\documentclass[usenatbib]{mn2e}
\usepackage{amsmath}
\usepackage{amssymb}
\usepackage{graphicx}
\usepackage[draft=false]{hyperref}
\usepackage{ifthen}  
\usepackage{rotating}

\newcounter{loRes}
\newcounter{extra}
\def\glc{{\sc Galacticus}}

\def\mnras{MNRAS}

\def\d{{\rm d}}

\newcounter{AGNDone}
\def\AGN{\ifthenelse{\equal{\arabic{AGNDone}}{0}}{active galactic nuclei (AGN)\setcounter{AGNDone}{1}}{AGN}}

\newcounter{BIEDone}
\def\BIE{\ifthenelse{\equal{\arabic{BIEDone}}{0}}{Bayesian Inference Engine (BIE)\setcounter{BIEDone}{1}}{BIE}}

\newcounter{CDMDone}
\def\CDM{\ifthenelse{\equal{\arabic{CDMDone}}{0}}{cold dark matter (CDM)\setcounter{CDMDone}{1}}{CDM}}

\newcounter{CGMDone}
\def\CGM{\ifthenelse{\equal{\arabic{CGMDone}}{0}}{circumgalactic medium (CGM)\setcounter{CGMDone}{1}}{CGM}}

\newcounter{ePSDone}
\def\ePS{\ifthenelse{\equal{\arabic{ePSDone}}{0}}{extended Press-Schechter (ePS)\setcounter{ePSDone}{1}}{ePS}}

\newcounter{HODDone}
\def\HOD{\ifthenelse{\equal{\arabic{HODDone}}{0}}{halo occupation distribution (HOD)\setcounter{HODDone}{1}}{HOD}}

\newcounter{ICMDone}
\def\ICM{\ifthenelse{\equal{\arabic{ICMDone}}{0}}{intracluster medium (ICM)\setcounter{ICMDone}{1}}{ICM}}

\newcounter{IGMDone}
\def\IGM{\ifthenelse{\equal{\arabic{IGMDone}}{0}}{intergalactic medium (IGM)\setcounter{IGMDone}{1}}{IGM}}

\newcounter{IMFDone}
\def\IMF{\ifthenelse{\equal{\arabic{IMFDone}}{0}}{initial mass function (IMF)\setcounter{IMFDone}{1}}{IMF}}

\newcounter{ISMDone}
\def\ISM{\ifthenelse{\equal{\arabic{ISMDone}}{0}}{interstellar medium (ISM)\setcounter{ISMDone}{1}}{ISM}}

\newcounter{ODEDone}
\def\ODE{\ifthenelse{\equal{\arabic{ODEDone}}{0}}{ordinary differential equation (ODE)\setcounter{ODEDone}{1}}{ODE}}
\def\ODEs{\ifthenelse{\equal{\arabic{ODEDone}}{0}}{ordinary differential equations (ODEs)\setcounter{ODEDone}{1}}{ODEs}}

\newcounter{PCADone}
\def\PCA{\ifthenelse{\equal{\arabic{PCADone}}{0}}{principal component analysis (PCA)\setcounter{PCADone}{1}}{PCA}}

\newcounter{PPCDone}
\def\PPC{\ifthenelse{\equal{\arabic{PPCDone}}{0}}{posterior predictive check (PPC)\setcounter{PPCDone}{1}}{PPC}}
\def\PPCs{\ifthenelse{\equal{\arabic{PPCDone}}{0}}{posterior predictive checks (PPCs)\setcounter{PPCDone}{1}}{PPCs}}

\newcounter{PPDDone}
\def\PPD{\ifthenelse{\equal{\arabic{PPDDone}}{0}}{posterior probability distribution\setcounter{PPDDone}{1}}{posterior probability distribution}}
\def\PPDs{\ifthenelse{\equal{\arabic{PPDDone}}{0}}{posterior probability distributions\setcounter{PPDDone}{1}}{posterior probability distributions}}

\newcounter{MCMCDone}
\def\MCMC{\ifthenelse{\equal{\arabic{MCMCDone}}{0}}{Markov Chain Monte Carlo (MCMC)\setcounter{MCMCDone}{1}}{MCMC}}

\newcounter{SAMDone}
\def\SAM{\ifthenelse{\equal{\arabic{SAMDone}}{0}}{semi-analytic model (SAM)\setcounter{SAMDone}{1}}{SAM}}
\def\SAMs{\ifthenelse{\equal{\arabic{SAMDone}}{0}}{semi-analytic models (SAMs)\setcounter{SAMDone}{1}}{SAMs}}

\newcounter{SDSSDone}
\def\SDSS{\ifthenelse{\equal{\arabic{SDSSDone}}{0}}{Sloan Digital Sky Survey (SDSS)\setcounter{SDSSDone}{1}}{SDSS}}

\newcounter{SEDDone}
\def\SED{\ifthenelse{\equal{\arabic{SEDDone}}{0}}{spectral energy distribution (SED)\setcounter{SEDDone}{1}}{SED}}
\def\SEDs{\ifthenelse{\equal{\arabic{SEDDone}}{0}}{spectral energy distributions (SEDs)\setcounter{SEDDone}{1}}{SEDs}}

\title[Building A Model of Galaxy Formation]{Building a Predictive Model of Galaxy Formation - I: Phenomenological Model Constrained to the \boldmath{$z=0$} Stellar Mass Function}
\author[Andrew J. Benson]{Andrew J. Benson\\
Carnegie Observatories, 813 Santa Barbara Street, Pasadena, CA 91101, USA.}

\begin{document}

\maketitle

\begin{abstract}
We constrain a highly simplified semi-analytic model of galaxy formation using the $z\approx 0$ stellar mass function of galaxies. Particular attention is paid to assessing the role of random and systematic errors in the determination of stellar masses, to systematic uncertainties in the model, and to correlations between bins in the measured and modeled stellar mass functions, in order to construct a realistic likelihood function. We derive constraints on model parameters and explore which aspects of the observational data constrain particular parameter combinations. We find that our model, once constrained, provides a remarkable match to the measured evolution of the stellar mass function to $z=1$, although fails dramatically to match the local galaxy HI mass function. Several ``nuisance parameters'' contribute significantly to uncertainties in model predictions. In particular, systematic errors in stellar mass estimate are the dominant source of uncertainty in model predictions at $z\approx 1$, with additional, non-negligble contributions arising from systematic uncertainties in halo mass functions and the residual uncertainties in cosmological parameters. Ignoring any of these sources of uncertainties could lead to viable models being erroneously ruled out. Additionally, we demonstrate that ignoring the significant covariance between bins the observed stellar mass function leads to significant biases in the constraints derived on model parameters. Careful treatment of systematic and random errors in the constraining data, and in the model being constrained, are crucial if this methodology is to be used to test hypotheses relating to the physics of galaxy formation.
\end{abstract}


\section{Introduction}\label{sec:Introduction}

Galaxy formation is a complex, nonlinear process, driven by the interplay of many different physical mechanisms \citep{benson_galaxy_2010}. Fortunately, there exists a wealth of observational data from which we can attempt to infer the physics of galaxy formation. Given the complexity of the problem, this inference necessarily proceeds via the use of models, which we use to estimate the statistical properties of the galaxy population given some set of assumptions about the underlying theory of galaxy formation. The goal of this process is to proceed from information (observational data), through knowledge (empirical descriptions of the galaxy population which summarize observational facts), to understanding (insight into the underlying physics).

Recently, there has been significant interest in the process of quantitatively constraining the parameters of \SAMs\ using Bayesian \MCMC\ \citep{henriques_monte_2009,henriques_tidal_2010,lu_bayesian_2011,lu_bayesian_2012,mutch_constraining_2013,lu_bayesian_2013}, model emulator \citep{bower_parameter_2010,gomez_characterizing_2012,gomez_dissecting_2013} methods, and particle swarm algorithms\footnote{Note that, unlike \protect\MCMC\ and model emulator techniques, the particle swarm method does not directly compute the full \protect\PPD\ of the model parameters. Instead the \protect\PPD\ must be approximate as a multivariate Gaussian by considering the behaviour of points in the vicinity of the maximum likelihood solution.} \citep{ruiz_calibration_2013}. This approach is crucial to facilitate the goal of model inference---that is, inferring the likelihood of a model given the data---and for making the reliable, quantitative predictions that are necessary in order to test models. To fully exploit the power of this approach it is crucial to address the following issues:
\begin{enumerate}
 \item construction of realistic likelihood functions;
 \item accounting for random and systematic errors in observations \emph{and} models;
 \item inferring the minimalistic viable model.
\end{enumerate}
The first issue requires careful assessment of the covariance in both observational and model determinations of the statistic (e.g. galaxy mass function) in question. The second issue requires a careful consideration of all sources of error in the observations, and model discrepancies (i.e. differences between the model and the modeled reality arising from approximations inherent in the modeling). Finally, the third point requires that we begin by studying simplistic models and gradually add complexity as guided by the data.

\subsection{Goals}

This is the first in a series of papers in which we will develop a well-constrained \SAM\ of galaxy formation based on the {\sc Galacticus} toolkit\footnote{\href{https://sites.google.com/site/galacticusmodel/}{\tt https://sites.google.com/site/galacticusmodel/}}. Specifically, we wish to constrain a model of galaxy formation in the standard $\Lambda$CDM cosmological model. We could, alternatively, choose to relax this assumption and explore non-flat cosmologies, with dark energy with equation of state $w\ne -1$, and with a variety of dark matter particle phenomenology (warm dark matter, self-interacting dark matter, etc.). We choose \emph{not} to do this for two reasons. First, such an exploration is currently computationally impractical\footnote{While \protect\glc\ is able to compute solutions for all of these variants on the standard cosmological model, the additional parameters and computational load introduced make it impractical for \protect\MCMC\ exploration at present.}. The second reason is that our understanding of galaxy formation remains sufficiently poor that we do not expect it to give strong insights into the nature of these extensions to the standard cosmological model. As such, we believe that these extensions should only be explored in studies such as this work if they are strongly motivated by other experiments. Currently, this is not the case \citep{frenk_dark_2012,hinshaw_nine-year_2013}.

Our goal is to construct a model of galaxy formation by starting with the simplest possible model and adding physics, guided at each step by constraints from the best available observations of key measures of the galaxy population. We intend to put a very significant emphasis on carefully assessing uncertainties, both random and systematic, in both observations and model, so as avoid being mislead by over-constraining the model, or constraining to biased data.

\subsection{Motivation for Simple Models}

The {\sc Galacticus} model can incorporate detailed modeling of galaxy formation physics \citep{benson_galacticus:_2012}. However, for both practical and methodological reasons, we choose in this work to run {\sc Galacticus} with a highly simplified model. Specifically, we run a model that contains minimal physics beyond that of dark matter structure formation and subhalo merging, instead relying entirely on empirical prescriptions.

Practically, such a simplified model runs much faster than a model with the complete complement of physics, making the \MCMC\ analysis tractable.

From the standpoint of methodology, we wish to build up a complete description of galaxy formation by gradually piecing together the components of a model, guided at each step by the best available observational data. Obviously, this must be a physical model of galaxy formation. However, our initial goal is simply to construct a prescription close enough to basic data upon which we can then begin to improve. Additionally, it is interesting to understand what the basic scaling properties of different processes must be in order to reproduce data. From a model inference perspective, we also wish to begin with a model that we \emph{expect} to be able to rule out. When the model is ruled out, we will switch one or more components to a more physical model, or add additional physics, until a viable model is recovered.

Our simple model is therefore constructed to encapsulate the following established general principles of galaxy formation theory \citep{benson_galaxy_2010}:
\begin{itemize}
 \item gas accretes into dark matter halos from the \IGM;
 \item that gas, initially diffusely distributed throughout the halo, must cool\footnote{Or it may already be cool \protect\citep{birnboim_virial_2003,kerevs_how_2005}---in any case it must collect in a high density region at the center of the halo where it can form a galaxy.} and condense to galactic densities;
 \item stars must form from this condensed gas;
 \item some material must be driven out of galaxies to limit star formation in lower mass halos;
 \item above some halo mass, some additional process must act to inhibit the condensation of gas \citep{benson_what_2003}.
\end{itemize}

Based on the success of similar prescriptions (e.g. \citealt{neistein_degeneracy_2010}), we expect that such a model, carefully constructed, should be able to match the constraint that we will apply in this work.

\subsection{Choice of Constraint}

We have chosen as our initial, sole constraint, the stellar mass function of galaxies measured in the local Universe, specifically the measurement by \cite{li_distribution_2009}. While there are more recent measures of the stellar mass function (which appeared after this study was begun, and some of which are explored in \S\ref{sec:Results}), the precise choice of dataset should not matter in the following sense. If all sources of random and systematic error affecting the data are correctly accounted for then constraints on the model parameters should be unbiased, and therefore consistent with constraints that would be derived from more recent measurements. Of course, more recent measurements might have smaller random and systematic errors, and thereby provide stronger constraints---this will be invaluable to future improvements in our modeling, but our primary goal here is to derive robust, unbiased constraints on model parameters.

The stellar mass function represents one of the most fundamental properties of the galaxy population that a viable theory of galaxy formation should be able to reproduce. There are, of course, other fundamental properties that that theory should also reproduce or predict (e.g. luminosity functions, gas content, star formation rates, the dependence of galaxy clustering on luminosity etc.). Our goal here is to select just a single, fundamental measure of the galaxy population. In future works in this series we will then explore the effects of adding additional constraints. Furthermore, many of these (e.g. luminosity functions and star formation rates) are closely tied to the stellar mass function.

The remainder of this paper is arranged as follows. In \S\ref{sec:Model} we describe the galaxy formation model that will be employed in this work. \S\ref{sec:Likelihood} describes how we construct the model likelihood function, including random and systematic errors in model and data. In \S\ref{sec:Results} we present results on the parameter constraints derived, and explore extrapolations of the model to other datasets. These results are discussed in \S\ref{sec:Discussion}, and we present our overall conclusions in \S\ref{sec:Conculsion}. We include several appendices addressing various technical aspects of our modeling. Appendix~\ref{app:HODConstraints} describes how we build a \HOD\ model of the observed stellar mass function (see also \S\ref{sec:ObservedCovariance}), Appendix~\ref{app:OptimalHaloMasses} describes an algorithm for selecting the optimal set of halo masses to simulate in a \SAM\ when constraining to a given data set, Appendix~\ref{app:Numerics} explores the numerical robustness and convergence of our \SAM\ code, Appendix~\ref{app:ModelCovariance} describes our model for the covariance matrix of the model galaxy stellar mass function, and, finally, Appendix~\ref{app:MergerTreeConstruction} explores the accuracy of the algorithm we employ for merger tree construction.

\section{Model}\label{sec:Model}

As stated in \S\ref{sec:Introduction}, our aim is to begin with a highly simplified, phenomenological model of galaxy formation which we will refine (and into which we will introduce more physics) as aspects of the model are ruled out by consideration of a variety of constraints. Therefore, we begin with a model which aims to compute only the distribution of galaxy masses (both stellar and gaseous) using simple scaling relations applied to dark matter halo merger trees.

We note that the model described below is not the simplest possible model. One could imagine a model which neglected the role of cooling and simply assumed a star formation rate proportional to the growth rate of halos. This would result in a stellar mass function that was a simple mapping of the halo(+subhalo) mass function to lower masses. It is well known that such a model is highly incompatible with observations \citep{benson_what_2003}. Furthermore, we could consider a model with no feedback. Such a model might plausibly fit the $z=0$ stellar mass function given a sufficiently flexible prescription for the star formation rate\footnote{For example, if the reduced specific star formation rate is arranged to be twice the halo growth rate, $\dot{M}_\star/M_\star = 2 \dot{M}_{\rm h}/M_{\rm h}$, then $M_\star \propto M_{\rm h}^2$ which will result in a flattening of the low-mass stellar mass function relative to the low-mass halo mass function, without any explicit inclusion of feedback. Indeed, the model of \protect\cite{lilly_phenomenological_2013} succeeds in matching the stellar mass function for precisely this reason. It should be noted however that this is not a physical model of star formation. As such, this phenomenological prescription for the star formation rate may well be encapsulating the net effect of the physics of star formation plus feedback.}.

Instead, we adopt a \emph{sufficiently simple} model, by which we mean one in which the various prescriptions are simple scaling relations, and the interconnections between the mass reservoirs are simple to understand. We include feedback specifically so that we can explore whether feedback is actually required by the constraints for such a sufficiently simple phenomenological model. This model can be run using v0.9.3 of \glc\ (revision 2284).

Our sufficiently simple model tracks the flow of mass between three reservoirs:
\begin{itemize}
 \item \emph{hot halo gas} defined as diffuse gas occupying the majority of the volume of the dark matter halo (a.k.a. \CGM, \ICM);
 \item \emph{ISM gas} defined as gas in the \ISM\ of a galaxy;
 \item \emph{stars} defined as mass which is locked up into long-lived (main sequence lifetimes $\gtrapprox 10$Gyr) stars.
\end{itemize}
The mass in each component for a given halo is denoted by $M_{\rm hot}$, $M_{\rm ISM}$, and $M_\star$ respectively.

Our model is fully specified by three classes of rule:
\begin{enumerate}
 \item dark matter merger trees;
 \item a set of \ODEs\ describing the evolution of the baryonic reservoirs in isolated galaxies;
 \item rules for the effects of halo and galaxy mergers on the baryonic reservoirs.
\end{enumerate}
We proceed to detail these three classes below.

\subsection{Merger Trees}

Merger trees are generated using the algorithm described by \cite{parkinson_generating_2008}, which reproduces the progenitor mass functions of halos in the Millennium Simulation. \cite{jiang_generating_2013} study several algorithms for constructing merger trees and compare their results with trees extracted from N-body simulations. They find that the \cite{parkinson_generating_2008} algorithm is the only algorithm which gives good agreement with the N-body merger trees for all statistics that they consider. Trees are generated backward in time starting from $z=0$, and are followed until each branch falls below a specified mass resolution, $M_{\rm res}$. We test for convergence in our results with respect to the numerical parameters controlling the construction of these trees (see Appendix~\ref{app:Numerics}). Throughout, we use the fitting formulae of \cite{kitayama_semianalytic_1996} to compute the virial density contrast for dark matter halos, $\Delta_{\rm h}$, and the critical overdensity for collapse, $\delta_{\rm c}$.

In the interests of computational speed, we allow the mass resolution to be a function of the final, $z=0$, mass of the tree, $M_0$. Specifically,
\begin{equation}
 M_{\rm res} = \hbox{max}(f_0 M_0,M_{\rm res,min}),
 \label{eq:TreeMassResolution}
\end{equation}
where $f_0=0.1$ and $M_{\rm res,min}=10^{8}M_\odot$. This approximation will introduce some error into the model calculation. We measure and account for this error as a model discrepancy term (see \S\ref{sec:VariableMassResolution}).

We must also specify the set of $z=0$ halo masses for which trees are to be built. These masses can be distributed arbitrarily in the range of required masses, providing the contribution of their galaxies to the stellar mass function is appropriately weighted. Given a set of $z=0$ halo masses, $M_i$, the weight to assign galaxies in tree $i$ is simply
\begin{equation}
w_i = \int_{\sqrt{M_{i+1}/M_i}}^{\sqrt{M_i/M_{i-1}}} \hbox{minmax}\left[\phi_{\rm min},\phi_{\rm max},{{\rm d} n \over {\rm d}\ln M}(M)\right] M {\rm d}M, 
\label{eq:MergerTreeWeight}
\end{equation}
where ${\rm d}n/{\rm d}\ln M$ is the halo mass function, for which we use the fitting function of \cite{tinker_toward_2008} with the appropriate value of $\Delta_{\rm h}$, and 
\begin{equation}
 \hbox{minmax}(a,b,x) = \left\{\begin{array}{ll}a & \hbox{ if } x < a \\ x & \hbox{ if } a \leq x \leq b\\ b & \hbox{ if } x > b,\end{array}\right.
\end{equation}
with $\phi_{\rm min}=10^{-7}$~Mpc$^{-3}$ and $\phi_{\rm max}=10^{-3}$~Mpc$^{-3}$. The choice of distribution of $z=0$ halo masses will affect both the precision of results obtained (since we construct the stellar mass function by summing over realizations of merger trees, the finite number of trees generated will lead to statistical fluctuations in the results), and the time taken for each model evaluation. In Appendix~\ref{app:OptimalHaloMasses} we explore three different distributions of $z=0$ halo masses and find that simply constructing a volume-limited sample (i.e. drawing masses at random from the $z=0$ halo mass function) is close to optimal.

\subsubsection{Cosmological Dependence}\label{sec:CosmologyDependence}

The statistics of the merger tree structure will depend on the cosmological parameters (via the critical overdensity for collapse, $\delta_{\rm c}$), which we will allow to vary in our analysis. Additionally, given the algorithm described above, the $z=0$ masses of the trees will depend both on the cosmological parameters and on systematic uncertainties in the halo mass function (see \S\ref{sec:HaloMassFunctionSystematics}). Unfortunately, this dependence can introduce discontinuities in the likelihood surface. Since tree construction involves binary branching a small change in a cosmological parameter can lead to large changes in the structure of one or more trees. (Similarly, changing the $z=0$ mass by a small amount can lead to a large change in the structure of a tree.) 

In principle this is not a problem, providing that the number of trees simulated is very large---the more trees simulated the smaller the discontinuities in the likelihood surface will be. In practice, however, the number of trees we can run (given computational constraints) is not that large, and the discontinuities in the likelihood surface are significant, making it difficult for an \MCMC\ algorithm to achieve convergence.

Therefore, we adopt the following approach. A single set of merger trees is generated using the maximum likelihood cosmological parameters from \cite[see \S\ref{sec:CosmologicalParameters} for a full discussion of how we utilize these cosmological parameter constraints]{hinshaw_nine-year_2013} and no systematic offsets in the halo mass function. These trees are stored to file, and are re-used by each model evaluation\footnote{Specifically, the masses and redshifts of each node in each tree is reused in model evaluations. This will lead to some difference in the time at which nodes are defined since the cosmological time-redshift relation will vary between model evaluations.}. The weight assigned to each tree (see eqn.~\ref{eq:MergerTreeWeight}) \emph{is} recomputed for each model evaluation using the cosmological parameters of that model.

This approach reduces variance between model evaluations---a point that we will address in \S\ref{sec:CosmologyIndepTrees}.

We note that, in future \MCMC\ applications where a very large number of trees can be used, there is no reason to use a fixed set of trees, even if the model is evaluated twice with exactly the same cosmological parameters. While this does make the likelihood surface stochastic this would not affect the assumption of detailed balance in the \MCMC\ algorithm, since the probability for the merger-tree process is Lebesgue measurable\footnote{We thank Martin Weinberg for clarifying this point.}, but would broaden the \PPD\ for the model parameters\footnote{This broadening is instead accounted for by including our estimate of the model uncertainty in the likelihood function as described in Appendix~\protect\ref{app:ModelCovariance}.}. This approach is desirable in principle, as it avoids any bias in the model posterior which may be introduced through the use of a single set of merger trees (i.e. any finite set of merger trees will be somewhat biased relative to the full ensemble).

We have tested this approach, building a unique set of merger trees each time the model is evaluated, and find that at present it significantly extends the convergence time for the \MCMC\ algorithm, making it currently impractical.

\subsection{Differential Evolution}

Galaxies in our model typically evolve differentially, except when this evolution is punctuated by merger events (see \S\ref{sec:Merging}). The set of \ODEs\ to be solved along each branch of every merger tree is described below.

\subsubsection{Accretion From The IGM}

Gas is assumed to accrete into halos from the \IGM\ at a rate proportional to the growth rate of the total halo mass, with the constant of proportionality being the universal baryon fraction. The exception to this rule occurs for halos deemed to be unable to accrete due to the the \IGM\ being heated to high temperatures as a result of the reionization of the Universe. We adopt a simple step-function model for this suppression of accretion due to reionization which has been shown to capture the essential features of more realistic models \citep{font_population_2011}. Specifically:
\begin{equation}
 \dot{M}_{\rm IGM} = \left\{ \begin{array}{ll} (\Omega_{\rm b} / \Omega_{\rm M}) \dot{M}_{\rm h, smooth} & \hbox{ if } V_{\rm h} > V_{\rm reion} \hbox{ or } z > z_{\rm reion}, \\ 0 & \hbox{ otherwise,} \end{array} \right.
\end{equation}
where $\Omega_{\rm b}$ and $\Omega_{\rm M}$ are the density parameters for baryons and total matter respectively, $\dot{M}_{\rm h, smooth}$ is the total rate\footnote{As determined from our merger trees.} of smooth accretion onto the halo\footnote{Hot gas will also be obtained from merging with resolved halos---see \S\protect\ref{sec:HaloMerging}.}, $V_{\rm h}$ is the halo virial velocity, $z$ is redshift, and $V_{\rm reion}$ and $z_{\rm reion}$ are parameters of the model.

\subsubsection{Cooling}

The cooling of gas\footnote{More specifically, since we do not include any physics here, this should be viewed as the \emph{transfer} of gas from the hot halo to the \protect\ISM. Whether or not this occurs via cooling is irrelevant.} from the hot halo to the \ISM\ of a galaxy is modeled with a simple redshift-dependent timescale\footnote{Here, and elsewhere, we introduce an explicit redshift-dependence into our model. It could be argued that this is unphysical since a galaxy should have no direct knowledge of the redshift (expect, perhaps, via a redshift-dependent background radiation field for example). However, we justify the inclusion of such a explicit dependence by arguing that the evolution of a galaxy may depend directly on its density, which is expected to scale with the density of the Universe---a redshift-dependent quantity. Our present, sufficiently simple model does not directly trace galaxy densities, so we believe inclusion of an explicit redshift dependence as a proxy is justified. Of course, our goal is to proceed toward more physical models in which any such density dependence is modeled directly.}, plus an exponential break\footnote{The exponential break is introduced since it is well-established that without this type of suppression it is impossible to reproduce the exponential decline in the abundances of high-mass galaxies \protect\citep{benson_what_2003}. We expect, therefore, that our \protect\MCMC\ analysis will show that the parameter $\mathcal{M}_{\rm cool}$ cannot be arbitrarily large.}:
\begin{eqnarray}
 \dot{M}_{\rm cool} &=& {M_{\rm hot} \over \tau_{\rm cool}} (1+z)^{-\alpha_{\rm cool}} \nonumber \\ 
& & \times \left( 1 + \exp\left[ {\log_{10}(M_{\rm halo}/\mathcal{M}_{\rm cool})] \over \Delta \log_{10} \mathcal{M}_{\rm cool}}\right]\right)^{-\beta_{\rm cool}},
 \label{eq:coolingRate}
\end{eqnarray}
where $\tau_{\rm cool}$, $\alpha_{\rm cool}$, $\beta_{\rm cool}$, $\mathcal{M}_{\rm cool}$, and $ \Delta \log_{10} \mathcal{M}_{\rm cool}$ are parameters of the model.

\subsubsection{Star Formation}

The rate of star formation is assumed to be proportional to the mass of the \ISM, and inversely proportional to a timescale that scales with both redshift and halo virial velocity as follows:
\begin{equation}
 \dot{M}_{\rm sf} = \hbox{min}\left({M_{\rm ISM} \over \tau_\star} (1+z)^{-\alpha_\star} \left[{V_{\rm h} \over 200~\hbox{km/s}}\right]^{-\beta_\star},{M_{\rm ISM} \over \tau_{\rm \star,min}}\right),
 \label{eq:starFormationRate}
\end{equation}
where $\tau_\star$, $\alpha_\star$, $\beta_\star$, and $\tau_{\rm \star,min}$ are parameters of the model (the final parameter being introduced to prevent timescales becoming arbitrarily small). Note that we do not explicitly consider recycling of material from stars. The star formation rate should therefore be viewed as the \emph{reduced} star formation rate (i.e. the rate of formation of long-lived stars).

\subsubsection{Feedback}

Feedback (i.e. outflow of material from the \ISM\ to the hot halo) is assumed to occur at a rate proportional to the star formation rate, with a constant of proportionality which scales with both redshift and halo virial velocity:
\begin{eqnarray}
 \dot{M}_{\rm wind} &=& \hbox{min}\left(f_{\rm wind} \dot{M}_{\rm sf} (1+z)^{\alpha_{\rm wind}} \left[{V_{\rm h} \over 200~\hbox{km/s}}\right]^{\beta_{\rm wind}},\right. \nonumber \\
 & & \left. {M_{\rm ISM} \over \tau_{\rm wind, min}} \right),
 \label{eq:OutflowModel}
\end{eqnarray}
where $f_{\rm wind}$, $\alpha_{\rm wind}$, $\beta_{\rm wind}$, and $\tau_{\rm wind, min}$ are parameters of the model (the final parameter being introduced to prevent arbitrarily high outflow rates from occurring).

\subsubsection{Network}

Finally, the rates described above are combined into the following network describing the net rates of change of the masses in our three reservoirs:
\begin{alignat}{9}
 \dot{M}_{\rm hot} &=& \dot{M}_{\rm IGM} &-& \dot{M}_{\rm cool} &+& \dot{M}_{\rm wind} &{}&                , \\
 \dot{M}_{\rm ISM} &=& {}                &+& \dot{M}_{\rm cool} &-& \dot{M}_{\rm wind} &-& \dot{M}_{\rm sf}, \\
 \dot{M}_\star     &=& {}                &{}&                   &{}&                   &+& \dot{M}_{\rm sf}.
\end{alignat}
Note that the only net source of mass is accretion from the \IGM. This network of \ODEs\ is integrated along each branch of each merger tree.

\subsubsection{Initial Conditions}

The initial conditions for differential evolution of our model are given by specifying the baryonic content of dark matter halos corresponding to the tip (i.e. progenitorless halo) of each branch in the merger tree. We assume that each branch-tip halo contains zero mass in \ISM\ and stellar mass, and a mass of hot gas equal to
\begin{equation}
 M_{\rm hot} = \left\{ \begin{array}{ll} (\Omega_{\rm b} / \Omega_{\rm M}) M_{\rm h} & \hbox{ if } V_{\rm h} > V_{\rm reion} \hbox{ or } z > z_{\rm reion}, \\ 0 & \hbox{ otherwise,} \end{array} \right.
\end{equation}
where $M_{\rm h}$ is the total mass of the halo.

\subsection{Punctuated Evolution}\label{sec:Merging}

In addition to the differential evolution of galaxies along each branch of the merger tree, we include instantaneous, punctuated evolution in response to mergers of both halos and galaxies, as described below.

\subsubsection{Halo Merging}\label{sec:HaloMerging}

When two halos merge (i.e. when a smaller halo first becomes a subhalo in a larger host halo) the hot halo gas from the new subhalo is assumed to be instantaneously removed\footnote{Presumably by ram pressure or other environmental forces, although the details are purposely not specified in this sufficiently simple model.} and is added to the hot halo gas of the host halo:
\begin{eqnarray}
 M_{\rm hot, host} &\rightarrow& M_{\rm hot, host} + M_{\rm hot, subhalo}; \nonumber \\
 M_{\rm hot, subhalo} &\rightarrow& 0.
\end{eqnarray}
Additionally, the subhalo is assigned a time-of-merging, $t_{\rm merge}$, at which it will undergo a galaxy merger with the central galaxy of its host halo. The time-of-merging is computed using $t_{\rm merge} = t + t_{\rm df}$ where $t$ is the time at which the halo becomes a subhalo, and $t_{\rm df}$ is the dynamical friction timescale computed using the fitting formula of \cite[][their eqn.~5]{jiang_fitting_2008}, accounting for both mass and orbit dependencies. The orbital parameters of each new subhalo are assumed to be precisely equal to the mode of the distribution found by \cite{benson_orbital_2005}, specifically $(V_{\rm r},V_\phi)=(0.90,0.75) V_{\rm h, host}$, where $V_{\rm h, host}$ is the virial velocity of the host halo.

\subsubsection{Galaxy Merging}

At time $t_{\rm merge}$, the subhalo is removed from the calculation. The masses of its baryonic reservoirs (\ISM\ and stars only, as its hot halo gas reservoir has already been removed) are simply added to those of the central galaxy of the subhalo's host halo:
\begin{eqnarray}
 M_{\rm ISM, host} &\rightarrow& M_{\rm ISM, host} + M_{\rm ISM, subhalo}, \nonumber \\
 M_{\rm ISM, subhalo} &\rightarrow& \diameter, \nonumber \\
 M_{\rm \star, host} &\rightarrow& M_{\rm \star, host} + M_{\rm \star, subhalo}, \nonumber \\
 M_{\rm \star, subhalo} &\rightarrow& \diameter,
\end{eqnarray}
where ``$\diameter$'' indicates that a reservoir no longer exists. We do not include in our model any enhancement in the star formation rate after a merger (i.e. there is no explicit ``starburst'' mode), beyond any enhancement arising from the addition of new gas to a galaxy.

\section{Likelihood}\label{sec:Likelihood}

Perhaps the most important aspect of our calculation is the construction of a realistic likelihood function for our model. In this section we detail the calculation of this likelihood function.

\subsection{Observed Stellar Mass Function and Covariance Matrix}\label{sec:ObservedCovariance}

We adopt the stellar mass function of galaxies at $z\approx 0.07$ measured from the \SDSS\ by \cite{li_distribution_2009} as the single constraint on our model. \cite{li_distribution_2009} report errors on their mass function which they derive from mock catalogs. However, to correctly evaluate the model likelihood we need to know the full covariance matrix of the observed mass function. We have developed a procedure, utilizing the framework of \cite{smith_how_2012} to estimate the covariance matrix of this mass function.

The framework of \cite{smith_how_2012} computes three contributions\footnote{There are, of course, additional random and systematic uncertainties in stellar mass estimates. These will be addressed in \S\protect\ref{sec:StellarMassRandoms} and \S\protect\ref{sec:StellarMassSystematics} respectively.} to the covariance matrix:
\begin{enumerate}
 \item a Poisson term due to the finite number of galaxies in each bin;
 \item a large-scale structure term arising from structure in the Universe on scales comparable to and larger than the volume surveyed;
 \item a ``halo'' term arising from the fact that galaxies are found in associations (i.e. groups and clusters).
\end{enumerate}
Calculating these three contributions for a given observational sample requires the following information (we refer the reader to \cite{smith_how_2012} for a full explanation):
\begin{enumerate}
 \item the survey geometry (i.e. the angular mask of the survey);
 \item the depth of the survey as a function of galaxy mass;
 \item knowledge of how galaxies populate dark matter halos, the so-called \HOD.
\end{enumerate}

For the angular mask, we make use of the catalog of random points within the survey footprint provided by the NYU-VAGC\footnote{Specifically, \href{http://sdss.physics.nyu.edu/lss/dr72/random/lss_random-0.dr72.dat}{http://sdss.physics.nyu.edu/lss/dr72/random/\ lss\_random-0.dr72.dat}.} (\citealt{blanton_new_2005}; see also \citealt{adelman-mccarthy_sixth_2008,padmanabhan_improved_2008}). \cite{li_distribution_2009} consider only the main, contiguous region and so we keep only those points which satisfy RA$>100^\circ$, RA$<300^\circ$, and RA$<247^\circ$ or $\delta< 51^\circ$. When the survey window function is needed, these points are used to determine which elements of a 3D grid fall within the window function.

To estimate the depth of the \cite{li_distribution_2009} sample as a function of galaxy stellar mass we make use of semi-analytic models in the Millennium Database. Specifically, we use the \SAM\ of \citeauthor{de_lucia_hierarchical_2007}~(\citeyear{de_lucia_hierarchical_2007}; specifically the {\tt millimil..DeLucia2006a} and {\tt millimil..DeLucia2006a\_sdss2mass} tables in the Millennium Database). For each snapshot in the database, we extract the stellar masses and observed-frame SDSS r-band absolute magnitudes (including dust extinction), and determine the median absolute magnitude as a function of stellar mass. Using the limiting apparent magnitude of the \cite{li_distribution_2009} sample, $r=17.6$, we infer the corresponding absolute magnitude at each redshift and, using our derived absolute magnitude--stellar mass relation, infer the corresponding stellar mass.

\begin{figure}
 \includegraphics[width=85mm,trim=0mm 0mm 0mm 4mm,clip]{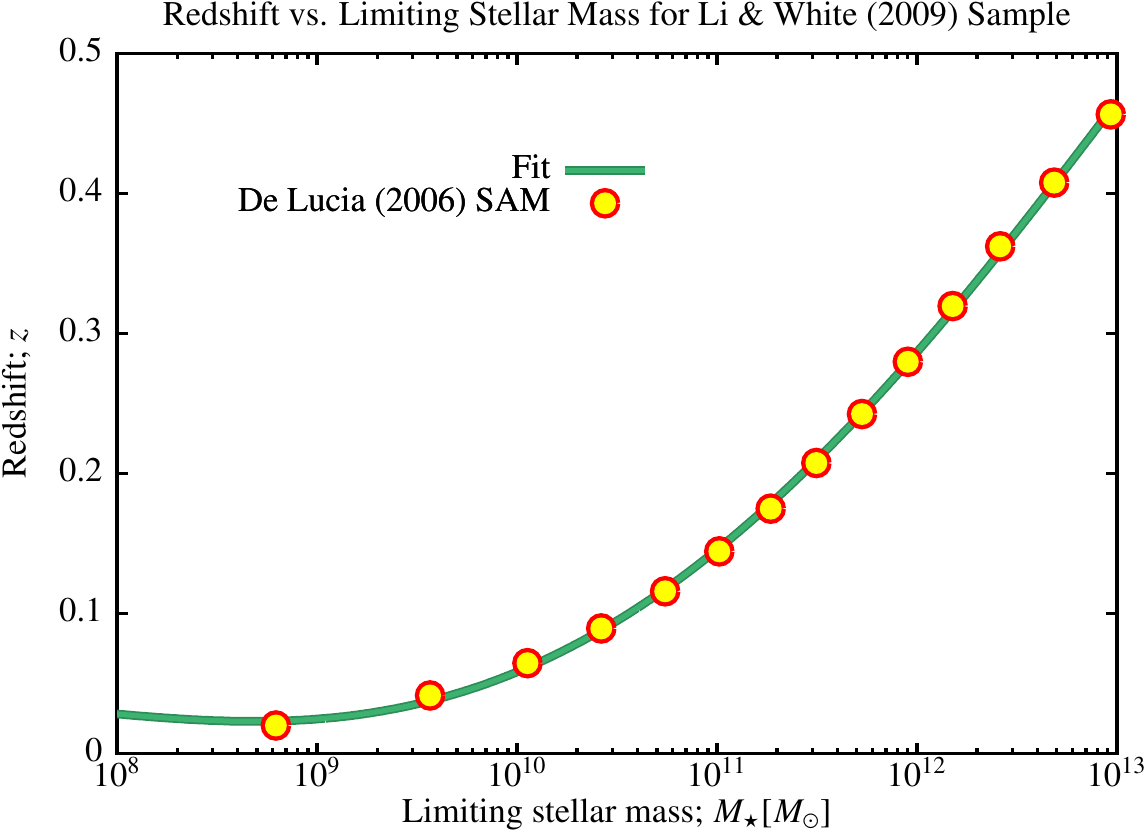}
 \caption{The maximum redshift at which a galaxy of given stellar mass can be detected in the sample of \protect\cite{li_distribution_2009}. Points show the results obtained using the \protect\cite{de_lucia_hierarchical_2007} model from the Millennium Database, while the lines shows a polynomial fit to these results (given in eqn.~\ref{eq:DepthPolynomial}).}
 \label{fig:SDSSDepthFit}
\end{figure}

The end result of this procedure is the limiting stellar mass as a function of redshift, accounting for k-corrections, evolution, and the effects of dust. Figure~\ref{fig:SDSSDepthFit} shows the resulting relation between stellar mass and the maximum redshift at which such a galaxy would be included in the sample. Points indicate measurements from the \SAM, while the line shows a polynomial fit:
\begin{eqnarray}
 z(M_\star) &=& -5.950 + 2.638 m - 0.4211 m^2 \nonumber \\ 
            & & + 2.852\times 10^{-2} m^3 - 6.783 \times 10^{-4} m^4,
 \label{eq:DepthPolynomial}
\end{eqnarray}
where $m= \log_{10}(M_\star/M_\odot)$. We use this polynomial fit to determine the depth of the sample as a function of stellar mass. We adopt a solid angle of $2.1901993$~sr \citep{percival_shape_2007} for the sample.

Computing the large-scale structure contribution to the covariance function requires integration of the non-linear matter power spectrum over the Fourier transform of the survey window function. We use the method of \cite{peacock_non-linear_1996} to determine the non-linear matter power spectrum, because of its simplicity and speed. We have checked that using a more accurate non-linear matter power spectrum (e.g. \citealt{lawrence_coyote_2010}) makes negligible difference to our results.

To find a suitable \HOD\ to describe the galaxies in the \cite{li_distribution_2009} sample we adopt the model of \cite{behroozi_comprehensive_2010}\footnote{One might ask, why not simply use the \protect\HOD\ from the \protect\SAM\ in this calculation of the data covariance matrix? This would indeed be the ideal solution, and one which we intend to investigate in future. For now, such an approach is practically impossible because it requires iteratively constraining the \protect\SAM\ to the data and updating the data covariance matrix on each iteration. This is too slow to be practical. However, planned improvements in the speed of data covariance matrix construction may allow this approach to be applied in future.}. This is an 11 parameter model which describes separately the numbers of satellite and central galaxies occupying a halo of given mass---the reader is referred to \cite{behroozi_comprehensive_2010} for a complete description of the functional form of this parametric \HOD\ (see also Appendix~\ref{app:OptimalHaloMasses}). 

To reproduce the mass function of \cite{li_distribution_2009}, $\phi^{\rm (observed)}$, using this \HOD\ we use the \BIE\ \citep{weinberg_computational_2013} to constrain the \HOD\ parameters. Details of this procedure are given in Appendix~\ref{app:HODConstraints}. We use a likelihood
\begin{equation}
 \ln \mathcal{L} = -{1\over 2} \Delta\cdot \mathcal{C}^{-1}\cdot \Delta^{\rm T} - {N \over 2} \ln(2\pi) - {\ln |\mathcal{C}| \over 2},
\end{equation}
where $N$ is the number of bins in the mass function, $\mathcal{C}$ is the covariance matrix of the observed mass function, and $\Delta_i = \phi_i^{\rm (HOD)} - \phi_i^{\rm (observed)}$, where $\phi^{\rm (HOD)}$ is the mass function computed from our \HOD\ model. Of course, it is precisely this covariance matrix, $\mathcal{C}$, that we are trying to compute. We therefore adopt an iterative approach as follows:
\begin{enumerate}
 \item make an initial estimate of the covariance matrix, assuming that only Poisson errors contribute (the covariance matrix is therefore diagonal, and the terms are easily computed from the measured mass function and the survey volume as a function of stellar mass);
 \item find the maximum likelihood parameters of the \HOD\ given the observed mass function and the current estimate of the covariance matrix;
 \item using this \HOD\ and the framework of \cite{smith_how_2012}, compute a new estimate of the covariance matrix, including all three contributions;
 \item repeat steps 2 and 3 until convergence in the covariance matrix is achieved.
\end{enumerate}
In practice we find that this procedure leads to an \HOD\ and covariance matrix which oscillate between two states in successive iterations. The differences in the covariance matrix are relatively small however, so we choose to conservatively adopt the covariance matrix with the larger values. In future, adding additional constraints to the \HOD\ (as described below) should help mitigate this problem.

\begin{figure}
 \includegraphics[width=85mm,trim=0mm 0mm 0mm 2.5mm,clip]{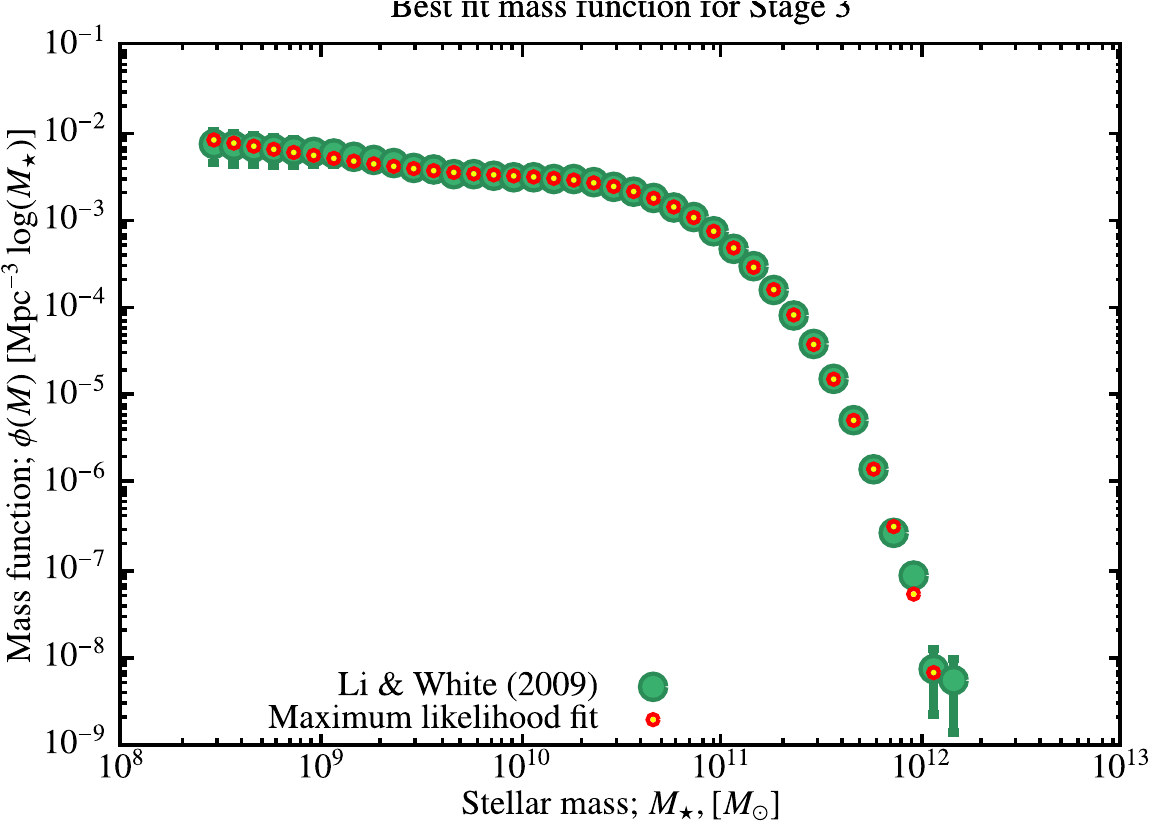}
 \caption{The maximum likelihood mass function obtained from our parametric \protect\HOD\ (yellow points), compared to the observed stellar mass function of \protect\citeauthor{li_distribution_2009}~(\citeyear{li_distribution_2009}; green points).}
 \label{fig:MaximumLikelihoodMassFunctionHOD}
\end{figure}

\begin{figure}
 \includegraphics[width=85mm,trim=0mm 0mm 0mm 0mm,clip]{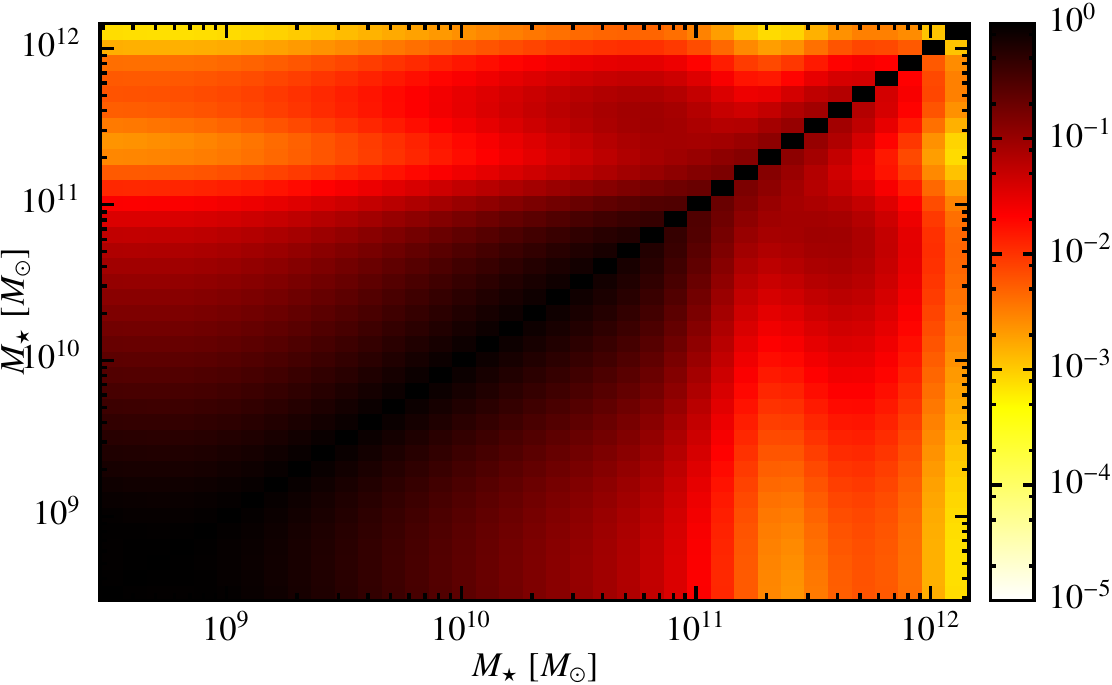}
 \caption{The correlation matrix of the observed galaxy stellar mass function of \protect\cite{li_distribution_2009}. Color indicates the strength of correlation between bins, according to the scale shown on the right.}
 \label{fig:CorrelationMatrixSDSS}
\end{figure}

The resulting maximum likelihood mass function is shown in Fig.~\ref{fig:MaximumLikelihoodMassFunctionHOD}, clearly illustrating that this parametric \HOD\ can produce an excellent match to the observed mass function. The resulting correlation matrix is shown in Fig.~\ref{fig:CorrelationMatrixSDSS}. As expected, at the higher masses the correlation matrix is dominated by the on-diagonal terms---arising from the Poisson fluctuations in the number of galaxies due to the scarcity of these massive systems. At lower masses the matrix has significant off-diagonal amplitude, indicating strong correlations between nearby bins, arising from both large-scale structure and halo contributions to the covariance. This structure significantly weakens the constraint arising from the low-mass end of the mass function. Also noticable  are regions of enhanced correlation peaking at roughly $(4\times10^{11}M_\odot,5\times10^{10}M_\odot)$. These arise from the ``halo'' term, descrbing the fact that galaxies are found in groups and clusters.

\begin{figure}
 \includegraphics[width=85mm,trim=0mm 0mm 0mm 2.5mm,clip]{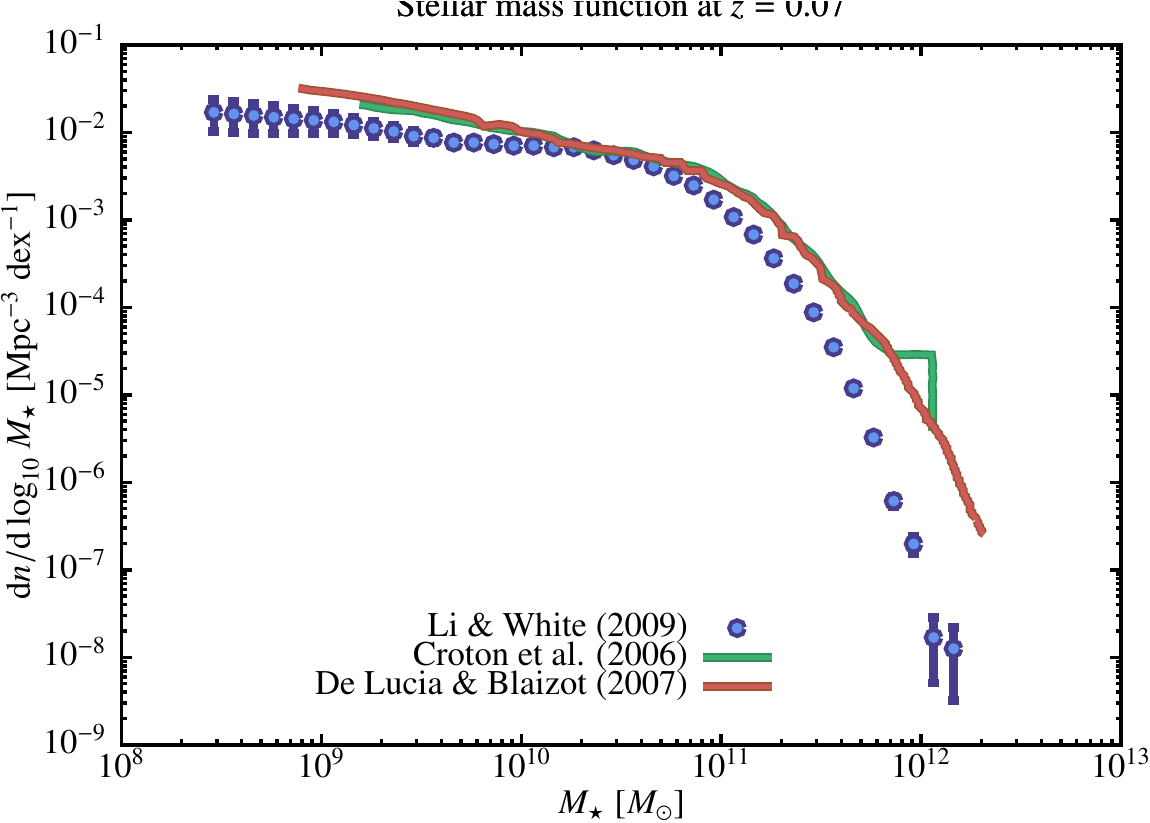}
 \caption{Comparison of three galaxy stellar mass functions. Blue points show the observationally determined stellar mass function from \protect\cite{li_distribution_2009}. The green line shows the model stellar mass function of \protect\citeauthor{croton_many_2006}~(\citeyear{croton_many_2006}; measured from Fig.~3 of \citealt{mutch_constraining_2013}), while the red lines shows the model stellar mass function of \citeauthor{de_lucia_hierarchical_2007}~(\citeyear{de_lucia_hierarchical_2007}; measured from Fig.~7 of \citealt{kitzbichler_high-redshift_2007}) which is a model very similar to that of \protect\cite{croton_many_2006}.}
 \label{fig:MassFunctionCrotonOverlay}
\end{figure}

\cite{li_distribution_2009} estimated errors on their mass function by generating 20 mock catalogs from the $z=0$ Millennium Database (utilizing the \SAM\ of \citealt{croton_many_2006}), and then measuring the variance of the resulting model stellar mass functions. This method should account for all of the sources of variance considered in our method. They report only the variance in each bin (i.e. the diagonal elements of the covariance matrix). We therefore compare the root-variance of the diagonal elements of our matrix with the errors reported by \cite{li_distribution_2009}. We find that our errors exceed those reported by \cite{li_distribution_2009} in the lowest mass bins by a factor of approximately 3. In this regime, the variance is dominated by the ``halo'' term, suggesting that this difference lies in the relative distribution of galaxies between halos in the two models. For the highest mass bins, our estimates are a factor 2--3 smaller than those of \cite{li_distribution_2009}. Both approaches to estimating the errors on the mass function rely on the assumption of an underlying model. However, for the highest mass bins the errors should be dominated by the Poisson term, and therefore depend only on the measured mass function and the survey volume. Figure~\ref{fig:MassFunctionCrotonOverlay} shows the \cite{croton_many_2006} stellar mass function overlaid on the \cite{li_distribution_2009} mass function. At high masses the \cite{croton_many_2006} model mass function exceeds that measured by \cite{li_distribution_2009}. This will lead to an overestimation of Poisson errors in the mass function in this regime. Our approach has the advantage that the underlying model used to determine the covariance is constrained to match the measured mass function to high precision. In any case, this illustrates one difficulty in determining covariances using models---the covariance is only as reliable as the model. As models are refined this should become less of a problem.

These techniques for covariance matrix estimation are applicable to mass functions and luminosity functions at any redshift, providing the survey geometry and depth are known. The only significant assumption is that the \HOD\ model is a good description of the true galaxy population\footnote{Since the parameters of this \protect\HOD\ model are uncertain (as is the survey depth to some degree) we should arguably include this uncertainty when constructing our likelihood function. This point is discussed further at the end of Appendix~\protect\ref{app:HODConstraints}.}. Here, we have self-calibrated the \HOD\ to match the mass function being studied. It is possible, however, to introduce additional constraints on the \HOD\ (e.g. the galaxy correlation function, or the evolution of the mass function with redshift) which would place tighter constraints on its parameters, or, possibly, rule it out as a viable description of the galaxy population. In the latter case, a more elaborate model would then be required. Indeed, given a complete galaxy formation model of the type studied in this work, it is possible to predict the \HOD\ directly from that model. In future, once our models are more tightly constrained, it may be possible to use them to predict the \HOD\ directly and use this to compute the covariance matrices of new observational datasets.

\subsection{Errors in Stellar Mass Estimates}

Stellar masses of galaxies are inferred from broadband photometry by fitting template \SEDs\ (see \cite{taylor_galaxy_2011} for a recent discussion and application of this methodology). This process includes many uncertainties, both random and systematic, due to photometric errors, stellar population synthesis, dust modeling, assumptions about possible star formation histories, and the \IMF. To correctly compare model and data we must take into account these errors. In our analysis we follow the approach of \cite{behroozi_comprehensive_2010}.

\subsubsection{Random Errors}\label{sec:StellarMassRandoms}

We assume that measured stellar masses in the \SDSS\ have random errors distributed as a Gaussian in $\log_{10}M_\star$ with a variance of $\sigma_\star^2=0.07^2$ \citep{behroozi_comprehensive_2010}. Therefore, when constructing the stellar mass function from a model galaxy population we convolve the mass function with a Gaussian of this variance.

\subsubsection{Systematic Errors}\label{sec:StellarMassSystematics}

Estimated stellar masses may also be systematically biased due to the limitations of the template fitting procedure. A detailed discussion of these sources of systematic error can be found in \cite{behroozi_comprehensive_2010}. We follow \cite{behroozi_comprehensive_2010} and assume that the systematic error in stellar mass can be modeled using:
\begin{equation}
 \log_{10}(M_\star) \rightarrow \log_{10}(M_\star) + \mu + \kappa \log_{10}\left({M_\star\over 10^{11.3}M_\odot}\right).
 \label{eq:StellarMassSystematic}
\end{equation}
Here, $\mu$ and $\kappa$ are unknown coefficients. We treat these as free parameters in our \MCMC\ analysis, and will marginalize over these parameters when determining the constraints on other parameters of the model. For priors on $\mu$ and $\kappa$, we follow \cite{behroozi_comprehensive_2010} and assume normal distributions with zero mean and dispersions of $0.27$ and $0.1$ respectively, truncated beyond $\pm 0.4$ and $\pm 0.15$ respectively.

Note that we are not fully including the \IMF\ assumed in estimating \SDSS\ stellar masses as a source of systematic uncertainty here. To first order, the choice of \IMF\ leads to a relatively simple uniform shift in $\log M_\star$. As such, if a different \IMF\ were assumed a simple offset in the parameters of our model would be required to accommodate this change. However, in detail changes in the \IMF\ will lead to changes in galaxy colours, potentially in a way not captured by our simple model for stellar mass systematics.  Furthermore, as we do not incorporate physics related to stellar evolution (such as metal enrichment, or calculation of stellar population luminosities) in our simple model, the \IMF\ need not be specified in our model. We intend to explore these points in greater detail in subsequent papers in this series.

\subsection{Systematic Errors in the Halo Mass Function}\label{sec:HaloMassFunctionSystematics}

Although the halo mass function has been determined to very high precision from cosmological N-body simulations (see, for example, \citealt{tinker_toward_2008}) for specific halo finding algorithms, there remain significant systematic uncertainties in this function arising from the choice of halo finding algorithm. For example, \cite{knebe_haloes_2011} report relative differences of up to 30\% in the halo mass function from a cross-comparison of 17 different halo finding algorithms. To account for this systematic uncertainty in the halo mass function, $n(M)$, we introduce a simple model of the systematic error in the halo mass function arising from the choice of halo finding algorithm. Specifically,
\begin{equation}
 n(M) \rightarrow n(M) [1+\alpha+\beta\log_{10}(M_{\rm halo}/10^{12}M_\odot)]
\label{eq:HaloSystematicModel}
\end{equation}
where $\alpha$ and $\beta$ are nuisance parameters to be included in our \MCMC\ analysis and marginalized over.

To determine priors on $\alpha$ and $\beta$, we make use of the relative residuals in halo mass functions determined by 14 halo finding algorithms reported in the lower panel of Fig.~17 of \cite{knebe_haloes_2011}. We fit eqn.~(\ref{eq:HaloSystematicModel}) to each line in that figure to determine the best-fit values of $\alpha$ and $\beta$, and then measure the means, $\langle \alpha\rangle$ and $\langle\beta\rangle$, and covariance, $\mathcal{C}$, of these parameters. We find
\begin{eqnarray}
 \langle \alpha \rangle &=& -0.0198, \nonumber \\
 \langle \beta  \rangle &=& +0.0099, \nonumber \\
 \mathcal{C} &=& \left( \begin{array}{rr}  0.0091 & -0.0047 \\ -0.0047 &  0.0032 \end{array} \right). \nonumber \\
\end{eqnarray}
The means are consistent with zero given the variances and sample size. This is expected as the residuals in the lower panel of Fig.~17 of \cite{knebe_haloes_2011} were defined relative to the mean halo mass function from the 14 halo finding algorithms considered. To incorporate this multi-dimensional Gaussian prior into our likelihood function we define two new parameters $H_1$ and $H_2$ which each have standard normal priors construct $\alpha$ and $\beta$ as:
\begin{eqnarray}
 \alpha &=& -0.01979 +0.09557 H_1 \nonumber \\
 \beta  &=& +0.00986 -0.04963 H_1 +0.02724 H_2.
\end{eqnarray}

\subsection{Construction of Model Mass Function}\label{sec:ModelCovariance}

To construct a model mass function, we generate a set of merger trees, and solve the system of \ODEs\ and rules given in \S\ref{sec:Model} to determine the stellar masses of model galaxies at $z=0.07$. For each galaxy, the weight of its merger tree and the stellar mass are shifted to the cosmological model assumed in the analysis of the data\footnote{That is, stellar masses are multiplied by the ratio of the square of the luminosity distance to $z=0.07$ in the data and model cosmologies, while the merger tree weight is multiplied by the ratio of $d_{\rm A}^2 {\rm d}r_{\rm c}/{\rm d}z$ in those two cosmologies, where $d_{\rm A}$ is angular diameter distance and $r_{\rm c}$ is comoving distance, and all terms are evaluated at $z=0.07$.}. Then the weight of the merger tree (eqn.~\ref{eq:MergerTreeWeight}) is distributed over a Gaussian of variance $\sigma_\star^2$ in $\log_{10}M_\star$,  centered on the stellar mass of the galaxy shifted for systematic errors according to eqn.~(\ref{eq:StellarMassSystematic}). This distribution is accumulated into bins which match those of the measured stellar mass function. That is:
\begin{equation}
 \phi_m = \sum_{j=1}^{N_{\rm t}} w_j \sum_{k=1}^{N_j} f_m(M^\prime_\star),
\end{equation}
where $\phi_m$ is the mass function in the $m^{\rm th}$ bin, $N_{\rm t}$ is the number of merger tree realizations, $N_j$ is the number of galaxies formed in tree $j$, $f_m(x)$ is the fraction of a lognormal of variance $\sigma_\star^2$ and mean $x$ that lies within the bounds of the $m^{\rm th}$ bin of the mass function, and $M^\prime_\star$ is the stellar mass of the model galaxy with systematic errors applied according to eqn.~(\ref{eq:StellarMassSystematic}). The covariance of the model mass function, $\mathcal{C}^{\rm (model)}$, which arises from the Monte Carlo nature of our model, is estimated as described in Appendix~\ref{app:ModelCovariance}, where we show that this model accurately reproduces the measured covariance of the model.

\subsection{Model Discrepancy}\label{sec:ModelDiscrepancy}

Our model is of course approximate. As such, even if the correct model parameters were known with infinite precision it would still not precisely reproduce the properties of observed galaxies. To mitigate this fact, we attempt to model the ``model discrepancies'' which arise as a consequence of this approximate nature.

We assume that model discrepancy can be approximated by a combination of systematic shifts in stellar mass and in abundance (or by inflation of the prior on systematic shifts in stellar mass and inflation of the model covariance matrix). Specifically, we model a systematic shift in stellar masses due to model discrepancy using the model given by eqn.~(\ref{eq:StellarMassSystematic}). Once the best-fitting values of $\mu$ and $\kappa$ have been found the remaining discrepancy is described as an offset plus multi-variate Gaussian covariance, $\mathcal{C}^{\rm (discrepancy)}$, in the mass function $\phi$. In general, we adopt a model in which
\begin{equation}
 \phi_i \rightarrow \zeta_i + \eta_i \phi_i,
 \label{eq:ModelDiscrepancy}
\end{equation}
where subscript $i$ refers to the $i^{\rm th}$ bin of the mass function, $\phi$, and $\zeta$ and $\eta$ are coefficients. That is, we consider both constant offset and multiplicative offset contributions to the discrepancy.

We will estimate model discrepancies using our model itself. Therefore, we must adopt some set of model parameters to use for this purpose. As we do not know the optimal set of parameters in advance (discerning this optimal set being the purpose of our study) we instead adopt a set of ``\emph{a priori}'' parameters, described in Appendix~\ref{app:Numerics}, for this purpose. This approach will be valid providing that the \emph{a priori} model parameters are sufficiently close to the maximum of the model posterior found by our \MCMC\ analysis. We will address this point in \S\ref{sec:ModelDiscrepancyResults}.

\subsubsection{Fixed Orbital Parameters}

As noted in \S\ref{sec:HaloMerging}, when computing the merging timescale of dark matter subhalos, we assign each subhalo orbital parameters equal to the mode of the measured cosmological distribution. That is, we ignore the scatter in orbital parameters. We do this to allow our models to converge more rapidly and reduce the CPU time required for calculations. This is, therefore, an approximation which we should account for using a model discrepancy term. 

To compute the model discrepancy, we run two calculations using our \emph{a priori} model: one with fixed orbital parameters, and a second with orbital parameters drawn from the full cosmological distribution. We then define
\begin{equation}
 \eta_i = \phi^{\rm (variable)}_i/\phi^{\rm (fixed)}_i,
\end{equation}
where $\phi^{\rm (variable)}_i$ and $\phi^{\rm (fixed)}_i$ are the stellar mass functions in the $i^{\rm th}$ in the model with variable and fixed orbital parameters respectively. The covariance, $\mathcal{C}^{\rm (discrepancy)}$, is set to the sum of the covariances of the two models\footnote{For all model discrepancy evaluations we run models much larger than those used in our \protect\MCMC\ analysis. Therefore, their covariance is a sub-dominant contribution to the total covariance, except where noted. Nevertheless, we include it in our estimate of the total covariance.}. We assume that the distribution of $\eta$ can be modeled as a multivariate log-normal distribution with this covariance. We find that $\eta_i$ is 100$\pm$1--2\%.

\subsubsection{(Lack of) Scatter in Merging Times}

\cite{jiang_fitting_2008} report a scatter between merger times computed from simulations and their fitting formula for merger timescale. This scatter is well described by a log-normal with width $\sigma\approx 0.4$. In the models run in our \MCMC\ analysis we ignore this scatter to reduce the CPU time needed for models to converge. We account for this approximation via a model discrepancy term.

To compute the model discrepancy, we run two calculations using our \emph{a priori} model: one with no scatter in merger times, and a second with $\sigma=0.4$ as recommended by \cite{jiang_fitting_2008}. We then define
\begin{equation}
 \eta_i = \phi^{\sigma=0.4}_i/\phi^{\sigma=0.0}_i,
\end{equation}
where $\phi^{\sigma=0.4}_i$ and $\phi^{\sigma=0.0}_i$ are the stellar mass functions in the $i^{\rm th}$ in the model with and without scatter respectively. The covariance, $\mathcal{C}^{\rm (discrepancy)}$, is set to the sum of the covariances of the two models. We assume that the distribution of $\eta$ can be modeled as a multivariate log-normal distribution with this covariance.  We find that $\eta_i$ is 100$\pm$1--5\% with the effect being largest in the highest mass bins.

\subsubsection{Monte Carlo-Generated Merger Trees}\label{sec:MonteCarloDiscrepancy}

Our merger trees are generated using a modification of the extended Press-Schechter formalism \citep{parkinson_generating_2008}. As such, they do not match perfectly the statistics of merger trees drawn from N-body simulations, although they are very close (see Appendix~\ref{app:MergerTreeConstruction}). We note that we are specifically \emph{not} assuming that the N-body trees represent the ``correct'' answer here. Merger tree construction algorithms give significantly different results \citep{srisawat_sussing_2013}, and the relatively small number of snapshots in the Millennium Simulation may limit the efficacy with which merger tree building algorithms can perform. Furthermore, \cite{jiang_generating_2013} note that the algorithm of \cite{parkinson_generating_2008} produces merger trees that differ from those extracted from N-body simulations by no more than the variation in N-body merger trees given different analysis methods. As such, we will treat this discrepancy as a source of covariance only (i.e. with no systematic offset) to reflect the fact that we do not know which type of tree (Monte Carlo or N-body) is more correct.

To quantify this model discrepancy we run our \emph{a priori} model on merger trees extracted from all 512 sub-volumes of the Millennium Simulation \citep{springel_simulations_2005}. We adjust the cosmological parameters and power spectrum assumed in our \emph{a priori} model to precisely match those used in the Millennium Simulation. We force the merger trees to be monotonically growing in mass (see Appendix~\ref{app:MergerTreeConstruction}). We then run the same model (i.e. still using the Millennium Simulation cosmology and power spectrum) using Monte Carlo merger trees, with the root masses of the trees at $z=0$ precisely matched to the masses of $z=0$ Millennium Simulation halos to remove any effects of variance due to finite volume. In this case, we re-sample the merger trees onto the set of times corresponding to the Millennium Simulation snapshots (for details of this re-sampling process, see \citealt{benson_convergence_2012}).

We parameterize the difference between the two mass functions using the systematic mass model of eqn.~(\ref{eq:StellarMassSystematic}), finding best-fit values of $\mu=-0.1025$, and $\kappa=-0.0225$. As there is no reason to assume that either Monte Carlo or N-body merger trees are ``better'', we do not shift the means of the priors on these parameters, but instead add the above values in quadrature to the variance of their Gaussian priors.

After applying this systematic shift to the Monte Carlo models, we assume a contribution to the covariance matrix of
\begin{equation}
 \mathcal{C}^{\rm (discrepancy)}_{ij} = [\phi^{\rm (N-body)}_i-\phi^{\rm (MC)}_i] [\phi^{\rm (N-body)}_j-\phi^{\rm (MC)}_j],
\end{equation}
where $\phi^{\rm (N-body)}_i$ and $\phi^{\rm (MC)}_i$ are the stellar mass functions in the $i^{\rm th}$ in the models using N-body and Monte Carlo merger trees respectively. For $\phi^{\rm (N-body)}_i$ we use the mass function estimated from the union of all 512 sub-volumes of the Millennium Simulation. We add to this covariance the sum of the internal covariances of the two models.

For the N-body merger trees, there is, additionally, some covariance arising from the finite volume of the Millennium Simulation. To estimate the total covariance in these models, including this finite volume source, we measure the covariance between the 512 sub-volumes of the Millennium Simulation. We begin by measuring the covariance between individual sub-volumes. We then proceed to pair sub-volumes and measure the covariance between pairs of sub-volumes. We repeat this process, doubling the number of sub-volumes combined each time, until we are combining 64 sub-volumes. From this we measure the dependence of the covariance matrix on the number of sub-volumes combined. We extrapolate this dependence to 512 sub-volumes to estimate the covariance in the full Millennium Simulation.

Over the power-law range of the mass function, we find that $\phi^{\rm (N-body)}_i$ and $\phi^{\rm (MC)}_i$ differ by typically $\sim1\%$. At the exponential cut-off, the N-body model predicts a significantly different mass function from the Monte Carlo merger trees, which makes the ratio of $\phi^{\rm (N-body)}_i$ to $\phi^{\rm (MC)}_i$ first decline to around $0.3$ before beginning to diverge to large values in the final few mass bins. This discrepancy is therefore large, but is nevertheless included. This is an aspect of the model discrepancy which warrants further study in future. In any case, the covariance introduced by this model discrepancy becomes large in the final few bins of the mass function, which will cause them to make only a small contribution to our likelihood.

\subsubsection{Variable Mass Resolution in Merger Trees}\label{sec:VariableMassResolution}

To permit rapid calculation of the model expectation, we use a variable mass resolution in our merger trees as described in eqn.~(\ref{eq:TreeMassResolution}). This is an approximation which affects the model estimate of the galaxy stellar mass function, which we account for using a model discrepancy term. 

To compute the model discrepancy, we run two calculations using our \emph{a priori} model: one with variable mass resolution as described in eqn.~(\ref{eq:TreeMassResolution}), and a second with fixed mass resolution ($M_{\rm res,min}=5\times 10^{9}M_\odot, f_0=0$). We assume that this discrepancy can be described by a constant offset in abundance (accounting for the missing halos in each tree) and a (possibly mass-dependent) systematic offset in stellar mass. We find that the systematic difference in these mass functions is well described by the systematic mass model of eqn.~(\ref{eq:StellarMassSystematic}) with $\mu=-0.0125$, $\kappa=0.165$. As such, we set the means of the priors on these parameters to these values. After applying this shift to the fixed mass resolution model, we then define
\begin{equation}
 \eta_i = \phi^{\rm (fixed)}_1/\phi^{\rm (variable)}_1,
\end{equation}
where $\phi^{\rm (fixed)}_i$ and $\phi^{\rm (variabl)}_i$ are the stellar mass functions in the $i^{\rm th}$ in the model with fixed and variable mass resolution respectively. The covariance, $\mathcal{C}^{\rm (discrepancy)}$, is set to the sum of the covariances of the two models. We assume that the distribution of $\eta$ can be modeled as a multivariate log-normal distribution with this covariance.

Note that we define the abundance discrepancy in all bins as equal to that in the lowest mass bin consistent with our assumption of a constant offset in abundance. This avoids the discrepancy term becoming uncomfortably large in the high mass bins. In these bins it is our opinion that the simple mass discrepancy model is insufficient to capture the details of this model discrepancy. Therefore, instead of applying a large correction, we include additional covariance to account for this failing of our discrepancy model. Specifically, any remaining offset, $\eta^\prime$, in the ratio $\phi^{\rm (fixed)}_i/\phi^{\rm (variable)}_i$ after applying the $\eta_i$ correction described above is used to add to the covariance matrix, $\mathcal{C}_{ij}^{\rm (discrepancy)} \rightarrow \mathcal{C}_{ij}^{\rm (discrepancy)} + \phi_i \phi_j \log(\eta_i^\prime) \log(\eta_j^\prime)$.

We find that $\eta_i$ is approximately 95\%. The additional covariance due to the limitations of our mass systematic model becomes large in the six highest mass bins, reducing the constraining power of the data in this regime. This is clearly a very large correction, but currently necessary due to practical limitations.

\subsubsection{Cosmology-independent Merger Trees}\label{sec:CosmologyIndepTrees}

As discussed in \S\ref{sec:CosmologyDependence}, we use a single set of merger trees in all model evaluations (to avoid introducing discontinuities in the likelihood surface). This means that cosmological dependence of merger tree structure is lost from the model. To account for this fact we introduce a model discrepancy term which increases the model covariance. To compute this term we generate a large number of model realizations with cosmological parameters and halo mass systematic parameters drawn from their respective priors (see \S\ref{sec:HaloMassFunctionSystematics} and \S\ref{sec:CosmologicalParameters}). We run one such set of models using a fixed set of merger trees (i.e. with trees generated using the maximum likelihood cosmological and systematics parameters), and a second set in which we generate a new set of trees for each model evaluation. We then construct the covariance matrix of the stellar mass function from each set of models. The difference between the covariance of the set with cosmology-dependent trees and those with cosmology-independent trees is taken as our model discrepancy term.

\subsection{Final Likelihood}

We assume that both the model and observed mass functions can be described as multivariate normal distributions, and so the model likelihood is finally evaluated as
\begin{equation}
 \ln \mathcal{L} = -{1\over 2} \Delta \mathcal{C}^{-1} \Delta^{\rm T} - {N \over 2} \ln(2\pi) - {\ln |\mathcal{C}| \over 2},
\end{equation}
where $N$ is the number of bins in the mass function, the covariance matrix is given by
\begin{equation}
 \mathcal{C} = \mathcal{C}^{\rm (observed)} + \mathcal{C}^{\rm (model)} + \mathcal{C}^{\rm (discrepancy)},
\end{equation}
where $\Delta_i = \zeta_i + \eta_i \phi_i^{\rm (model)} + \epsilon_i - \phi_i^{\rm (observed)}$ (recalling that $\zeta_i$ and $\eta_i$ describe model discrepancies---see eqn.~\ref{eq:ModelDiscrepancy}).

\subsection{MCMC Procedure}

In the following sections we describe the \MCMC\ procedure used to find constraints on model parameters.

\subsubsection{Parameter Priors}

Table~\ref{tb:ParameterPriors} summarizes the priors that we adopt for our model parameters. In this table, $U(x_{\rm l},x_{\rm h})$ indicates a uniform prior on the parameter over the range $[x_{\rm l},x_{\rm u}]$ and zero outside of that range. $U_{\ln{}}(x_{\rm l},x_{\rm h})$ indicates a uniform prior in the logarithm of the parameter over the range $[x_{\rm l},x_{\rm u}]$ and zero outside of that range. $N(\mu,\sigma^2,x_{\rm l},x_{\rm h})$ indicates a normal prior with mean $\mu$, variance $\sigma^2$, and truncated to zero outside the range $[x_{\rm l},x_{\rm u}]$.

Priors on composite cosmological parameters, $C_{1\ldots6}$, and halo mass function systematics, $H_{1\ldots2}$, are standard normals by construction (see \S\ref{sec:CosmologicalParameters} and \S\ref{sec:HaloMassFunctionSystematics} respectively), while the priors on the parameters of our model for systematics in stellar masses are described in \S\ref{sec:StellarMassSystematics}. For all other parameters we adopt relatively uninformative priors, as follows:
\begin{description}
 \item[{\boldmath $\alpha_{\rm cool}$}] We expect cooling to be more rapid in the past (densities are higher so cooling rates will be boosted and freefall timescales will be shorter). We therefore allow this parameter to range from a very negative value of $-10$ to $+1$ (to allow for the possibility that cooling actually becomes slightly less efficient at higher redshifts).

 \item[{\boldmath $\beta_{\rm cool}$}] This parameter controls the sharpness cut-off in cooling efficiency. Given that the mass function cuts off sharply at high mass we expect this parameter should be positive and probably greater than unity. We set a range $0$ to $8$.

 \item[{\boldmath $\Delta \log_{10}\mathcal{M}_{\rm cool}$}] Given the sharpness of the cut-off in cooling efficiency, we expect that this parameter will have to be relatively small, of order unity. We therefore limit it to the range $0.3$ to $3.0$.

 \item[{\boldmath $\tau_{\rm cool}$}] Timescales for cooling (or, more precisely, for infall) at the present day should be no shorter than typical halo dynamical times (which are of order 1~Gyr), but could be much longer if heating processes (perhaps from outflows or \AGN) are at work. We therefore adopt a uniform prior in the logarithm of this parameter between 1~Gyr and $3\times 10^3$~Gyr.

 \item[{\boldmath $\mathcal{M}_{\rm cool}$}] This parameter is intended to model the mass scale at which cooling transitions from efficient to inefficient. Based on studies such as \cite{benson_what_2003} we expect this scale must lie slightly above the halo mass associated with $L_*$ galaxies. If the cut-off in efficient cooling is associated with the transition from cold-mode to hot-mode cooling then we would expect $\mathcal{M}_{\rm cool}$ to be of order a few times $10^{11}M_\odot$ \citep{kerevs_how_2005}. To be conservative we therefore limit this parameter to the range $10^{10}$ to $3\times10^{13}M_\odot$.

 \item[{\boldmath $\alpha_\star$}] We do not have a strong intuition for the expected range for this parameter, nor a good way to infer it more directly from other data. Therefore, our prior is based on our experience of trial runs of our model. It will be important therefore to check whether the posterior on this parameter lies well within the prior or not.

 \item[{\boldmath $\beta_\star$}] As for $\alpha_\star$ our prior here is based on trial runs of our model.

 \item[{\boldmath $\tau_{\rm \star, min}$}]  The minimum timescale for star formation is limited to be sufficiently short that it will not affect the model results, while ensuring that model integration does not become unreasonably slow.

 \item[{\boldmath $\tau_\star$}] Typical specific star formation rates for star forming galaxies are of order $0.1$ to $1.0$~Gyr$^{-1}$ at $z\approx 0$ \citep{brinchmann_physical_2004}. Assuming a typical gas fraction of order 10\% for galaxies with rotation speeds of approximately $200$~km/s at the present day, this translate to a star formation timescale of around $0.1$ to $1.0$~Gyr. As such, we set the prior on the star formation timescale to lie in the range $0.03$ to $1$~Gyr.

 \item[{\boldmath $\alpha_{\rm wind}$}] It is not immediately obvious how wind mass loadings should scale with redshift, but we expect that outflows may be less strong at higher redshifts since galaxies will be denser (and so more energy may be radiated away rather than coupled to an outflow). Therefore, we allow a broad range for this parameter extending from very negative values to slightly above zero.

 \item[{\boldmath $\beta_{\rm wind}$}] We expect outflows to be stronger in systems of lower virial velocity, which suggests this parameter should be negative. We therefore allow it to range from a very negative value (more negative than any plausible physical outflow model) to zero.

 \item[{\boldmath $f_{\rm wind}$}] \cite{martin_demographics_2012} report mass-loading factors of order unity. Since we wish to also explore the possibility of models with minimal outflow we let this parameter range from a very small value to unity.

 \item[{\boldmath $\tau_{\rm wind, min}$}] The minimum timescale for outflows is limited to be sufficiently short that it will not affect the model results, while ensuring that model integration does not become unreasonably slow.

 \item[{\boldmath $f_{\rm df}$}] Merging timescales should be well calibrated through our use of the \cite{jiang_fitting_2008} fitting formula. However, we allow for the possibility for variations in this timescale (due to approximations in how merging is defined, effects of baryonic physics on merging, etc.) by adopting a prior that is uniform in the logarithm of this parameter between $0.1$ and $10$.

 \item[{\boldmath $V_{\rm reion}$}] The effects of reionization on the accretion of gas into halos has been well studied. For example, \cite{okamoto_mass_2008} find that accretion is suppressed in systems below a characteristic velocity of roughly 25~km/s. \cite{font_population_2011} explore the same, simple truncation model for the effects of reionization that we adopt in this work. They find that the results of \cite{okamoto_mass_2008} can be reproduced in such a model with $V_{\rm reion}=34$km/s. Previous studies \citep{gnedin_effect_2000} had reported much stronger effects from reionization. To explore the full range of possibilities we therefore, we adopt a prior that is uniform between 30 and 60~km/s.
\end{description}

\begin{table*}
 \begin{tabular}{llll@{(}r@{, }r@{}l@{ }r@{}l@{ }r@{)}}
  \hline
  {\bf Parameter} & {\bf Units} & {\bf Description} & \multicolumn{7}{l}{\bf Prior} \\
  \hline
  $\alpha_{\rm cool}$ & -- & Exponent of $1/(1+z)$ in cooling rate & $U$&$-10$&$+1$&&&& \\
  $\beta_{\rm cool}$ & -- & Exponent of cut-off mass-scale for efficient cooling & $U$&$0$&$+8$&&&& \\
  $\Delta \log_{10}\mathcal{M}_{\rm cool}$ & -- & Width of cut-off for efficient cooling & $U_{\ln{}}$&$3\times10^{-1}$&$3\times10^1$&&&& \\
  $\tau_{\rm cool}$ & Gyr & Timescale for cooling &  $U_{\ln{}}$&$1$&$3\times10^3$ &&&&\\
  $\mathcal{M}_{\rm cool}$ & $M_\odot$ & Cut-off mass-scale for efficient cooling &  $U_{\ln{}}$&$10^{10}$&$3\times10^{13}$&&&& \\
  $\alpha_\star$ & -- &  Exponent of $(1+z)$ in star formation rate &  $U$&$0$&$+4$&&&& \\
  $\beta_\star$ & -- &  Exponent of virial velocity in star formation rate &  $U$&$-8$&$-2$&&&& \\
  $\tau_{\rm \star, min}$ & Gyr &  Minimum timescale for star formation &  $U_{\ln{}}$&$10^{-5}$&$10^{-1})$&&&& \\
  $\tau_\star$ & Gyr &  Timescale for star formation &  $U_{\ln{}}$&$3\times10^{-2}$&$1$&&&& \\
  $\alpha_{\rm wind}$ & -- &  Exponent of $(1+z)$ in wind outflow rate &  $U$&$-14$&$+1$&&&& \\
  $\beta_{\rm wind}$ & -- &  Exponent of virial velocity in wind outflow rate &  $U$&$-20$&$0$&&&& \\
  $f_{\rm wind}$ & -- &  Normalization of wind outflow rate &  $U_{\ln{}}$&$10^{-5}$&$1$&&&& \\
  $\tau_{\rm wind, min}$ & Gyr &  Minimum timescale for wind outflow rate & $U_{\ln{}}$&$10^{-3}$&$10^{-1}$&&&& \\
  $f_{\rm df}$ & -- & Multiplier for subhalo merging times & $U_{\ln{}}$&$10^{-1}$&$10^{+1}$&&&& \\
  $V_{\rm reion}$ & km/s & Velocity scale for suppression of accretion from the \IGM\ & $U$&$30$&$60$&&&& \\
  $C_{1\ldots6}$ & -- &  Cosmological parameter combination & $N$&$0$&$1$&,&$-\infty$&,&$+\infty$ \\
  $\mu$ & -- &  Coefficient in stellar mass systematic (eqn.~\protect\ref{eq:StellarMassSystematic}) & $N$&$-0.0125$&$0.0833$&,&$-0.5$&,&$+0.5$ \\
  $\kappa$ & -- &  Coefficient in stellar mass systematic (eqn.~\protect\ref{eq:StellarMassSystematic}) & $N$&$0.165$&$0.0105$&,&$-0.32$&,&$+0.32$ \\
  $H_{1\ldots2}$ & -- &  Halo mass function systematic error parameter combination & $N$&$0$&$1$&,&$-\infty$&,&$+\infty$ \\
  \hline
 \end{tabular}
 \caption{Priors on parameters. $U(x_{\rm l},x_{\rm h})$ is uniform in the parameter over the range $[x_{\rm l},x_{\rm u}]$ and is zero outside of that range. $U_{\ln{}}(x_{\rm l},x_{\rm h})$ is uniform in the logarithm of the parameter over the range $x_{\rm l}$--$x_{\rm u}$ and is zero outside of that range. $N(\mu,\sigma^2,x_{\rm l},x_{\rm h})$ is a normal distribution with mean $\mu$, variance $\sigma^2$, and truncated to zero outside the range $[x_{\rm l},x_{\rm u}]$.}
 \label{tb:ParameterPriors}
\end{table*}

\subsubsection{Cosmological Parameters}\label{sec:CosmologicalParameters}

We include six cosmological parameters in our \MCMC\ analysis. Even though the values of these parameters are well constrained by combinations of cosmic microwave background, supernovae, Cepheid, lensing, and baryon acoustic oscillation experiments \citep[see][for a review]{planck_collaboration_planck_2013} some uncertainty in their values remain (particularly for $\sigma_8$), and we wish to assess the impact of these uncertainties on the model posterior and any predictions from that posterior. Therefore, we make use of the WMAP 9-year data \citep{hinshaw_nine-year_2013} to define priors on the combination of parameters $(H_0, \Omega_{\rm M}, \Omega_{\rm b}, \sigma_8, n_{\rm s}, \tau)$, and assume a flat universe such that $\Omega_\Lambda=1-\Omega_{\rm M}$. Using the \MCMC\ chains provided by the WMAP collaboration\footnote{Specifically the chains combining constraints from WMAP-9, SPT \citep{keisler_measurement_2011}, ACT \citep{das_atacama_2011}, SNLS3 \citep{sullivan_snls3:_2011}, BAO \citep{beutler_6df_2011,padmanabhan_2_2012,anderson_clustering_2012,blake_wigglez_2012}, and $H_0$ \citep{riess_3_2011,freedman_carnegie_2012}: \protect\href{http://lambda.gsfc.nasa.gov/data/map/dr5/dcp/chains/wmap_lcdm_wmap9_spt_act_snls3_chains_v5.tar.gz}{\tt http://lambda.gsfc.nasa.gov/data/map/dr5/dcp/chains/\ wmap\_lcdm\_wmap9\_spt\_act\_snls3\_chains\_v5.tar.gz}} we measure the covariance matrix of these six parameters and model their prior as a multivariate normal distribution. Specifically, we define six meta-parameters, $C_1\ldots6$, each with a standard normal prior and construct the cosmological parameters from these with the required covariances as follows:
\begin{eqnarray}
H_0 &=& 69.55 + 0.4011 C_1 - 0.5585 C_2 + 0.003770 C_3 \nonumber \\ 
    & & + 0.3875 C_4 \hbox{ km/s/Mpc}, \\
\Omega_{\rm M} &=& \hbox{min} \left( [0.1371 + 0.0004757 C_1 + 0.001892 C_2]\right.\nonumber \\
               & &  \left. \times \left[H_0 / 100~\hbox{km/s/Mpc}\right]^{-2},1 \right), \\
\Omega_\Lambda &=& 1-\Omega_{\rm M}, \\
\Omega_{\rm b} &=& (0.02225 + 0.0003305 C_1) \nonumber \\
               & & \times (H_0/100~\hbox{km/s/Mpc})^{-2}, \\
\sigma_8 &=& 0.8178 + 0.003817 C_1 + 0.007931 C_2 \nonumber \\
         & & + 0.01002 C_3 + 0.001584 C_4 + 0.0029312 C_5 \nonumber \\
         & & + 0.001727 C_6, \\
n_{\rm s} &=&  0.9616 + 0.005302 C_1 - 0.001576 C_2 \nonumber \\
          & & + 0.0009843 C_3 + 0.0006839 C_4 + 0.005735 C_5, \\
\tau &=& \hbox{max}(0.08155 + 0.001718 C_1 - 0.001858 C_2 \nonumber \\
     & & + 0.01179 C_3,0).
\end{eqnarray}

The redshift of reionization, $z_{\rm reion}$, is then found by solving the equation
\begin{equation}
 \tau = \int_0^{z_{\rm reion}} n_{\rm e}(z) {{\rm d} l \over {\rm d} z} {\rm d} z,
\end{equation}
where $n_{\rm e}(z)$ is the electron density at redshift $z$ assuming a fully-ionized universe.

\subsubsection{MCMC}\label{sec:MCMC}

We use the \BIE\ to perform a differential evolution \MCMC\ simulation \citep{terr_braak_markov_2006} to constrain the parameters of our model. We use 288 parallel chains for our main calculation. The state of each chain is initialized by drawing at random from the prior distributions. At each step of the simulation a proposed state, $S_i^\prime$, for each chain, $i$, is constructed by selecting at random (without replacement) two other chains, $m$ and $n$, and finding
\begin{equation}
 S_i^\prime = S_i + \gamma (S_m - S_n) + \epsilon,
\end{equation}
where $\gamma$ is a parameter chosen to keep the acceptance rate of proposed states sufficiently high, and $\epsilon$ is a random vector each component of which is drawn from a Cauchy distribution with median zero and width parameter set equal to 0.1\% of the width of the prior\footnote{The width of the prior is taken to be the maximum minus minimum values for uniform priors and the standard deviation (or the upper minus lower limit if that is larger) for normal priors.} to ensure that the chains are positively recurrent. For a multivariate normal likelihood function in $N$ dimensions the optimal value of $\gamma$ is $\gamma_0=2.38/\sqrt{N}$ \citep{terr_braak_markov_2006}. The proposed state is accepted with probability $P$ where
\begin{equation}
 P = \left\{ \begin{array}{ll} 1 & \hbox{ if } \mathcal{L}(S_i^\prime) > \mathcal{L}(S_i), \\ \mathcal{L}(S_i^\prime)/\mathcal{L}(S_i) & \hbox{ otherwise.} \end{array} \right.
\end{equation}

We begin our \MCMC\ simulation at a high temperature of $T=128$. That is, log-likelihoods are divided through by $T$ to smooth the likelihood surface and make it easier for the chains to rapidly explore the surface. The simulation is allowed to progress until the chains have converged on the heated posterior distribution as judged by the Gelman-Rubin statistic, $\hat{R}$ \citep{gelman_a._inference_1992}, after outlier chains (identified using the Grubb's outlier test \citep{grubbs_procedures_1969,stefansky_rejecting_1972} with significance level $\alpha=0.05$) have been discarded. Specifically, we declare convergence when $\hat{R}=1.2$. The final state of the chains is then used as the starting point for a new simulation with lower temperature. We repeat this process, reducing the temperature each time, until we reach $T=1$. Once the chains have converged for $T=1$ we let them run to generate a sufficiently large sampling of the converged, unheated posterior. We find that this gradual cooling is an effective way to achieve convergence onto the final posterior. Table~\ref{tb:tempering} lists the temperatures and corresponding values of $\gamma$ used.

\begin{table}
 \caption{The sequence of temperatures and corresponding values of $\gamma$ used in our \protect\MCMC\ simulations.}
\label{tb:tempering}
\begin{center}
\begin{tabular}{rr@{.}l}
\hline
{\bf Temperature} & \multicolumn{2}{c}{{\boldmath $\gamma$}} \\
 \hline
 128 & 0&3365 \\
  32 & 0&3365 \\
   8 & 0&1743 \\
   2 & 0&1200 \\
   1 & 0&1000 \\
\hline
\end{tabular}
\end{center}
\end{table}

\section{Results}\label{sec:Results}

Before beginning the \MCMC\ algorithm, we check that the numerical parameters of our model are sufficient to give robust, converged results. The results of this numerical study are presented in Appendix~\ref{app:Numerics}.

\subsection{Parameter Constraints}

\begin{figure*}
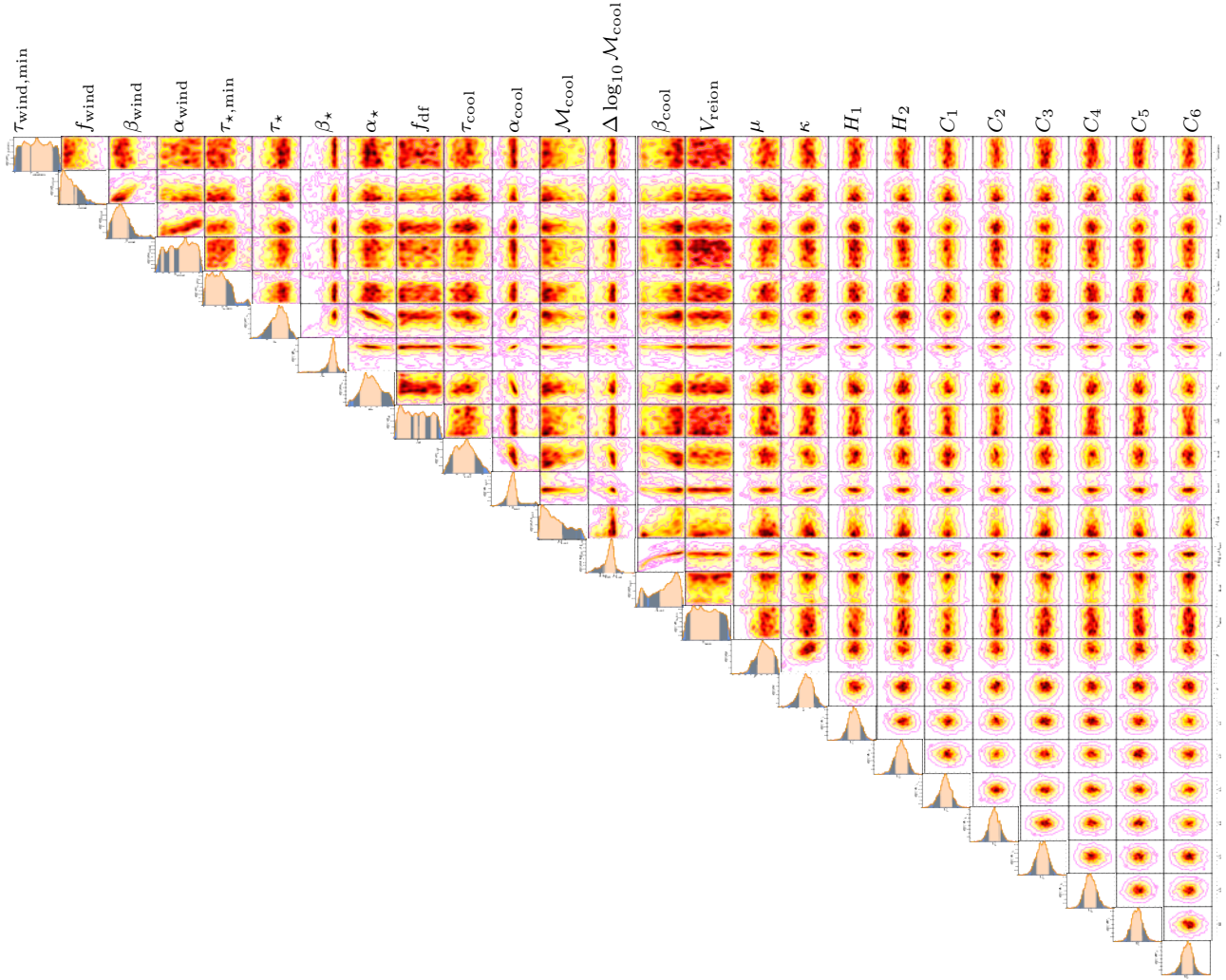

  \renewcommand{\arraystretch}{0}
\begin{tabular}{l@{}l@{}l@{}l@{}l@{}l@{}l@{}l@{}l@{}l@{}l@{}l@{}l@{}l@{}l@{}l@{}l@{}l@{}l@{}l@{}l@{}l@{}l@{}l@{}l@{}}
\multicolumn{1}{c}{\begin{sideways}\,$\tau_{\rm wind, min}$\end{sideways}}&
\multicolumn{1}{c}{\begin{sideways}\,$f_{\rm wind}$\end{sideways}}&
\multicolumn{1}{c}{\begin{sideways}\,$\beta_{\rm wind}$\end{sideways}}&
\multicolumn{1}{c}{\begin{sideways}\,$\alpha_{\rm wind}$\end{sideways}}&
\multicolumn{1}{c}{\begin{sideways}\,$\tau_{\rm \star, min}$\end{sideways}}&
\multicolumn{1}{c}{\begin{sideways}\,$\tau_\star$\end{sideways}}&
\multicolumn{1}{c}{\begin{sideways}\,$\beta_\star$\end{sideways}}&
\multicolumn{1}{c}{\begin{sideways}\,$\alpha_\star$\end{sideways}}&
\multicolumn{1}{c}{\begin{sideways}\,$f_{\rm df}$\end{sideways}}&
\multicolumn{1}{c}{\begin{sideways}\,$\tau_{\rm cool}$\end{sideways}}&
\multicolumn{1}{c}{\begin{sideways}\,$\alpha_{\rm cool}$\end{sideways}}&
\multicolumn{1}{c}{\begin{sideways}\,$\mathcal{M}_{\rm cool}$\end{sideways}}&
\multicolumn{1}{c}{\begin{sideways}\,$\Delta \log_{10}\mathcal{M}_{\rm cool}$\end{sideways}}&
\multicolumn{1}{c}{\begin{sideways}\,$\beta_{\rm cool}$\end{sideways}}&
\multicolumn{1}{c}{\begin{sideways}\,$V_{\rm reion}$\end{sideways}}&
\multicolumn{1}{c}{\begin{sideways}\,$\mu$\end{sideways}}&
\multicolumn{1}{c}{\begin{sideways}\,$\kappa$\end{sideways}}&
\multicolumn{1}{c}{\begin{sideways}\,$H_1$\end{sideways}}&
\multicolumn{1}{c}{\begin{sideways}\,$H_2$\end{sideways}}&
\multicolumn{1}{c}{\begin{sideways}\,$C_1$\end{sideways}}&
\multicolumn{1}{c}{\begin{sideways}\,$C_2$\end{sideways}}&
\multicolumn{1}{c}{\begin{sideways}\,$C_3$\end{sideways}}&
\multicolumn{1}{c}{\begin{sideways}\,$C_4$\end{sideways}}&
\multicolumn{1}{c}{\begin{sideways}\,$C_5$\end{sideways}}&
\multicolumn{1}{c}{\begin{sideways}\,$C_6$\end{sideways}}\\
\end{tabular}

 \caption{The \protect\PPD\ of parameters in our model constrained to fit the observed stellar mass function of \protect\cite{li_distribution_2009}. Panels on the diagonal show the marginalized \protect\PPD\ for each individual parameter, while off-diagonal panels show the \protect\PPD\ for pairs of parameters, marginalized over all other parameters.}
 \label{fig:PosteriorTriangle}
\end{figure*}

\begin{table*}
 \caption{Values of model parameters at the indicated percentages of their 1-D marginalized cumulative \protect\PPDs.}
 \label{tb:Intervals}
 \begin{center}
   \begin{tabular}{lr@.lr@.lr@.lr@.lr@.lr@.lr@.l}
 \hline
  & \multicolumn{14}{c}{\bf Cumulative probability} \\
 {\bf Parameter} &
 \multicolumn{2}{c}{\bf 0.27\%} & 
 \multicolumn{2}{c}{\bf 4.55\%} & 
 \multicolumn{2}{c}{\bf 31.73\%} & 
 \multicolumn{2}{c}{\bf 50.00\%} & 
 \multicolumn{2}{c}{\bf 68.27\%} & 
 \multicolumn{2}{c}{\bf 95.45\%} & 
 \multicolumn{2}{c}{\bf 99.73\%}
 \\
 \hline
    $\tau_{\rm wind, min}$ & $0$&$845 \times 10^{-3}$ & $0$&$117 \times 10^{-2}$ & $0$&$417 \times 10^{-2}$ & $0$&$940 \times 10^{-2}$ & $0$&$0214$ & $0$&$0781$ & $0$&$111$ \\
$f_{\rm wind}$ & $0$&$715 \times 10^{-5}$ & $0$&$117 \times 10^{-4}$ & $0$&$428 \times 10^{-4}$ & $0$&$120 \times 10^{-3}$ & $0$&$428 \times 10^{-3}$ & $0$&$0113$ & $0$&$180$ \\
$\beta_{\rm wind}$ & $-20$&$3$ & $-19$&$2$ & $-16$&$0$ & $-14$&$4$ & $-12$&$6$ & $-6$&$09$ & $-0$&$376$ \\
$\alpha_{\rm wind}$ & $-14$&$3$ & $-13$&$0$ & $-8$&$55$ & $-5$&$72$ & $-3$&$49$ & $0$&$192$ & $1$&$29$ \\
$\tau_{\rm \star, min}$ & $0$&$746 \times 10^{-5}$ & $0$&$122 \times 10^{-4}$ & $0$&$615 \times 10^{-4}$ & $0$&$192 \times 10^{-3}$ & $0$&$556 \times 10^{-3}$ & $0$&$0107$ & $0$&$105$ \\
$\tau_\star$ & $0$&$0306$ & $0$&$0673$ & $0$&$173$ & $0$&$242$ & $0$&$321$ & $0$&$611$ & $0$&$884$ \\
$\beta_\star$ & $-7$&$20$ & $-5$&$03$ & $-3$&$90$ & $-3$&$67$ & $-3$&$48$ & $-2$&$78$ & $-2$&$07$ \\
$\alpha_\star$ & $-0$&$0526$ & $0$&$487$ & $1$&$55$ & $2$&$00$ & $2$&$47$ & $3$&$67$ & $4$&$02$ \\
$f_{\rm df}$ & $0$&$0836$ & $0$&$114$ & $0$&$344$ & $0$&$830$ & $1$&$94$ & $7$&$50$ & $10$&$9$ \\
$\tau_{\rm cool}$ & $0$&$940$ & $2$&$13$ & $13$&$7$ & $38$&$2$ & $91$&$5$ & $8$&$58 \times 10^{2}$ & $2$&$76 \times 10^{3}$ \\
$\alpha_{\rm cool}$ & $-9$&$55$ & $-7$&$50$ & $-5$&$71$ & $-5$&$19$ & $-4$&$72$ & $-0$&$359$ & $1$&$02$ \\
$\mathcal{M}_{\rm cool}$ & $7$&$03 \times 10^{9}$ & $1$&$07 \times 10^{10}$ & $3$&$72 \times 10^{10}$ & $1$&$10 \times 10^{11}$ & $4$&$24 \times 10^{11}$ & $1$&$23 \times 10^{13}$ & $2$&$85 \times 10^{13}$ \\
$\Delta \log_{10}\mathcal{M}_{\rm cool}$ & $0$&$307$ & $0$&$547$ & $2$&$07$ & $2$&$77$ & $3$&$39$ & $6$&$19$ & $14$&$6$ \\
$\beta_{\rm cool}$ & $0$&$0505$ & $0$&$669$ & $3$&$80$ & $5$&$34$ & $6$&$46$ & $7$&$78$ & $8$&$24$ \\
$V_{\rm reion}$ & $29$&$0$ & $31$&$4$ & $39$&$2$ & $44$&$3$ & $49$&$8$ & $58$&$3$ & $60$&$6$ \\
$\mu$ & $-0$&$167$ & $0$&$652 \times 10^{-2}$ & $0$&$186$ & $0$&$253$ & $0$&$327$ & $0$&$467$ & $0$&$510$ \\
$\kappa$ & $-0$&$313$ & $-0$&$159$ & $-0$&$896 \times 10^{-2}$ & $0$&$0401$ & $0$&$0903$ & $0$&$209$ & $0$&$297$ \\
$C_1$ & $-2$&$69$ & $-1$&$66$ & $-0$&$467$ & $0$&$0357$ & $0$&$535$ & $1$&$80$ & $2$&$76$ \\
$C_2$ & $-2$&$88$ & $-1$&$73$ & $-0$&$458$ & $0$&$0440$ & $0$&$524$ & $1$&$67$ & $2$&$56$ \\
$C_3$ & $-2$&$79$ & $-1$&$83$ & $-0$&$495$ & $-0$&$0652$ & $0$&$362$ & $1$&$54$ & $2$&$60$ \\
$C_4$ & $-2$&$99$ & $-1$&$83$ & $-0$&$645$ & $-0$&$186$ & $0$&$274$ & $1$&$50$ & $2$&$51$ \\
$C_5$ & $-2$&$79$ & $-1$&$73$ & $-0$&$503$ & $0$&$199 \times 10^{-2}$ & $0$&$478$ & $1$&$64$ & $2$&$60$ \\
$C_6$ & $-2$&$71$ & $-1$&$69$ & $-0$&$557$ & $-0$&$0774$ & $0$&$423$ & $1$&$72$ & $2$&$83$ \\
$H_1$ & $-2$&$86$ & $-1$&$78$ & $-0$&$559$ & $-0$&$0416$ & $0$&$441$ & $1$&$69$ & $3$&$04$ \\
$H_2$ & $-2$&$89$ & $-1$&$77$ & $-0$&$476$ & $0$&$0156$ & $0$&$463$ & $1$&$59$ & $2$&$66$ \\

 \hline
   \end{tabular}
 \end{center}
\end{table*}

Figure~\ref{fig:PosteriorTriangle} shows 1-D and 2-D marginalized \PPDs\ for  all 25 parameters in our analysis, while Table~\ref{tb:Intervals} lists parameter values at various percentages of their 1-D marginalized cumulative \PPDs. The posteriors of the ``nuisance'' parameters (i.e. those describing systematic uncertainties in stellar mass, the halo mass function, and cosmological parameters) are dominated by their priors (indicating that the data do not provide significant additional constraint on these quantities). While the majority of the model parameters are rather poorly constrained, a handful of them are strongly constrained by the observed stellar mass function.

\begin{figure*}
 \begin{tabular}{cc}
  \includegraphics[width=85mm,trim=0mm 0mm 0mm 2.5mm,clip]{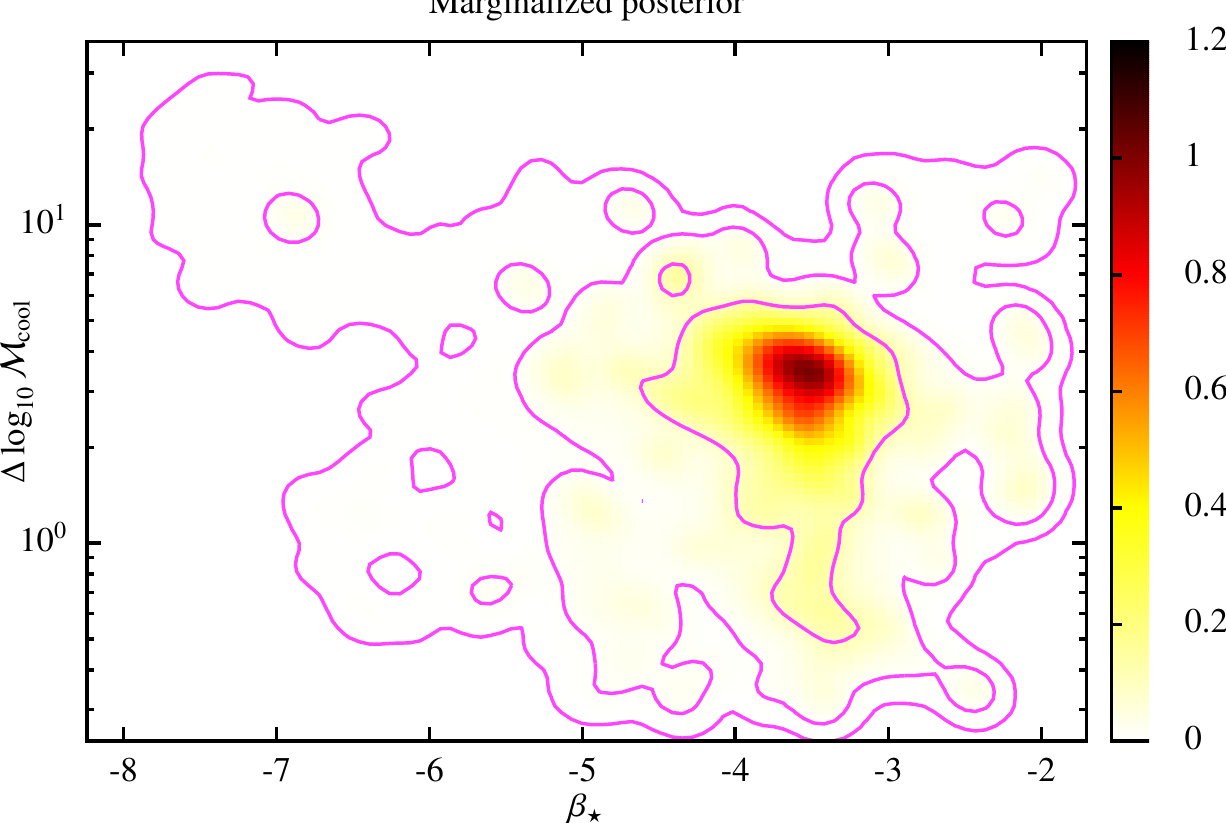} & \includegraphics[width=85mm,trim=0mm 0mm 0mm 2.5mm,clip]{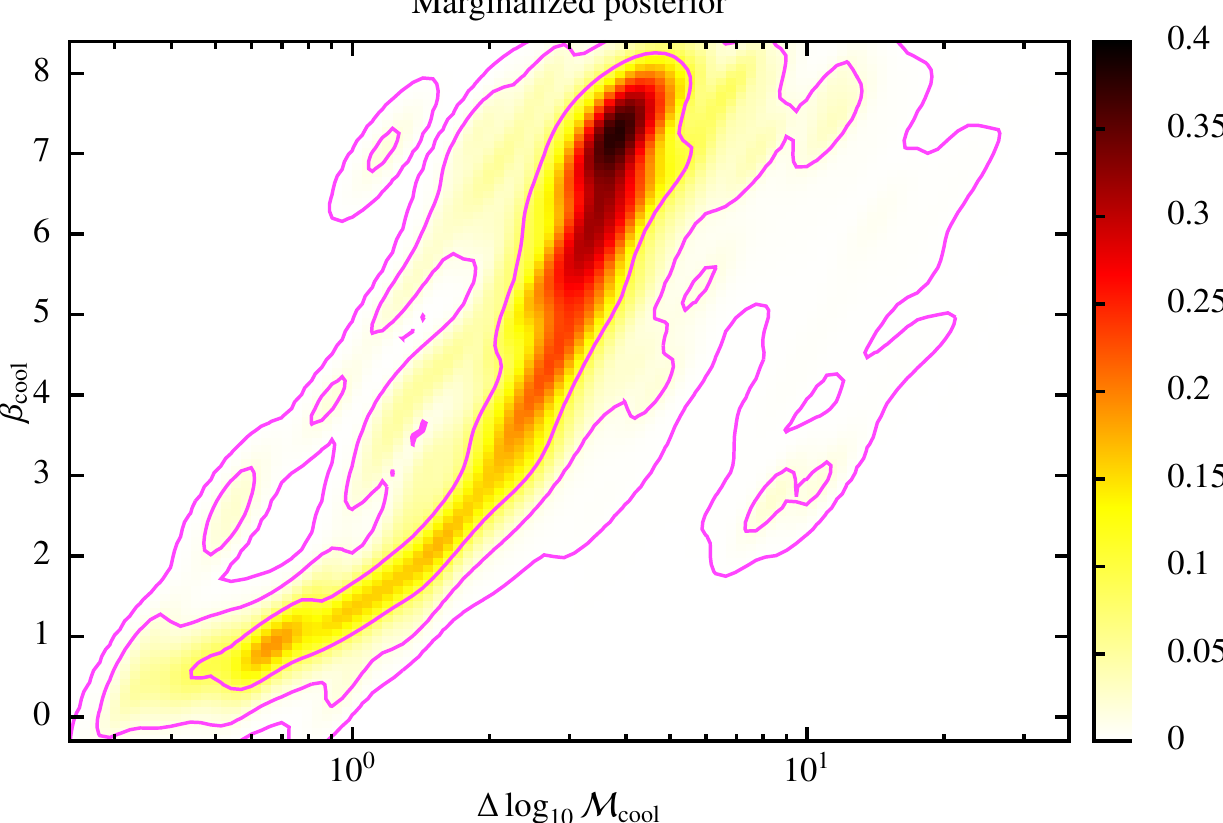}
 \end{tabular}
 \caption{\emph{Left panel:} The joint \protect\PPD\ of $\beta_\star$ (the exponent of halo virial velocity appearing in the star formation rate; eqn.~\protect\ref{eq:starFormationRate}) and $\Delta \log_{10}\mathcal{M}_{\rm cool}$ (the width in the logarithm of halo mass of the cut-off in cooling efficiency; eqn.~\protect\ref{eq:coolingRate}). \emph{Right panel:} The joint \protect\PPD\ of $\Delta \log_{10}\mathcal{M}_{\rm cool}$ and $\beta_{\rm cool}$ (the exponent of the cut-off factor in cooling efficiency; eqn.~\protect\ref{eq:coolingRate}).}
 \label{fig:PosteriorInteresting}
\end{figure*}

Figure~\ref{fig:PosteriorInteresting} shows enlarged versions of two cases where the posterior is well constrained. In the left hand panel we plot the marginalized posterior for $\beta_\star$ and $\Delta \log_{10}\mathcal{M}_{\rm cool}$. These parameters are both well constrained by the data (in the sense that their posteriors are much narrower than their priors), and are uncorrelated. The right hand panels shows the joint posterior of $\Delta \log_{10}\mathcal{M}_{\rm cool}$ and $\beta_{\rm cool}$. This pair of parameters is jointly well constrained, but highly correlated.

Finally, the posteriors of the parameters $\alpha_\star$ and $\beta_\star$---for which we did not have good pre-existing knowledge from which to set priors---are peaked well inside the ranges allowed by their priors. As such, our choice of prior did not strongly affect the results\footnote{It is always possible that an island of high-likelihood exists outside the range of the priors that we set.}.

We will discuss implications of the posterior in \S\ref{sec:Discussion}.

\subsection{Posterior Predictive Checks}\label{sec:PosteriorPredictiveChecks}

As described by \cite{lu_bayesian_2012}, \PPCs\ \citep{gilks_w._r._markov_1995,gelman_a._bayesian_2013} are a powerful means by which to check that the model family characterized by the \PPD\ is a viable description of the observed data. We follow the procedure of \cite{lu_bayesian_2012} to assess the \PPC, and the reader is directed to that work for a complete description. Specifically, we adopt a test-statistic similar to that of \citeauthor{lu_bayesian_2012}~(\citeyear{lu_bayesian_2012}; eqn.~6):
\begin{equation}
 \mathcal{T}_l = \mathcal{T}({\bf y}_l) = \Delta_l\cdot\mathcal{C}^{-1}_\searrow\cdot\Delta_l^{\rm T},
\end{equation}
where $\Delta_l = {\bf y}_l - \bar{\bf y}$, ${\bf y}_l$ is the $l^{\rm th}$ model mass function sampled from our converged \MCMC\ chains, $\bar{\bf y}$ is the mean model mass function over the \MCMC\ chains, and $\mathcal{C}_\searrow$ is the diagonalized covariance matrix\footnote{We use the diagonalized covariance matrix here as it is more robust---we find that using the full covariance matrix estimated from this sample of 10,000 model realizations can lead to imprecisions in the matrix inversion to get $\mathcal{C}^{-1}$ leading to wildly varying test statistics. Future studies might avoid this problem (and so be able to employ the full covariance matrix) by running the \MCMC\ chains for longer thereby allowing a much larger number of indepdent model realizations to be generated.} of the set of model results. We compute the same quantity for the observational data
\begin{equation}
  \mathcal{T}^\prime = \Delta^\prime \cdot\mathcal{C}^{-1}_\searrow\cdot\Delta^{\prime \rm T},
\end{equation}
where $\Delta^\prime = {\bf y}^\prime - \bar{\bf y}$, and ${\bf y}^\prime$ is the observed mass function. The Bayesian $p$-value is then
\begin{equation}
 \hat{p}_{\rm B} = {1\over L} \sum_{l=1}^L I_{\mathcal{T}_l \ge \mathcal{T}^\prime}.
\end{equation}

\begin{figure*}
 \vspace{+18mm}
 \begin{tabular}{cc}
   \vspace{-18mm} $p=\protect\input{plots/posteriorPredictiveChecks/sdssStellarMassFunctionZ0.07_testStatistic_pValue.txt}$ &  \\ 
 \includegraphics[width=85mm,trim=0mm 0mm 0mm 2.5mm,clip]{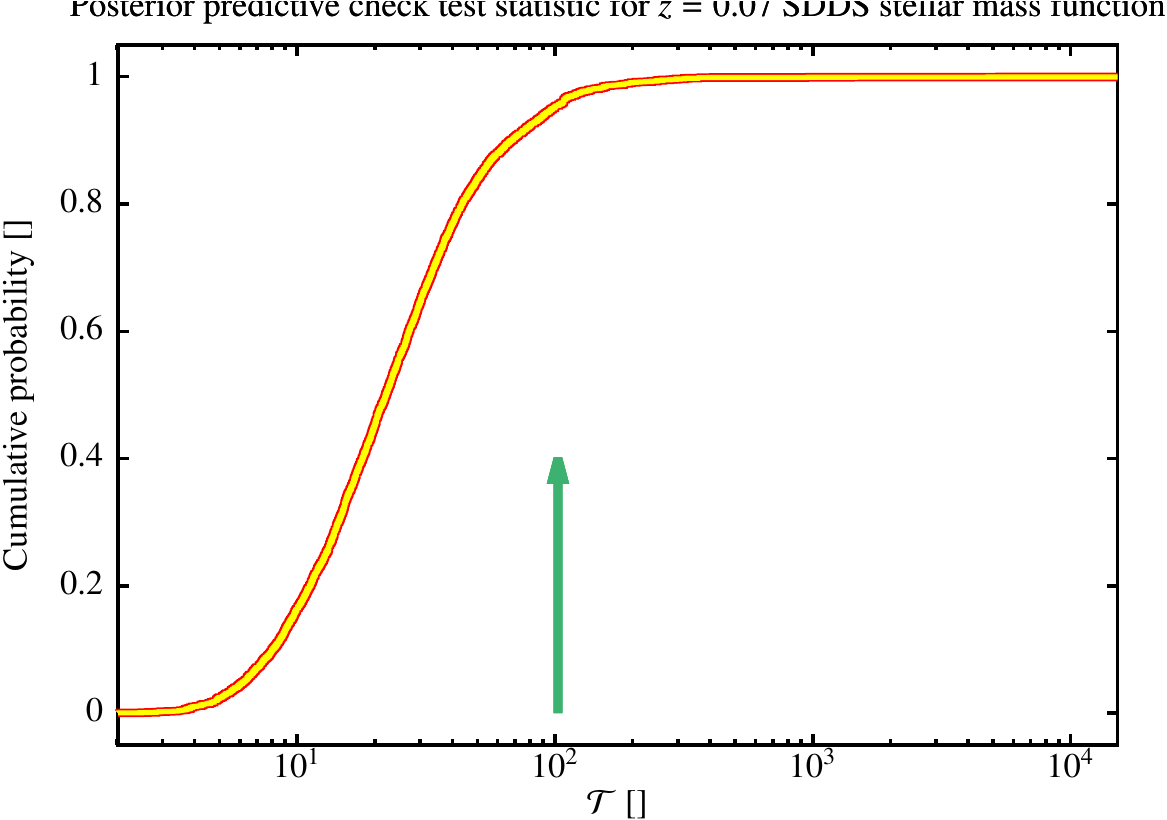} &
   \includegraphics[width=85mm,trim=0mm 0mm 0mm 2.5mm,clip]{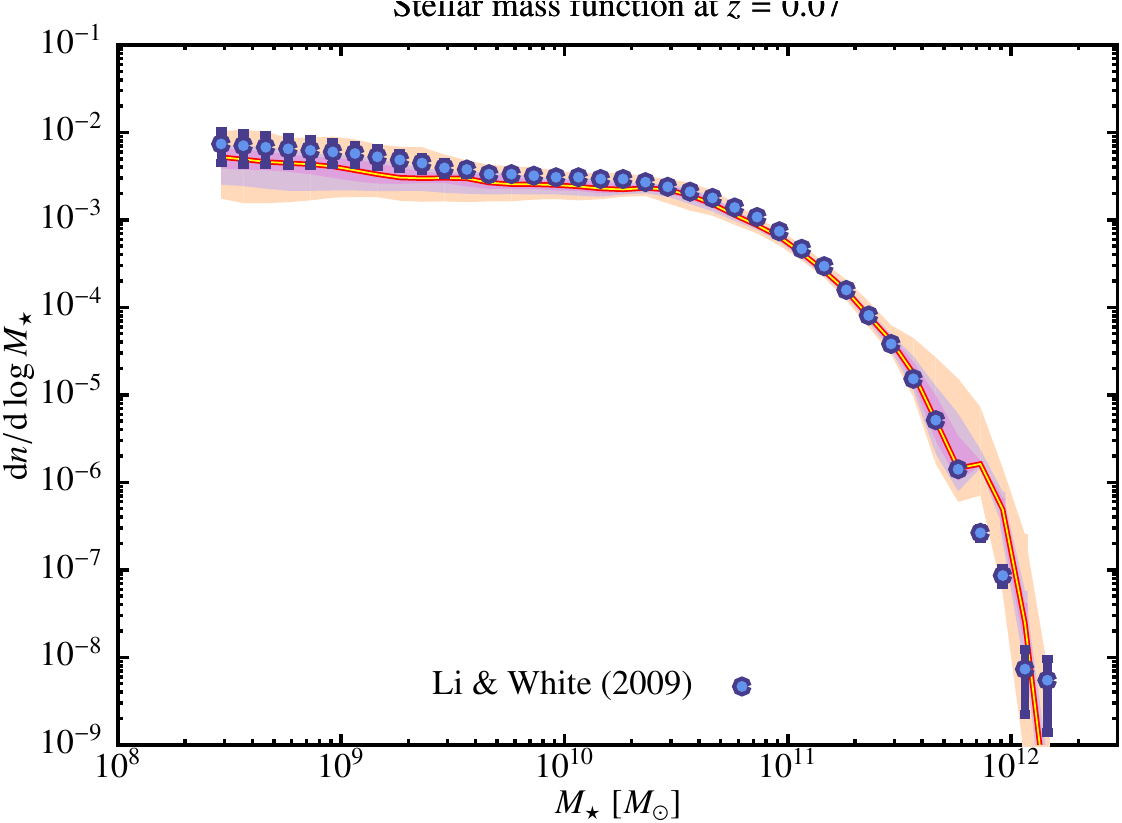}
\end{tabular}
 \caption{\emph{Left panel:} The cumulative probability of the test statistic used in our \protect\PPC\ analysis for models sampled from the converged \protect\MCMC\ chains, $\mathcal{T}_l$ (yellow line). The green arrow shows the same test statistic computed for the data, $\mathcal{T}^\prime$. The resulting $p$-value (also indicated in the panel) is $\protect\input{plots/posteriorPredictiveChecks/sdssStellarMassFunctionZ0.07_testStatistic_pValue.txt}$. \emph{Right panel:} The stellar mass function of galaxies at $z\approx 0.07$. Blue points with error bars show the mass function measured from the \protect\SDSS\ by \protect\cite{li_distribution_2009}. Error bars on these points are the square-roots of the diagonal elements of the covariance matrix constructed as described in \S\protect\ref{sec:ObservedCovariance}. The shaded regions enclose the inner 68.26\%, 95.44\%, and 99.74\% (1, 2, and 3-``$\sigma$'' respectively) of the distribution of models sampled from the converged \protect\MCMC\ chains, while the yellow line indicates the median of those models.}
 \label{fig:PosteriorPredictiveCheck}
\end{figure*}

The left-hand panel of Figure~\ref{fig:PosteriorPredictiveCheck} shows the cumulative probability of the test statistic, $\mathcal{T}$, for the model (yellow line) and indicates the test statistic of the data by the green arrow, showing that $\mathcal{T}^\prime$ lies in the tail of the distribution of model test statistics ($\hat{p}_{\rm B} = 0.045
$). The model is therefore only a very marginally good description of the data (as measured by this particular test statistic). The right-hand panel of Figure~\ref{fig:PosteriorPredictiveCheck} shows the observed stellar mass function (blue points) along with the distribution of model mass functions drawn from the \PPD. Specifically, we sample 10,000 points from the converged \MCMC\ chains (excluding outlier chains) and evaluate the model mass function for each point. We then find the median of the distribution of model mass functions in each mass bin. This is shown by the yellow line. Additionally, shaded regions enclose the inner 68.26\%, 95.44\%, and 99.74\% (1, 2, and 3-``$\sigma$'' respectively) of the distribution of model mass functions. The model clearly closely follows the data, but noticeably lies systematically below the data at the low-mass end (e.g. roughly 85\% of models like below the data points). The knee and high-mass end of the mass function are well-matched---the ``bump'' in the model mass function just below $10^{12}M_\odot$ arises from the small number of model merger trees used to construct the mass function in this regime.

\subsection{Effects of Uncertainties}

\begin{figure}
 \begin{tabular}{c}
  \includegraphics[width=85mm,trim=0mm 0mm 0mm 0mm,clip]{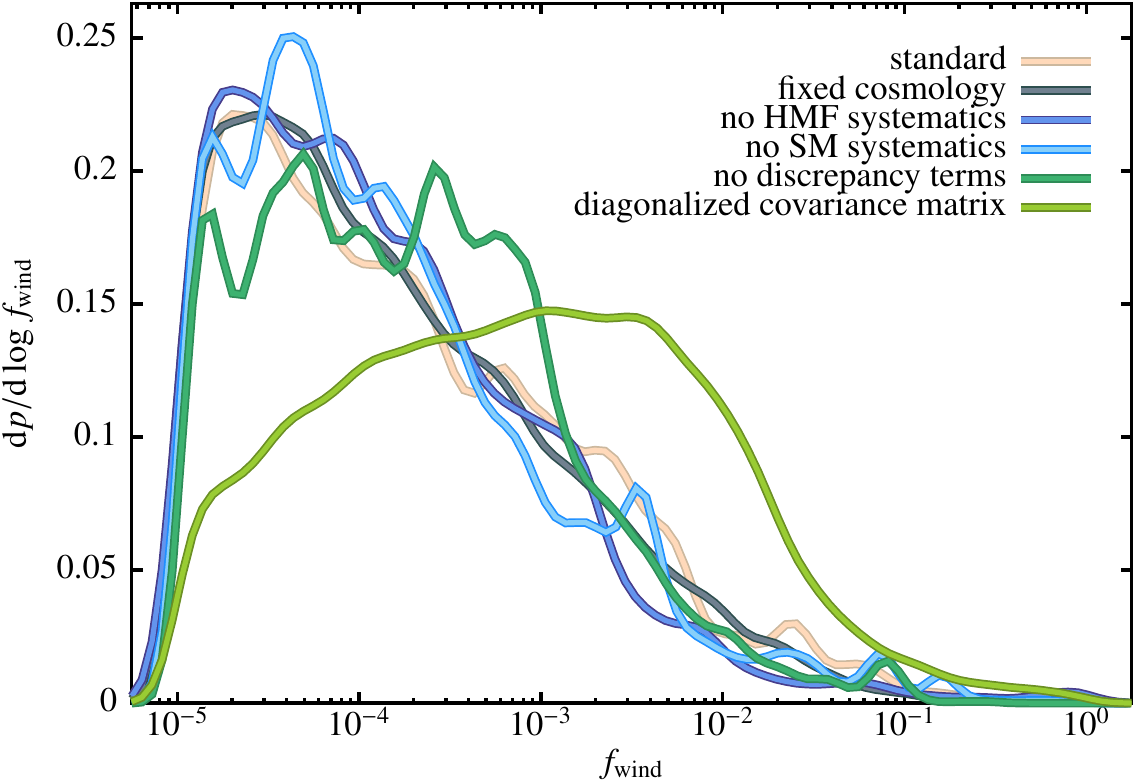} \\
  \includegraphics[width=85mm,trim=0mm 0mm 0mm 0mm,clip]{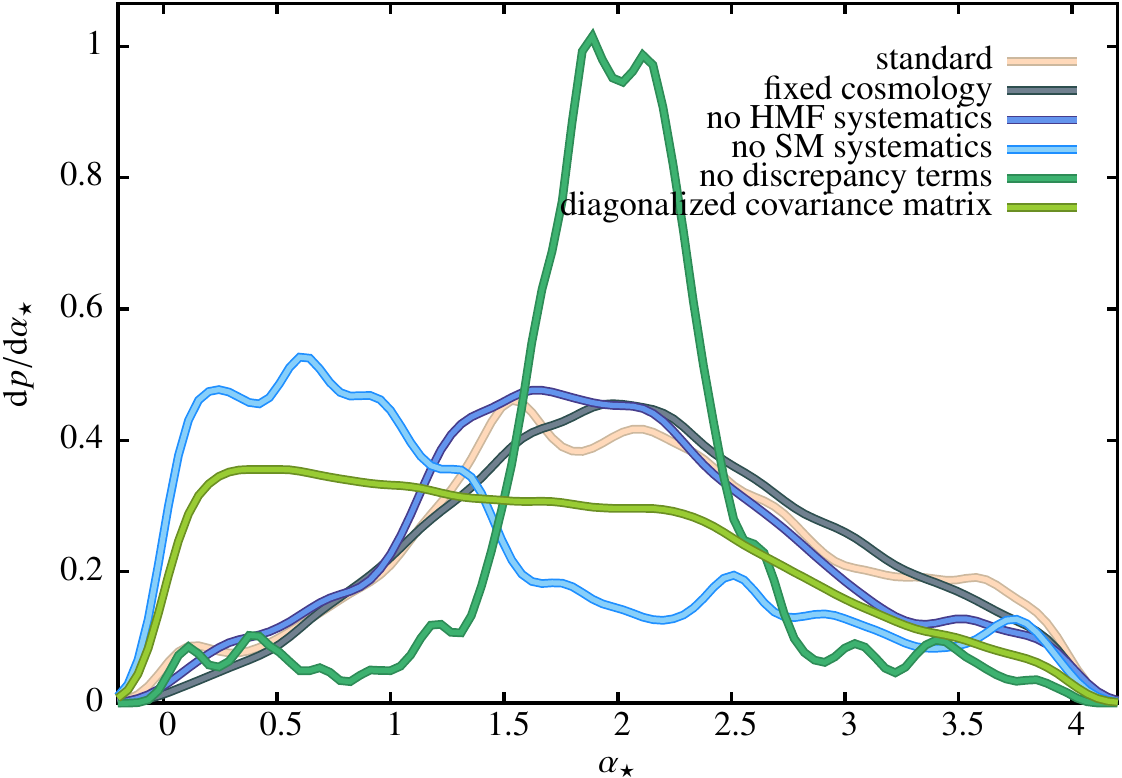} \\
  \includegraphics[width=85mm,trim=0mm 0mm 0mm 0mm,clip]{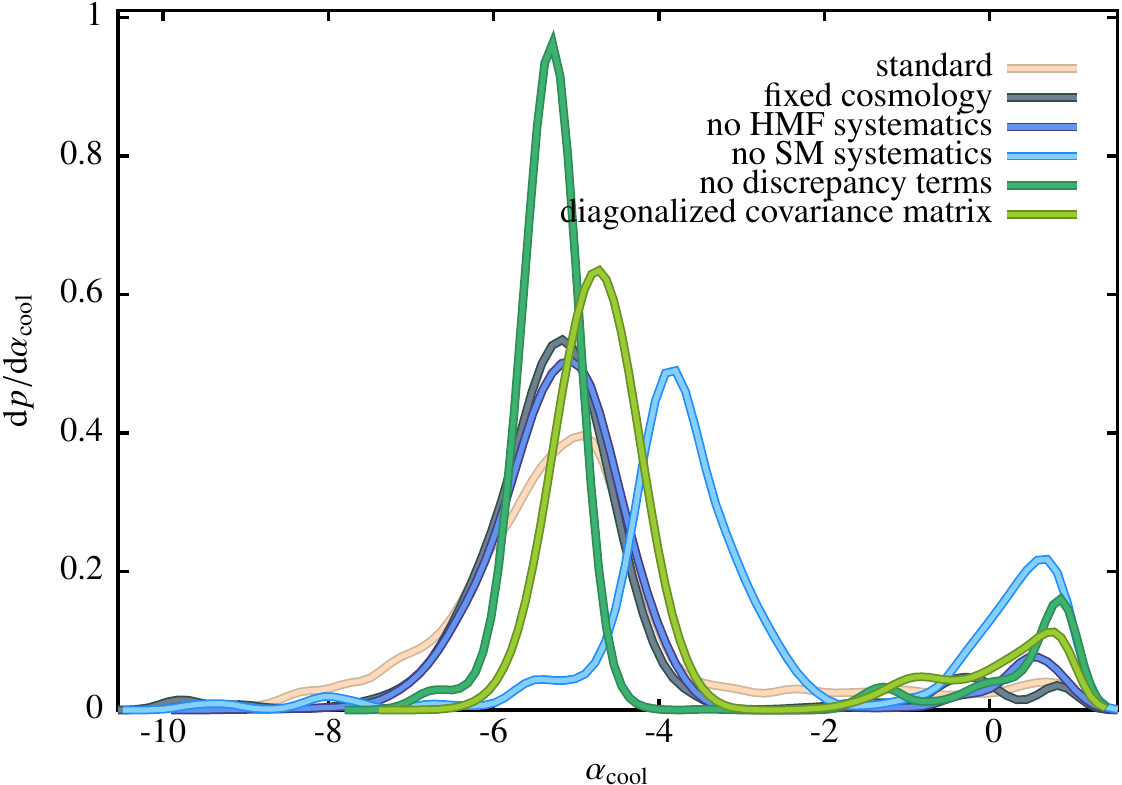}
 \end{tabular}
 \caption{Marginalized \protect\PPDs\ on two parameters (\emph{left panel:} $f_{\rm wind}$; \emph{right panel:} $\alpha_\star$) from our \protect\MCMC\ chains. Line colors correspond to different modeling assumptions. From top to bottom: ``standard'' is our full calculation; ``fixed cosmology'' fixes cosmological parameters to the maximum likelihood values of their priors (see \S\protect\ref{sec:noCosmologyResults}); ``no HMF systematics'' fixes halo mass function systematics parameters to the maximum likelihood values of their priors (see \S\protect\ref{sec:noHMFSystematicsResults}); ``no SM systematics'' fixes stellar mass systematics parameters to the maximum likelihood values of their priors (see \S\protect\ref{sec:noSMSytematicsResults}); ``no discrepancy terms'' removes all model discrepancy terms (see \S\protect\ref{sec:ModelDiscrepancyResults}); finally, ``diagonalized covariance matrix'' diagonalizes the covariance matrix before computing model likelihoods (see \S\protect\ref{sec:DiagonalizedCovariance})}
 \label{fig:PosteriorComparison}
\end{figure}

To assess the importance of the various uncertainties that we account for in our analysis we perform additional \MCMC\ simulations in which we successively remove each source of uncertainty. These simulations were performed in the same manner as described in \S\ref{sec:MCMC}, except that 144 parallel chains per simulation were used and the chains were initialized from the posterior of the previous simulation\footnote{Since in each successive simulation we are removing nuisance parameters this means that the chains still begin in an overdispersed state as required by the Gelman-Rubin convergence criterion.}. Throughout this discussion we will refer to the marginalized \PPDs\ shown in Fig.~\ref{fig:PosteriorComparison} which illustrate the range of responses of \PPDs\ to these uncertainties.

\subsubsection{Cosmological Parameters}\label{sec:noCosmologyResults}

We first perform an \MCMC\ simulation in which the cosmological parameters $C_{1\ldots6}$ are fixed to zero, such that all cosmological parameters are fixed at their maximum likelihood values from WMAP-9. We find that the \PPDs\ of model parameters are not significantly affected by fixing cosmological parameters to the maximum likelihood values\footnote{This does not imply that this would be the case for other constraints of course. In particular, it is likely that certain parameters (e.g. $\sigma_8$) have a much stronger effect at higher redshifts.}, as can be seen in all three panels of Fig.~\ref{fig:PosteriorComparison}. Consequently, the model remains a viable description of the data even when cosmological parameters are held fixed.

\subsubsection{Halo Mass Function Systematics}\label{sec:noHMFSystematicsResults}

Keeping the cosmological parameters fixed, we next fix the halo mass function systematic parameters, $H_{1\ldots2}$ to zero and repeat our \MCMC\ analysis. The effect of these nuisance parameters on the \PPDs\ is also quite small. There is a small change in the \PPD\ of the $f_{\rm wind}$ parameter (see top panel of Fig.~\ref{fig:PosteriorComparison}) in which the probability is supressed at high values of $f_{\rm wind}$, and a similar effect on the \PPD\ of $\alpha_\star$ (middle panel of Fig.~\ref{fig:PosteriorComparison}). In general, we find that inclusion of halo mass function systematic uncertainties lead to small increases in the widths of \PPDs\, but no significant bias.

\subsubsection{Stellar Mass Systematics}\label{sec:noSMSytematicsResults}

We next remove stellar mass systematics by fixing $\mu$ and $\kappa$ at the maximum likelihood values of their priors\footnote{Note that this means $\mu=-0.0125$ and $\kappa=0.165$ since we retain the discrepancy term from \S\protect\ref{sec:VariableMassResolution}.} and repeating our \MCMC\ analysis. We find that the inclusion of these systematics can dramatically alter the \PPD\ for some parameters, while leaving others unaffacted. For example, in Fig.~\ref{fig:PosteriorComparison} it is clear that the \PPD\ of $f_{\rm wind}$ is unaffected by stellar mass systematics, while the \PPD\ of $\alpha_\star$ is strongly affected, with the peak of the \PPD\ shifting from $2$ to $0.5$. We note that, in our full analysis, the peaks of the \PPDs\ for thse parameters are located at $\mu\approx0.22$ and $\kappa\approx 0.035$ respectively. These are both far from the mean of their priors, so it is not surprising to find a large shift in the \PPDs\ of other parameters once $\mu$ and $\kappa$ are no longer allowed to vary freely.

\subsubsection{Model Discrepancies}\label{sec:ModelDiscrepancyResults}

Next, we set all model discrepancy terms to zero (such that the stellar mass systematic parameters are now fixed at $\mu=0$, $\kappa=0$) and repeat our \MCMC\ analysis. We find that the peaks of the \PPDs\ now return close to their locations for our full analysis. However, the \PPDs\ of several model parameters are now \emph{much} narrower than in our full analysis. This indicates the cumulative effect of removing all sources of model and data systematic uncertainties. For example, in the middle panel of Fig.~\ref{fig:PosteriorComparison} we see that the width of the \PPD\ for $\alpha_\star$ is reduced by a factor of approximately 2 compared to the full analysis. This highlights the importance of the inclusion of nuisance parameters to model systematic uncertainties in deriving robust constraints on model parameters (and, therefore, for being able to make robust predictions).

\begin{figure}
 \includegraphics[width=85mm,trim=0mm 0mm 0mm 2.5mm,clip]{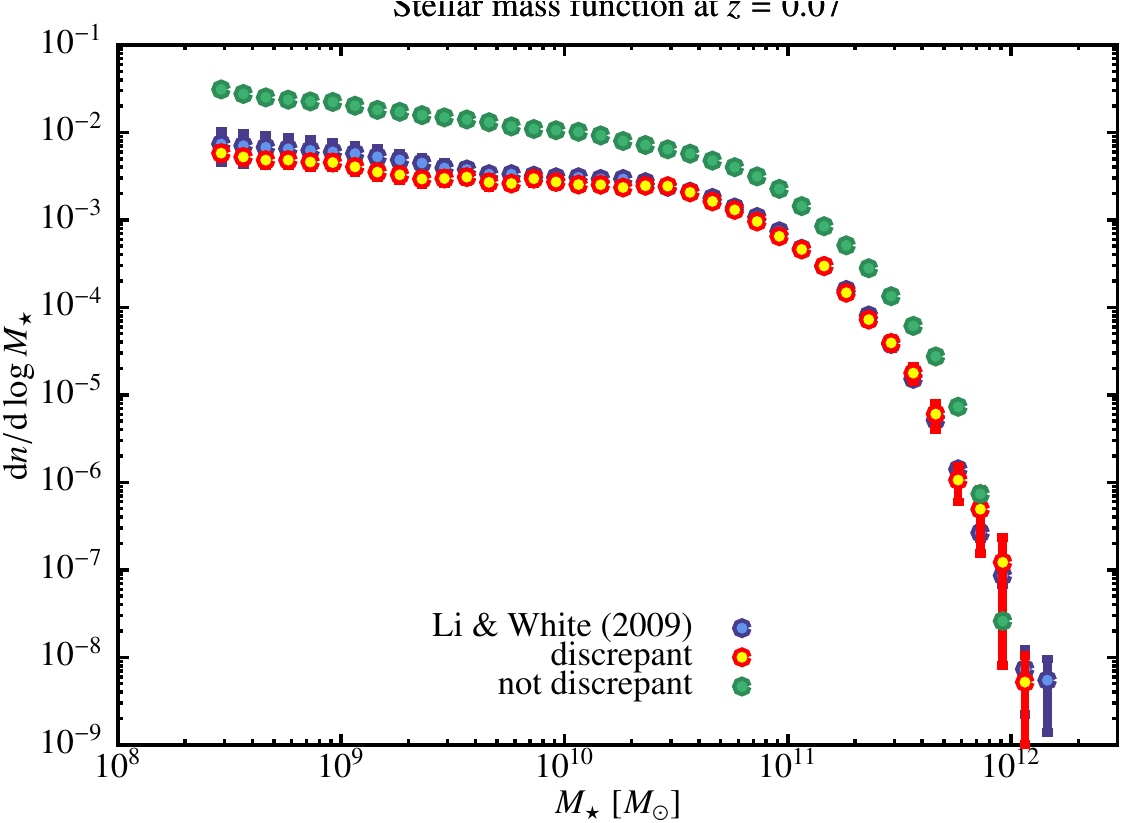}
 \caption{The maximum likelihood model evaluated including model discrepancy terms (yellow points), together with the same model with discrepancies removed (green points).}
 \label{fig:ModelDiscrepancyCompare}
\end{figure}

It is also instructive to check whether our parameterizations of the model discrepancy terms remain valid for the maximum likelihood model. This is not guaranteed to be the case---the model discrepancy terms were computed using our \emph{a priori} model which turns out to be significantly different from the final maximum likelihood model. Therefore, in Figure~\ref{fig:ModelDiscrepancyCompare} we plot the stellar mass function data (blue points), our maximum likelihood model computed exactly as in our \MCMC\ analysis and with model discrepancy correction terms applied (yellow points). Additionally, we re-run the maximum likelihood model, but now remove all model discrepancies (i.e. we use distributions of virial orbits and merging times, and use a fixed merger tree resolution---note that we do not remove the discrepancy relating to the use of \cite{parkinson_generating_2008} merger trees rather than N-body-derived merger trees, but this causes no offset in the model mass function), and do not include model discrepancy corrections when evaluating the mass function. The result is shown by the green points, which are clearly significantly offset from the model including discrepancies. This clearly shows that our parameterizations of model discrepancies---calibrated using our \emph{a priori} model---do not well describe the model discrepancies in our maximum likelihood model. 

In future work, we can use the maximum likelihood model from this work to assess model discrepancies for further \MCMC\ analysis. This should result in more accurate parameterization of model discrepancies. However, the fundamental problem illustrated by Figure~\ref{fig:ModelDiscrepancyCompare} is not resolved by this fact---most likely we will be evaluating model discrepancies for some other observable for which the present maximum likelihood model may not be the best description.

This suggests, for future studies, an iterative procedure. That is, model discrepancies are determined using the current maximum likelihood model. These are then used in a quick search of parameter space to find a new maximum likelihood model. If this is sufficiently different from the previous maximum likelihood model (where ``sufficiently different'' here means that the model discrepancy terms differ substantially), then it is used to evaluate new model discrepancies. This process is repeated until convergence on the model discrepancies is reached.

\subsubsection{Diagonalized Covariance}\label{sec:DiagonalizedCovariance}

Finally, we explore the importance of off-diagonal elements of the covariance matrices by simply setting all off-diagonal elements to zero and repeating our \MCMC\ analysis\footnote{A similar study of the importance of correlated errors was carried out by \protect\cite{lu_bayesian_2011}, although using a phenomenological model of the covariance matrix.}. This throws away any information about correlations between bins in the stellar mass function (both observed and modeled). As we have shown in Fig.~\ref{fig:CorrelationMatrixSDSS}, these correlations are significant, and so we may expect diagonalizing the covariance matrix to significantly affect our results. This is in fact the case, as can be seen in all three panels of Fig.~\ref{fig:PosteriorComparison}. Comparing the ``no discrepancy terms'' and ``diagonalized covariance matrix'' terms we see that the posterior is substantially shifted in all three cases when correlations between bins in the stellar mass function are ignored. In particular, for the parameter $\alpha_{\rm cool}$ (bottom panel of Fig.~\ref{fig:PosteriorComparison}) the shift is, approximating the \PPD\ as Gaussian, slightly over $1\sigma$.

\subsection{Predictions}

Using the \PPD\ of our model, we can explore predictions for other observables. Given the phenomenological nature of the model described in \S\ref{sec:Model} we in fact should consider these to be ``extrapolations'' rather than ``predictions''. True ``predictions'' will only arise from a model with a more physical basis.

When comparing the \PPC\ of the model posterior with these datasets we must be careful to account for the errors in the observed data. This is not a problem when carrying out \protect\PPC\ on the dataset used to constrain the model because the model posterior implicitly accounts for the observational errors on that dataset. When comparing to other datasets though it is possible that those datasets may have much larger errors. Computing the test statistic at only the measured values of the data would then likely lead to an extreme value for the test statistic, suggesting inconsistency with the model. 

To avoid this problem, we perturb each model realization by a random realization of the errors in the observed data, using the covariance matrix (assumed to be diagonal) of the data to perturb around the true values. The distribution of the test statistic, and the associated $p$-value are then computed in the usual way.

\begin{figure*}
\vspace{+9mm}
 \begin{tabular}{cc}
    \vspace{-9mm} $p=\protect\input{plots/posteriorPredictiveChecks/hiMassFunctionZ0.00_testStatistic_pValue.txt}$ &  \\ 
 \includegraphics[width=85mm,trim=0mm 0mm 0mm 2.5mm,clip]{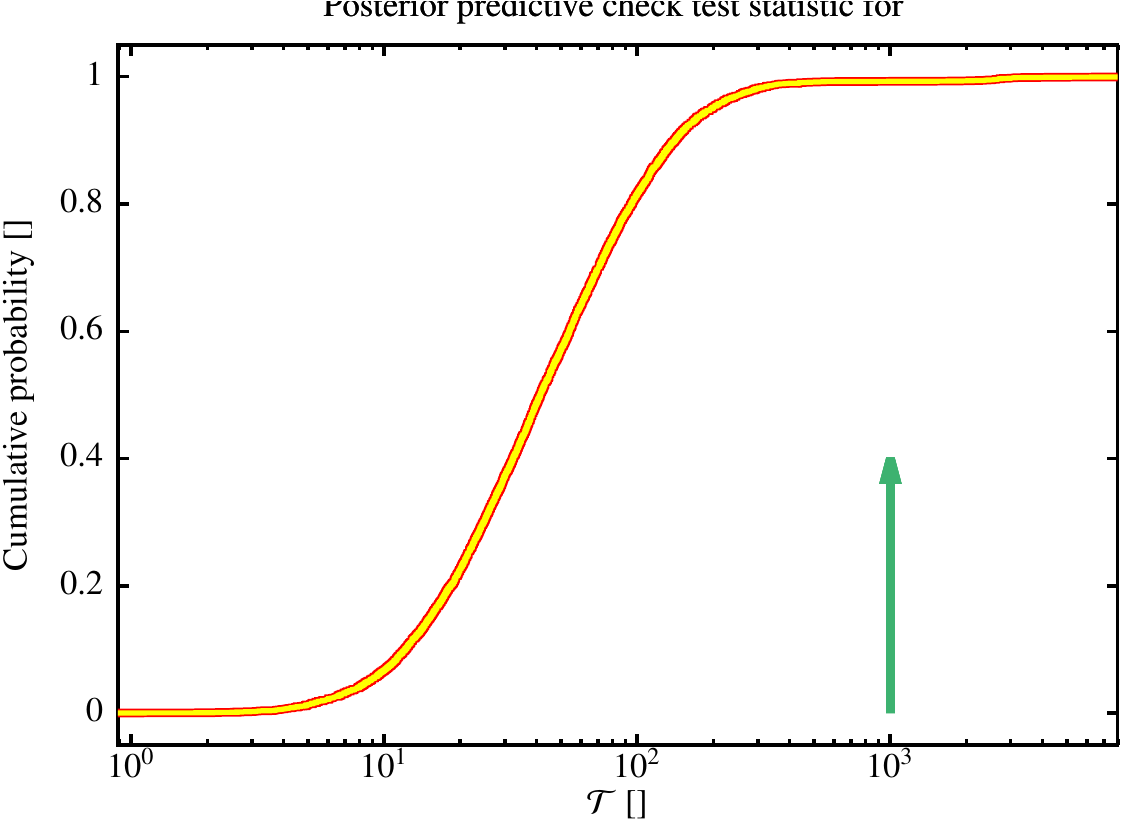} &
  \includegraphics[width=85mm,trim=0mm 0mm 0mm 2.5mm,clip]{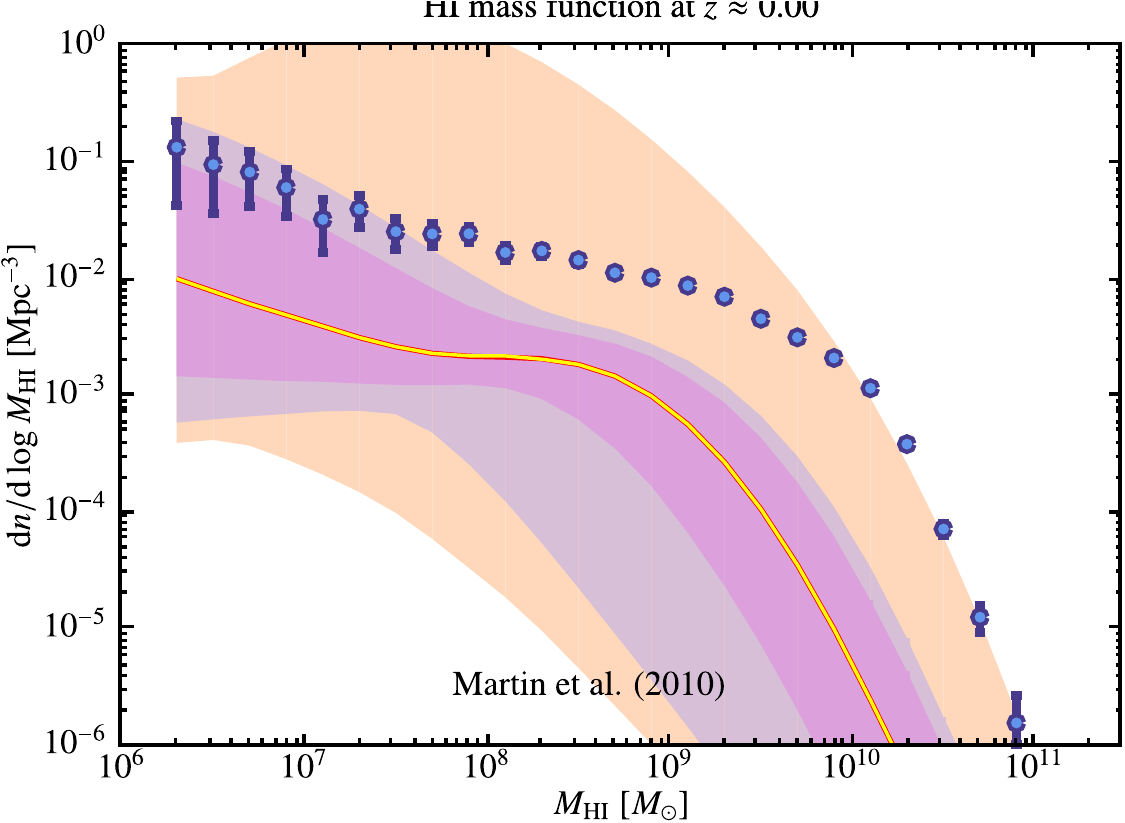}
 \end{tabular}
 \caption{\emph{Left panel:} The cumulative probability distribution of the test statistic used in our \protect\PPC\ analysis for the HI galaxy mass function at $z=0$ using models sampled from the converged \protect\MCMC\ chains, $\mathcal{T}_l$ (yellow line). The green arrow shows the same test statistic computed for the data, $\mathcal{T}^\prime$. The resulting $p$-value (also showin in the panel) is $\protect\input{plots/posteriorPredictiveChecks/hiMassFunctionZ0.00_testStatistic_pValue.txt}$. \emph{Right panel:} The $z=0$ HI galaxy mass function of \protect\cite{martin_arecibo_2010} compared to the extrapolation from our model. Shaded regions enclose the inner 68.26\%, 95.44\%, and 99.74\% (1, 2, and 3-``$\sigma$'' respectively) of the distribution of models sampled from the converged \protect\MCMC\ chains, while the yellow line indicates the median of those models.}
 \label{fig:PosteriorPredictiveCheckHI}
\end{figure*}

Figure~\ref{fig:PosteriorPredictiveCheckHI} compares our model to the galaxy HI mass function measured from the ALFALFA survey \citep{martin_arecibo_2010}, while Figure~\ref{fig:PosteriorPredictiveCheckPRIMUS0100} compares our model to stellar mass functions out to $z\approx 1$ from the PRIMUS survey \citep{moustakas_primus:_2013}.

\begin{figure*}
 \begin{tabular}{cc}
  \vspace{-6mm}\hspace{-16mm} $p=\protect\input{plots/posteriorPredictiveChecks/primusStellarMassFunctionZ0.100_testStatistic_pValue.txt}$ & $z=0.100$ \\ 
  \includegraphics[width=80mm,trim=0mm 0mm 0mm 2.5mm,clip]{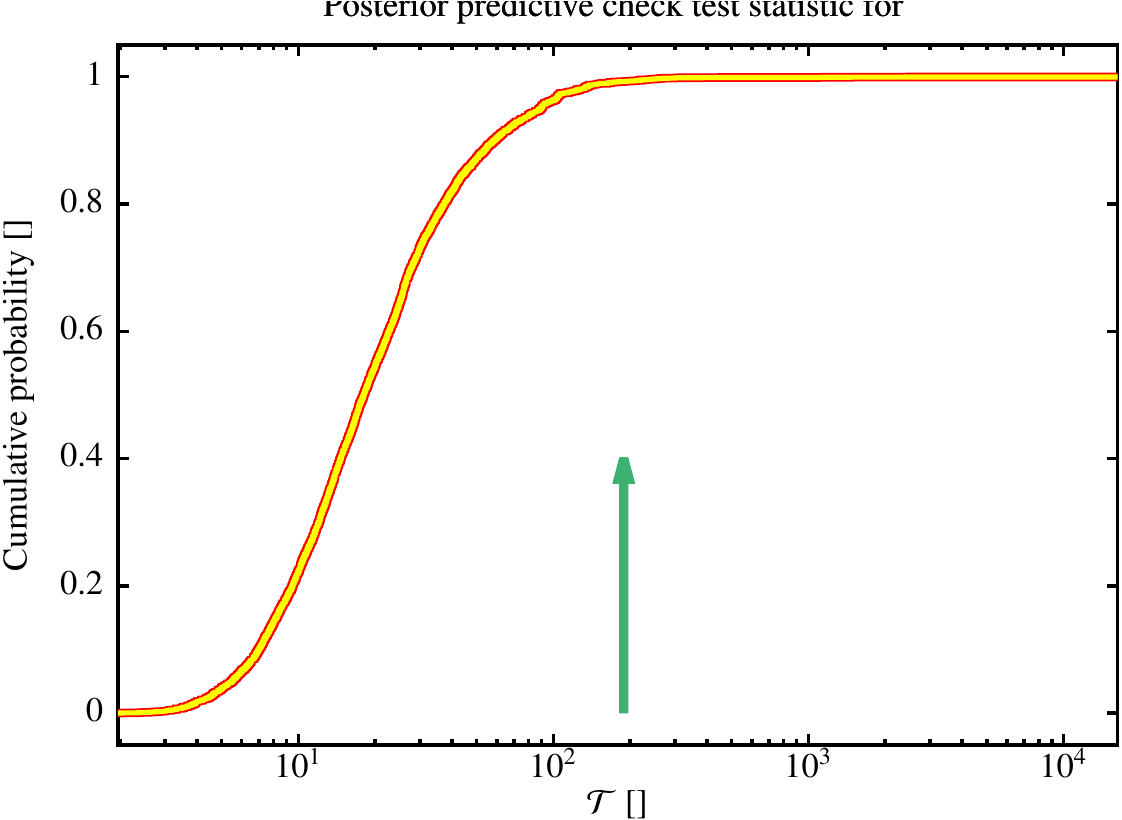} &
  \includegraphics[width=80mm,trim=0mm 0mm 0mm 2.5mm,clip]{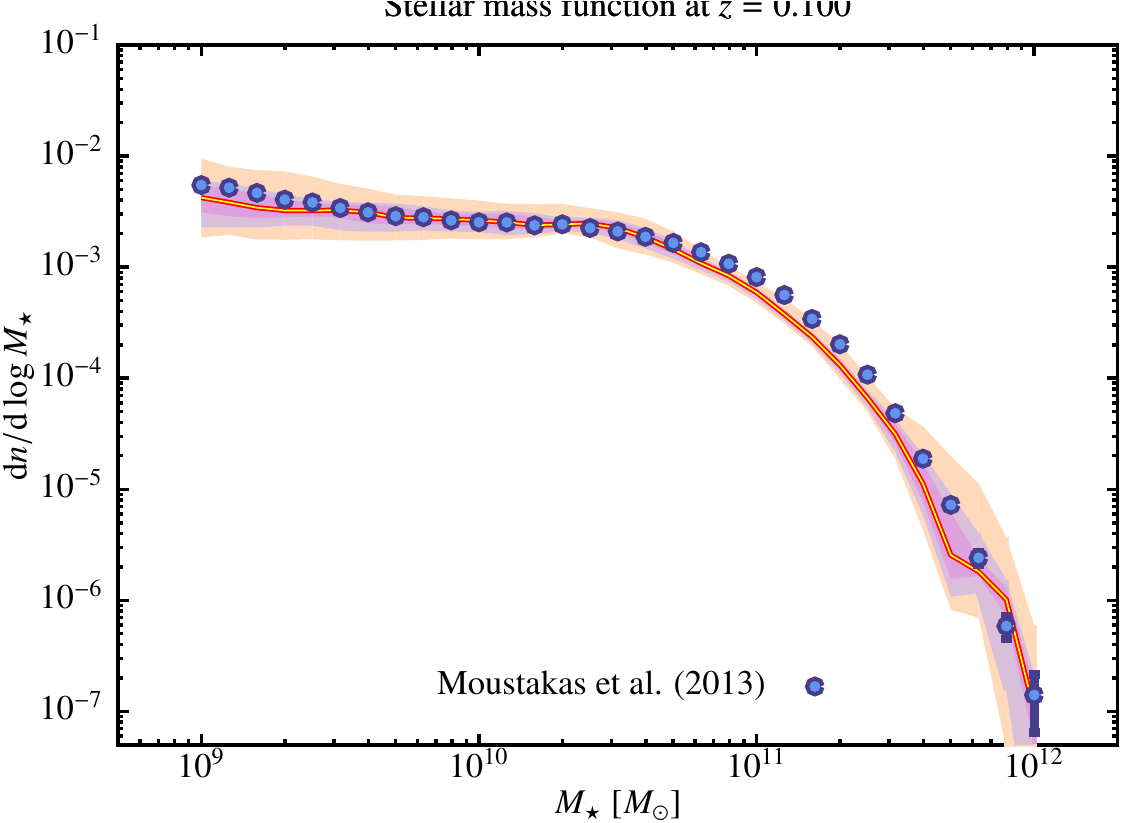} \\
   \vspace{-6mm} $p=\protect\input{plots/posteriorPredictiveChecks/primusStellarMassFunctionZ0.250_testStatistic_pValue.txt}$ & $z=0.250$ \\ 
\includegraphics[width=80mm,trim=0mm 0mm 0mm 2.5mm,clip]{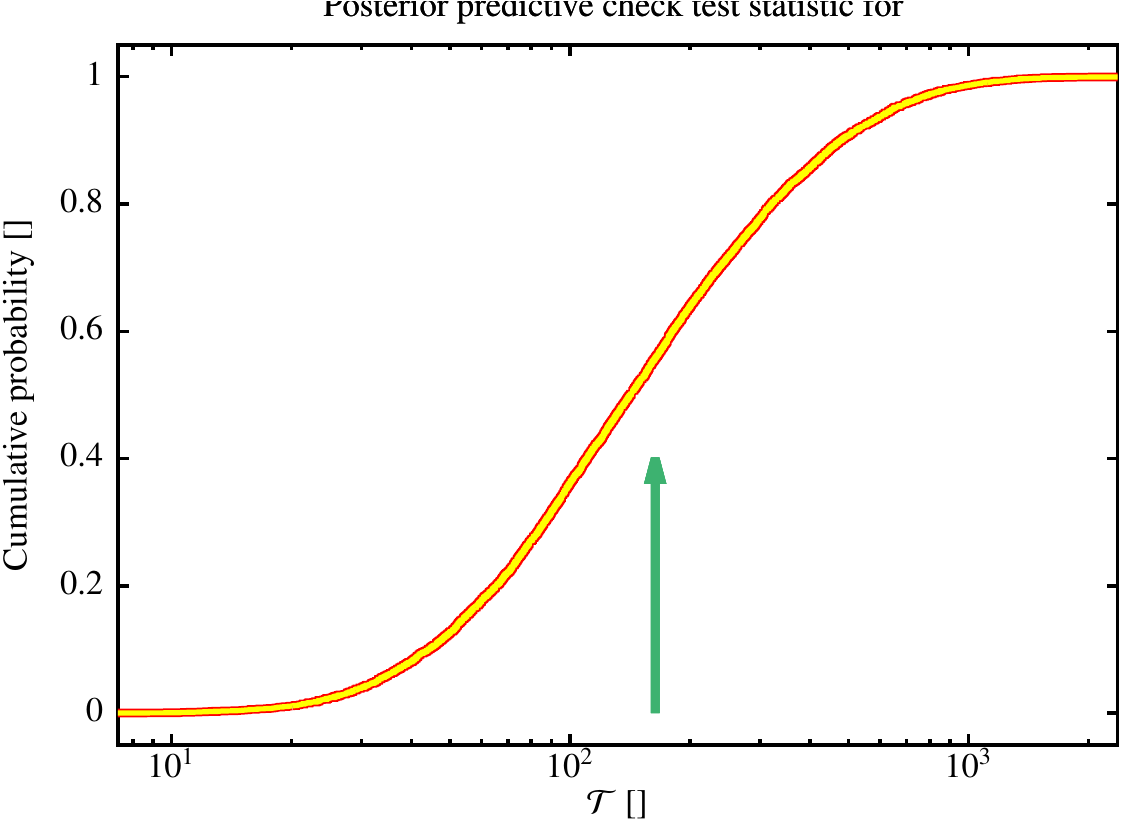} &
  \includegraphics[width=80mm,trim=0mm 0mm 0mm 2.5mm,clip]{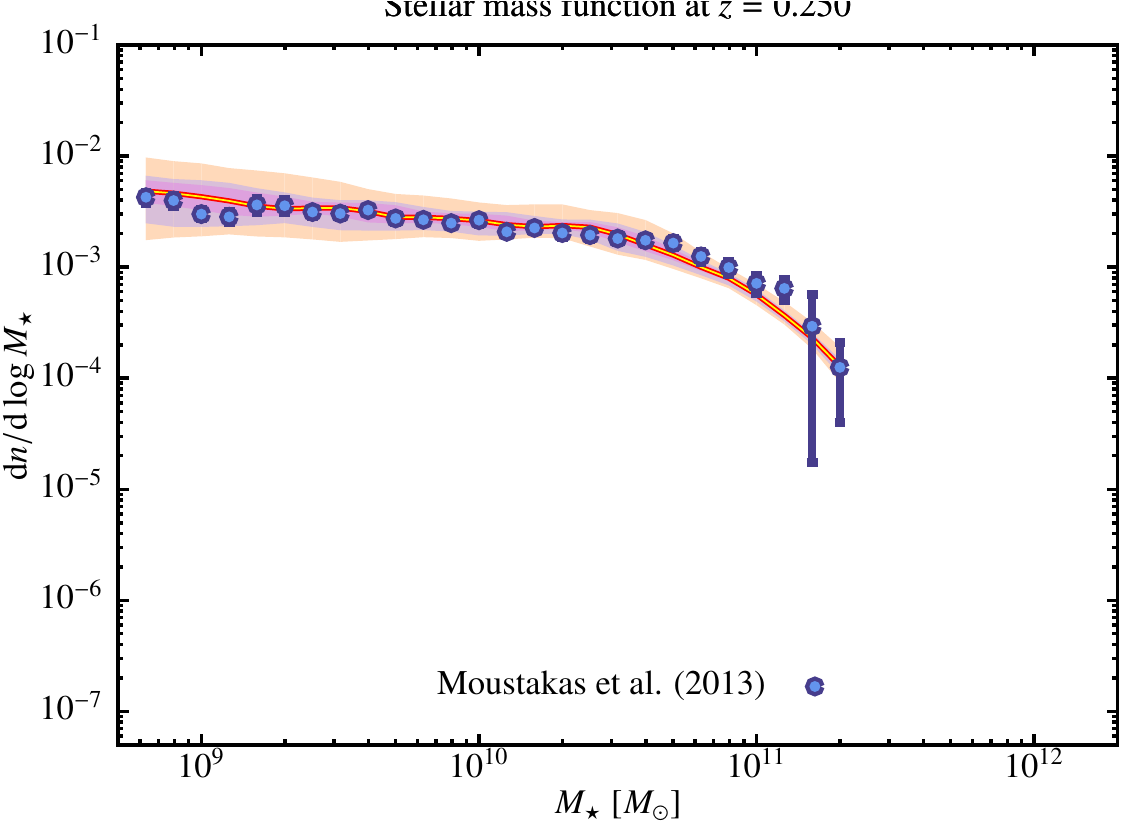} \\
  \vspace{-6mm} $p=\protect\input{plots/posteriorPredictiveChecks/primusStellarMassFunctionZ0.350_testStatistic_pValue.txt}$ & $z=0.350$ \\ 
 \includegraphics[width=80mm,trim=0mm 0mm 0mm 2.5mm,clip]{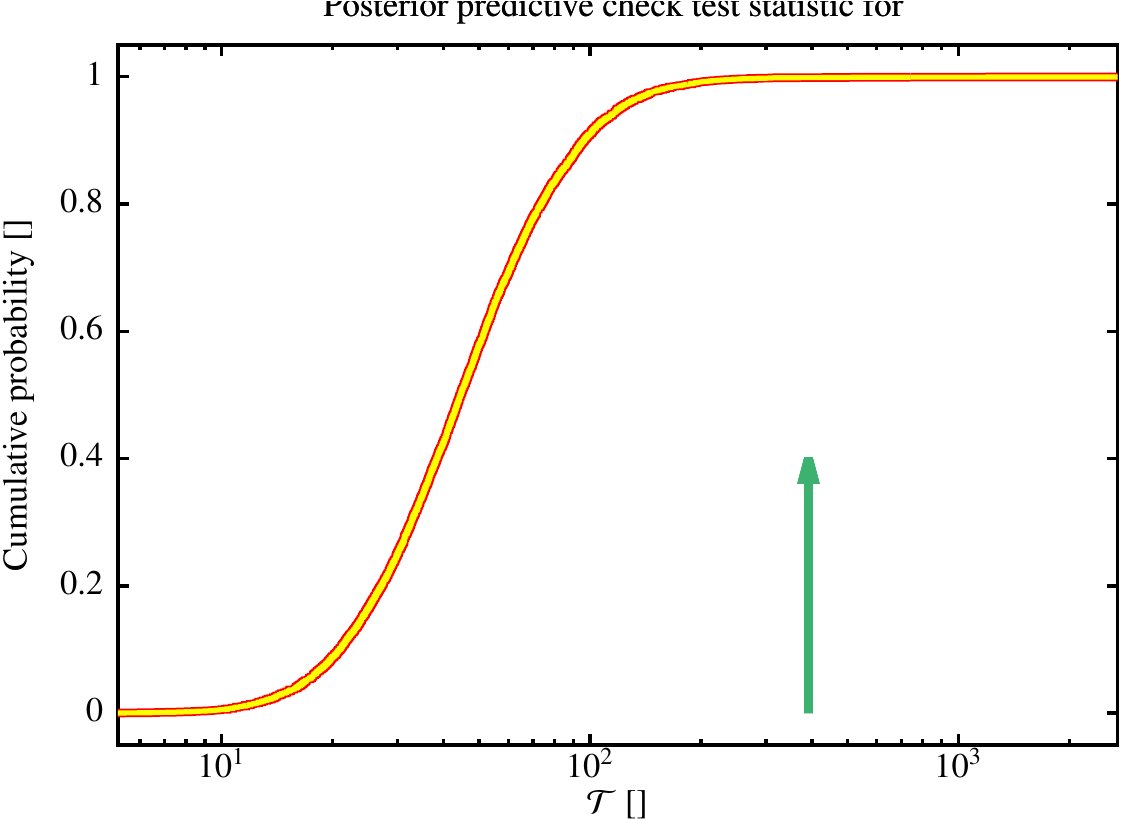} &
  \includegraphics[width=80mm,trim=0mm 0mm 0mm 2.5mm,clip]{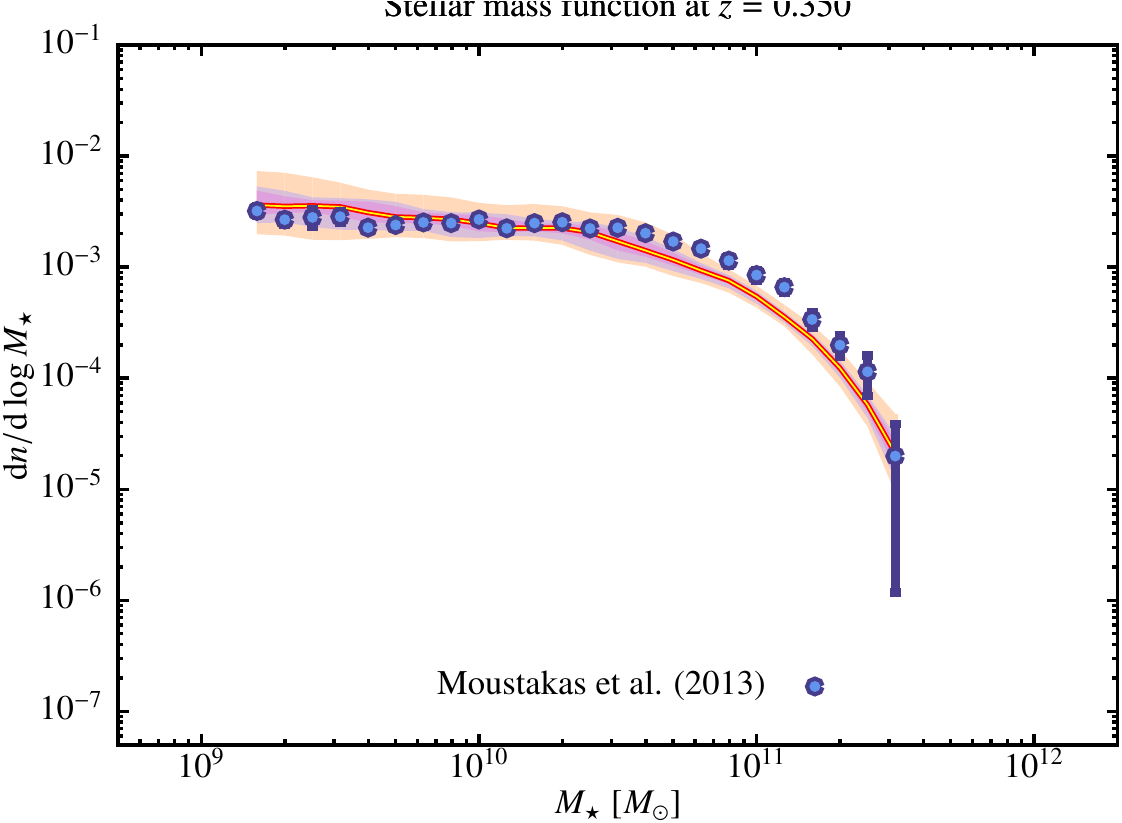} \\
  \vspace{-6mm} $p=\protect\input{plots/posteriorPredictiveChecks/primusStellarMassFunctionZ0.450_testStatistic_pValue.txt}$ & $z=0.450$ \\ 
  \includegraphics[width=80mm,trim=0mm 0mm 0mm 2.5mm,clip]{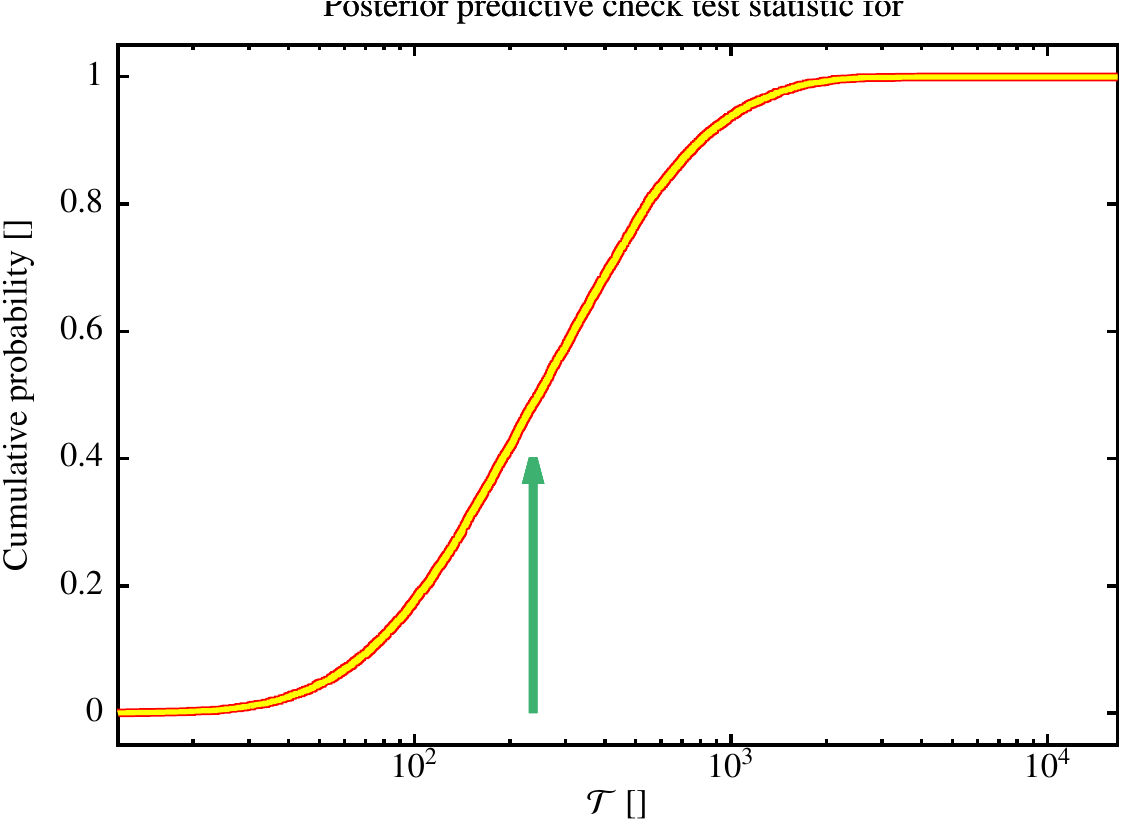} &
  \includegraphics[width=80mm,trim=0mm 0mm 0mm 2.5mm,clip]{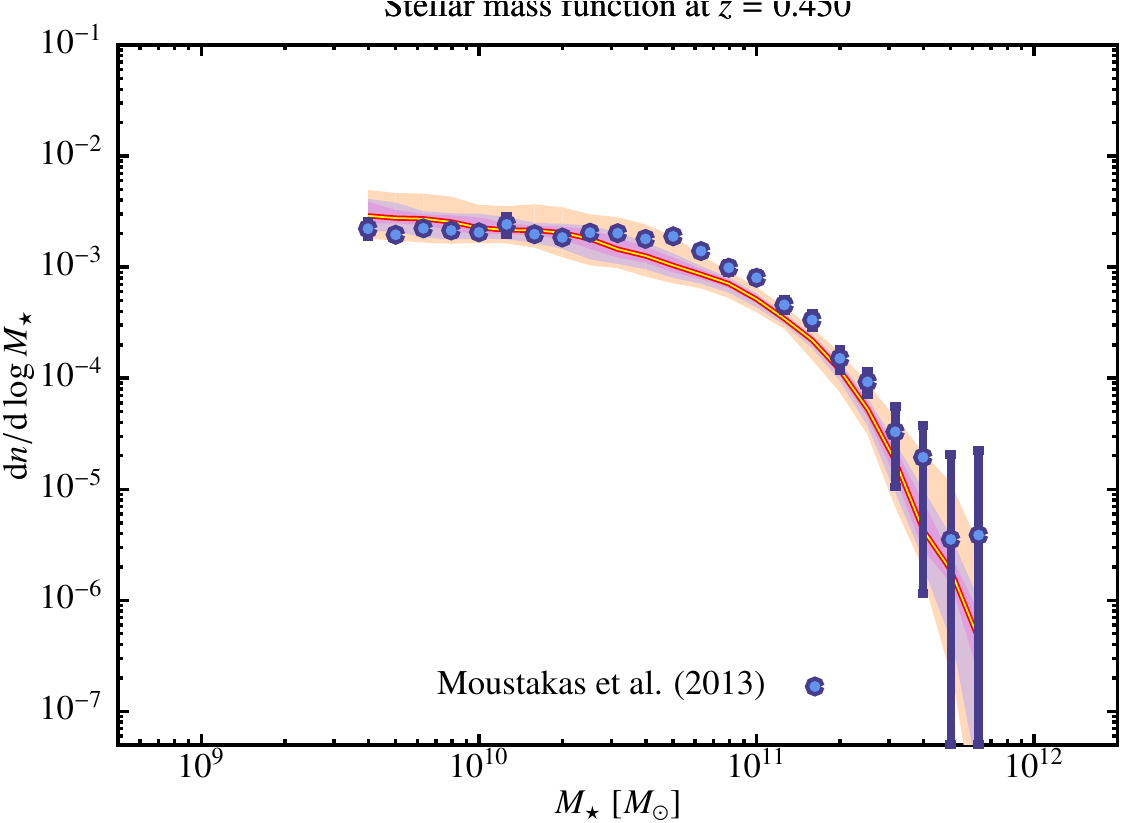} \\
 \end{tabular}
 \caption{\emph{Left panel:} The cumulative probability distribution of the test statistic used in our \protect\PPC\ analysis for models sampled from the converged \protect\MCMC\ chains, $\mathcal{T}_l$ (yellow line), for stellar mass functions measured from the PRIMUS survey \protect\citep{moustakas_primus:_2013} at the redshifts indicated in each panel. Where visible, the green arrow shows the same test statistic computed for the data, $\mathcal{T}^\prime$. The resulting $p$-value is $\protect\input{plots/posteriorPredictiveChecks/primusStellarMassFunctionZ0.100_testStatistic_pValue.txt}$, $\protect\input{plots/posteriorPredictiveChecks/primusStellarMassFunctionZ0.250_testStatistic_pValue.txt}$, $\protect\input{plots/posteriorPredictiveChecks/primusStellarMassFunctionZ0.350_testStatistic_pValue.txt}$, $\protect\input{plots/posteriorPredictiveChecks/primusStellarMassFunctionZ0.450_testStatistic_pValue.txt}$. \emph{Right panel:} Stellar mass functions from the PRIMUS survey at redshifts as indicated in the panels (blue points). The shaded regions enclose the inner 68.26\%, 95.44\%, and 99.74\% (1, 2, and 3-``$\sigma$'' respectively) of the distribution of models sampled from the converged \protect\MCMC\ chains, while the yellow line indicates the median of those models.}
 \label{fig:PosteriorPredictiveCheckPRIMUS0100}
\end{figure*}

\begin{figure*}
 \begin{tabular}{cc}
  \vspace{-7mm}\hspace{-12mm} $p=\protect\input{plots/posteriorPredictiveChecks/primusStellarMassFunctionZ0.575_testStatistic_pValue.txt}$ & $z=0.575$ \\ 
 \includegraphics[width=80mm,trim=0mm 0mm 0mm 2.5mm,clip]{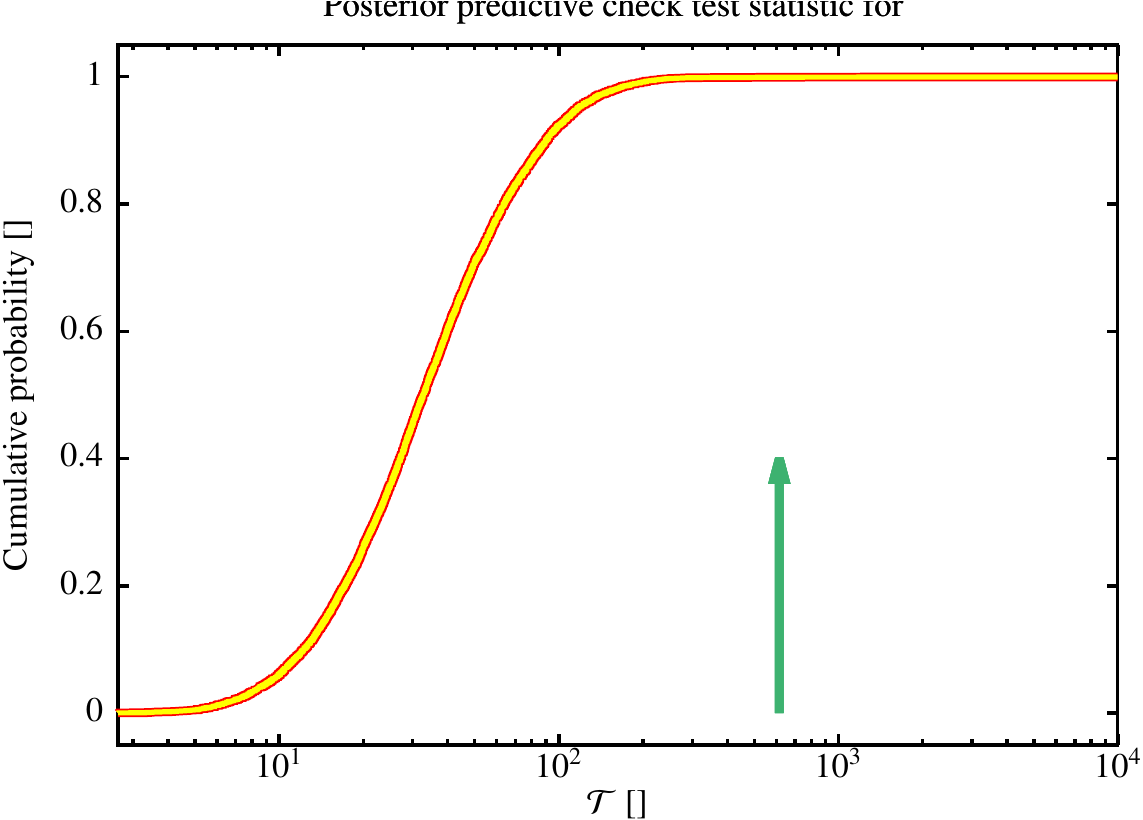} &
  \includegraphics[width=80mm,trim=0mm 0mm 0mm 2.5mm,clip]{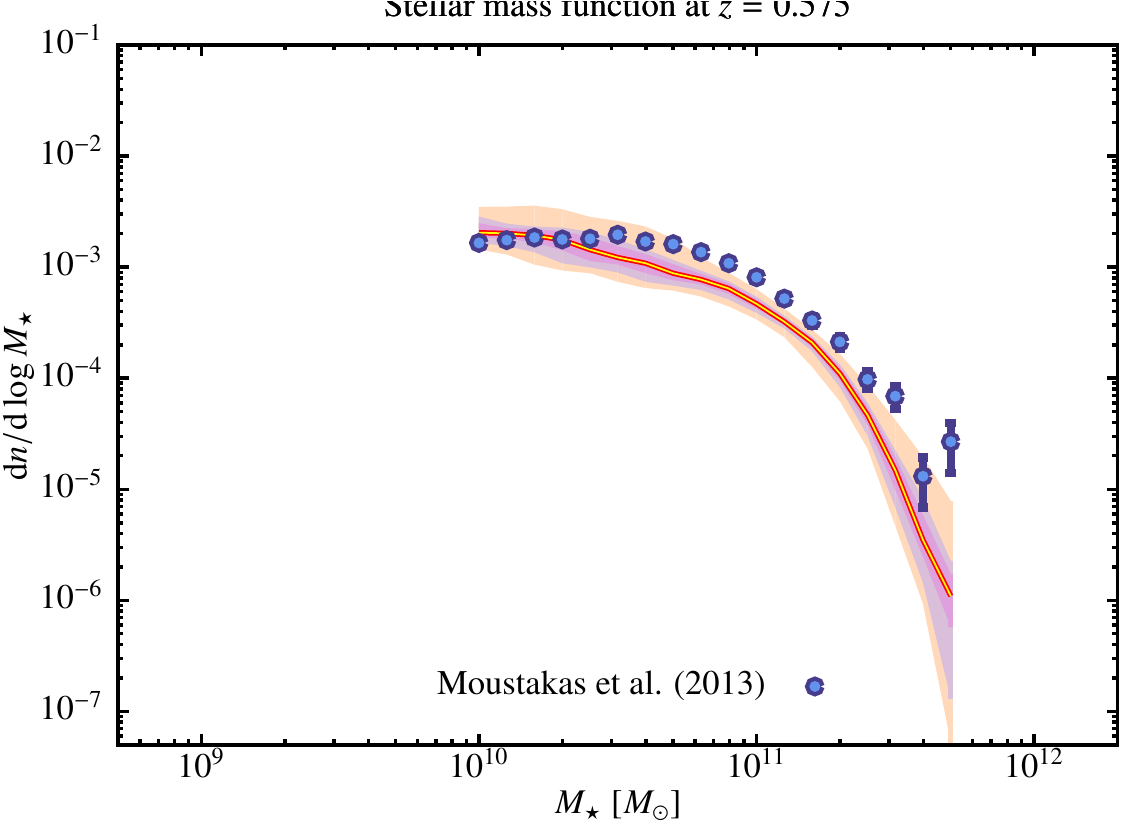} \\
  \vspace{-6mm} $p=\protect\input{plots/posteriorPredictiveChecks/primusStellarMassFunctionZ0.725_testStatistic_pValue.txt}$ & $z=0.725$ \\ 
 \includegraphics[width=80mm,trim=0mm 0mm 0mm 2.5mm,clip]{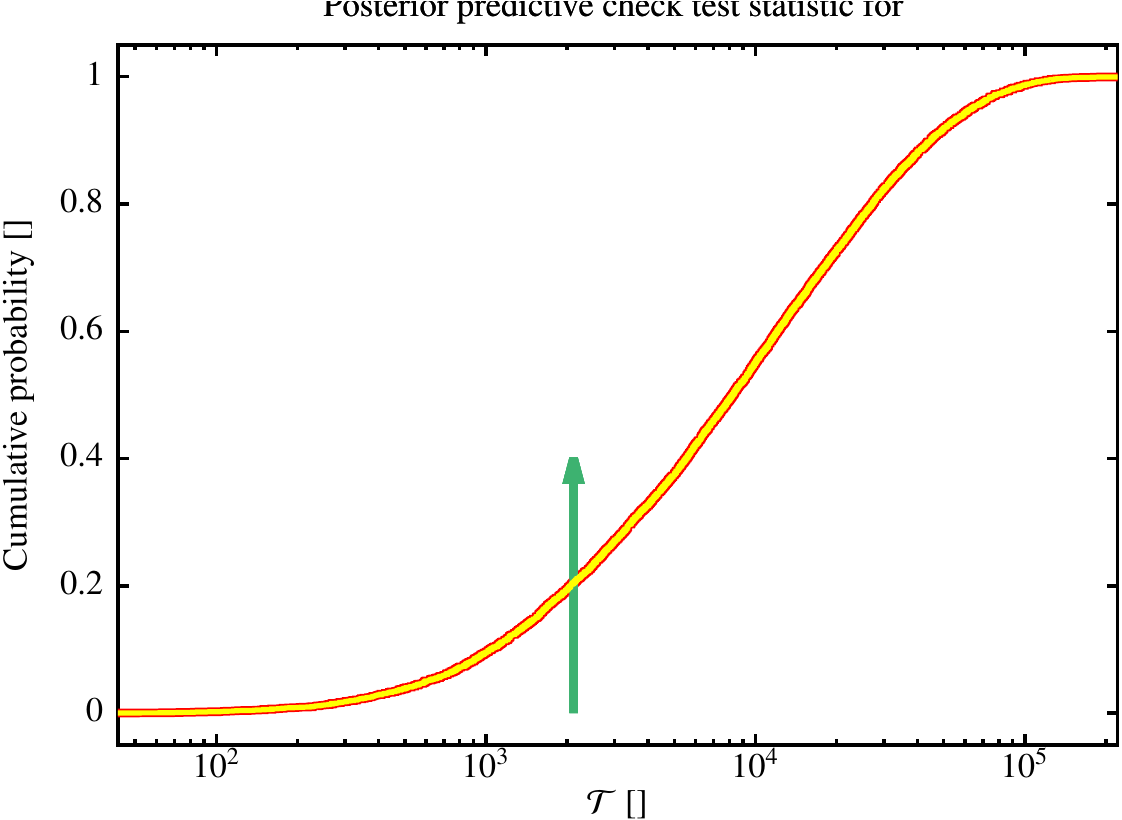} &
  \includegraphics[width=80mm,trim=0mm 0mm 0mm 2.5mm,clip]{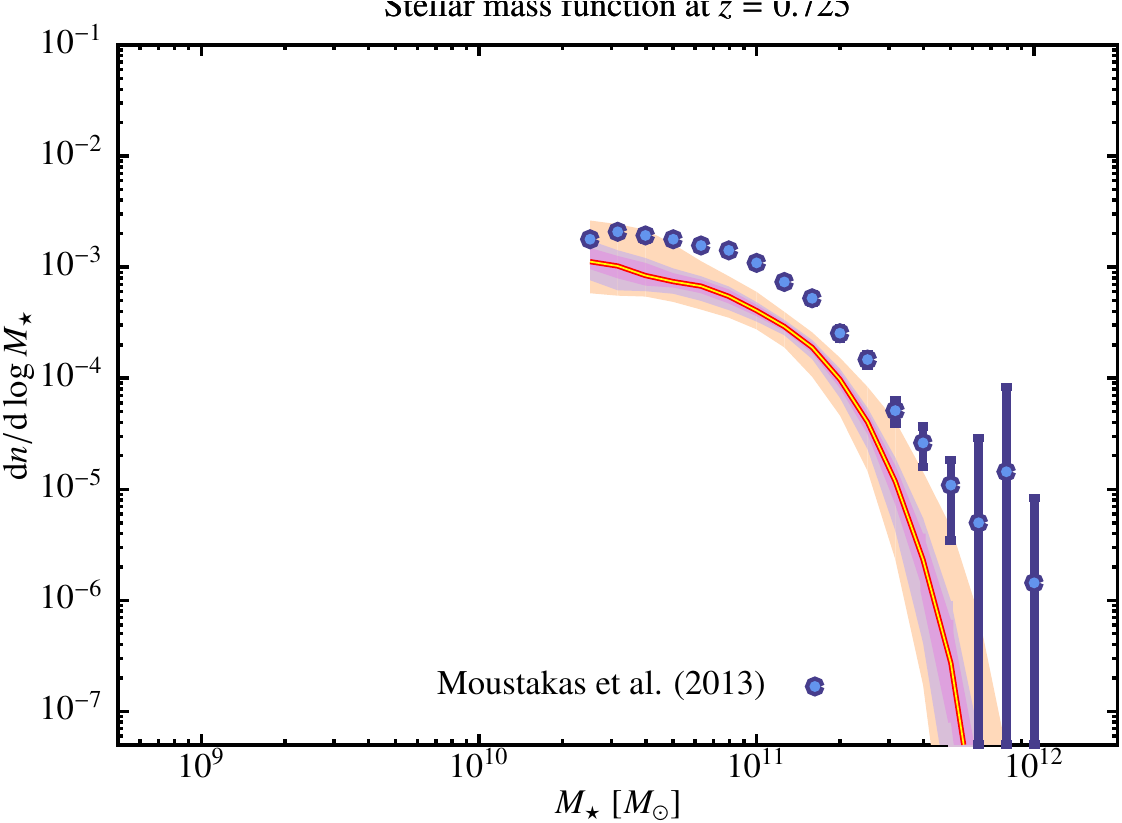} \\
  \vspace{-6mm} $p=\protect\input{plots/posteriorPredictiveChecks/primusStellarMassFunctionZ0.900_testStatistic_pValue.txt}$ & $z=0.900$ \\ 
 \includegraphics[width=80mm,trim=0mm 0mm 0mm 2.5mm,clip]{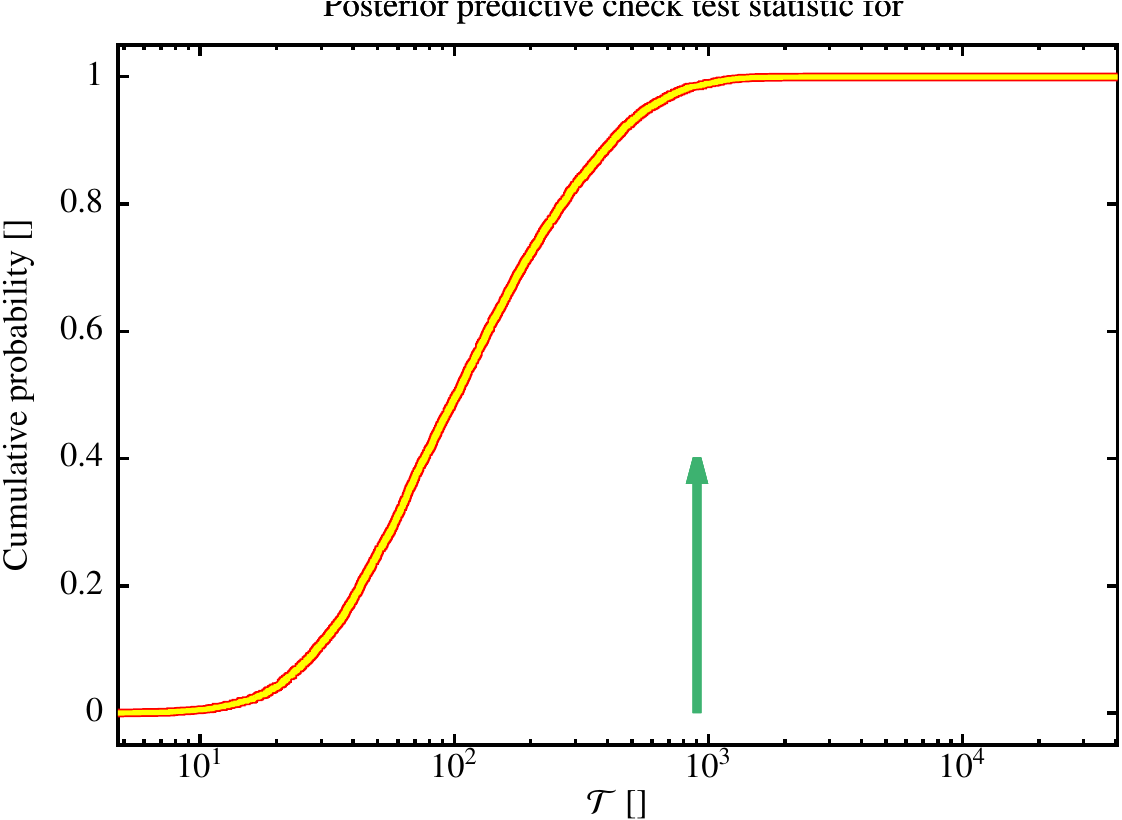} &
  \includegraphics[width=80mm,trim=0mm 0mm 0mm 2.5mm,clip]{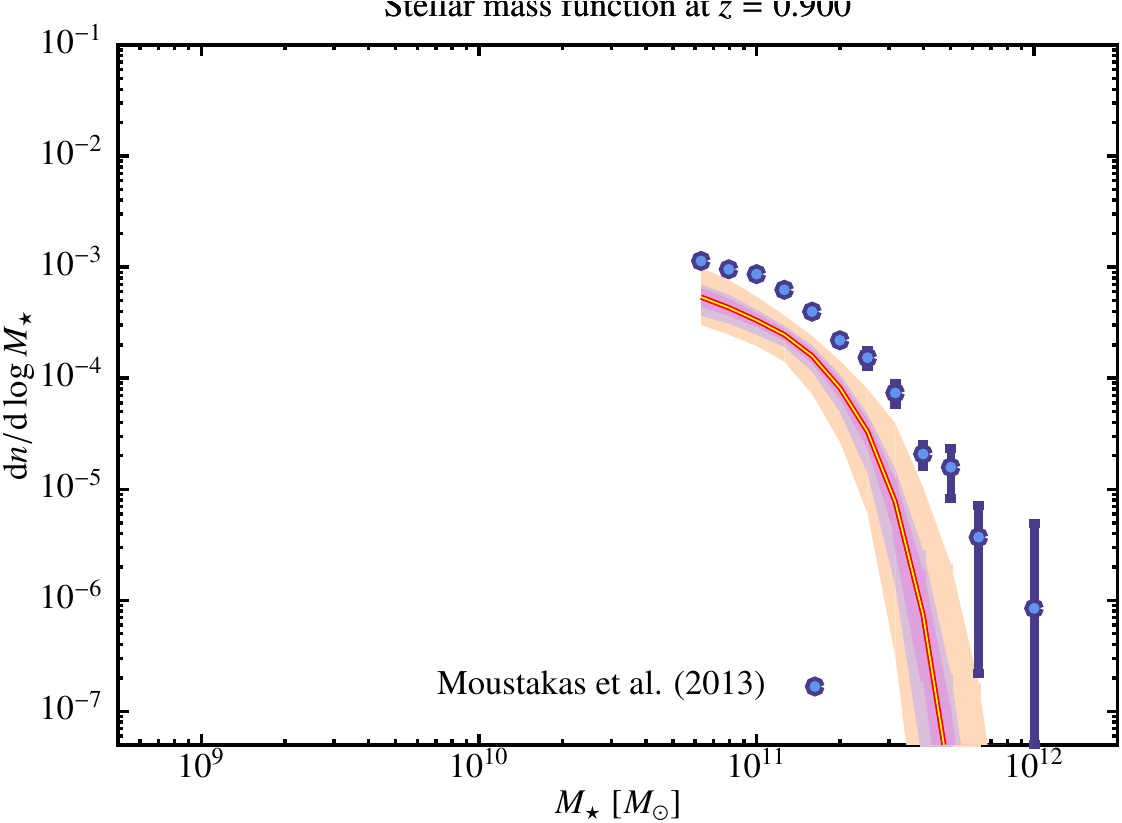} \\
 \end{tabular}
 \addtocounter{figure}{-1}
 \caption{\emph{(cont.)} $p$-values for these redshifts are $\protect\input{plots/posteriorPredictiveChecks/primusStellarMassFunctionZ0.575_testStatistic_pValue.txt}$, $\protect\input{plots/posteriorPredictiveChecks/primusStellarMassFunctionZ0.725_testStatistic_pValue.txt}$, and $\protect\input{plots/posteriorPredictiveChecks/primusStellarMassFunctionZ0.900_testStatistic_pValue.txt}$.}
\end{figure*}

\section{Discussion}\label{sec:Discussion}

\begin{table*}
 \caption{Results of a \protect\PCA\ analysis of the model posterior using the ``projection pursuit'' methodology of \protect\cite{bower_parameter_2010}. Rows list principal components, arranged in order of least to most contribution to the variance in the posterior (as listed in the second column). For each principal component, model parameters which contribute significantly to the variance of the component (greater than 10\%) are listed, along with the fraction of the variance in the component that they contribute.}
 \label{tb:ProjectionPursuit}
 \begin{center}
  \begin{tabular}{llr@{\,}lr@{\,}lr@{\,}lr@{\,}l}
\hline
{\bf Projection} & {\bf Variance} & \multicolumn{8}{c}{\bf Contributions}\\
\hline
{\bf 0} & $0.13 \times 10^{-2}$ & $0.56$ & $\alpha_{\rm cool}$; & $0.26$ & $\tau_{\rm cool}$;\\
{\bf 1} & $0.16 \times 10^{-2}$ & $0.73$ & $\Delta \log_{10}\mathcal{M}_{\rm cool}$;\\
{\bf 2} & $0.26 \times 10^{-2}$ & $0.39$ & $\beta_{\rm wind}$; & $0.30$ & $\beta_\star$;\\
{\bf 3} & $0.69 \times 10^{-2}$ & $0.57$ & $\beta_\star$;\\
{\bf 4} & $0.95 \times 10^{-2}$ & $0.56$ & $\tau_\star$; & $0.21$ & $\alpha_\star$; & $0.10$ & $\beta_\star$;\\
{\bf 5} & $0.21 \times 10^{-1}$ & $0.30$ & $\tau_{\rm cool}$;\\
{\bf 6} & $0.38 \times 10^{-1}$ & $0.41$ & $f_{\rm wind}$; & $0.15$ & $\beta_{\rm wind}$;\\
{\bf 7} & $0.41 \times 10^{-1}$ & $0.78$ & $\tau_{\rm \star, min}$; & $0.11$ & $f_{\rm wind}$;\\
{\bf 8} & $0.59 \times 10^{-1}$ & $0.39$ & $\alpha_\star$;\\
{\bf 9} & $0.72 \times 10^{-1}$ & $0.33$ & $V_{\rm reion}$; & $0.12$ & $\alpha_{\rm wind}$;\\
{\bf 10} & $0.76 \times 10^{-1}$ & $0.44$ & $V_{\rm reion}$; & $0.31$ & $\tau_{\rm wind, min}$;\\
{\bf 11} & $0.76 \times 10^{-1}$ & $0.33$ & $f_{\rm df}$; & $0.15$ & $\mathcal{M}_{\rm cool}$; & $0.12$ & $\tau_{\rm cool}$; & $0.11$ & $\alpha_\star$;\\
{\bf 12} & $0.86 \times 10^{-1}$ & $0.43$ & $\tau_{\rm wind, min}$; & $0.38$ & $\mathcal{M}_{\rm cool}$;\\
{\bf 13} & $0.90 \times 10^{-1}$ & $0.49$ & $\alpha_{\rm wind}$; & $0.20$ & $f_{\rm df}$; & $0.15$ & $\beta_{\rm wind}$;\\
{\bf 14} & $0.10$ & $0.46$ & $\beta_{\rm cool}$;\\
{\bf 15} & $0.20$ & $0.87$ & $\mu$;\\
{\bf 16} & $0.81$ & $0.85$ & $\kappa$;\\
{\bf 17} & $0.84$ & $0.35$ & $C_3$; & $0.21$ & $C_6$; & $0.17$ & $C_2$;\\
{\bf 18} & $0.89$ & $0.40$ & $C_1$; & $0.39$ & $H_2$;\\
{\bf 19} & $0.95$ & $0.23$ & $H_2$; & $0.23$ & $C_3$; & $0.12$ & $H_1$;\\
{\bf 20} & $0.96$ & $0.41$ & $C_4$; & $0.33$ & $H_1$;\\
{\bf 21} & $0.98$ & $0.81$ & $C_5$; & $0.11$ & $H_1$;\\
{\bf 22} & $1.1$ & $0.13$ & $C_1$; & $0.11$ & $H_1$;\\
{\bf 23} & $1.1$ & $0.35$ & $C_2$; & $0.23$ & $H_1$;\\
{\bf 24} & $1.2$ & $0.27$ & $C_1$;\\
\hline
\end{tabular}

 \end{center}
\end{table*}

Bearing in mind the phenomenological nature of our current model, it would be unreasonable to draw too much inference about the nature of galaxy formation physics from the \PPD\ of Figure~\ref{fig:PosteriorTriangle}. Nevertheless, there are a few interesting facts to be gleaned from the \PPD. To elucidate these points we apply the ``projection pursuit'' methodology of \cite{bower_parameter_2010} to identify the linear combinations of model parameters which contribute \emph{least} to the variance of the \PPD. We follow \cite{bower_parameter_2010} in scaling each parameter by the width of its prior when carrying out the \PCA\ analysis. Specifically, for uniform priors we scale parameters by the difference between the upper and lower limits of the prior, while for normal priors we scale by the root variance. Table~\ref{tb:ProjectionPursuit} shows the results of this analysis.

\begin{figure*}
 \begin{tabular}{cc}
  \includegraphics[width=85mm,trim=0mm 0mm 0mm 2.5mm,clip]{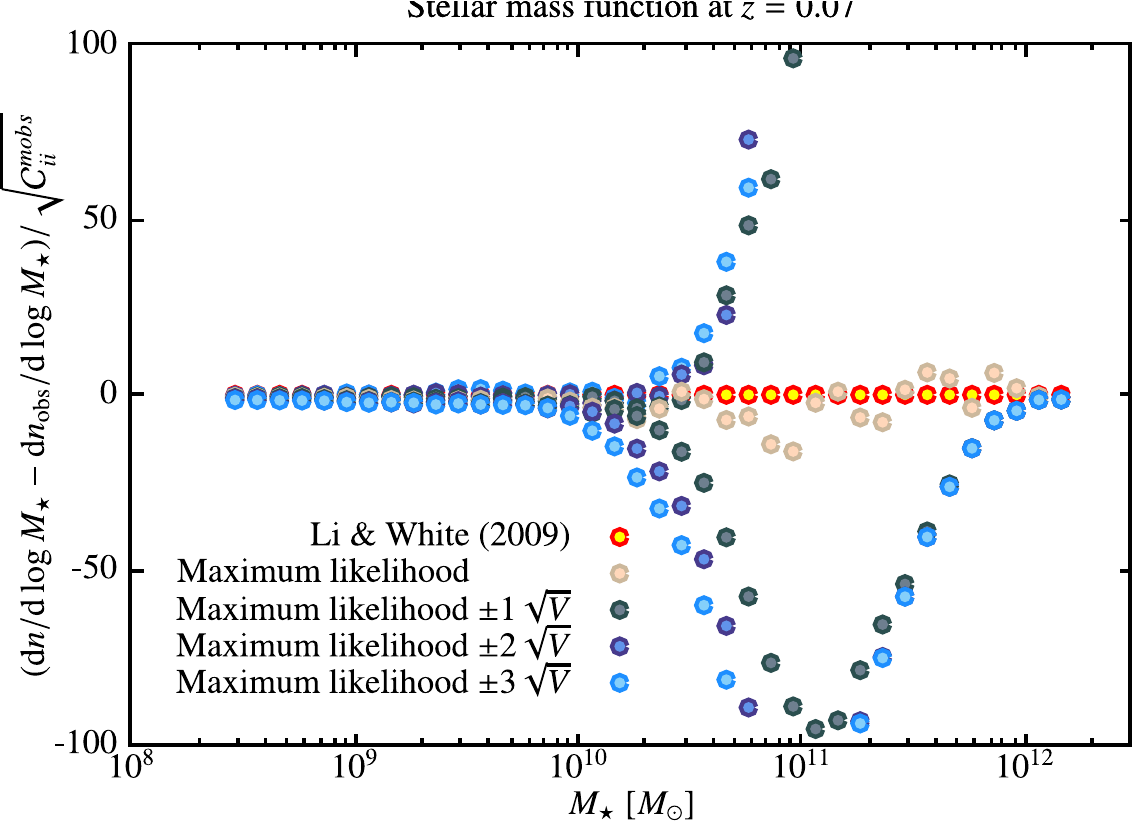} &
  \includegraphics[width=85mm,trim=0mm 0mm 0mm 2.5mm,clip]{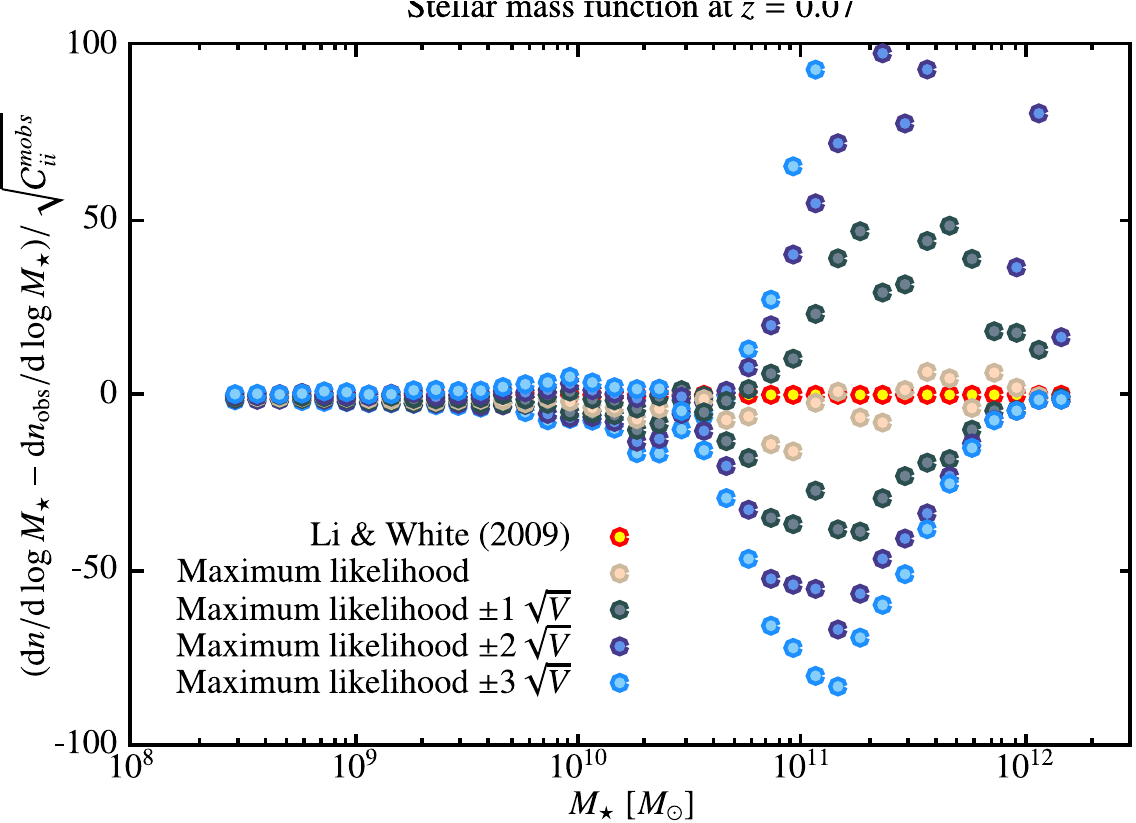}
 \end{tabular}
 \caption{Models perturbed away from the maximum likelihood model along the directions of projections 0 and 2 (from Table~\protect\ref{tb:ProjectionPursuit}) by 0, $\pm1$, $\pm2$, and $\pm3$ times the root variance of the projection. The points show the offset of the model mass function from the observed mass function normalized by the root-variance of each of the observed mass function.}
 \label{fig:ProjectionModels}
\end{figure*}

From the \PPD\ and projection pursuit analysis we find:
\begin{itemize}
 \item The exponent of the virial velocity dependence of the star formation efficiency, $\beta_\star$, is tightly constrained to the range $[-4.05,-3.24]$ (68.3\% confidence interval) with maximum likelihood value $-3.59$. As can be seen in Table~\ref{tb:ProjectionPursuit}, $\beta_\star$ is the dominant contributor to the second principal projection (which has an additional contribution from $\beta_{\rm cool}$). We explore what drives this tight constraint on this parameter combination by perturbing the maximum likelihood model along the direction of this principal projection\footnote{When performing this perturbation we include contributions from all parameters to this principal projection, not just the dominant contributors listed in Table~\protect\ref{tb:ProjectionPursuit}.} in parameter space by $\pm1$, 2, and $3 \sqrt{V}$ where $V$ is the variance of the principal projection\footnote{Of course, we remove the scaling of the parameters used in the \protect\PCA\ analysis when constructing the perturbation.}. The results of this study are shown in the right-hand panel of Figure~\ref{fig:ProjectionModels}, in which we plot the model stellar mass function relative to the observed mass function and normalized by the root-variance of the observed mass function. This indicates that it is the ``knee'' and high-mass cut-off of the mass function which is strongly affected by this parameter combination (and, therefore, which is driving the tight constraint on these parameters). Note in particular that the perturbation changes sign at around $2\times 10^{11}M_\odot$---the constraint is driven primarily by the shape of the knee/cut-off region of the mass function.

 \item The very flat slope of the stellar mass function at low masses leads to a preference for very negative exponents in the virial velocity dependence of the outflow rate (see eqn.~\ref{eq:OutflowModel}), $\beta_{\rm wind}$. Values as low as $\beta_{\rm wind}=-14.95$ are preferred (even more negative values are allowed, and are limited mostly by the imposed prior), although values $\beta_{\rm wind}<-0.57$ are not ruled out at greater than $99.7\%$ confidence. This illustrates the difficulty of obtaining a sufficiently flat galaxy mass function given the very steep slope of the halo mass function.

Previous \SAMs\ have also found the need for a strong dependence of outflow mass loading on halo virial velocity. For example, \cite{cole_recipe_1994} and \cite{bower_parameter_2010} required values $\beta_{\rm wind}=-5.5$ and $-3.2$ respectively, while \cite{lu_bayesian_2013} find $\beta_{\rm wind}\approx 6$ is required to produce slopes shallow enough to match their constraints. There is a correlation between $\beta_{\rm wind}$ the normalization of the outflow rate, $f_{\rm wind}$, and the redshift exponent of outflow, $\alpha_{\rm wind}$, (projection 9 in Table~\ref{tb:ProjectionPursuit}) such that less negative values of $\beta_{\rm wind}$ require larger values of the normalization. This is fortunate as physical models of outflows typically predict values $\beta_{\rm wind}$ in the range $-2.7$ to $-1$ \citep{murray_maximum_2005,oppenheimer_feedback_2010,creasey_how_2013,lagos_dynamical_2013}, while observations indicate mass loading factors for galaxies of stellar mass $\sim 10^{10}M_\odot$ of order unity \citep{martin_demographics_2012}. In this current, sufficiently simple, model, outflows can never actually remove gas from a halo (instead simply returning it to the hot component within the halo). An extended model in which outflows can actually leave the halo (returning only if the halo later grows substantially in mass) might reduce or remove the need for such a strong scaling with halo virial velocity.

 \item The exponent of the cooling timescale, $\alpha_{\rm cool}$, is tightly constrained to lie in the range $-6.12$ to $-4.35$ at 68.3\% confidence with maximum likelihood value $-5.12$. As shown in Table~\ref{tb:ProjectionPursuit}, $\alpha_{\rm cool}$ is the dominant contributor to the projection with least variance, with other significant contributions from the cooling timescale, $\tau_{\rm cool}$, and the exponent of the cut-off in cooling efficiency, $\beta_{\rm cool}$. In the left-hand panel of Figure~\ref{fig:ProjectionModels} we show the effect of perturbing the maximum likelihood model along this parameter projection. Not surprisingly, these perturbations affect the high mass end of the mass function, shifting it systematically above or below the observed mass function. This parameter combination directly affects the masses of high-mass galaxies.

 \item Our model includes a parameterization of a cut-off in cooling efficiency in high mass halos. We find that a relatively low mass cut-off is preferred (maximum likelihood value of $10^{10.1}M_\odot$, with 68.3\% of the posterior lying between $10^{10.0}$ and $10^{11.6}M_\odot$), although the posterior extends to cut-off masses as high as $10^{13.4}M_\odot$ at $99.7\%$ confidence. This cut-off has often been associated with \AGN\ feedback, or the transition from ``cold-mode'' to ``hot-mode'' cooling. Our posterior on the cut-off mass is consistent with the expected mass scales for these processes \citep{birnboim_virial_2003,kerevs_how_2005}, but is too broad to place a meaningful constraint.

 \item The combination of the width of the cut-off in cooling efficiency, $\Delta\log_{10}\mathcal{M}_{\rm cool}$, and the exponent of the cut-off, $\beta_{\rm cool}$, is, however, very well constrained. Cooling must transition from being efficient to inefficient over about 1 order of magnitude in halo mass if $\beta_{\rm cool}$ is around unity. Broader transitions are allowed if the cut-off is made sharper by raising the exponent, $\beta_{\rm cool}$---a result of the strong correlation between $\Delta\log_{10}\mathcal{M}_{\rm cool}$ and $\beta_{\rm cool}$.
\end{itemize}

\begin{figure}
  \includegraphics[width=85mm,trim=0mm 0mm 0mm 2.5mm,clip]{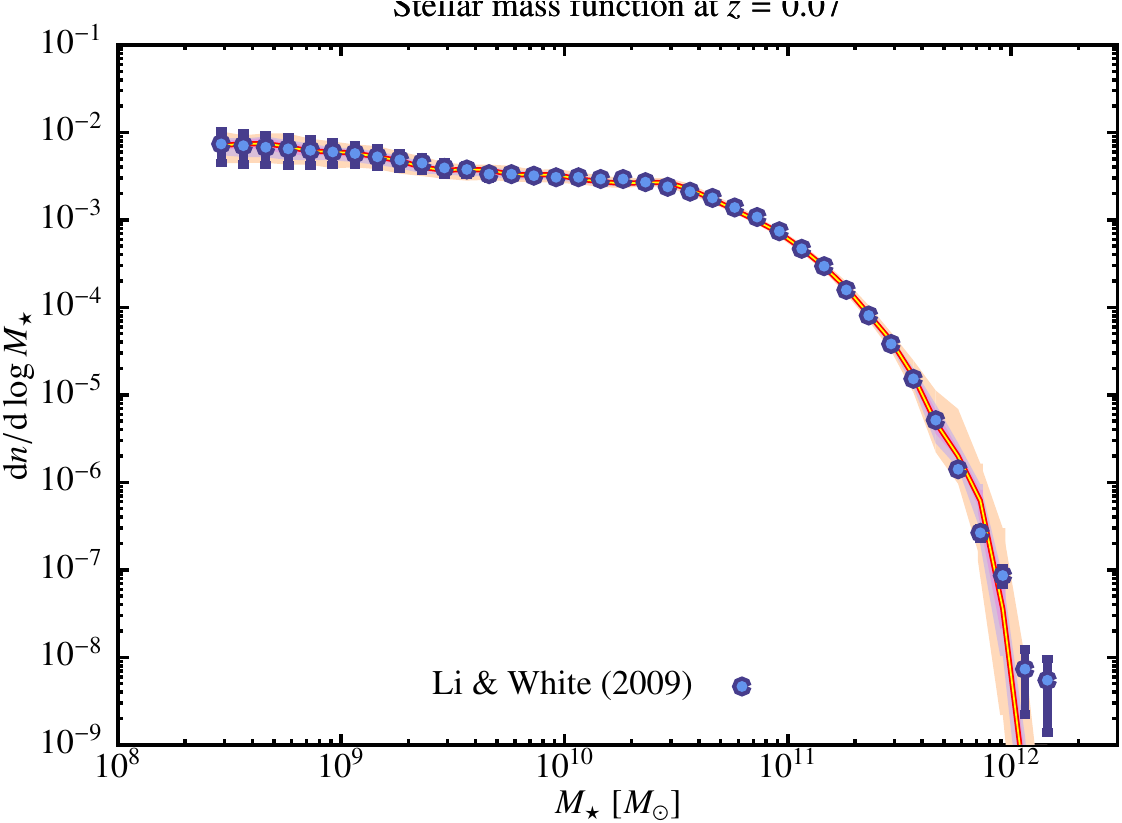}
 \caption{The stellar mass function of galaxies at $z\approx 0.07$ when the model is constrained using a diagonalized covariance matrix. Blue points with error bars show the mass function measured from the \protect\SDSS\ by \protect\cite{li_distribution_2009}. Error bars on these points are the square-roots of the diagonal elements of the covariance matrix constructed as described in \S\protect\ref{sec:ObservedCovariance}. The shaded regions enclose the inner 68.26\%, 95.44\%, and 99.74\% (1, 2, and 3-``$\sigma$'' respectively) of the distribution of models sampled from the converged \protect\MCMC\ chains, while the yellow line indicates the median of those models.}
 \label{fig:DiagonalizedMassFunction}
\end{figure}

As mentioned in \S\ref{sec:PosteriorPredictiveChecks} the median model mass function drawn from the \PPD\ lies systematically below the data at the low-mass end of the stellar mass function. This is a consequence of two facts. First, the data show a weak, but significant change in slope at around $M_\star\approx 4\times10^9M_\odot$. Our model contains nothing which would allow it to reproduce such a feature (i.e. in this regime the model is essentially self-similar). The second fact is that the covariance matrix of the data has strong off-diagonal correlation at low-masses (a consequence of large scale structure contributions to the covariance). Combined, these facts mean that the model can achieve a high likelihood with a constant slope at low masses because the low-mass data points effectively move up and down in unison, rather than as independent points. Colloquially speaking, the model lies roughly ``1-$\sigma$'' below each of the low mass points, but this corresponds to a likelihood consistent with being ``1-$\sigma$'' below a single point. Figure~\ref{fig:DiagonalizedMassFunction} shows the model stellar mass function computed when the covariance matrix used in our likelihood calculation is diagonalized. In this case the model is forced to go through all points in the low mass part of the mass function, effectively overfitting the data. This illustrates the importance of constructing the model likelihood using the full covariance matrix of the data, rather than just a diagonalized approximation to that matrix. Similar effects have been seen by \cite{lu_bayesian_2013} when constraining their model to fit the HI mass function of galaxies.

\begin{figure*}
 \begin{center}
 \begin{tabular}{cc}
 \vspace{-6mm} $z=0.900$ & $z=0.9000$ \\ 
 \includegraphics[width=80mm,trim=0mm 0mm 0mm 2.5mm,clip]{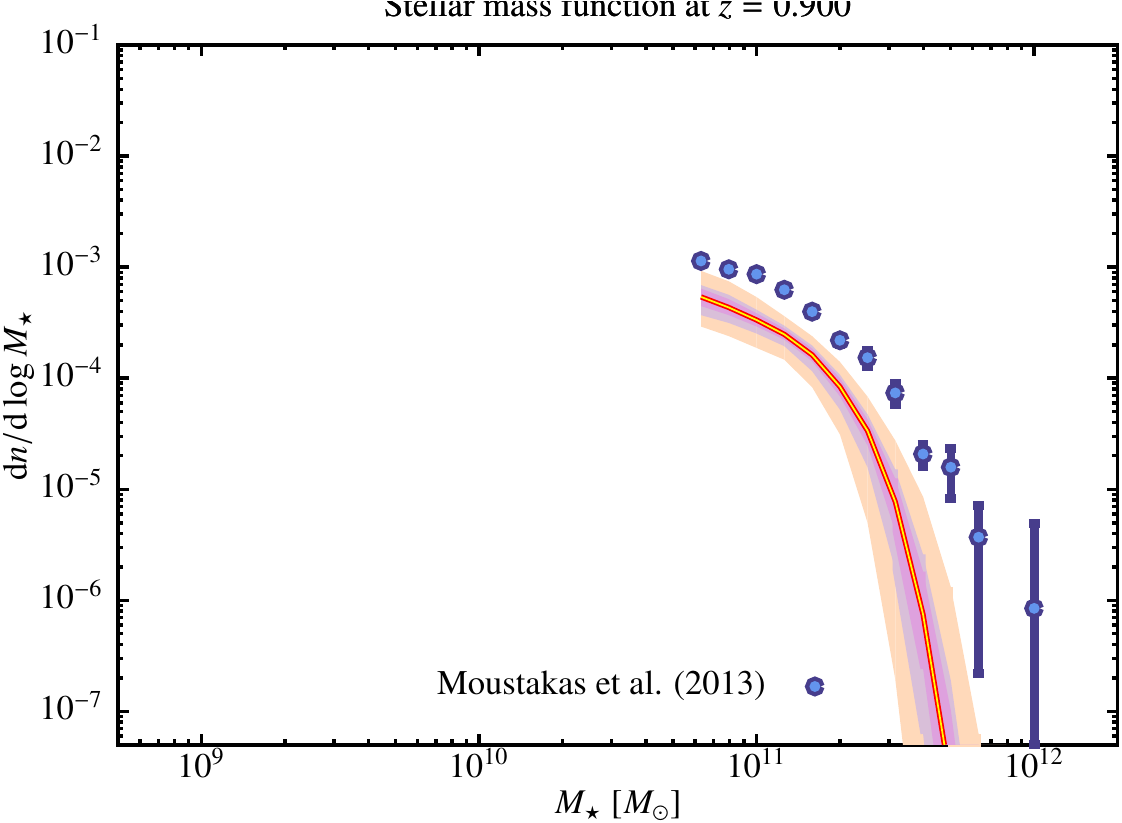} &
 \includegraphics[width=80mm,trim=0mm 0mm 0mm 2.5mm,clip]{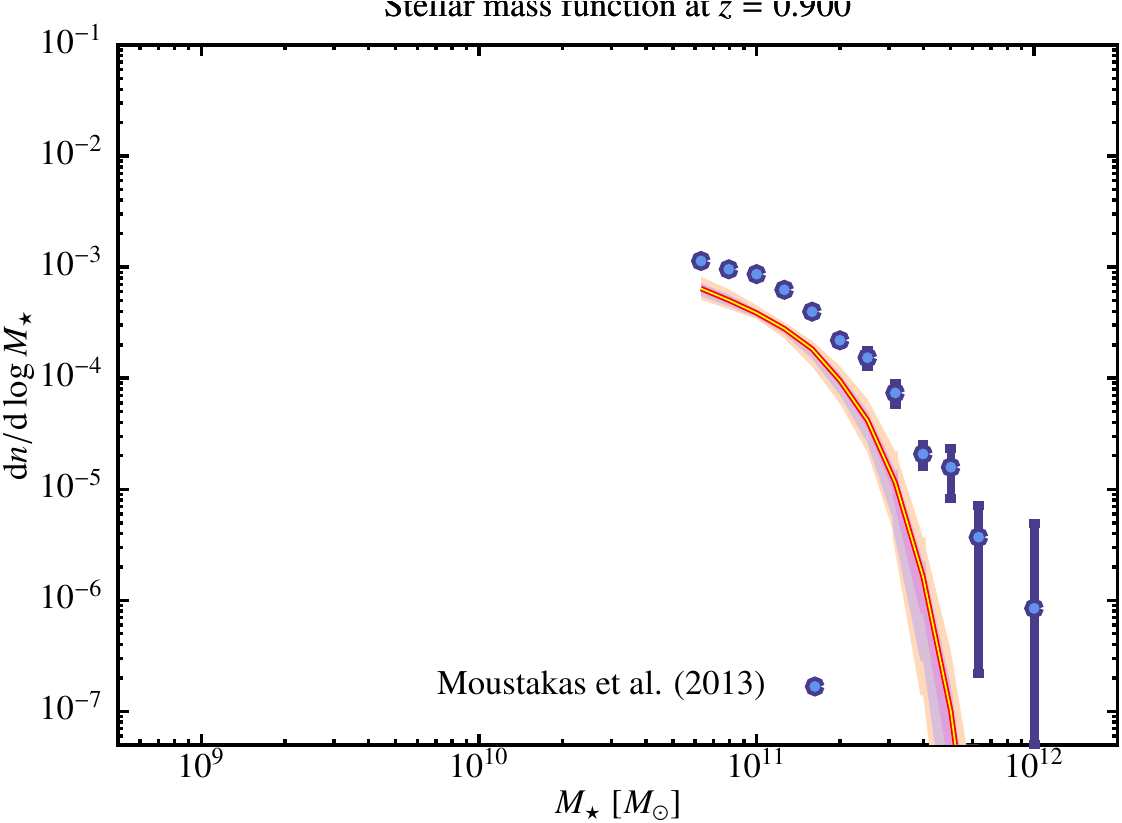}
 \end{tabular}
 \end{center}
 \caption{Stellar mass function from the PRIMUS survey at $z=0.9$ (blue points). The shaded regions enclose the inner 68.26\%, 95.44\%, and 99.74\% (1, 2, and 3-``$\sigma$'' respectively) of the distribution of models sampled from the converged \protect\MCMC\ chains, while the yellow line indicates the median of those models. \emph{Left panel:} Models in which cosmological parameters were held fixed at their maximum likelihood values. \emph{Right panel:} Models in which all systematic uncertainties and model discrepancy terms were set to zero.} 
 \label{fig:noCosmology:PRIMUS}
\end{figure*}

When our model is extrapolated to higher redshifts we find that, while it is formally not a good description of the stellar mass function out to $z\approx 1$, given that we have not yet performed a detailed analysis of random and systematic errors in those mass functions (as we have done for $z\approx 0$), the model is remarkably successful in capturing the evolution of the shape and normalization of the stellar mass function, even though it is constrained only by $z\approx 0$ data\footnote{After the \protect\MCMC\ calculations for this work were underway, \protect\cite{bernardi_massive_2013} published a new measurement of the SDSS galaxy stellar mass function, based on a new analysis of the photometry. They conclude that the galaxy magnitudes used by \protect\cite{li_distribution_2009} underestimate the total luminosities of the brightest galaxies. As such, \protect\cite{bernardi_massive_2013} find that the high-mass end of the mass function should be boosted relative to \protect\cite{li_distribution_2009}. If our model were fit to the result of \protect\cite{bernardi_massive_2013} instead of to those of \protect\cite{li_distribution_2009} we might therefore expect improved agreement between our model and $z\approx 1$ stellar mass functions. This will be explored in the next paper in this series.}. For example, at $z=0.9$ the 99.7\% confidence interval of the model comes close to several of the data points. We can compare this to the model extrapolation if we, for example, keep cosmological parameters fixed at their maximum likelihood values rather than letting them vary with the WMAP 9-year priors. The left panel of Fig.~\ref{fig:noCosmology:PRIMUS} shows the $z=0.9$ stellar mass function in this case. The range spanned by the model extrapolations is clearly much narrower than that shown in Figure~\ref{fig:PosteriorPredictiveCheckPRIMUS0100} (in which cosmological parameters were allowed to vary). Particularly at high redshifts, the remaining uncertainties in cosmological parameters contribute significantly to the uncertainty in theoretical models constrained by $z=0$ data. Even bigger changes are apparent if we set all systematic uncertainty parameters to zero and ignore model discrepancies, as shown in the right panel of Fig.~\ref{fig:noCosmology:PRIMUS}. In this case the range of model predictions at $z=0.9$ is dramatically reduced (although no significant bias is introduced), clearly indicating the importance of systematic uncertainty analysis when extrapolating models constrained by one dataset to make predictions for another.

\begin{figure}
 \begin{center}
  \includegraphics[width=80mm,trim=0mm 0mm 0mm 2.5mm,clip]{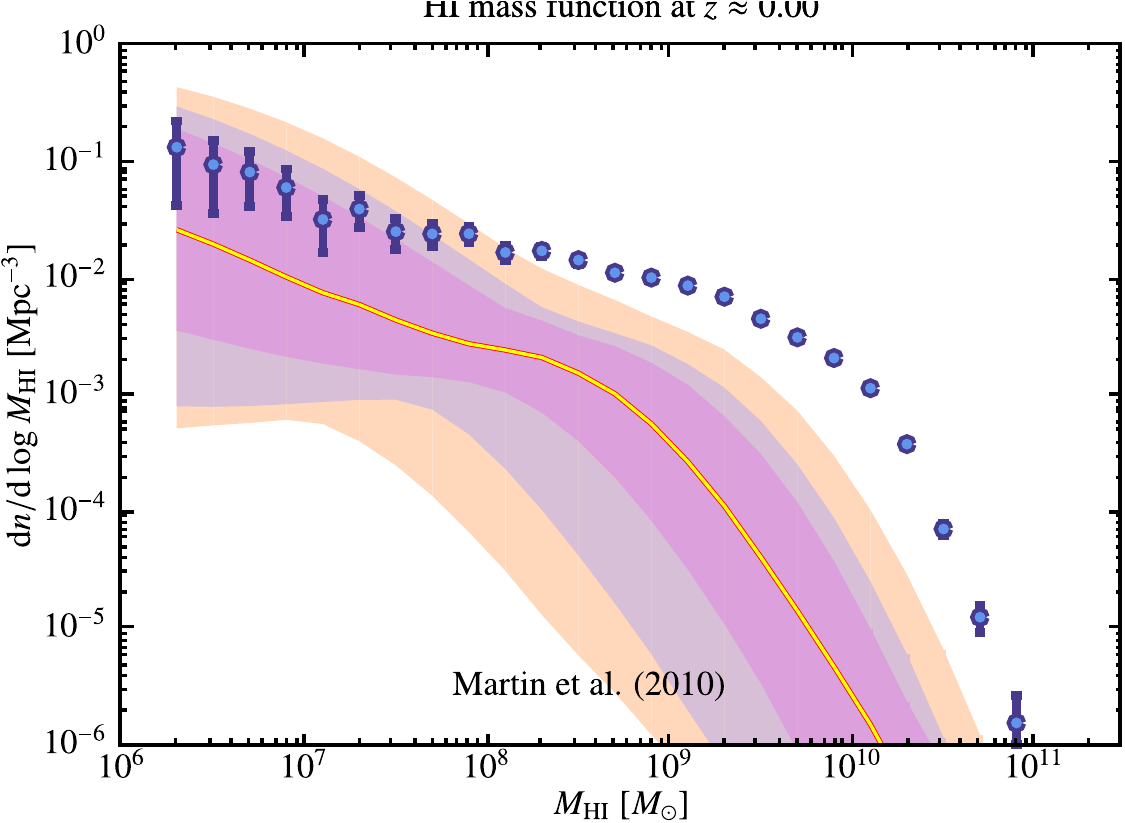}
 \end{center}
 \caption{The $z=0$ HI galaxy mass function of \protect\cite{martin_arecibo_2010} compared to the extrapolation from our model when all systematic uncertainty parameters are set to zero and model discrepancy terms are ignored. Shaded regions enclose the inner 68.26\%, 95.44\%, and 99.74\% (1, 2, and 3-``$\sigma$'' respectively) of the distribution of models sampled from the converged \protect\MCMC\ chains, while the yellow line indicates the median of those models.} 
 \label{fig:noSystematics:ALFALFA}
\end{figure}

In comparing our model with the HI galaxy mass function from the ALFALFA survey \citep{martin_arecibo_2010} it is clear that the model predictions are both highly unconstrained\footnote{This indicates that, within the structure of our model, it is possible to obtain the correct stellar mass function from galaxy populations with vastly different gas content.}, and, in most instances, dramatically underpredict the mass function. Our test statistic indicates an extremely poor agreement between model and data, as expected. This disagreement between model and HI mass function suggests that adding the HI mass function as a constraint on our model should provide significant additional constraining power, although additional freedom may have to be introduced into the model to allow a good fit to be found. We again caution that these results should not be considered final as we have not yet accounted for systematics in the determination of HI masses, nor have we assessed how model discrepancy terms would impact the HI mass function. In particular, the very coarse resolution of our merger trees might be expected to substantially affect the predicted HI mass function. These issues will be explored in the next paper in this series. Once again, the inclusion of systematic uncertainties in our modelling strongly affects the distributions of model HI mass functions, as can be seen in Fig.~\ref{fig:noSystematics:ALFALFA} in which we show the distribution of model HI mass functions when systematics and model discrepancies are ignored (compare with Fig.~\ref{fig:PosteriorPredictiveCheckHI}).

\subsection{Comparison with previous works}

Several previous works have applied \MCMC\ or similar techniques to \SAMs\ with the goal of constraining the parameters of these models. In the following we discuss several of these works and highlight the similarities and differences with the current work.

\cite{henriques_monte_2009} explored a 6-dimensional model parameter space using \MCMC\ techniques. They find that they are able to obtain reasonable fits to a combination of K-band luminosity function, B$-$V colours, and the black hole mass function all at $z\approx 0$, but find that this requires very strong feedback from supernovae, as do we in the present work. Importantly, \cite{henriques_monte_2009} highlight a discrepancy between their model constrained to the K-band luminosity function and the stellar mass function, which they take to suggest that a possible 0.3~dex systematic in stellar mass determinations may exist. Such systematics are modeled directly in our calculation. A systematic of 0.3~dex is entirely consistent with our derived posterior distributions for the parameters $\mu$ and $\kappa$ of our stellar mass systematics model. Importantly, the constraints on all of our other parameters correctly take into account the full range of possible systematics. \cite{henriques_monte_2009} also acknowledge the importance of systematics in assessing the validity of their model---in the current work we are able to address this issue directly. Continuations of this work \citep{guo_dwarf_2011,henriques_simulations_2013} explores constraints derived from luminosity and stellar mass functions over the redshift range $0<z<3$. \cite{henriques_simulations_2013} emply an 11-parameter model and find that the efficiency of reaccretion of ejected material must explicitly vary with cosmic epoch in order to fit the constraints. Once constrained in this way, their model reproduces quite well the evolution of galaxy abundances, the colours of $z=0$ dwarf galaxies, and the clustering of galaxies across a wide range of masses. Following \cite{henriques_monte_2009}, \cite{henriques_simulations_2013} attempt to address the presence of sytematic errors in datasets by using the scatter between multiple observational determinations of a particular constraint. This approach has the advantage that it does not require assumptions about the sources of systematics to be made. However, as noted by \cite{henriques_simulations_2013} it does require arbitrary judgements about the quality of different datasets to be made and, furthermore, will not account for systematics in common between datasets. As a result, \cite{henriques_simulations_2013} conclude that ``\emph{As a result, formal levels of agreement between theory and observation should be treated with caution.}'' Clearly, consideration of how to treat systematic uncertainties is crucial for progress in this area.

\cite{bower_parameter_2010} explored the 16-dimensional parameter space of their model using a model emulator technique, constraining their model to fit $z\approx 0$ b$_{\rm J}$ and K-band luminosity functions. They found, as do we, that a wide range\footnote{Although still small compared to the prior range of parameter space.} of parameter space is compatible with the observational data. \cite{bower_parameter_2010} take into account three model discrepancy terms. The first arises from their choice to exclude eight ``inactive parameters'' (parameters to which the results of the model are only weakly dependent) from their model emulator---this discrepancy accounts for the additional variance that those parameters would introduce were they included in the emulation. Since we treat all parameters as active in the present work we have no comparable discrepancy. The second discrepancy included by \cite{bower_parameter_2010} accounts for the finite volume of the simulation volume from which they extract their merger trees. This is accounted for in the present work when modeling the differences between Monte Carlo and N-body merger trees\footnote{Additionally, in this work, we account for the finite number of merger trees used in our model evaluation by directly modeling the model covariance matrix; see Appendix~\ref{app:ModelCovariance}.}; see \S\ref{sec:MonteCarloDiscrepancy}. Finally, \cite{bower_parameter_2010} include a discrepancy term to account for the structural deficiencies of their model. This is comparable to the discrepancies that we introduce in \S\ref{sec:ModelDiscrepancy}. \cite{bower_parameter_2010} adopt a rather different approach to estimating this discrepancy, namely exploring a range of model parameter space and judging which models they would consider to be adequate fits to the data based on previous experience. While this approach has its advantages compared to our own (namely being applicable even in cases where the ``correct'' answer cannot be modeled), it is difficult to apply to constraints for which previous intuition is lacking.

\cite{mutch_constraining_2013} constrain a 6-dimensional \SAM\ to simultaneously fit $z\approx0$ and $z\approx 0.83$ galaxy stellar mass functions. They also find that extremely strong supernova feedback is required to match these datasets, and note a minor tension between the model and data when fitting both redshifts together. \cite{mutch_constraining_2013} suggest that this tension may be removed by more accurate estimation of random and systematic errors in the data. In this work, we have shown that these errors do in fact make a significant difference to the constraints on models, and may help reconcile the tension found by \cite{mutch_constraining_2013}. We note, however, that our own models remain formally inconsistent with the high redshift stellar mass functions explored here, with the caveat that we have so far applied a rigorous error analysis only to the $z\approx0$ stellar mass function. Once a similar analysis is applied to the higher redshift data it remains possible that our model will be consistent with all data considered in this work. 

Finally, \cite{lu_bayesian_2013} constrain a 17-dimensional \SAM\ to fit a K-band luminosity function and HI galaxy mass function. Once again, a requirement for very strong supernova driven feedback is found, and that a very strong dependence on halo mass (or virial velocity) is required to produce a sufficiently shallow faint end slope (consistent with our own findings regarding the slope of the stellar mass function at low masses). \cite{lu_bayesian_2013} employ \PPCs\ to assess whether their model is a viable description of the data, finding that it is not (the data lies well into the tails of the distribution of model results), similar to our own findings. This suggests that the detailed form of these particular datasets requires more refined models (possibly containing additional physics). 

To summarize, a consistent picture is beginning to emerge from studies of this sort, with conclusions including the need for very strong supernova feedback, and the inability of current models to describe existing data with the accuracy required by observational uncertainties. In this work we have extended previous studies by taking into account the full range of model and data errors---both random (including covariances) and systematic---and sources of model discrepancy. While this results in a very high (25) dimensional model and requires additional modeling to assess observational and theoretical errors we have demonstrated that such an approach is feasible given current computational abilities, and is important if predictions from such models are to be used to robustly test galaxy formation theory.

\section{Conclusions}\label{sec:Conculsion}

We have performed a careful study of how to constrain a semi-analytic model of galaxy formation to a single observational dataset---specifically the stellar mass function of galaxies in the nearby Universe. We have paid particular attention to assessing the covariance matrices of data and model, both random and systematic errors in the data, and discrepancies in the model itself. Combining these multiple sources of uncertainty we derive a robust \PPD\ on the model parameters.

We find that, of these various sources of uncertainty, cosmological parameters and halo mass function systematics are sufficiently well constrained by other experiments to not significantly affect the specific observations to which we constrain our model in this work. However, the remaining uncertainties in cosmological parameters \emph{do} play a crucial role in setting the uncertainty of theoretical model expectations for predictions at higher redshifts. Systematic uncertainties in stellar mass estimates are the largest source of uncertainty in our model expectations for higher redshifts. Understanding these systematic errors in stellar mass, and how they vary with redshift, will be crucial for robustly constraing and testing models in future. Ignoring this contribution to the model uncertainty could result in viable models being erroneously ruled out\footnote{Of course, overestimating uncertainties can result in the opposite problem, namely forgiving models that are inadequate. This should be avoided to, but we believe that erring on the size of larger uncertainties is the more conservative way to proceed. Eventually, we would like to use these model frameworks to test specific hypotheses relating to aspects of galaxy formation physics. We would prefer to wrongly accept a bad model (which can always be ruled out later with the addition of better data) then to wrongly exclude a correct model which may then be no longer considered even as better data becomes available.}. We also find that ignoring correlations between bins in the stellar mass function when computing model likelihoods (i.e. using only the diagonal terms in the covariance matrix) can substantially bias the resulting \PPDs.

Given the phenomenological nature of our current model we refrain from drawing strong conclusions about the physics of galaxy formation. However, we find that the very flat slope of the galaxy stellar mass function is particularly challenging to reproduce within the framework of our model, requiring a very strong dependence of outflow efficiency on halo virial velocity. Furthermore, we find that the cooling of gas in halos must transition from efficient to inefficient over around one order of magnitude in halo mass (of less, depending on the functional form of the transition). 

We compare our model, constrained to match the stellar mass function of galaxies in the local Universe, to other datasets. We find that it succeeds in matching the evolution of the galaxy stellar mass function out to $z\approx 1$---a remarkable fact given that the model was not constrained by any high-redshift data. This suggests that the evolution of the stellar mass function is encoded in the present day mass function plus basic cosmological structure formation\footnote{At least, that is, given the particular structure of our model.}. However, we find that our model is not in agreement with the galaxy HI mass function in the local Universe, dramatically under-predicting the HI masses of galaxies. We note that \cite{lu_bayesian_2013} were also unable to find a model which came close to the observed HI mass function, although in their case the model dramatically over-predicted the observations. We caution that this failure should not be taken too seriously as yet---we have not yet accounted for systematic uncertainties in HI mass determinations, nor for the limitations of the model in predicting HI mass. Furthermore, it is possible that a mode exists in the model parameter space which adequately describes both the stellar and HI mass functions, yet has too low likelihood to be populated when the stellar mass function alone is used as a constraint \citep{lu_bayesian_2012,lu_bayesian_2013}. Nevertheless, this failing will be our focus in a future work.

Both \cite{mutch_constraining_2013} and \cite{lu_bayesian_2013} find significant tensions between their \SAMs\ and observational data (the evolution of the luminosity function, and $z=0$ luminosity function plus HI mass function respectively). These results illustrate the power of \MCMC\ methods to uncover failings of models, as noted in particular by \cite{lu_bayesian_2013}. These conclusions are clearly of great importance for advancing our understanding of galaxy formation. Given this importance, we believe that it is crucial to carefully determine the systematic and random uncertainties of both models and data, and to account for these uncertainties when assessing likelihoods to provide accurate and robust constraints on model parameters. We will therefore revisit these additional constraints in the next paper in this series, including a careful analysis of random and systematic errors in those datasets.

Our current model relies almost entirely on phenomenological treatments of baryonic physics. This limits the insight that can be obtained from such models. Our goal is to progress swiftly to incorporating detailed physical models. To do this, it is crucial that any physical model (e.g. of star formation) be embedded within a model which gives a broadly correct description of the global properties of galaxies, such as their masses, sizes, densities, and dark matter halos. Otherwise, it will be difficult, if not impossible, to judge whether the failing of a model is due to an inadequate physical model, or simply due to that model being embedded within a galaxy population which looks nothing like that which is actually observed.

As such, the goal of the next paper in this series will be to construct a model which gets the basic structural properties (masses, sizes, densities, and halo masses) of galaxies correct in the local Universe and, where possible, out to moderate redshifts ($z\approx1$). This will provide a sound basis on which the develop physical prescriptions. With such a model in hand it will become possible to augment the physics of the model and make quantitative inference about the validity of our understanding of galaxy formation physics.

Computational speed remains a serious issue for this type of study, in which tens of millions of model evaluations are required. Therefore, in the next paper in this series, we will also explore methods to mitigate this computational demand as will be required to make further studies of more realistic models tractable.

\section*{Acknowledgments}

We thank Richard Bower, Shaun Cole, Darren Croton, John Helly, Neal Katz, Yu Lu, Martin Weinberg, and Martin White for invaluable conversations. We would also like to thank Martin Weinberg for developing, sharing and providing assistance with the \BIE\ package, Alexander Knebe for providing data on halo mass functions, and the VIRGO Consortium for making availabel data from the ``MillGas'' simulation. We acknowledge the hospitality of the International Centre for Radio Astronomy Research at the University of Western Australia, the Aspen Center for Physics (NSF grant \#1066293), and the Institute for Computational Cosmology, at the University of Durham where parts of this work were completed.

Funding for the Sloan Digital Sky Survey (SDSS) has been provided by the Alfred P. Sloan Foundation, the Participating Institutions, the National Aeronautics and Space Administration, the National Science Foundation, the U.S. Department of Energy, the Japanese Monbukagakusho, and the Max Planck Society. The SDSS Web site is http://www.sdss.org/.

The SDSS is managed by the Astrophysical Research Consortium (ARC) for the Participating Institutions. The Participating Institutions are The University of Chicago, Fermilab, the Institute for Advanced Study, the Japan Participation Group, The Johns Hopkins University, Los Alamos National Laboratory, the Max-Planck-Institute for Astronomy (MPIA), the Max-Planck-Institute for Astrophysics (MPA), New Mexico State University, University of Pittsburgh, Princeton University, the United States Naval Observatory, and the University of Washington. 

\bibliographystyle{mn2e}
\bibliography{galacticusConstraints1Accented}

\begin{thebibliography}{}
\makeatletter
\def\mn@urlcharsother{%
\let\do\@makeother
\do\$\do\&\do\#\do\^\do\_\do\%\do\~}
\def\mn@doi{\begingroup
\mn@urlcharsother
\@ifnextchar[%
{\mn@doi@}
{\mn@doi@[]}}
\def\mn@doi@[#1]#2{%
\def\@tempa{#1}%
\ifx\@tempa\@empty
\href{http://dx.doi.org/#2}{doiXX:#2}%
\else
\href{http://dx.doi.org/#2}{#1}%
\fi
\endgroup
}
\def\mn@eprint#1#2{%
\mn@eprint@#1:#2::\@nil}
\def\mn@eprint@arXiv#1{\href{http://arxiv.org/abs/#1}{{\tt arXiv:#1}}}
\def\mn@eprint@dblp#1{\href{http://dblp.uni-trier.de/rec/bibtex/#1.xml}{dblp:#1}}
\def\mn@eprint@#1:#2:#3:#4\@nil{%
\def\@tempa{#1}%
\def\@tempb{#2}%
\def\@tempc{#3}%
\ifx\@tempc\@empty
\let\@tempc\@tempb
\let\@tempb\@tempa
\fi
\ifx\@tempb\@empty
\def\@tempb{arXiv}%
\fi
\@ifundefined{mn@eprint@\@tempb}
{\@tempb:\@tempc}
{\expandafter\expandafter\csname
  mn@eprint@\@tempb\endcsname\expandafter{\@tempc}}%
}

\bibitem[\protect\citeauthoryear{Adelman-{McCarthy} et~al.,}{Adelman-{McCarthy}
  et~al.}{2008}]{adelman-mccarthy_sixth_2008}
Adelman-{McCarthy} J.~K.,  et~al., 2008, \mn@doi [ApJS] {10.1086/524984;}, 175,
  297

\bibitem[\protect\citeauthoryear{Anderson et~al.,}{Anderson
  et~al.}{2012}]{anderson_clustering_2012}
Anderson L.,  et~al., 2012, \mn@doi [MNRAS] {10.1111/j.1365-2966.2012.22066.x},
  427, 3435

\bibitem[\protect\citeauthoryear{Behroozi, Conroy \& Wechsler}{Behroozi
  et~al.}{2010}]{behroozi_comprehensive_2010}
Behroozi P.~S.,  Conroy C.,    Wechsler R.~H.,  2010, ApJ, 717, 379

\bibitem[\protect\citeauthoryear{Benson}{Benson}{2005}]{benson_orbital_2005}
Benson A.~J.,  2005, {\mnras}, 358, 551

\bibitem[\protect\citeauthoryear{Benson}{Benson}{2010}]{benson_galaxy_2010}
Benson A.~J.,  2010, \mn@doi [Physics Reports] {10.1016/j.physrep.2010.06.001},
  495, 33

\bibitem[\protect\citeauthoryear{Benson}{Benson}{2012}]{benson_galacticus:_2012}
Benson A.~J.,  2012, NewA, 17, 175

\bibitem[\protect\citeauthoryear{Benson, Bower, Frenk, Lacey, Baugh \&
  Cole}{Benson et~al.}{2003}]{benson_what_2003}
Benson A.~J.,  Bower R.~G.,  Frenk C.~S.,  Lacey C.~G.,  Baugh C.~M.,    Cole
  S.,  2003, ApJ, 599, 38

\bibitem[\protect\citeauthoryear{Benson, Borgani, De~Lucia, Boylan-Kolchin \&
  Monaco}{Benson et~al.}{2012}]{benson_convergence_2012}
Benson A.~J.,  Borgani S.,  De~Lucia G.,  Boylan-Kolchin M.,    Monaco P.,
  2012, \mn@doi [MNRAS] {10.1111/j.1365-2966.2011.20002.x;}, 419, 3590

\bibitem[\protect\citeauthoryear{Bernardi, Meert, Sheth, Vikram,
  Huertas-Company, Mei \& Shankar}{Bernardi
  et~al.}{2013}]{bernardi_massive_2013}
Bernardi M.,  Meert A.,  Sheth R.~K.,  Vikram V.,  Huertas-Company M.,  Mei S.,
     Shankar F.,  2013, \mn@doi [MNRAS] {10.1093/mnras/stt1607}, 436, 697

\bibitem[\protect\citeauthoryear{Beutler et~al.,}{Beutler
  et~al.}{2011}]{beutler_6df_2011}
Beutler F.,  et~al., 2011, \mn@doi [MNRAS] {10.1111/j.1365-2966.2011.19250.x},
  416, 3017

\bibitem[\protect\citeauthoryear{Birnboim \& Dekel}{Birnboim \&
  Dekel}{2003}]{birnboim_virial_2003}
Birnboim Y.,  Dekel A.,  2003, MNRAS, 345, 349

\bibitem[\protect\citeauthoryear{Blake et~al.,}{Blake
  et~al.}{2012}]{blake_wigglez_2012}
Blake C.,  et~al., 2012, \mn@doi [MNRAS] {10.1111/j.1365-2966.2012.21473.x},
  425, 405

\bibitem[\protect\citeauthoryear{Blanton et~al.,}{Blanton
  et~al.}{2005}]{blanton_new_2005}
Blanton M.~R.,  et~al., 2005, \mn@doi [AJ] {10.1086/429803;}, 129, 2562

\bibitem[\protect\citeauthoryear{Bower, Vernon, Goldstein, Benson, Lacey,
  Baugh, Cole \& Frenk}{Bower et~al.}{2010}]{bower_parameter_2010}
Bower R.~G.,  Vernon I.,  Goldstein M.,  Benson A.~J.,  Lacey C.~G.,  Baugh
  C.~M.,  Cole S.,    Frenk C.~S.,  2010, \mn@doi [MNRAS]
  {10.1111/j.1365-2966.2010.16991.x;}, 407, 2017

\bibitem[\protect\citeauthoryear{Brinchmann, Charlot, White, Tremonti,
  Kauffmann, Heckman \& Brinkmann}{Brinchmann
  et~al.}{2004}]{brinchmann_physical_2004}
Brinchmann J.,  Charlot S.,  White S. D.~M.,  Tremonti C.,  Kauffmann G.,
  Heckman T.,    Brinkmann J.,  2004, \mn@doi [MNRAS]
  {10.1111/j.1365-2966.2004.07881.x}, 351, 1151

\bibitem[\protect\citeauthoryear{Cole, Aragon-Salamanca, Frenk, Navarro \&
  Zepf}{Cole et~al.}{1994}]{cole_recipe_1994}
Cole S.,  Aragon-Salamanca A.,  Frenk C.~S.,  Navarro J.~F.,    Zepf S.~E.,
  1994, MNRAS, 271, 781

\bibitem[\protect\citeauthoryear{Creasey, Theuns \& Bower}{Creasey
  et~al.}{2013}]{creasey_how_2013}
Creasey P.,  Theuns T.,    Bower R.~G.,  2013, \mn@doi [MNRAS]
  {10.1093/mnras/sts439}, 429, 1922

\bibitem[\protect\citeauthoryear{Croton et~al.,}{Croton
  et~al.}{2006}]{croton_many_2006}
Croton D.~J.,  et~al., 2006, MNRAS, 365, 11

\bibitem[\protect\citeauthoryear{Das et~al.,}{Das
  et~al.}{2011}]{das_atacama_2011}
Das S.,  et~al., 2011, \mn@doi [ApJ] {10.1088/0004-637X/729/1/62}, 729, 62

\bibitem[\protect\citeauthoryear{De~Lucia \& Blaizot}{De~Lucia \&
  Blaizot}{2007}]{de_lucia_hierarchical_2007}
De~Lucia G.,  Blaizot J.,  2007, \mn@doi [MNRAS]
  {10.1111/j.1365-2966.2006.11287.x;}, 375, 2

\bibitem[\protect\citeauthoryear{Font et~al.,}{Font
  et~al.}{2011}]{font_population_2011}
Font A.~S.,  et~al., 2011, \mn@doi [MNRAS] {10.1111/j.1365-2966.2011.19339.x},
  417, 1260

\bibitem[\protect\citeauthoryear{Freedman, Madore, Scowcroft, Burns, Monson,
  Persson, Seibert \& Rigby}{Freedman et~al.}{2012}]{freedman_carnegie_2012}
Freedman W.~L.,  Madore B.~F.,  Scowcroft V.,  Burns C.,  Monson A.,  Persson
  S.~E.,  Seibert M.,    Rigby J.,  2012, \mn@doi [ApJ]
  {10.1088/0004-637X/758/1/24}, 758, 24

\bibitem[\protect\citeauthoryear{Frenk \& White}{Frenk \&
  White}{2012}]{frenk_dark_2012}
Frenk C.~S.,  White S. D.~M.,  2012, \mn@doi [Annalen der Physik]
  {10.1002/andp.201200212;}, 524, 507

\bibitem[\protect\citeauthoryear{Gelman \& Rubin}{Gelman \&
  Rubin}{1992}]{gelman_a._inference_1992}
Gelman A.,  Rubin D.~B.,  1992, Statistical Science, 7, 457

\bibitem[\protect\citeauthoryear{Gelman, Carlin, Ster \& Rubin}{Gelman
  et~al.}{2013}]{gelman_a._bayesian_2013}
Gelman A.,  Carlin J.~B.,  Ster H.~S.,    Rubin D.~B.,  2013, Bayesian Data
  Analysis, 3rd edn.
Chapman and {Hall/CRC}, Boca Raton, {FL}

\bibitem[\protect\citeauthoryear{Gilks}{Gilks}{1995}]{gilks_w._r._markov_1995}
Gilks W.~R.,  1995, Markov Chain Monte Carlo in Practice: Interdisciplinary
  Statistics, 1st edn.
Chapman and {Hall/CRC}

\bibitem[\protect\citeauthoryear{Gnedin}{Gnedin}{2000}]{gnedin_effect_2000}
Gnedin N.~Y.,  2000, \mn@doi [ApJ] {10.1086/317042}, 542, 535

\bibitem[\protect\citeauthoryear{G\'omez, Coleman-Smith, {O'Shea}, Tumlinson \&
  Wolpert}{G\'omez et~al.}{2012}]{gomez_characterizing_2012}
G\'omez F.~A.,  Coleman-Smith C.~E.,  {O'Shea} B.~W.,  Tumlinson J.,    Wolpert
  R.~L.,  2012, \mn@doi [ApJ] {10.1088/0004-637X/760/2/112;}, 760, 112

\bibitem[\protect\citeauthoryear{G\'omez, Coleman-Smith, {O'Shea}, Tumlinson \&
  Wolpert}{G\'omez et~al.}{2013}]{gomez_dissecting_2013}
G\'omez F.~A.,  Coleman-Smith C.~E.,  {O'Shea} B.~W.,  Tumlinson J.,    Wolpert
  R.~L.,  2013, arXiv:1311:2587

\bibitem[\protect\citeauthoryear{Grubbs}{Grubbs}{1969}]{grubbs_procedures_1969}
Grubbs F.,  1969, Technometrics, 11, 1

\bibitem[\protect\citeauthoryear{Guo et~al.,}{Guo
  et~al.}{2011}]{guo_dwarf_2011}
Guo Q.,  et~al., 2011, \mn@doi [MNRAS] {10.1111/j.1365-2966.2010.18114.x}, 413,
  101

\bibitem[\protect\citeauthoryear{Helly, Cole, Frenk, Baugh, Benson \&
  Lacey}{Helly et~al.}{2003}]{helly_galaxy_2003}
Helly J.~C.,  Cole S.,  Frenk C.~S.,  Baugh C.~M.,  Benson A.,    Lacey C.,
  2003, MNRAS, 338, 903

\bibitem[\protect\citeauthoryear{Henriques \& Thomas}{Henriques \&
  Thomas}{2010}]{henriques_tidal_2010}
Henriques B. M.~B.,  Thomas P.~A.,  2010, \mn@doi [MNRAS]
  {10.1111/j.1365-2966.2009.16151.x;}, 403, 768

\bibitem[\protect\citeauthoryear{Henriques, Thomas, Oliver \&
  Roseboom}{Henriques et~al.}{2009}]{henriques_monte_2009}
Henriques B. M.~B.,  Thomas P.~A.,  Oliver S.,    Roseboom I.,  2009, \mn@doi
  [MNRAS] {10.1111/j.1365-2966.2009.14730.x;}, 396, 535

\bibitem[\protect\citeauthoryear{Henriques, White, Thomas, Angulo, Guo, Lemson
  \& Springel}{Henriques et~al.}{2013}]{henriques_simulations_2013}
Henriques B. M.~B.,  White S. D.~M.,  Thomas P.~A.,  Angulo R.~E.,  Guo Q.,
  Lemson G.,    Springel V.,  2013, \mn@doi [MNRAS] {10.1093/mnras/stt415},
  431, 3373

\bibitem[\protect\citeauthoryear{Hinshaw et~al.,}{Hinshaw
  et~al.}{2013}]{hinshaw_nine-year_2013}
Hinshaw G.,  et~al., 2013, \mn@doi [ApJS] {10.1088/0067-0049/208/2/19}, 208, 19

\bibitem[\protect\citeauthoryear{Jiang \& van~den Bosch}{Jiang \& van~den
  Bosch}{2013}]{jiang_generating_2013}
Jiang F.,  van~den Bosch F.~C.,  2013, arXiv:1311:5225

\bibitem[\protect\citeauthoryear{Jiang, Jing, Faltenbacher, Lin \& Li}{Jiang
  et~al.}{2008}]{jiang_fitting_2008}
Jiang C.~Y.,  Jing Y.~P.,  Faltenbacher A.,  Lin W.~P.,    Li C.,  2008, {ApJ},
  675, 1095

\bibitem[\protect\citeauthoryear{Jiang, Helly, Cole \& Frenk}{Jiang
  et~al.}{2014}]{jiang_n-body_2014}
Jiang L.,  Helly J.~C.,  Cole S.,    Frenk C.~S.,  2014, \mn@doi [MNRAS]
  {10.1093/mnras/stu390}, 440, 2115

\bibitem[\protect\citeauthoryear{Keisler et~al.,}{Keisler
  et~al.}{2011}]{keisler_measurement_2011}
Keisler R.,  et~al., 2011, \mn@doi [ApJ] {10.1088/0004-637X/743/1/28}, 743, 28

\bibitem[\protect\citeauthoryear{Kere\v{s}, Katz, Weinberg \& Dav\'e}{Kere\v{s}
  et~al.}{2005}]{kerevs_how_2005}
Kere\v{s} D.,  Katz N.,  Weinberg D.~H.,    Dav\'e R.,  2005, MNRAS, 363, 2

\bibitem[\protect\citeauthoryear{Kitayama \& Suto}{Kitayama \&
  Suto}{1996}]{kitayama_semianalytic_1996}
Kitayama T.,  Suto Y.,  1996, \mn@doi [ApJ] {DOI: 10.1086/177797; eprintid:
  arXiv:astro-ph/9604141}, 469, 480

\bibitem[\protect\citeauthoryear{Kitzbichler \& White}{Kitzbichler \&
  White}{2007}]{kitzbichler_high-redshift_2007}
Kitzbichler M.~G.,  White S. D.~M.,  2007, \mn@doi [MNRAS]
  {10.1111/j.1365-2966.2007.11458.x;}, 376, 2

\bibitem[\protect\citeauthoryear{Knebe et~al.,}{Knebe
  et~al.}{2011}]{knebe_haloes_2011}
Knebe A.,  et~al., 2011, \mn@doi [MNRAS] {10.1111/j.1365-2966.2011.18858.x;},
  415, 2293

\bibitem[\protect\citeauthoryear{Kravtsov, Berlind, Wechsler, Klypin,
  Gottl\"ober, Allgood \& Primack}{Kravtsov et~al.}{2004}]{kravtsov_dark_2004}
Kravtsov A.~V.,  Berlind A.~A.,  Wechsler R.~H.,  Klypin A.~A.,  Gottl\"ober
  S.,  Allgood B.,    Primack J.~R.,  2004, \mn@doi [ApJ] {10.1086/420959;},
  609, 35

\bibitem[\protect\citeauthoryear{Lagos, Lacey \& Baugh}{Lagos
  et~al.}{2013}]{lagos_dynamical_2013}
Lagos C. d.~P.,  Lacey C.~G.,    Baugh C.~M.,  2013, \mn@doi [MNRAS]
  {10.1093/mnras/stt1696}, 436, 1787

\bibitem[\protect\citeauthoryear{Lawrence, Heitmann, White, Higdon, Wagner,
  Habib \& Williams}{Lawrence et~al.}{2010}]{lawrence_coyote_2010}
Lawrence E.,  Heitmann K.,  White M.,  Higdon D.,  Wagner C.,  Habib S.,
  Williams B.,  2010, \mn@doi [ApJ] {DOI: 10.1088/0004-637X/713/2/1322;
  eprintid: arXiv:0912.4490}, 713, 1322

\bibitem[\protect\citeauthoryear{Leauthaud et~al.,}{Leauthaud
  et~al.}{2012}]{leauthaud_new_2012}
Leauthaud A.,  et~al., 2012, \mn@doi [ApJ] {10.1088/0004-637X/744/2/159;}, 744,
  159

\bibitem[\protect\citeauthoryear{Li \& White}{Li \&
  White}{2009}]{li_distribution_2009}
Li C.,  White S. D.~M.,  2009, MNRAS, 398, 2177

\bibitem[\protect\citeauthoryear{Lilly, Peng, Carollo \& Renzini}{Lilly
  et~al.}{2013}]{lilly_phenomenological_2013}
Lilly S.~J.,  Peng Y.,  Carollo M.,    Renzini A.,  2013, in The Intriguing
  Life of Massive Galaxies, Proceedings of the International Astronomical
  Union, {IAU} Symposium. eprint: {arXiv:1302.4450}, pp 141--150,
  \mn@doi{10.1017/S1743921313004535}, \url
  {http://adsabs.harvard.edu/abs/2013IAUS..295..141L}

\bibitem[\protect\citeauthoryear{Lu, Mo, Weinberg \& Katz}{Lu
  et~al.}{2011}]{lu_bayesian_2011}
Lu Y.,  Mo H.~J.,  Weinberg M.~D.,    Katz N.,  2011, \mn@doi [MNRAS]
  {10.1111/j.1365-2966.2011.19170.x;}, 416, 1949

\bibitem[\protect\citeauthoryear{Lu, Mo, Katz \& Weinberg}{Lu
  et~al.}{2012}]{lu_bayesian_2012}
Lu Y.,  Mo H.~J.,  Katz N.,    Weinberg M.~D.,  2012, \mn@doi [MNRAS]
  {10.1111/j.1365-2966.2012.20435.x;}, 421, 1779

\bibitem[\protect\citeauthoryear{Lu, Mo, Lu, Katz \& Weinberg}{Lu
  et~al.}{2013}]{lu_bayesian_2013}
Lu Y.,  Mo H.~J.,  Lu Z.,  Katz N.,    Weinberg M.~D.,  2013, arXiv:1311:0047

\bibitem[\protect\citeauthoryear{Martin, Papastergis, Giovanelli, Haynes,
  Springob \& Stierwalt}{Martin et~al.}{2010}]{martin_arecibo_2010}
Martin A.~M.,  Papastergis E.,  Giovanelli R.,  Haynes M.~P.,  Springob C.~M.,
    Stierwalt S.,  2010, \mn@doi [ApJ] {10.1088/0004-637X/723/2/1359}, 723,
  1359

\bibitem[\protect\citeauthoryear{Martin, Shapley, Coil, Kornei, Bundy, Weiner,
  Noeske \& Schiminovich}{Martin et~al.}{2012}]{martin_demographics_2012}
Martin C.~L.,  Shapley A.~E.,  Coil A.~L.,  Kornei K.~A.,  Bundy K.,  Weiner
  B.~J.,  Noeske K.~G.,    Schiminovich D.,  2012, \mn@doi [ApJ]
  {10.1088/0004-637X/760/2/127}, 760, 127

\bibitem[\protect\citeauthoryear{Moustakas et~al.,}{Moustakas
  et~al.}{2013}]{moustakas_primus:_2013}
Moustakas J.,  et~al., 2013, \mn@doi [ApJ] {10.1088/0004-637X/767/1/50}, 767,
  50

\bibitem[\protect\citeauthoryear{Murray, Quataert \& Thompson}{Murray
  et~al.}{2005}]{murray_maximum_2005}
Murray N.,  Quataert E.,    Thompson T.~A.,  2005, \mn@doi [ApJ]
  {10.1086/426067}, 618, 569

\bibitem[\protect\citeauthoryear{Mutch, Poole \& Croton}{Mutch
  et~al.}{2013}]{mutch_constraining_2013}
Mutch S.~J.,  Poole G.~B.,    Croton D.~J.,  2013, \mn@doi [MNRAS]
  {10.1093/mnras/sts182;}, 428, 2001

\bibitem[\protect\citeauthoryear{Neistein \& Weinmann}{Neistein \&
  Weinmann}{2010}]{neistein_degeneracy_2010}
Neistein E.,  Weinmann S.~M.,  2010, \mn@doi [MNRAS]
  {10.1111/j.1365-2966.2010.16656.x;}, 405, 2717

\bibitem[\protect\citeauthoryear{Okamoto, Gao \& Theuns}{Okamoto
  et~al.}{2008}]{okamoto_mass_2008}
Okamoto T.,  Gao L.,    Theuns T.,  2008, \mn@doi [MNRAS]
  {10.1111/j.1365-2966.2008.13830.x}, 390, 920

\bibitem[\protect\citeauthoryear{Oppenheimer, Dav\'e, Kereš, Fardal, Katz,
  Kollmeier \& Weinberg}{Oppenheimer et~al.}{2010}]{oppenheimer_feedback_2010}
Oppenheimer B.~D.,  Dav\'e R.,  Kereš D.,  Fardal M.,  Katz N.,  Kollmeier
  J.~A.,    Weinberg D.~H.,  2010, \mn@doi [MNRAS]
  {10.1111/j.1365-2966.2010.16872.x}, 406, 2325

\bibitem[\protect\citeauthoryear{Padmanabhan et~al.,}{Padmanabhan
  et~al.}{2008}]{padmanabhan_improved_2008}
Padmanabhan N.,  et~al., 2008, \mn@doi [ApJ] {10.1086/524677;}, 674, 1217

\bibitem[\protect\citeauthoryear{Padmanabhan, Xu, Eisenstein, Scalzo, Cuesta,
  Mehta \& Kazin}{Padmanabhan et~al.}{2012}]{padmanabhan_2_2012}
Padmanabhan N.,  Xu X.,  Eisenstein D.~J.,  Scalzo R.,  Cuesta A.~J.,  Mehta
  K.~T.,    Kazin E.,  2012, \mn@doi [MNRAS]
  {10.1111/j.1365-2966.2012.21888.x}, 427, 2132

\bibitem[\protect\citeauthoryear{Parkinson, Cole \& Helly}{Parkinson
  et~al.}{2008}]{parkinson_generating_2008}
Parkinson H.,  Cole S.,    Helly J.,  2008, {\mnras}, 383, 557

\bibitem[\protect\citeauthoryear{Peacock \& Dodds}{Peacock \&
  Dodds}{1996}]{peacock_non-linear_1996}
Peacock J.~A.,  Dodds S.~J.,  1996, \mn@doi [MNRAS] {eprintid:
  arXiv:astro-ph/9603031}, 280, L19

\bibitem[\protect\citeauthoryear{Percival et~al.,}{Percival
  et~al.}{2007}]{percival_shape_2007}
Percival W.~J.,  et~al., 2007, ApJ, 657, 645

\bibitem[\protect\citeauthoryear{{Planck Collaboration} et~al.,}{{Planck
  Collaboration} et~al.}{2013}]{planck_collaboration_planck_2013}
{Planck Collaboration} et~al., 2013, arXiv:1303:5076

\bibitem[\protect\citeauthoryear{Riess et~al.,}{Riess
  et~al.}{2011}]{riess_3_2011}
Riess A.~G.,  et~al., 2011, \mn@doi [ApJ] {10.1088/0004-637X/730/2/119}, 730,
  119

\bibitem[\protect\citeauthoryear{Ruiz et~al.,}{Ruiz
  et~al.}{2013}]{ruiz_calibration_2013}
Ruiz A.~N.,  et~al., 2013, arXiv:1310:7034

\bibitem[\protect\citeauthoryear{Smith}{Smith}{2012}]{smith_how_2012}
Smith R.~E.,  2012, \mn@doi [MNRAS] {10.1111/j.1365-2966.2012.21745.x}, 426,
  531

\bibitem[\protect\citeauthoryear{Springel et~al.,}{Springel
  et~al.}{2005}]{springel_simulations_2005}
Springel V.,  et~al., 2005, \mn@doi [Nature] {10.1038/nature03597}, 435, 629

\bibitem[\protect\citeauthoryear{Srisawat et~al.,}{Srisawat
  et~al.}{2013}]{srisawat_sussing_2013}
Srisawat C.,  et~al., 2013, \mn@doi [MNRAS] {10.1093/mnras/stt1545}

\bibitem[\protect\citeauthoryear{Stefansky}{Stefansky}{1972}]{stefansky_rejecting_1972}
Stefansky W.,  1972, Technometrics, 14, 469

\bibitem[\protect\citeauthoryear{Sullivan et~al.,}{Sullivan
  et~al.}{2011}]{sullivan_snls3:_2011}
Sullivan M.,  et~al., 2011, \mn@doi [ApJ] {10.1088/0004-637X/737/2/102}, 737,
  102

\bibitem[\protect\citeauthoryear{Taylor et~al.,}{Taylor
  et~al.}{2011}]{taylor_galaxy_2011}
Taylor E.~N.,  et~al., 2011, \mn@doi [MNRAS]
  {10.1111/j.1365-2966.2011.19536.x;}, 418, 1587

\bibitem[\protect\citeauthoryear{Terr~Braak}{Terr~Braak}{2006}]{terr_braak_markov_2006}
Terr~Braak C. J.~F.,  2006, \mn@doi [Stat Comput] {10.1007/s11222-006-8769-1},
  16, 239

\bibitem[\protect\citeauthoryear{Tinker, Kravtsov, Klypin, Abazajian, Warren,
  Yepes, Gottl\"ober \& Holz}{Tinker et~al.}{2008}]{tinker_toward_2008}
Tinker J.,  Kravtsov A.~V.,  Klypin A.,  Abazajian K.,  Warren M.,  Yepes G.,
  Gottl\"ober S.,    Holz D.~E.,  2008, ApJ, 688, 709

\bibitem[\protect\citeauthoryear{Weinberg}{Weinberg}{2013}]{weinberg_computational_2013}
Weinberg M.~D.,  2013, \mn@doi [MNRAS] {10.1093/mnras/stt1132}, 434, 1736

\makeatother
\end{thebibliography}

\appendix

\section{Constraints on HOD Model Parameters}\label{app:HODConstraints}

In computing the covariance matrix of the \cite{li_distribution_2009} stellar mass function we make use of a parametric form for the \HOD\ given by \cite{behroozi_comprehensive_2010}. The reader is referred to that paper for a full specification of the parametric form (see also Appendix~\ref{app:OptimalHaloMasses}). We constrain 11 parameters in the \HOD\ using the \BIE, using priors as geiven in Table~\ref{tb:HODParameterPriors}.

We use the differential evolution algorithm of \cite{terr_braak_markov_2006} as our \MCMC\ sampler, with $\gamma(\equiv0.1\gamma_0)=0.0507$ where $\gamma_0=2.38/\sqrt{N_{\rm dim}}$ is the optimal fraction for a multidimensional Gaussian posterior \citep{terr_braak_markov_2006} and $N_{\rm dim}$ is the dimension of the parameter space. This results in a reasonable acceptance rate in the chains. Every 10 steps, we set $\gamma=1$ to allow chains to swap modes. 

Technically, to ensure that the chains are positively recurrent in the differential evolution algorithm it is necessary to add a random element to each proposed step in the \MCMC\ algorithm \citep{terr_braak_markov_2006}. This random element should have small variance, but long tails to large values. For this purpose we draw the random components of the proposed step from Cauchy distributions with widths, $\epsilon$, as given in Table~\ref{tb:HODParameterPriors}.

\begin{figure*}
  \renewcommand{\arraystretch}{0}
\begin{tabular}{l@{}l@{}l@{}l@{}l@{}l@{}l@{}l@{}l@{}l@{}l@{}}
\multicolumn{1}{c}{\begin{sideways}\,$\alpha$\end{sideways}}&
\multicolumn{1}{c}{\begin{sideways}\,$\log_{10}(M_1/M_\odot)$\end{sideways}}&
\multicolumn{1}{c}{\begin{sideways}\,$\log_{10}(M_{\star,0}/M_\odot)$\end{sideways}}&
\multicolumn{1}{c}{\begin{sideways}\,$\beta$\end{sideways}}&
\multicolumn{1}{c}{\begin{sideways}\,$\delta$\end{sideways}}&
\multicolumn{1}{c}{\begin{sideways}\,$\gamma$\end{sideways}}&
\multicolumn{1}{c}{\begin{sideways}\,$\sigma_{\log M_\star}$\end{sideways}}&
\multicolumn{1}{c}{\begin{sideways}\,$B_{\rm cut}$\end{sideways}}&
\multicolumn{1}{c}{\begin{sideways}\,$B_{\rm sat}$\end{sideways}}&
\multicolumn{1}{c}{\begin{sideways}\,$\beta_{\rm cut}$\end{sideways}}&
\multicolumn{1}{c}{\begin{sideways}\,$\beta_{\rm sat}$\end{sideways}}\\
\ifthenelse{\equal{\arabic{loRes}}{0}}{
\includegraphics[scale=0.1]{plots/dataCovariance/triangle_0.pdf}\vspace{-1.9pt}&\raisebox{1.9pt}{\includegraphics[scale=0.1]{plots/dataCovariance/triangle_0_1.pdf}}&\raisebox{1.9pt}{\includegraphics[scale=0.1]{plots/dataCovariance/triangle_0_2.pdf}}&\raisebox{1.9pt}{\includegraphics[scale=0.1]{plots/dataCovariance/triangle_0_3.pdf}}&\raisebox{1.9pt}{\includegraphics[scale=0.1]{plots/dataCovariance/triangle_0_4.pdf}}&\raisebox{1.9pt}{\includegraphics[scale=0.1]{plots/dataCovariance/triangle_0_5.pdf}}&\raisebox{1.9pt}{\includegraphics[scale=0.1]{plots/dataCovariance/triangle_0_6.pdf}}&\raisebox{1.9pt}{\includegraphics[scale=0.1]{plots/dataCovariance/triangle_0_7.pdf}}&\raisebox{1.9pt}{\includegraphics[scale=0.1]{plots/dataCovariance/triangle_0_8.pdf}}&\raisebox{1.9pt}{\includegraphics[scale=0.1]{plots/dataCovariance/triangle_0_9.pdf}}&\raisebox{1.9pt}{\includegraphics[scale=0.1]{plots/dataCovariance/triangle_0_10.pdf}}\\
&\hspace{-2.4pt}\includegraphics[scale=0.1]{plots/dataCovariance/triangle_1.pdf}\vspace{-2.2pt}&\raisebox{2.2pt}{\includegraphics[scale=0.1]{plots/dataCovariance/triangle_1_2.pdf}}&\raisebox{2.2pt}{\includegraphics[scale=0.1]{plots/dataCovariance/triangle_1_3.pdf}}&\raisebox{2.2pt}{\includegraphics[scale=0.1]{plots/dataCovariance/triangle_1_4.pdf}}&\raisebox{2.2pt}{\includegraphics[scale=0.1]{plots/dataCovariance/triangle_1_5.pdf}}&\raisebox{2.2pt}{\includegraphics[scale=0.1]{plots/dataCovariance/triangle_1_6.pdf}}&\raisebox{2.2pt}{\includegraphics[scale=0.1]{plots/dataCovariance/triangle_1_7.pdf}}&\raisebox{2.2pt}{\includegraphics[scale=0.1]{plots/dataCovariance/triangle_1_8.pdf}}&\raisebox{2.2pt}{\includegraphics[scale=0.1]{plots/dataCovariance/triangle_1_9.pdf}}&\raisebox{2.2pt}{\includegraphics[scale=0.1]{plots/dataCovariance/triangle_1_10.pdf}}\\
&&\hspace{-3.9pt}\includegraphics[scale=0.1]{plots/dataCovariance/triangle_2.pdf}\vspace{-2.2pt}&\raisebox{2.2pt}{\includegraphics[scale=0.1]{plots/dataCovariance/triangle_2_3.pdf}}&\raisebox{2.2pt}{\includegraphics[scale=0.1]{plots/dataCovariance/triangle_2_4.pdf}}&\raisebox{2.2pt}{\includegraphics[scale=0.1]{plots/dataCovariance/triangle_2_5.pdf}}&\raisebox{2.2pt}{\includegraphics[scale=0.1]{plots/dataCovariance/triangle_2_6.pdf}}&\raisebox{2.2pt}{\includegraphics[scale=0.1]{plots/dataCovariance/triangle_2_7.pdf}}&\raisebox{2.2pt}{\includegraphics[scale=0.1]{plots/dataCovariance/triangle_2_8.pdf}}&\raisebox{2.2pt}{\includegraphics[scale=0.1]{plots/dataCovariance/triangle_2_9.pdf}}&\raisebox{2.2pt}{\includegraphics[scale=0.1]{plots/dataCovariance/triangle_2_10.pdf}}\\
&&&\hspace{-2.8pt}\includegraphics[scale=0.1]{plots/dataCovariance/triangle_3.pdf}\vspace{-2.1pt}&\raisebox{2.1pt}{\includegraphics[scale=0.1]{plots/dataCovariance/triangle_3_4.pdf}}&\raisebox{2.1pt}{\includegraphics[scale=0.1]{plots/dataCovariance/triangle_3_5.pdf}}&\raisebox{2.1pt}{\includegraphics[scale=0.1]{plots/dataCovariance/triangle_3_6.pdf}}&\raisebox{2.1pt}{\includegraphics[scale=0.1]{plots/dataCovariance/triangle_3_7.pdf}}&\raisebox{2.1pt}{\includegraphics[scale=0.1]{plots/dataCovariance/triangle_3_8.pdf}}&\raisebox{2.1pt}{\includegraphics[scale=0.1]{plots/dataCovariance/triangle_3_9.pdf}}&\raisebox{2.1pt}{\includegraphics[scale=0.1]{plots/dataCovariance/triangle_3_10.pdf}}\\
&&&&\hspace{-2.8pt}\includegraphics[scale=0.1]{plots/dataCovariance/triangle_4.pdf}\vspace{-1.9pt}&\raisebox{1.9pt}{\includegraphics[scale=0.1]{plots/dataCovariance/triangle_4_5.pdf}}&\raisebox{1.9pt}{\includegraphics[scale=0.1]{plots/dataCovariance/triangle_4_6.pdf}}&\raisebox{1.9pt}{\includegraphics[scale=0.1]{plots/dataCovariance/triangle_4_7.pdf}}&\raisebox{1.9pt}{\includegraphics[scale=0.1]{plots/dataCovariance/triangle_4_8.pdf}}&\raisebox{1.9pt}{\includegraphics[scale=0.1]{plots/dataCovariance/triangle_4_9.pdf}}&\raisebox{1.9pt}{\includegraphics[scale=0.1]{plots/dataCovariance/triangle_4_10.pdf}}\\
&&&&&\hspace{-2.4pt}\includegraphics[scale=0.1]{plots/dataCovariance/triangle_5.pdf}\vspace{-2.1pt}&\raisebox{2.1pt}{\includegraphics[scale=0.1]{plots/dataCovariance/triangle_5_6.pdf}}&\raisebox{2.1pt}{\includegraphics[scale=0.1]{plots/dataCovariance/triangle_5_7.pdf}}&\raisebox{2.1pt}{\includegraphics[scale=0.1]{plots/dataCovariance/triangle_5_8.pdf}}&\raisebox{2.1pt}{\includegraphics[scale=0.1]{plots/dataCovariance/triangle_5_9.pdf}}&\raisebox{2.1pt}{\includegraphics[scale=0.1]{plots/dataCovariance/triangle_5_10.pdf}}\\
&&&&&&\hspace{-2.8pt}\includegraphics[scale=0.1]{plots/dataCovariance/triangle_6.pdf}\vspace{-2.2pt}&\raisebox{2.2pt}{\includegraphics[scale=0.1]{plots/dataCovariance/triangle_6_7.pdf}}&\raisebox{2.2pt}{\includegraphics[scale=0.1]{plots/dataCovariance/triangle_6_8.pdf}}&\raisebox{2.2pt}{\includegraphics[scale=0.1]{plots/dataCovariance/triangle_6_9.pdf}}&\raisebox{2.2pt}{\includegraphics[scale=0.1]{plots/dataCovariance/triangle_6_10.pdf}}\\
&&&&&&&\hspace{-4.1pt}\includegraphics[scale=0.1]{plots/dataCovariance/triangle_7.pdf}\vspace{-2.3pt}&\raisebox{2.3pt}{\includegraphics[scale=0.1]{plots/dataCovariance/triangle_7_8.pdf}}&\raisebox{2.3pt}{\includegraphics[scale=0.1]{plots/dataCovariance/triangle_7_9.pdf}}&\raisebox{2.3pt}{\includegraphics[scale=0.1]{plots/dataCovariance/triangle_7_10.pdf}}\\
&&&&&&&&\hspace{-3.7pt}\includegraphics[scale=0.1]{plots/dataCovariance/triangle_8.pdf}\vspace{-2.1pt}&\raisebox{2.1pt}{\includegraphics[scale=0.1]{plots/dataCovariance/triangle_8_9.pdf}}&\raisebox{2.1pt}{\includegraphics[scale=0.1]{plots/dataCovariance/triangle_8_10.pdf}}\\
&&&&&&&&&\hspace{-3.3pt}\includegraphics[scale=0.1]{plots/dataCovariance/triangle_9.pdf}\vspace{-2.1pt}&\raisebox{2.1pt}{\includegraphics[scale=0.1]{plots/dataCovariance/triangle_9_10.pdf}}\\
&&&&&&&&&&\hspace{-3.3pt}\includegraphics[scale=0.1]{plots/dataCovariance/triangle_10.pdf}\vspace{-2.1pt}\\
}
{
\includegraphics[scale=0.1]{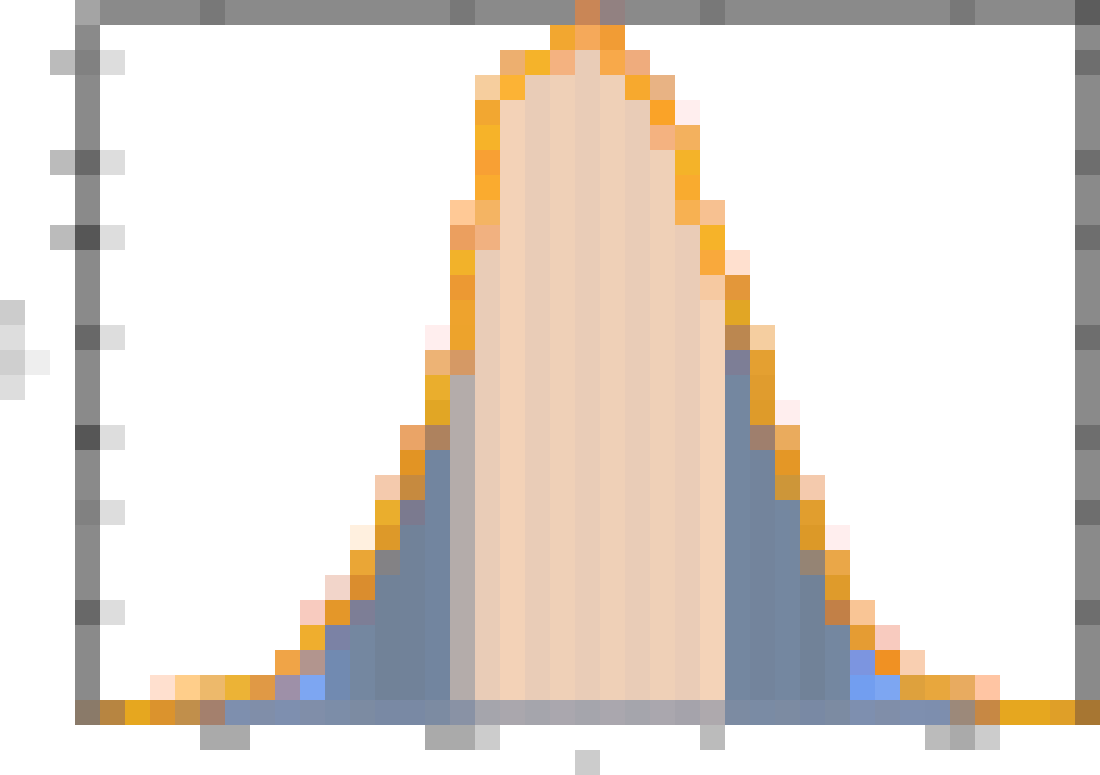}\vspace{-1.9pt}&\raisebox{1.9pt}{\includegraphics[scale=0.1]{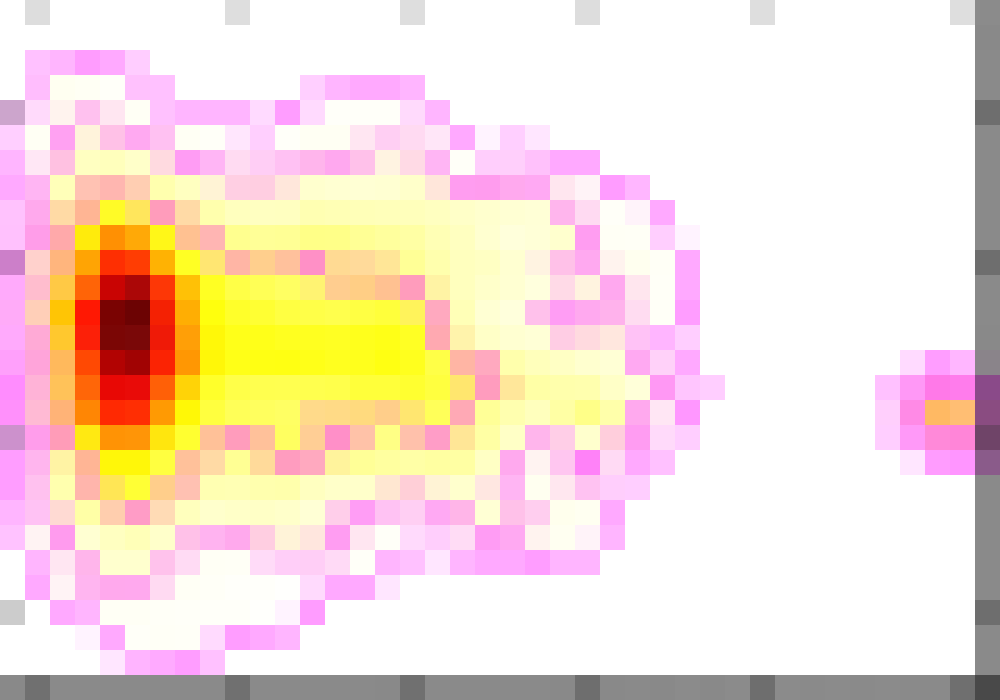}}&\raisebox{1.9pt}{\includegraphics[scale=0.1]{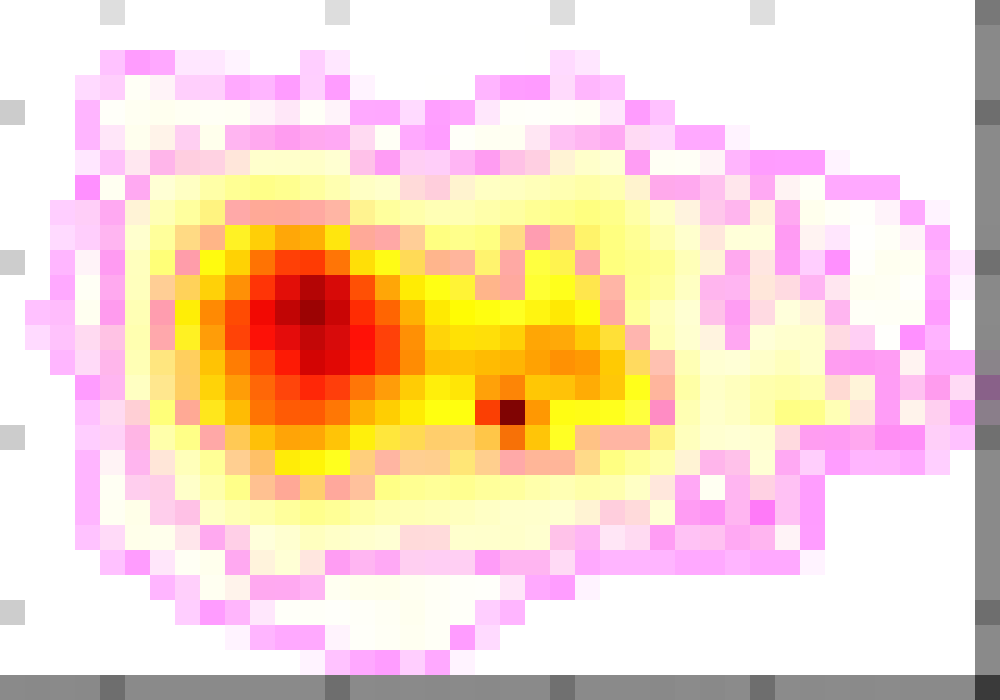}}&\raisebox{1.9pt}{\includegraphics[scale=0.1]{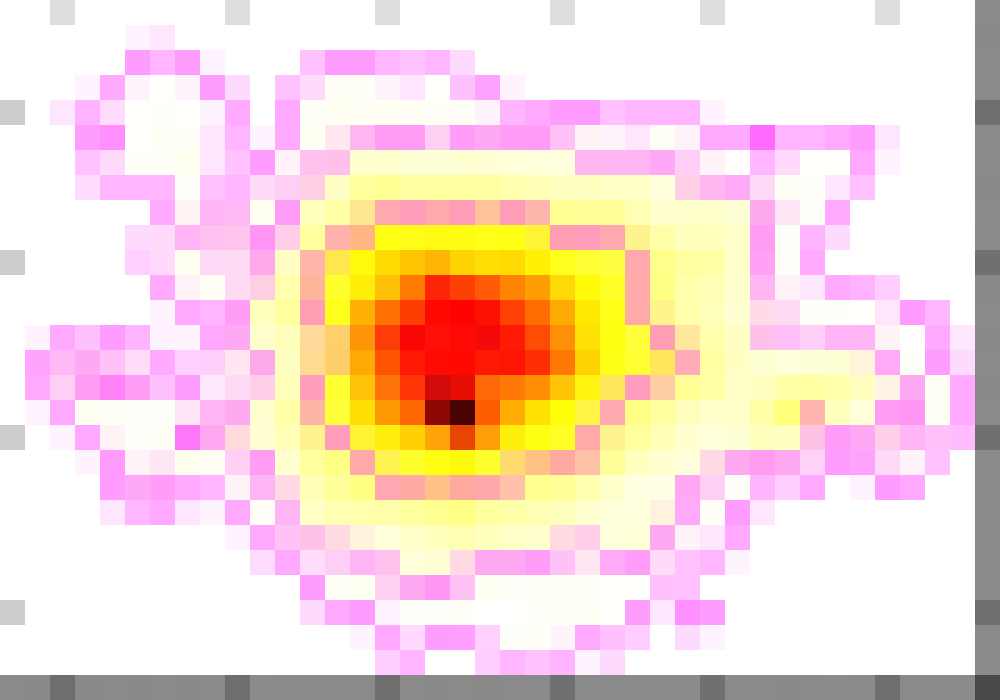}}&\raisebox{1.9pt}{\includegraphics[scale=0.1]{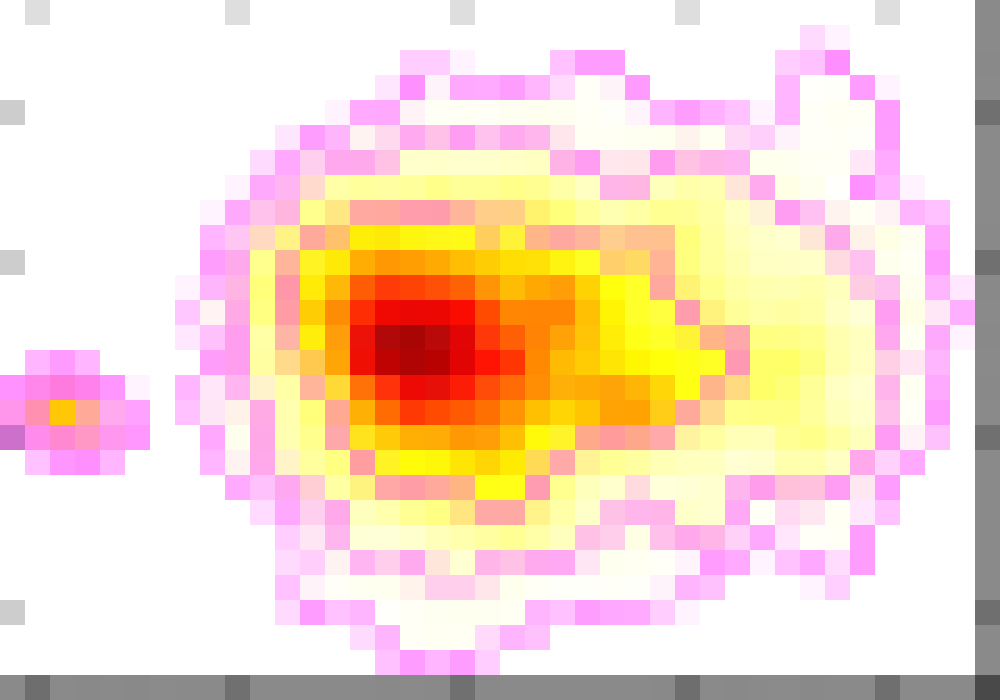}}&\raisebox{1.9pt}{\includegraphics[scale=0.1]{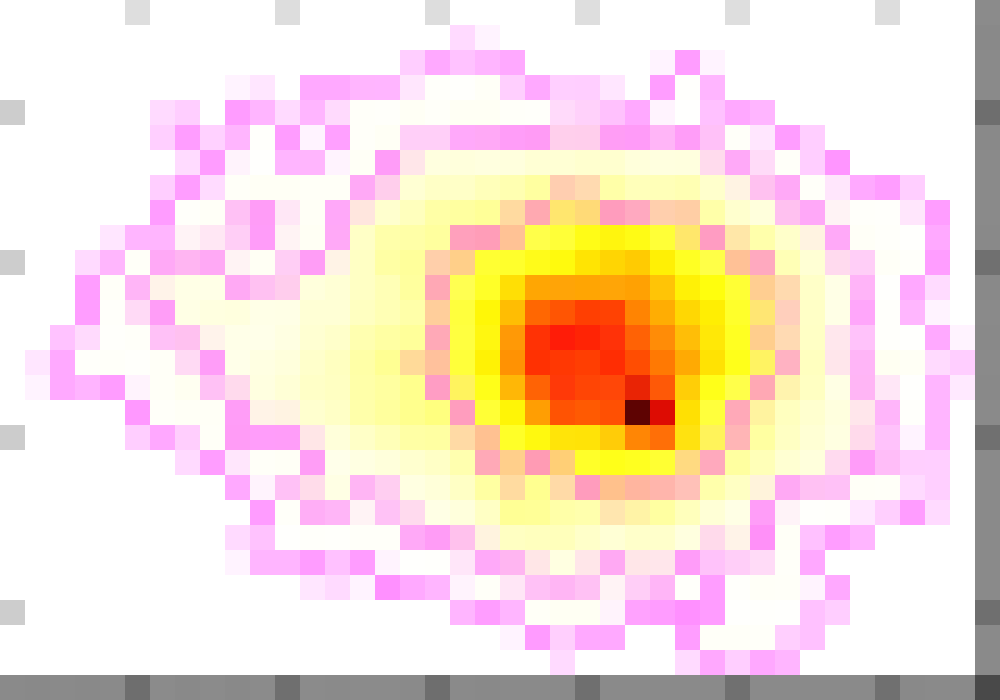}}&\raisebox{1.9pt}{\includegraphics[scale=0.1]{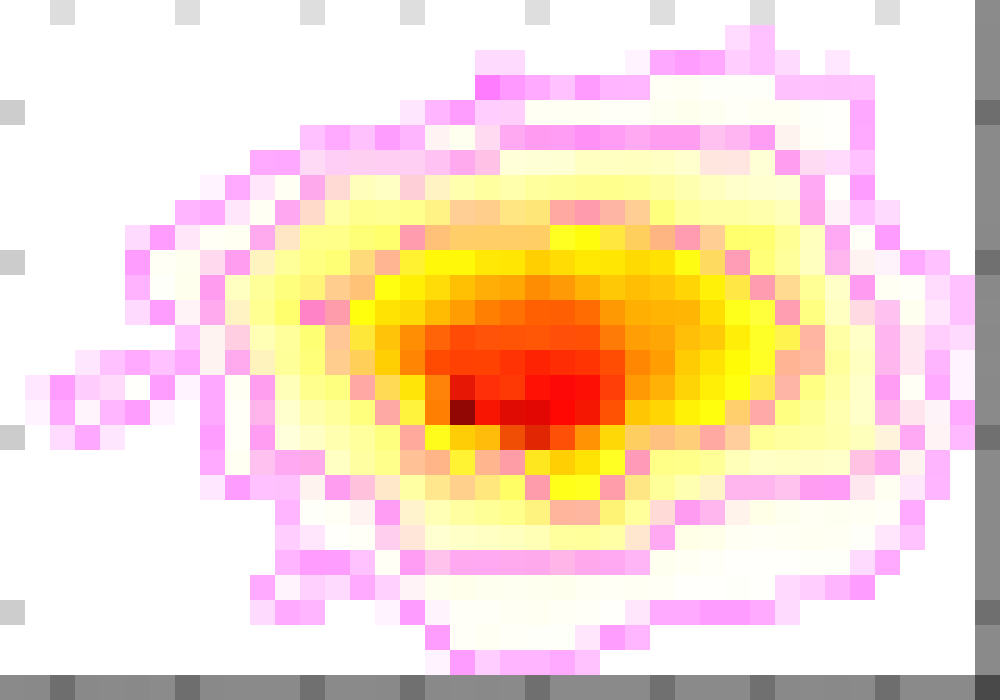}}&\raisebox{1.9pt}{\includegraphics[scale=0.1]{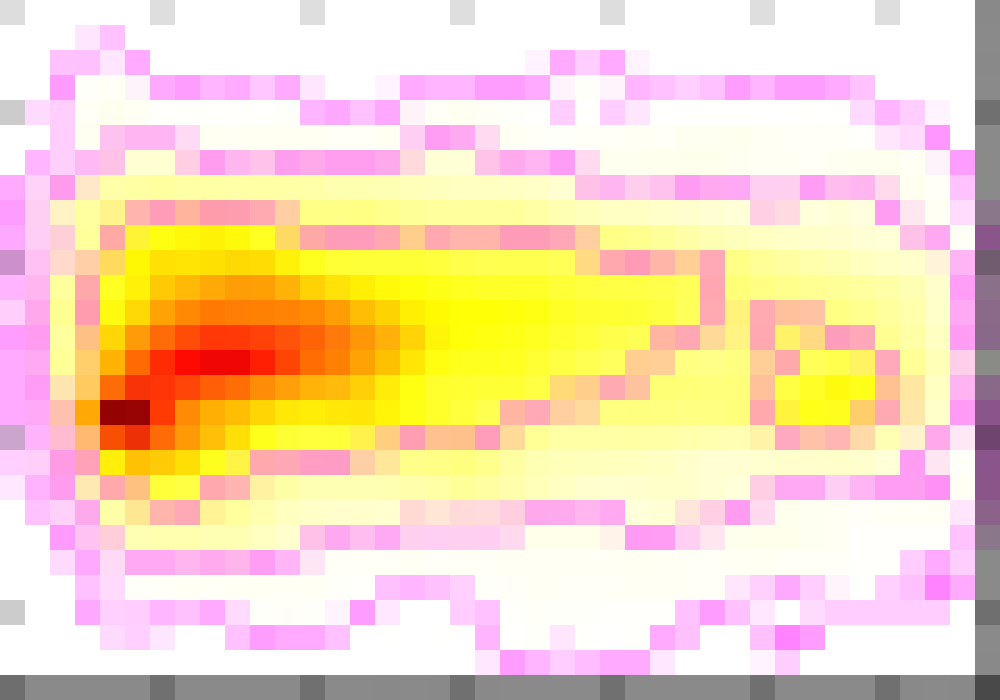}}&\raisebox{1.9pt}{\includegraphics[scale=0.1]{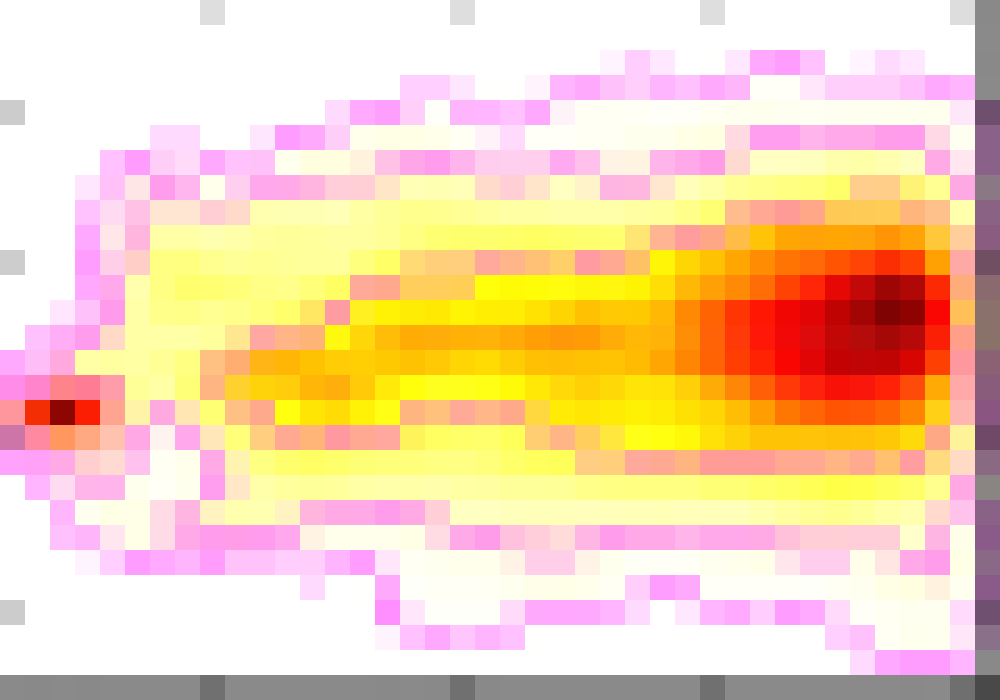}}&\raisebox{1.9pt}{\includegraphics[scale=0.1]{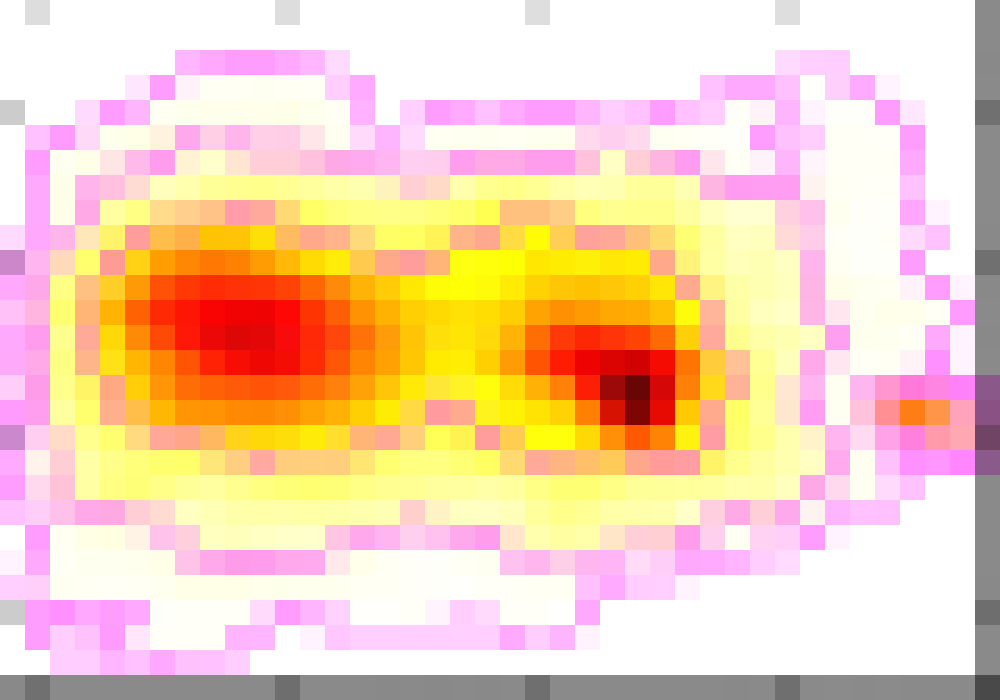}}&\raisebox{1.9pt}{\includegraphics[scale=0.1]{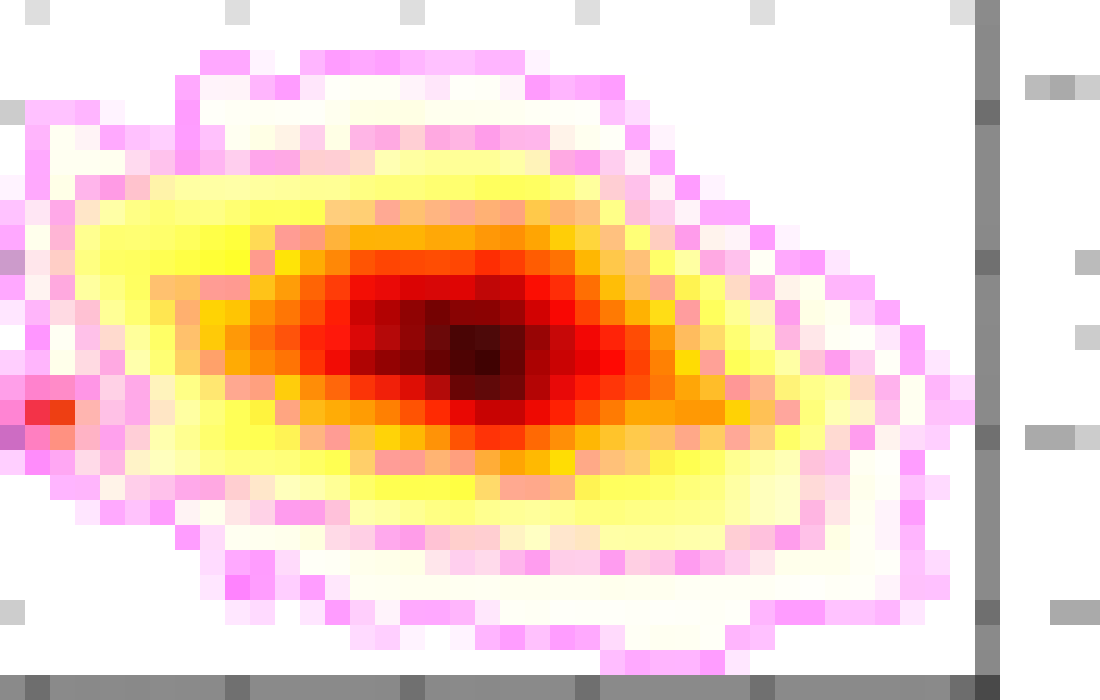}}\\
&\hspace{-2.4pt}\includegraphics[scale=0.1]{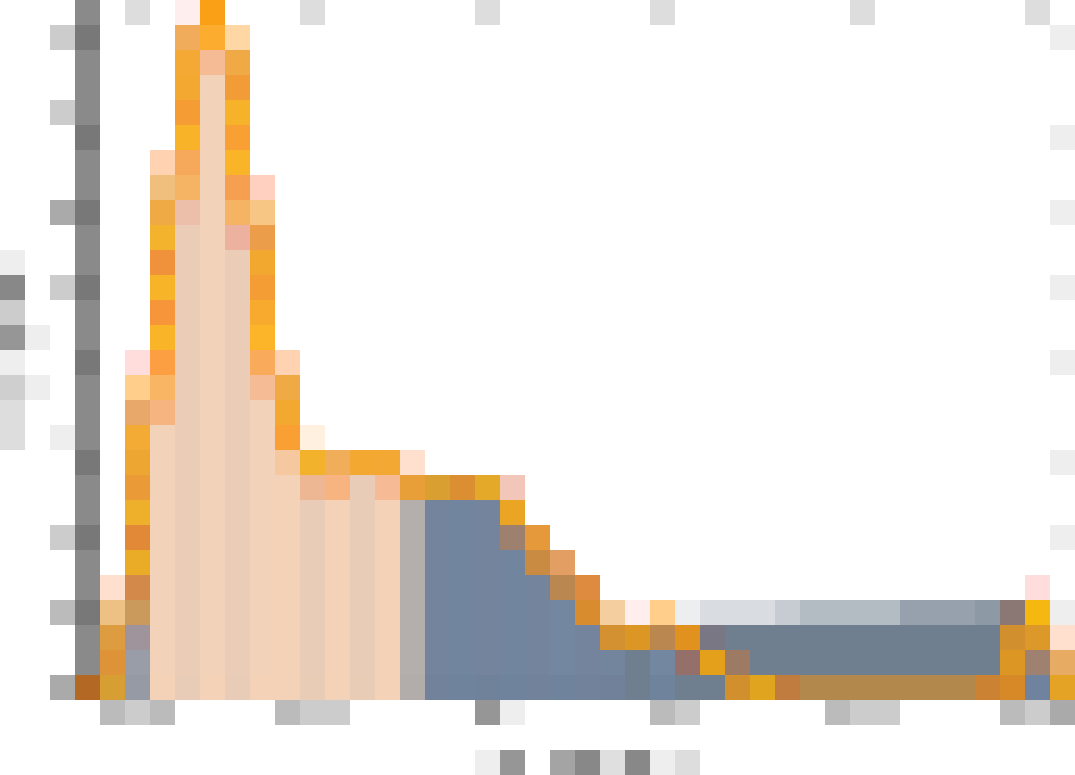}\vspace{-2.2pt}&\raisebox{2.2pt}{\includegraphics[scale=0.1]{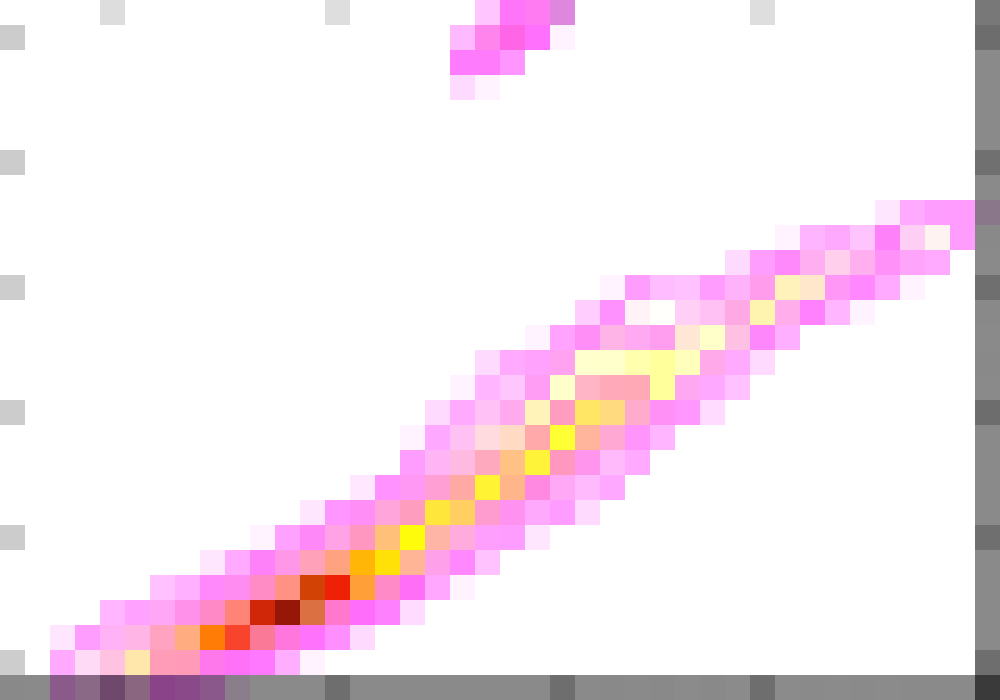}}&\raisebox{2.2pt}{\includegraphics[scale=0.1]{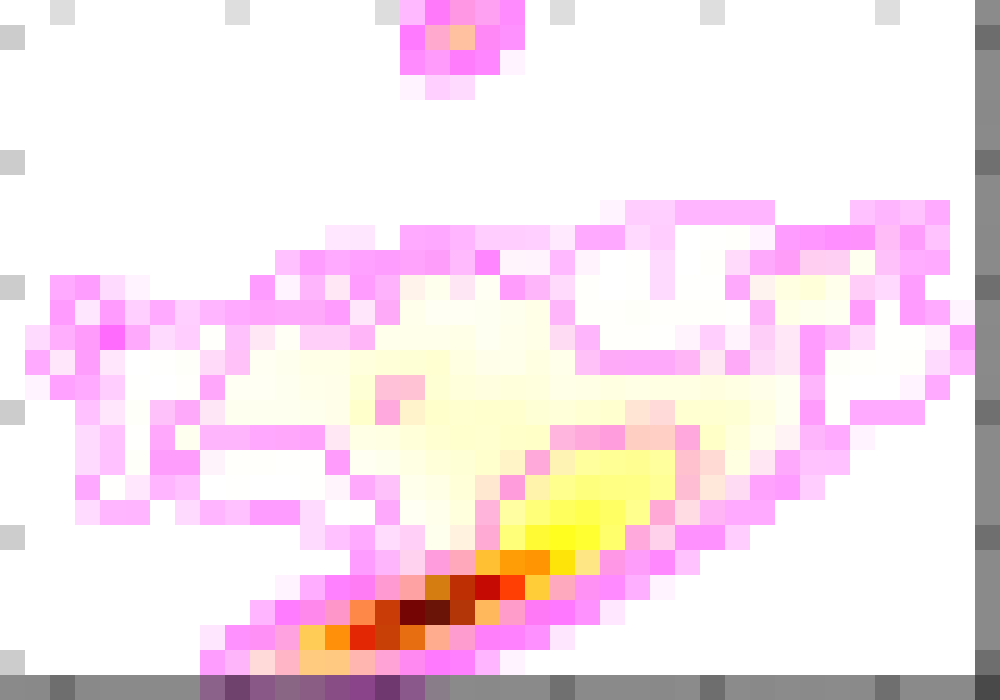}}&\raisebox{2.2pt}{\includegraphics[scale=0.1]{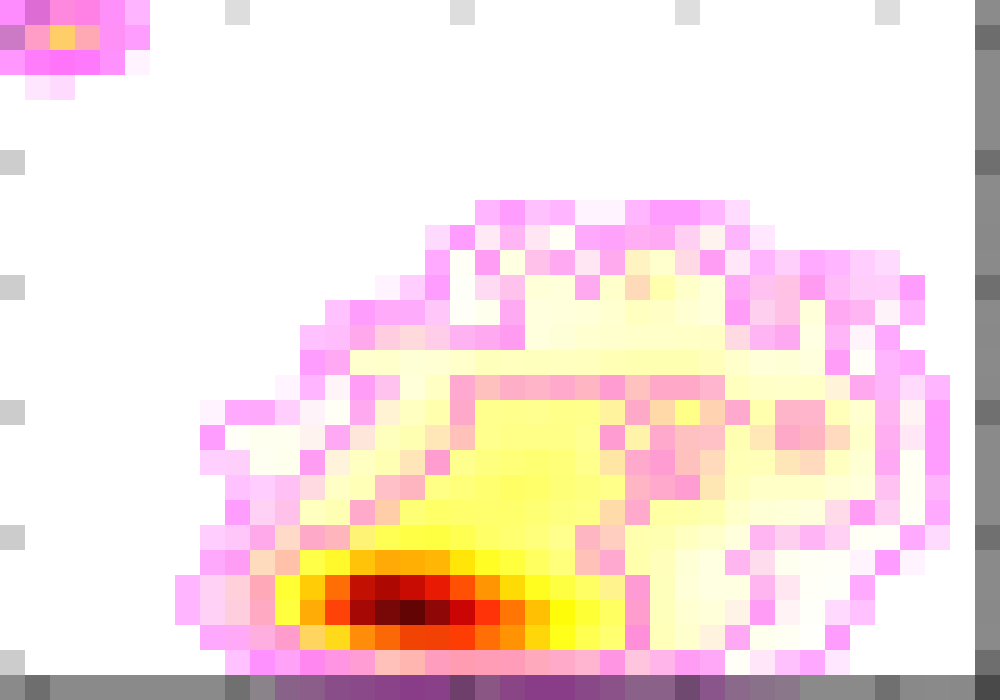}}&\raisebox{2.2pt}{\includegraphics[scale=0.1]{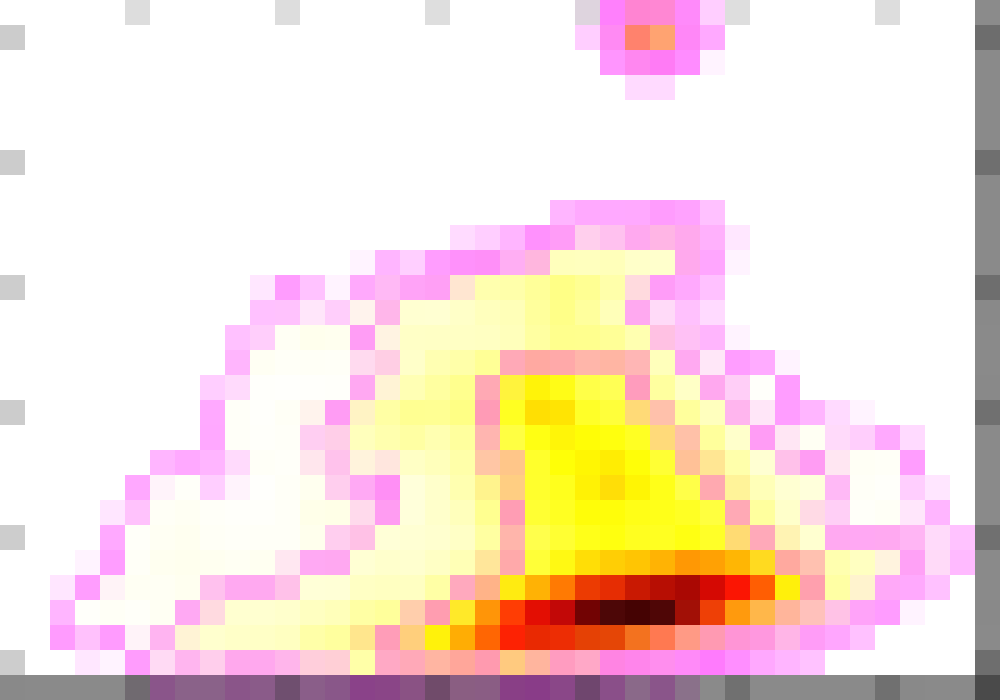}}&\raisebox{2.2pt}{\includegraphics[scale=0.1]{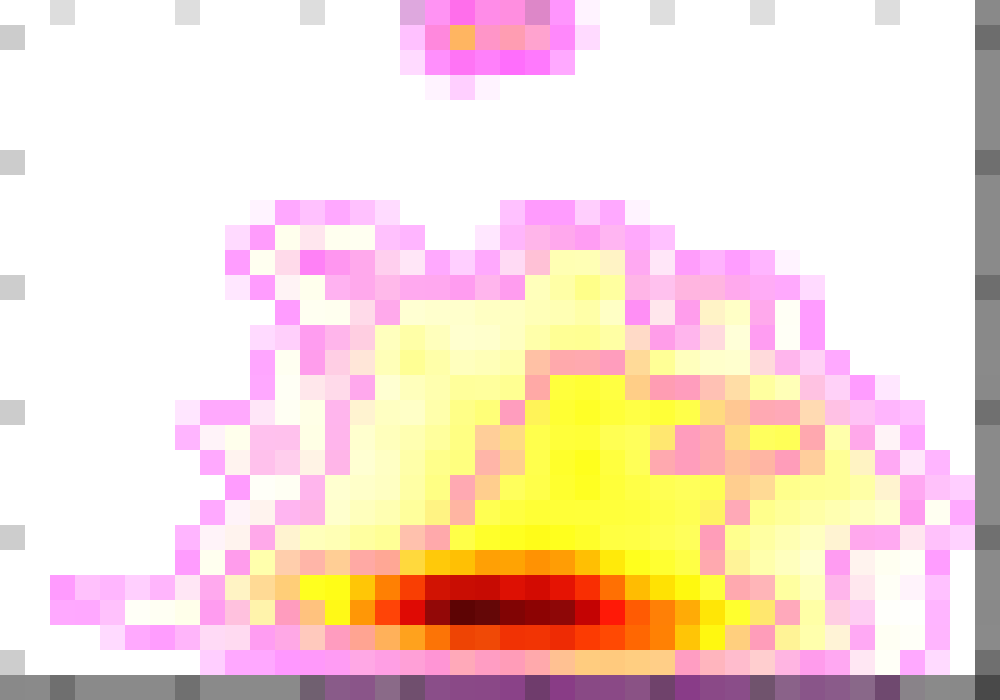}}&\raisebox{2.2pt}{\includegraphics[scale=0.1]{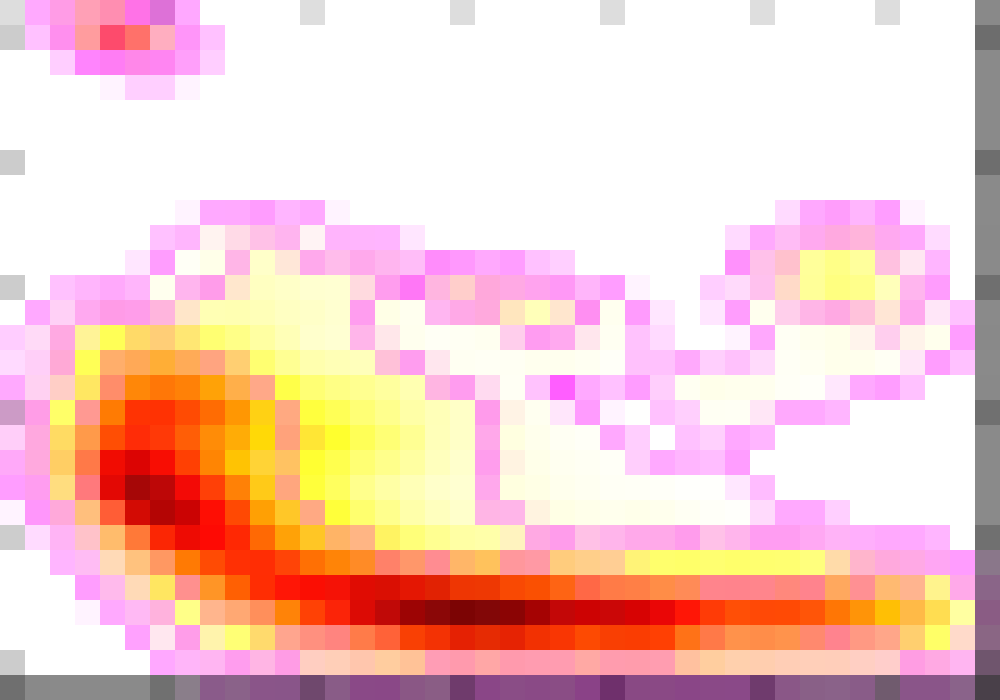}}&\raisebox{2.2pt}{\includegraphics[scale=0.1]{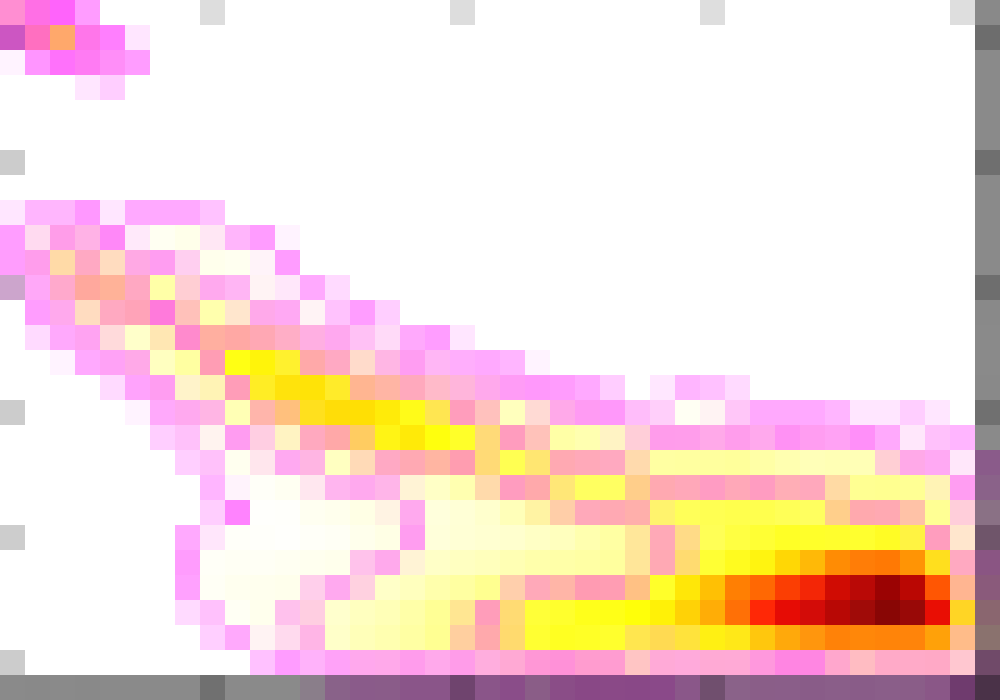}}&\raisebox{2.2pt}{\includegraphics[scale=0.1]{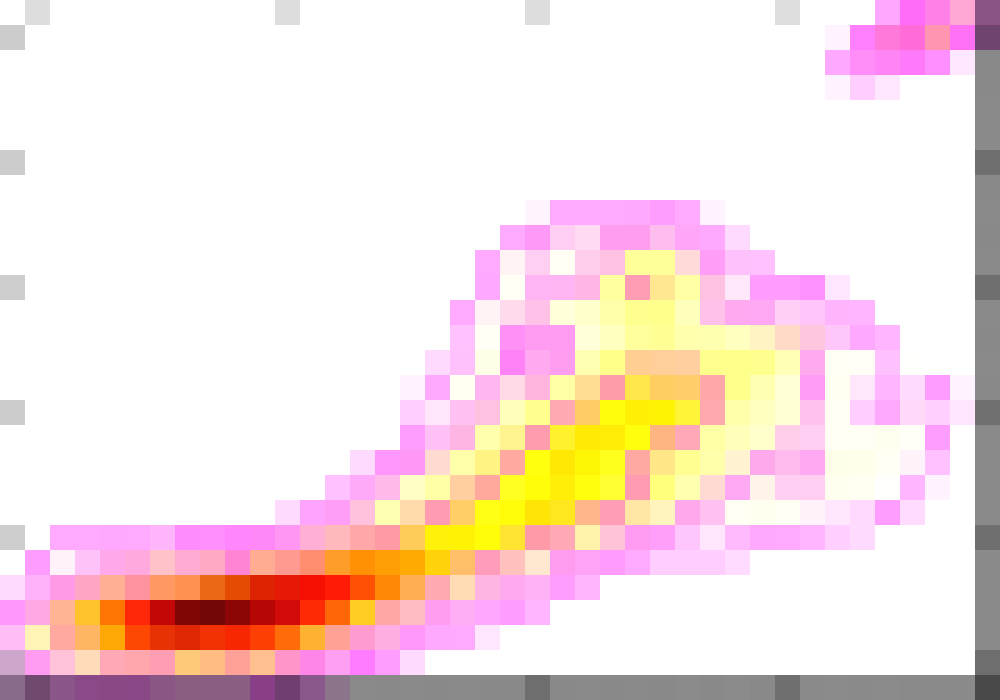}}&\raisebox{2.2pt}{\includegraphics[scale=0.1]{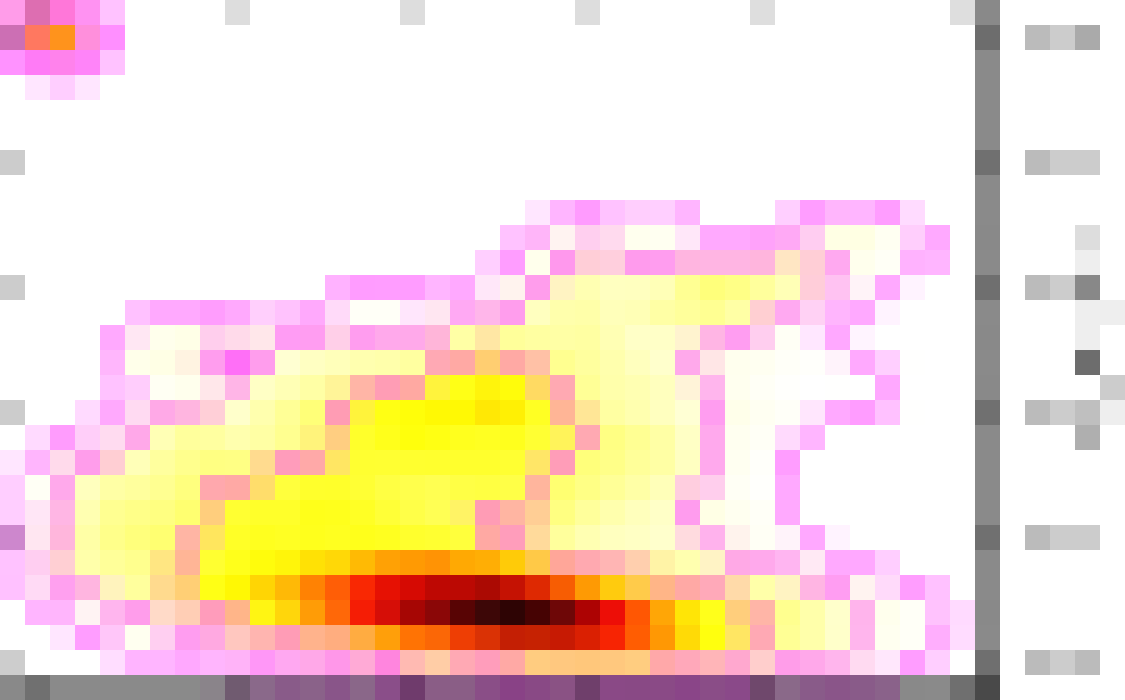}}\\
&&\hspace{-3.9pt}\includegraphics[scale=0.1]{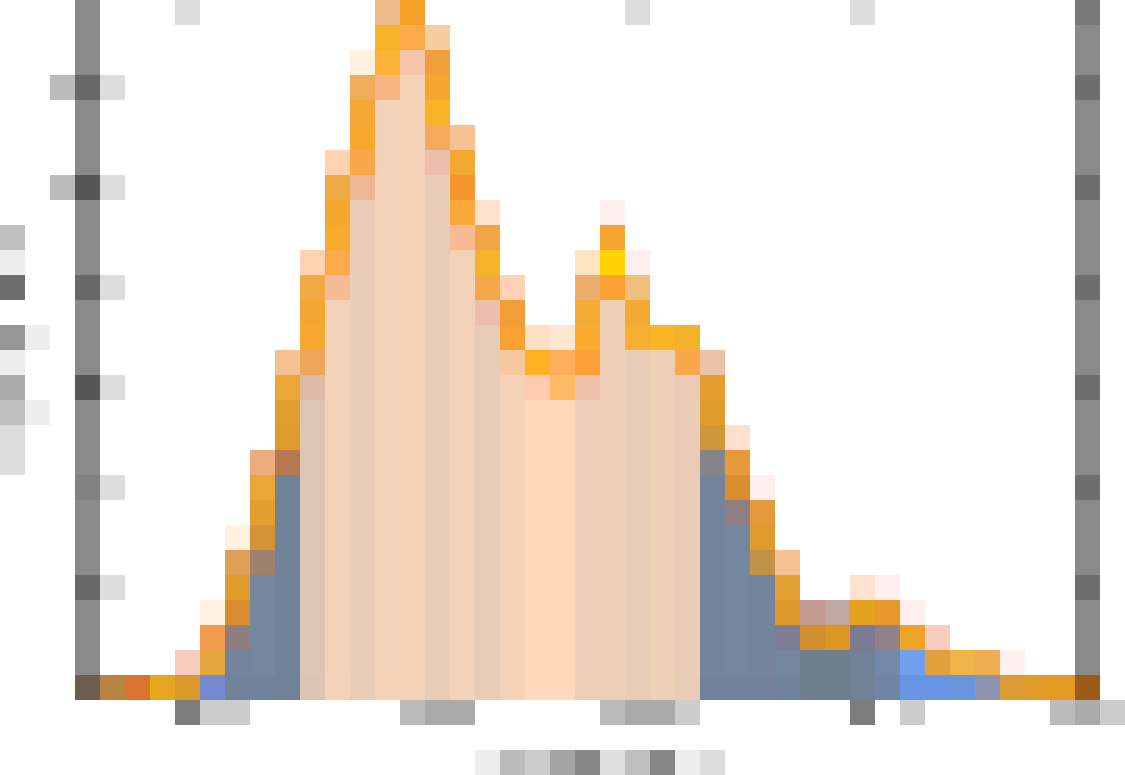}\vspace{-2.2pt}&\raisebox{2.2pt}{\includegraphics[scale=0.1]{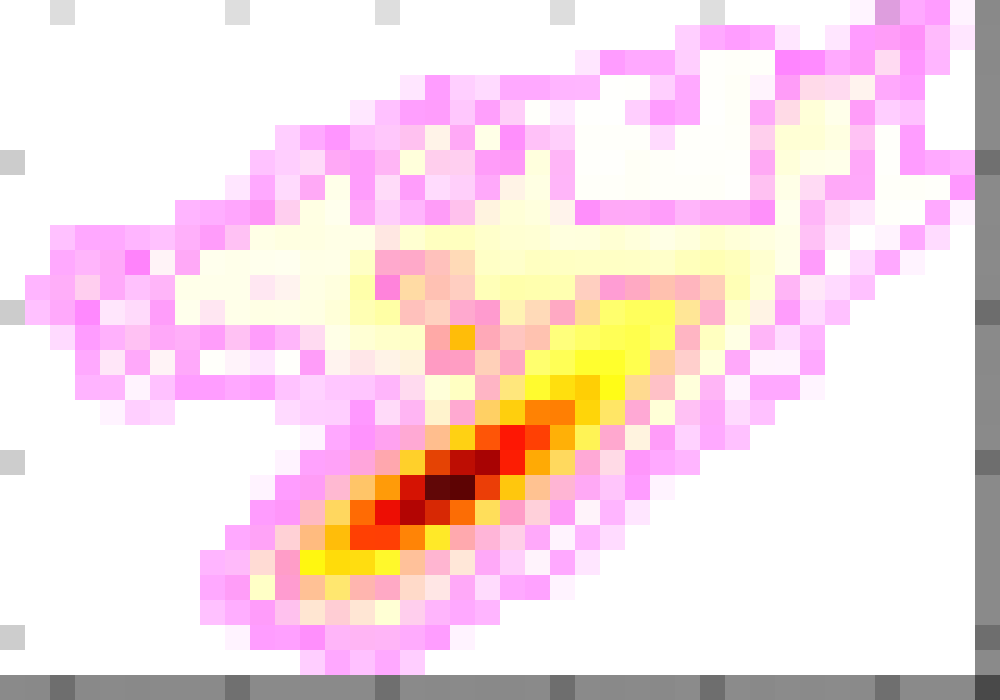}}&\raisebox{2.2pt}{\includegraphics[scale=0.1]{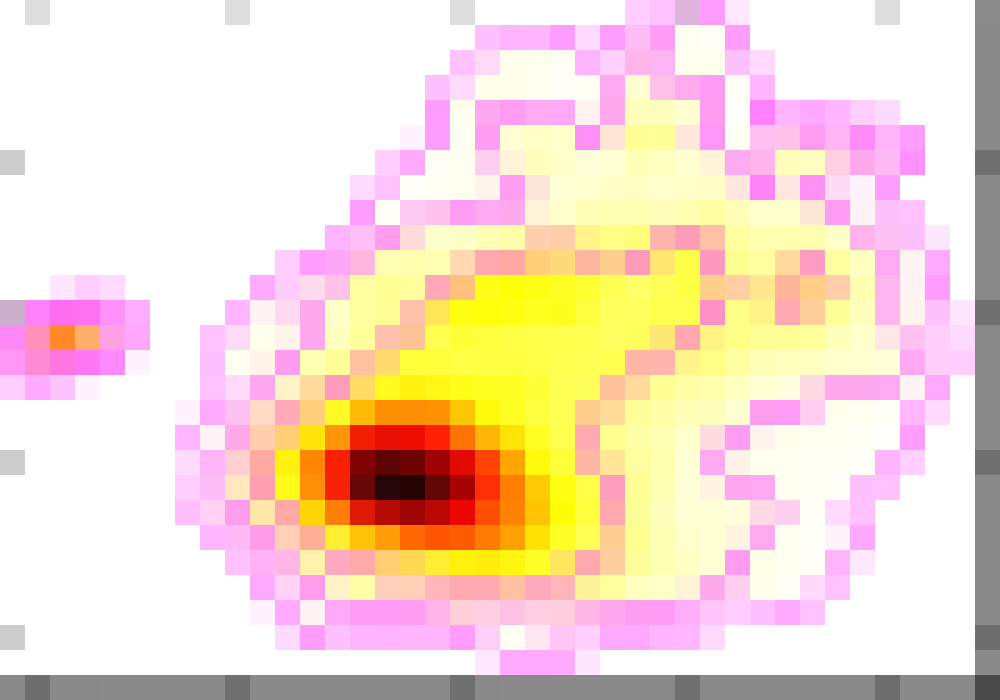}}&\raisebox{2.2pt}{\includegraphics[scale=0.1]{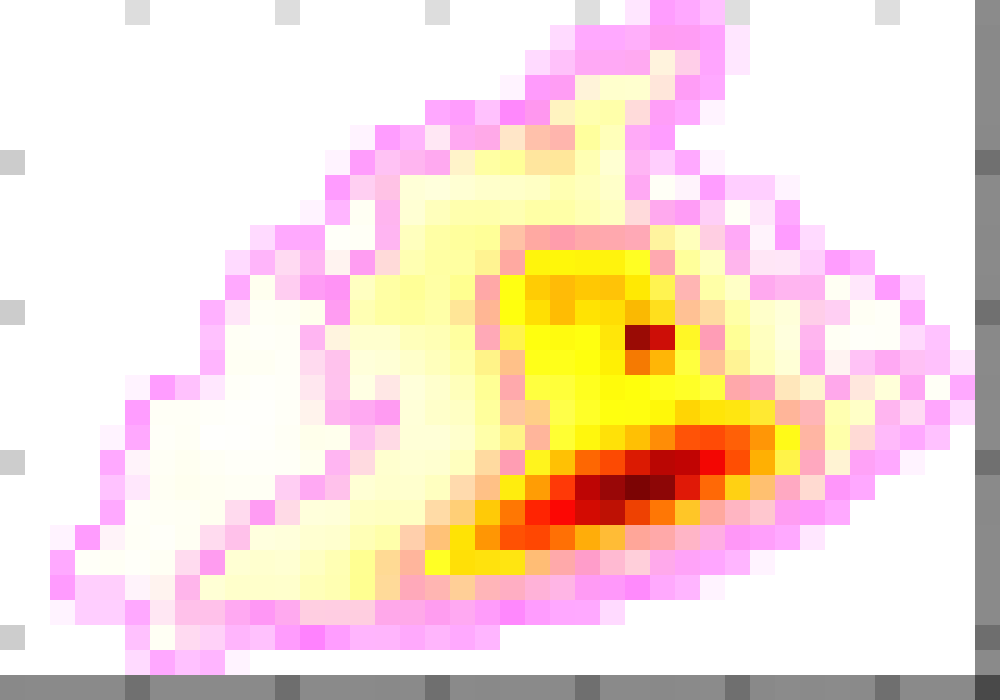}}&\raisebox{2.2pt}{\includegraphics[scale=0.1]{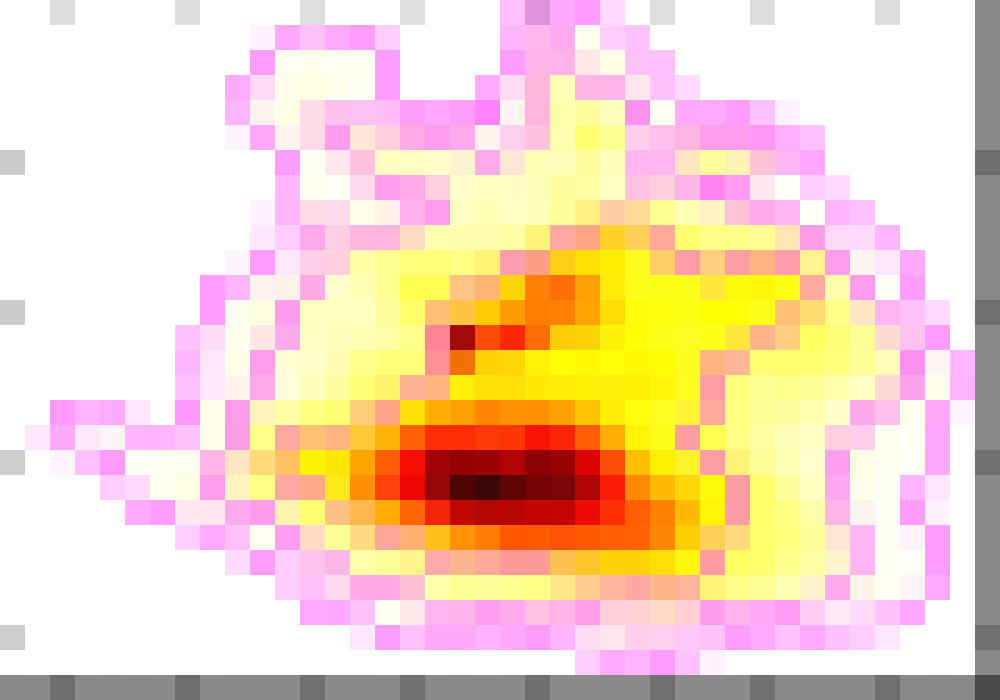}}&\raisebox{2.2pt}{\includegraphics[scale=0.1]{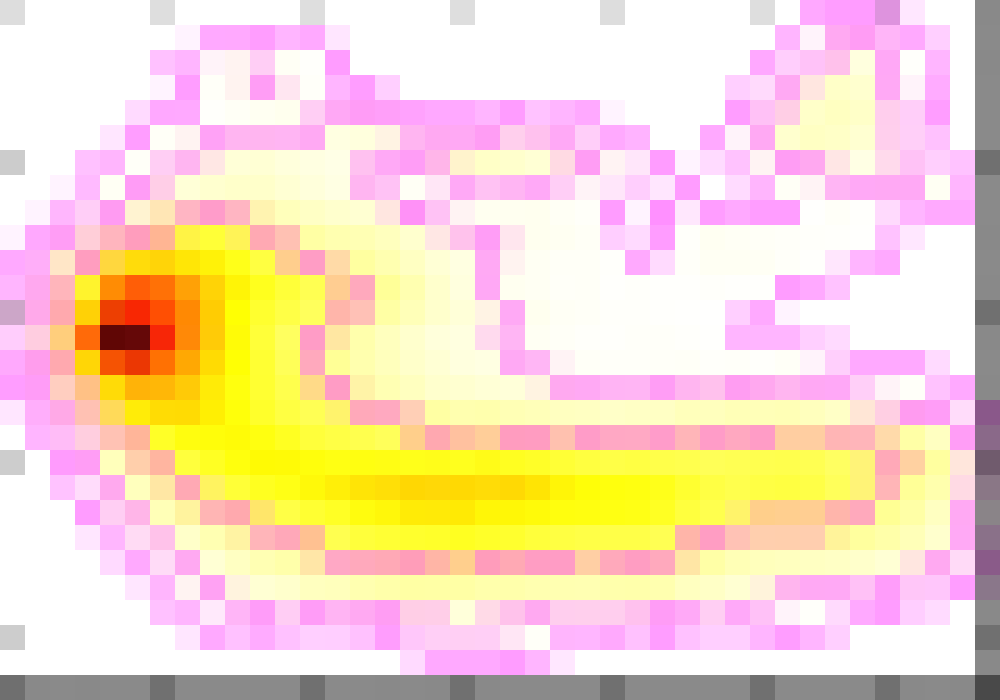}}&\raisebox{2.2pt}{\includegraphics[scale=0.1]{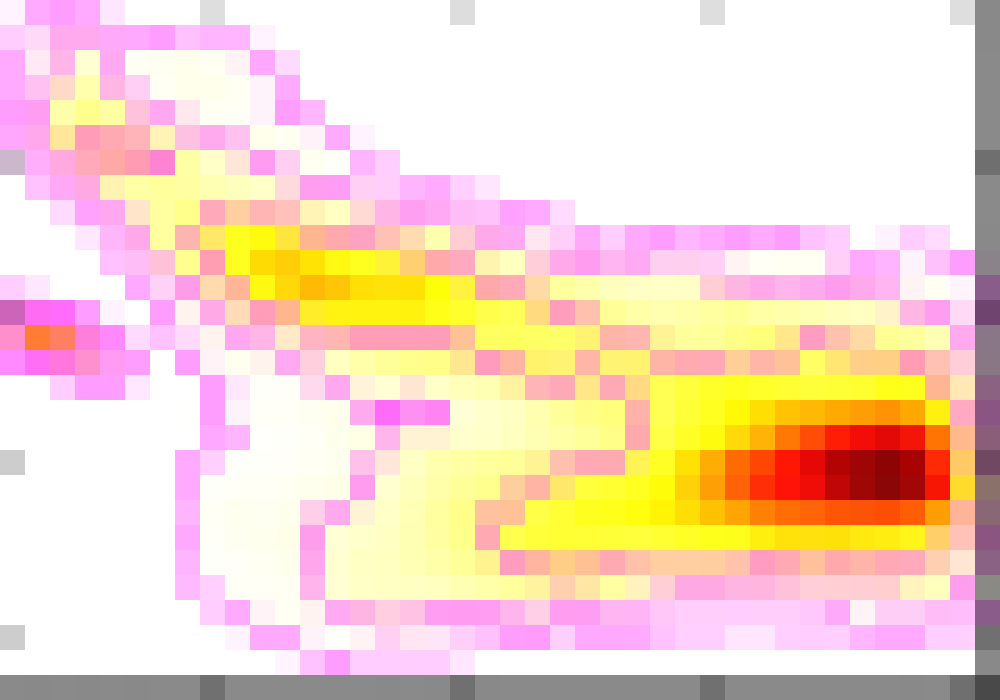}}&\raisebox{2.2pt}{\includegraphics[scale=0.1]{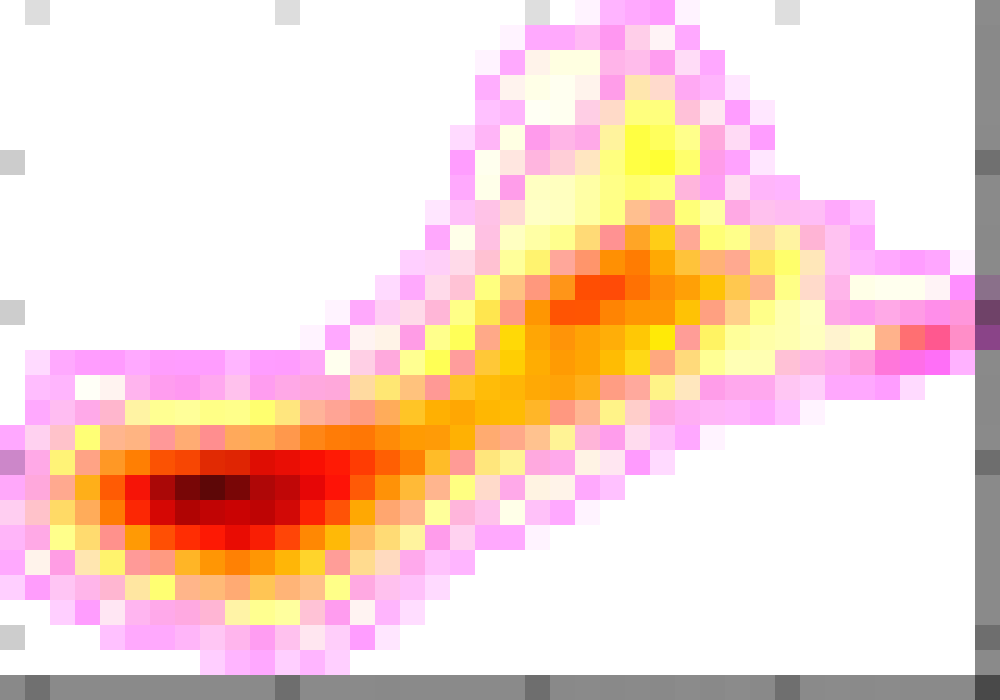}}&\raisebox{2.2pt}{\includegraphics[scale=0.1]{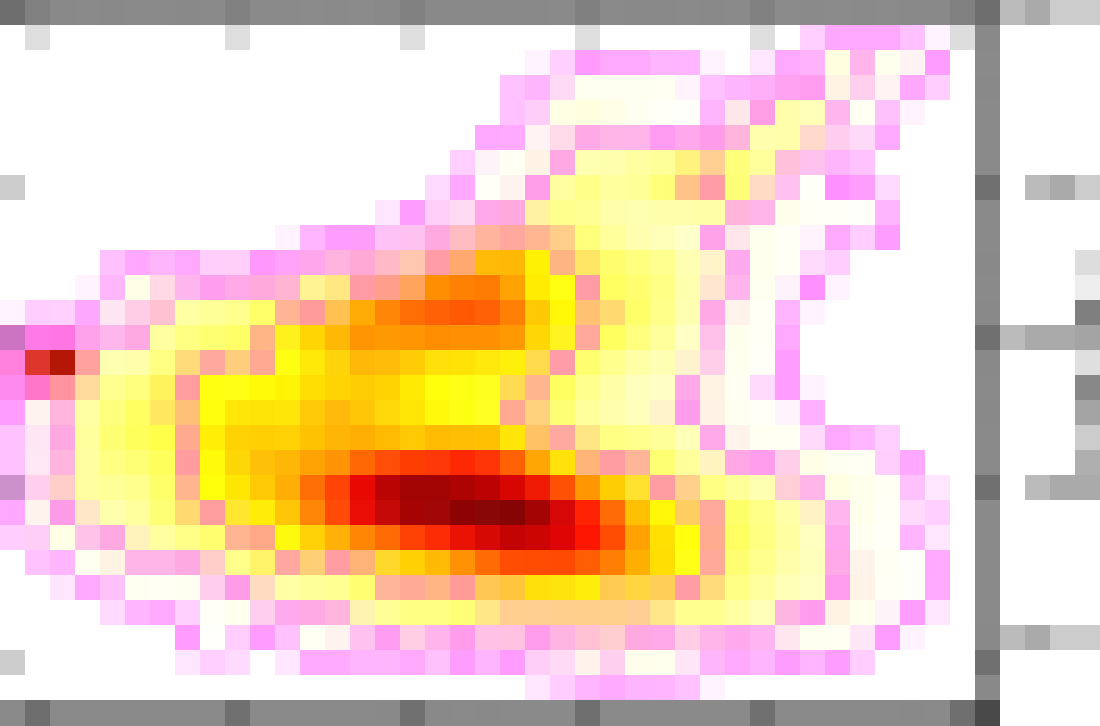}}\\
&&&\hspace{-2.8pt}\includegraphics[scale=0.1]{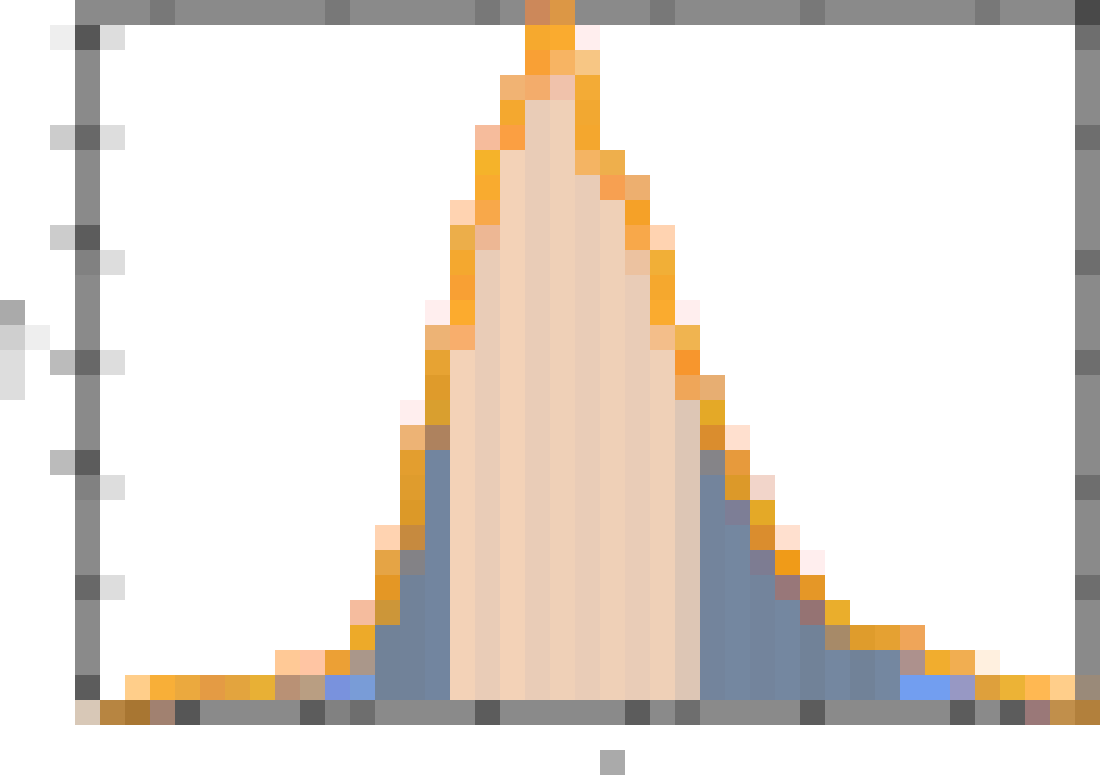}\vspace{-2.1pt}&\raisebox{2.1pt}{\includegraphics[scale=0.1]{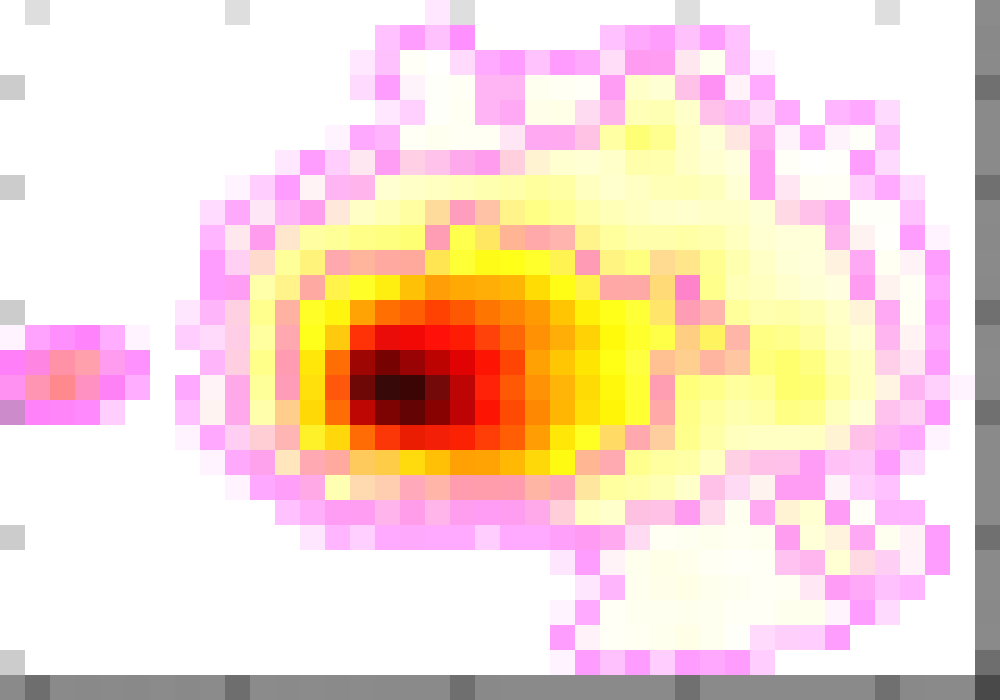}}&\raisebox{2.1pt}{\includegraphics[scale=0.1]{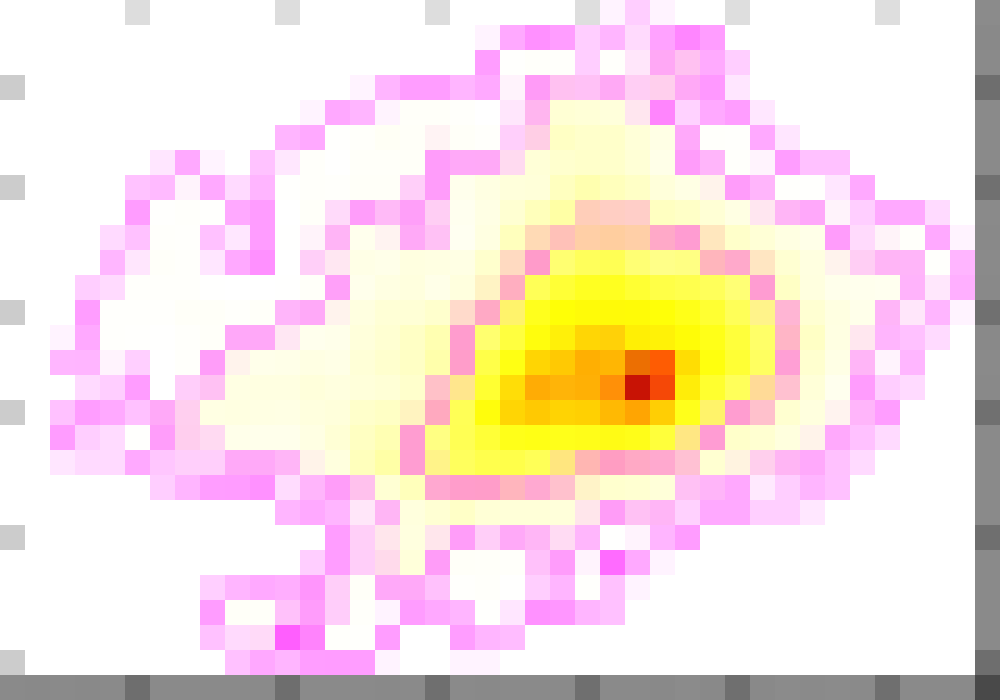}}&\raisebox{2.1pt}{\includegraphics[scale=0.1]{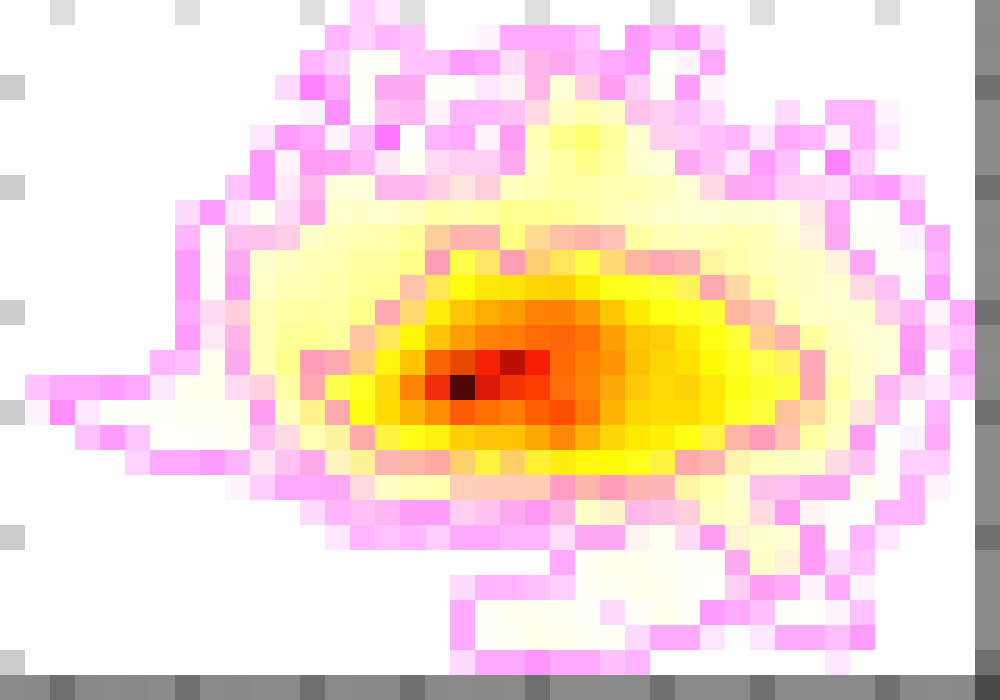}}&\raisebox{2.1pt}{\includegraphics[scale=0.1]{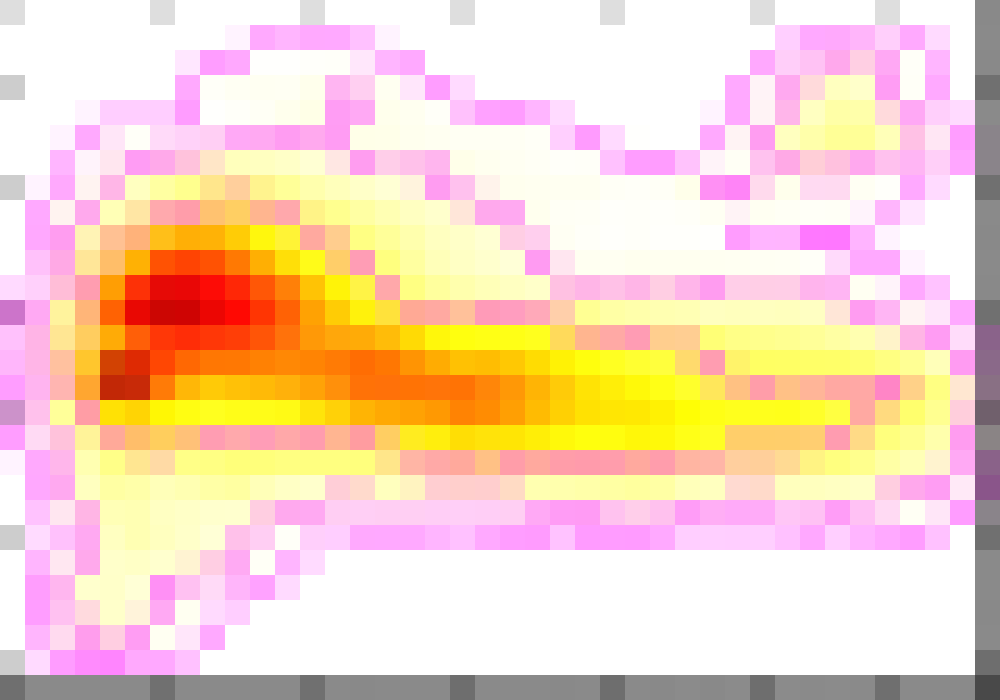}}&\raisebox{2.1pt}{\includegraphics[scale=0.1]{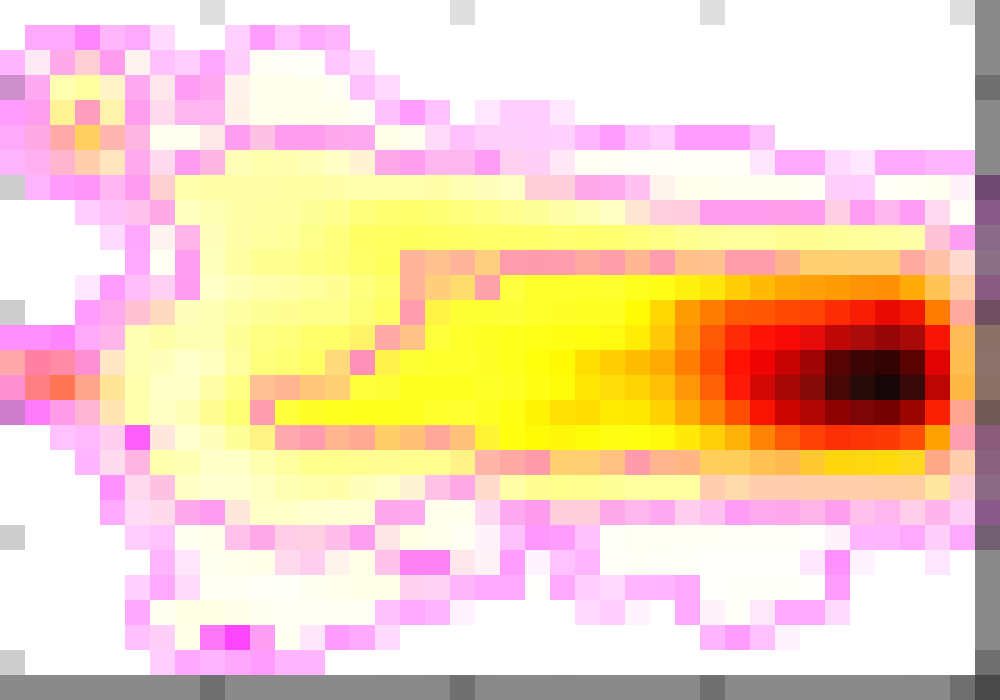}}&\raisebox{2.1pt}{\includegraphics[scale=0.1]{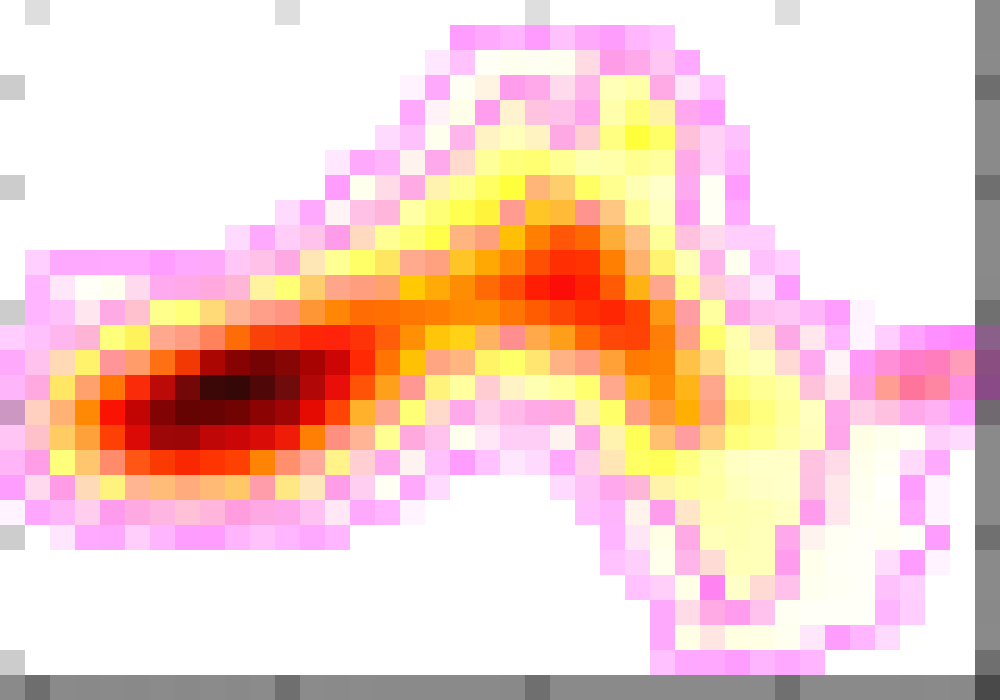}}&\raisebox{2.1pt}{\includegraphics[scale=0.1]{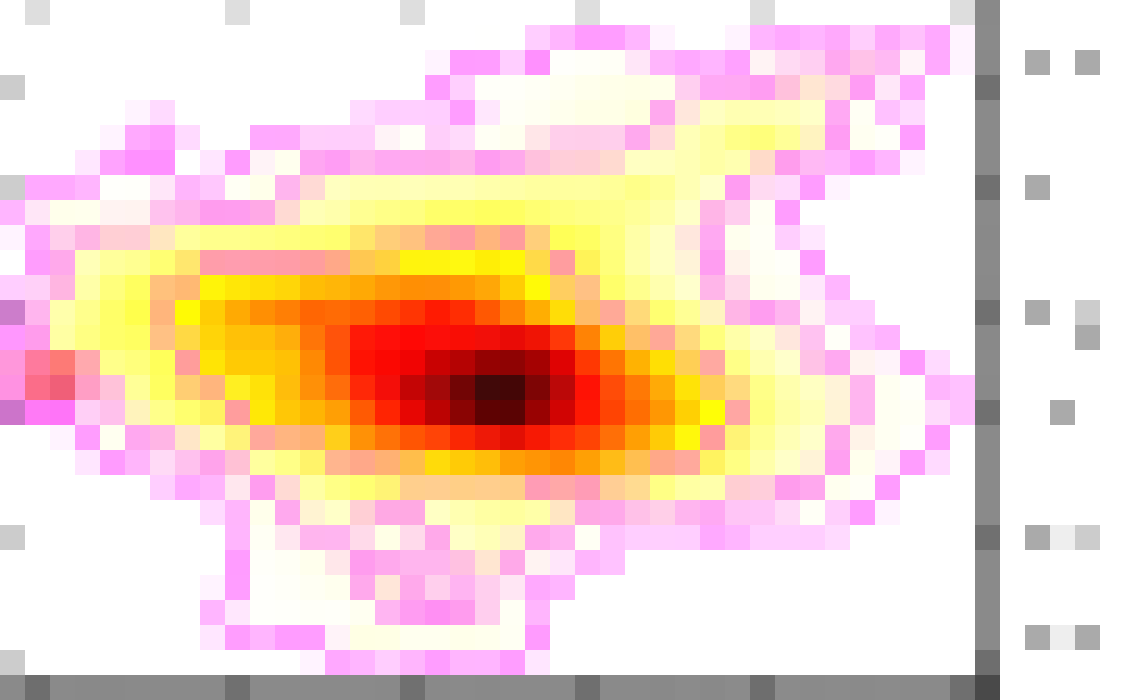}}\\
&&&&\hspace{-2.8pt}\includegraphics[scale=0.1]{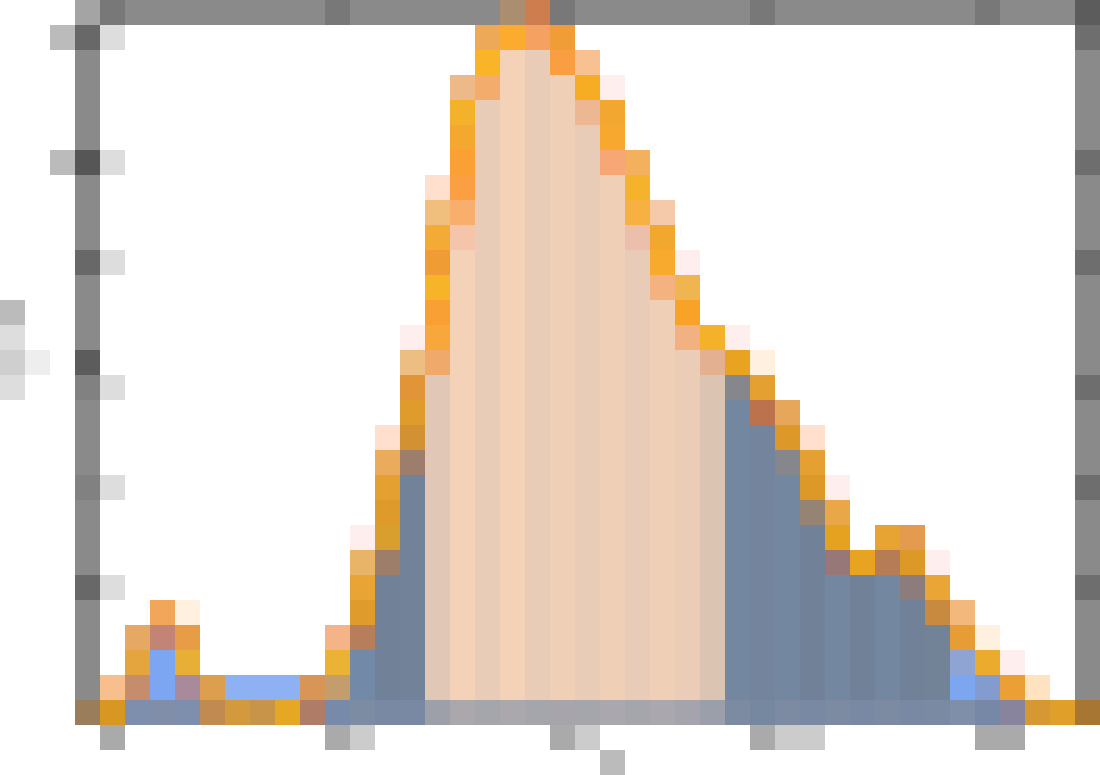}\vspace{-1.9pt}&\raisebox{1.9pt}{\includegraphics[scale=0.1]{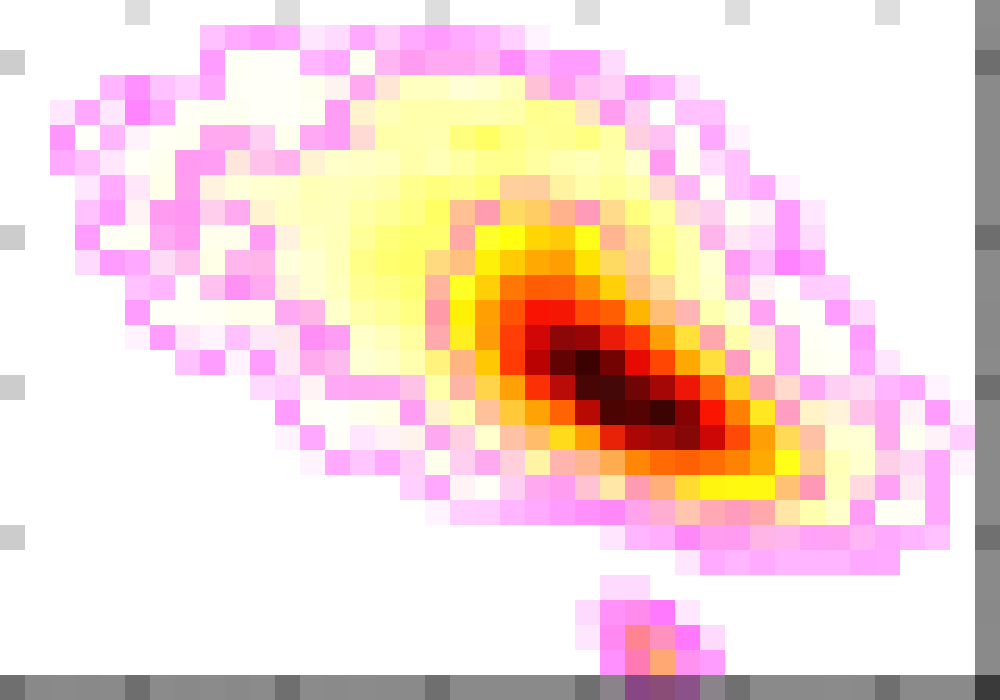}}&\raisebox{1.9pt}{\includegraphics[scale=0.1]{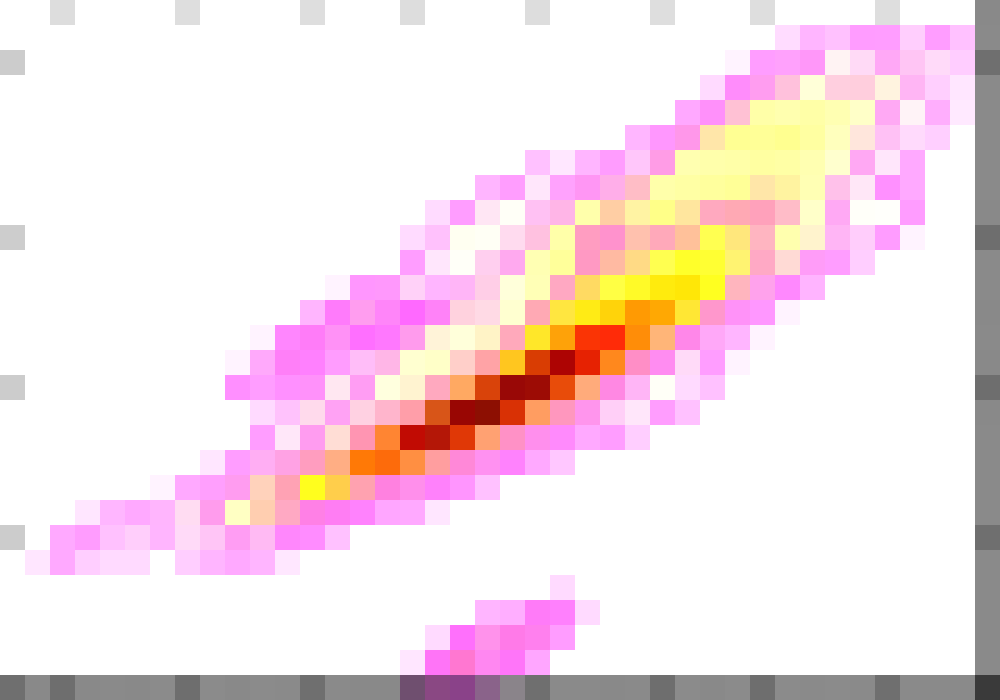}}&\raisebox{1.9pt}{\includegraphics[scale=0.1]{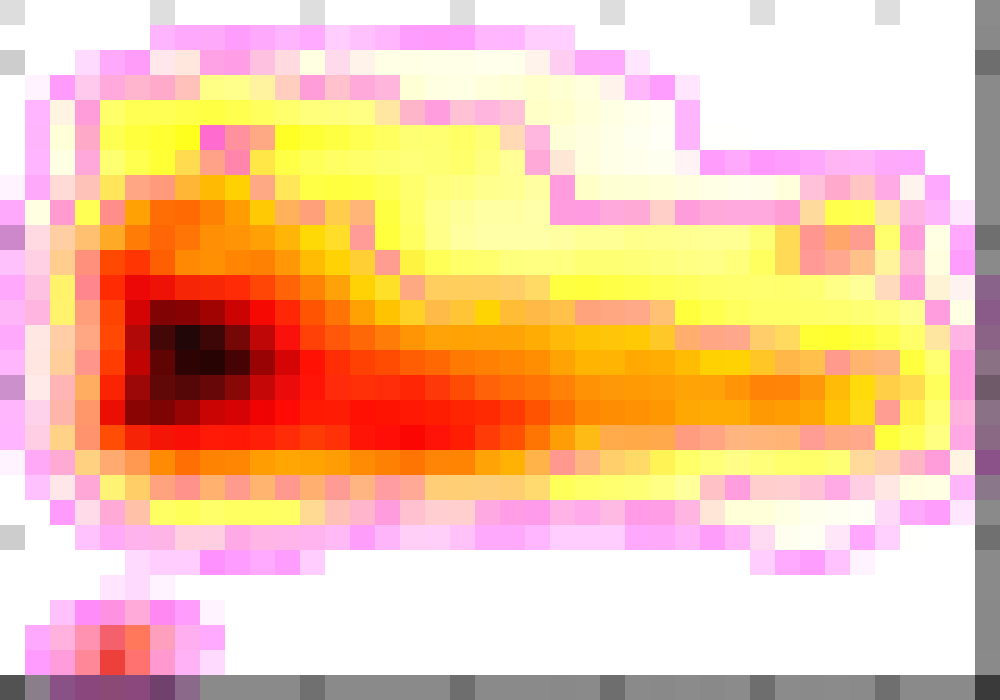}}&\raisebox{1.9pt}{\includegraphics[scale=0.1]{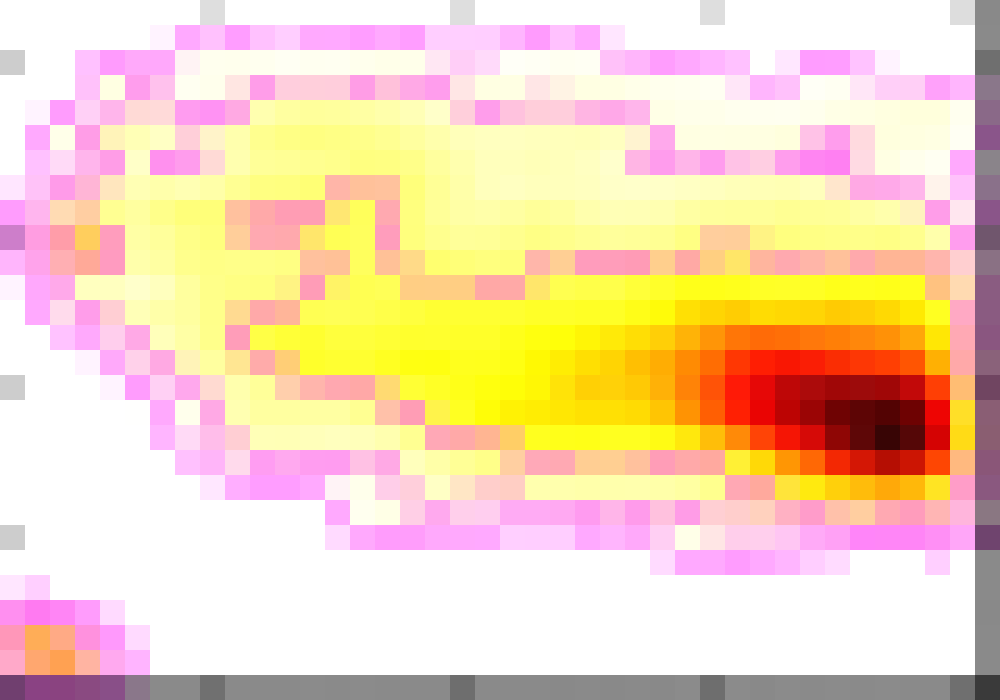}}&\raisebox{1.9pt}{\includegraphics[scale=0.1]{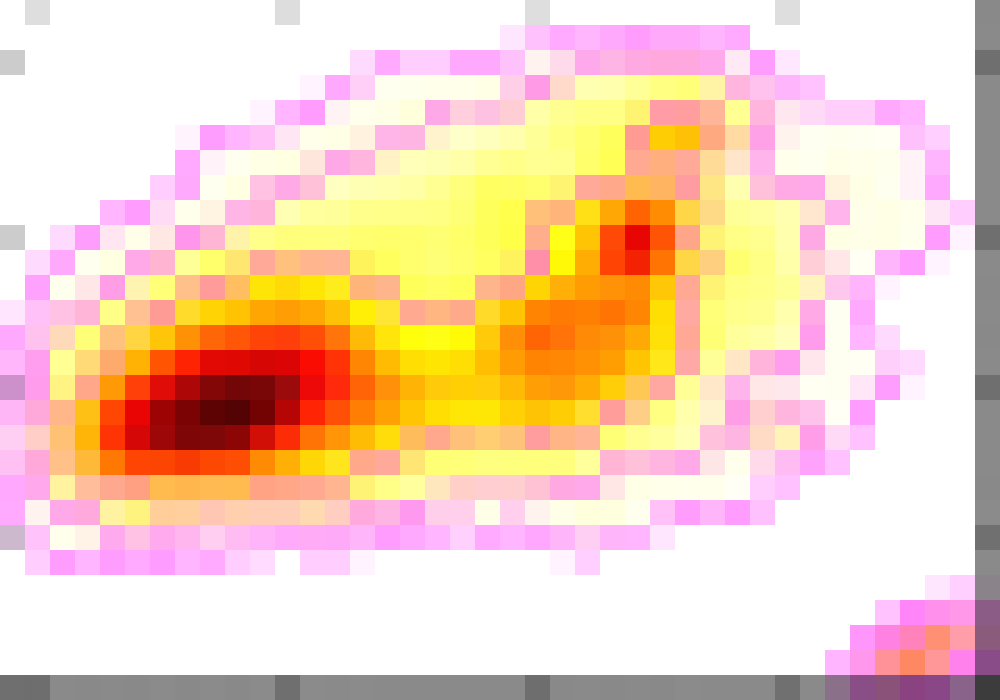}}&\raisebox{1.9pt}{\includegraphics[scale=0.1]{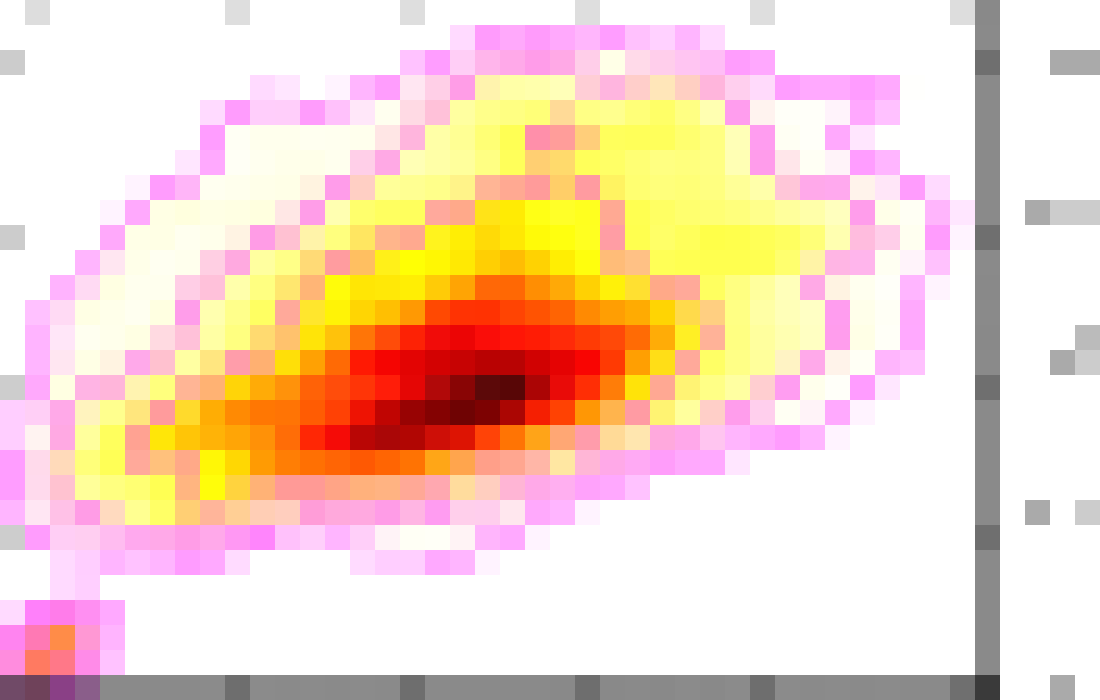}}\\
&&&&&\hspace{-2.4pt}\includegraphics[scale=0.1]{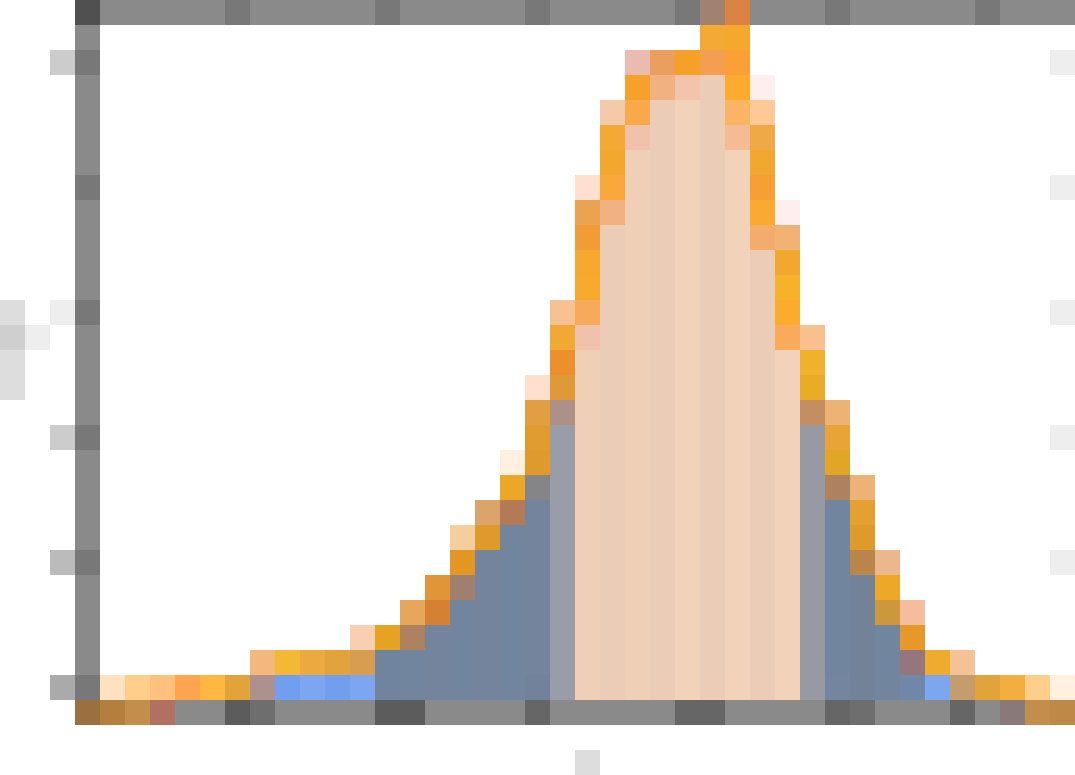}\vspace{-2.1pt}&\raisebox{2.1pt}{\includegraphics[scale=0.1]{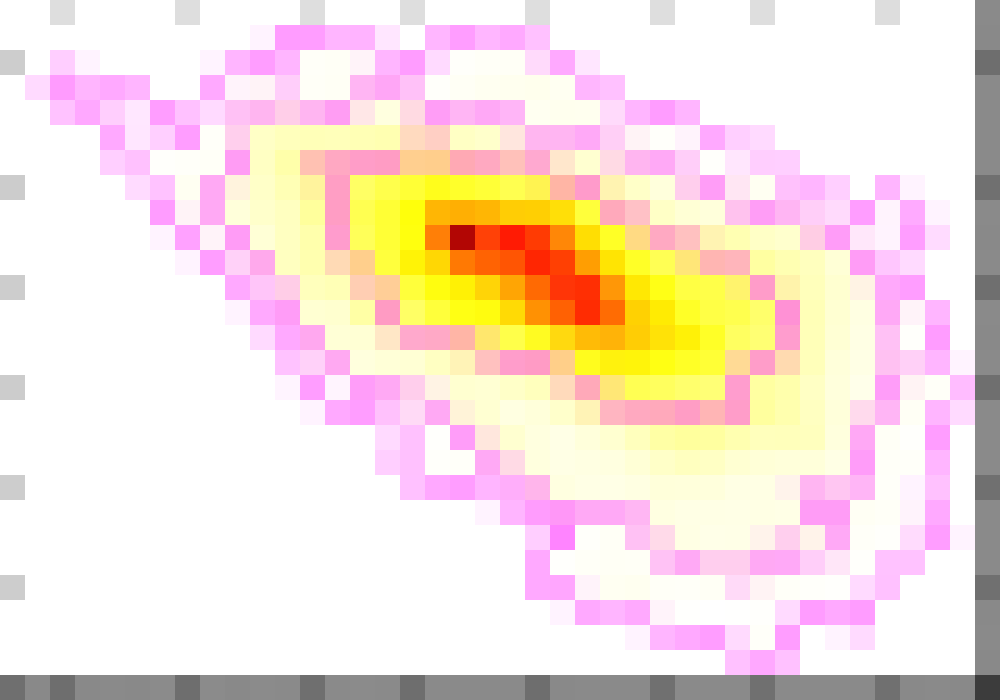}}&\raisebox{2.1pt}{\includegraphics[scale=0.1]{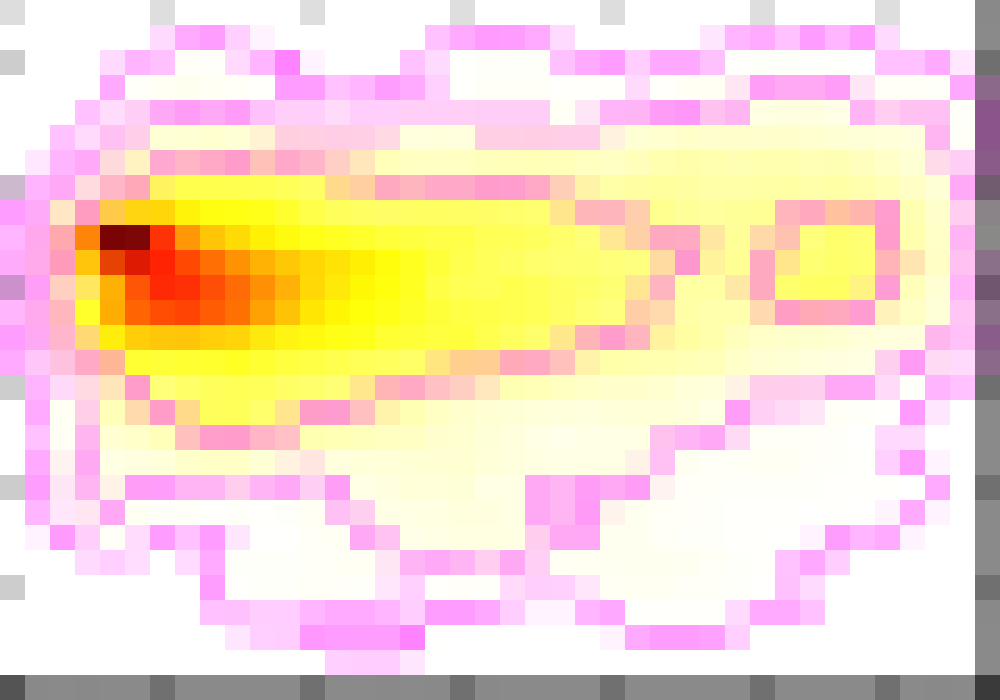}}&\raisebox{2.1pt}{\includegraphics[scale=0.1]{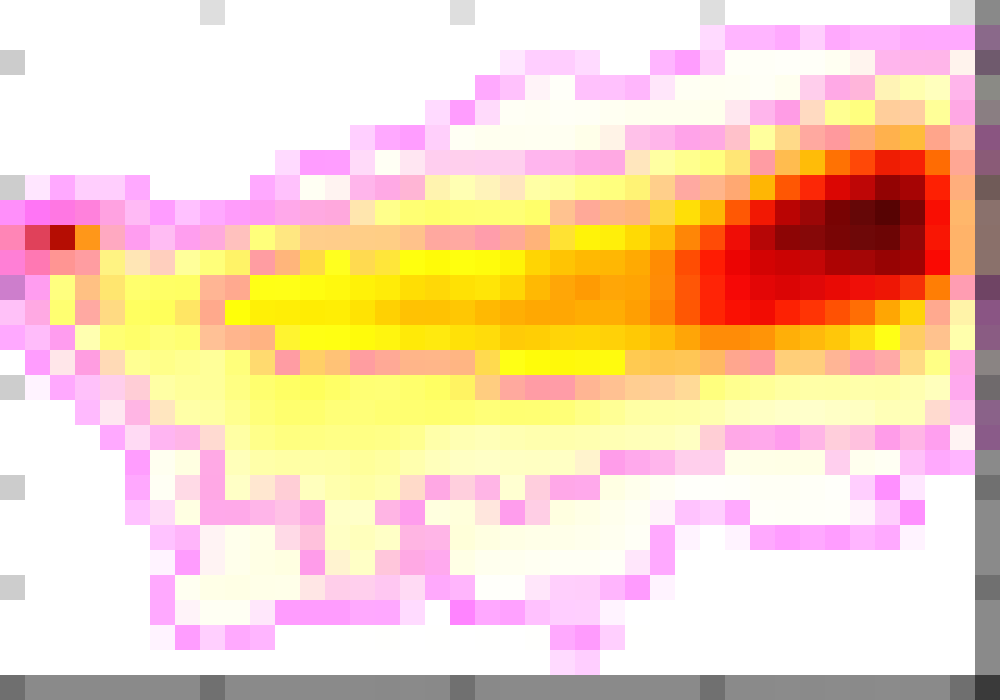}}&\raisebox{2.1pt}{\includegraphics[scale=0.1]{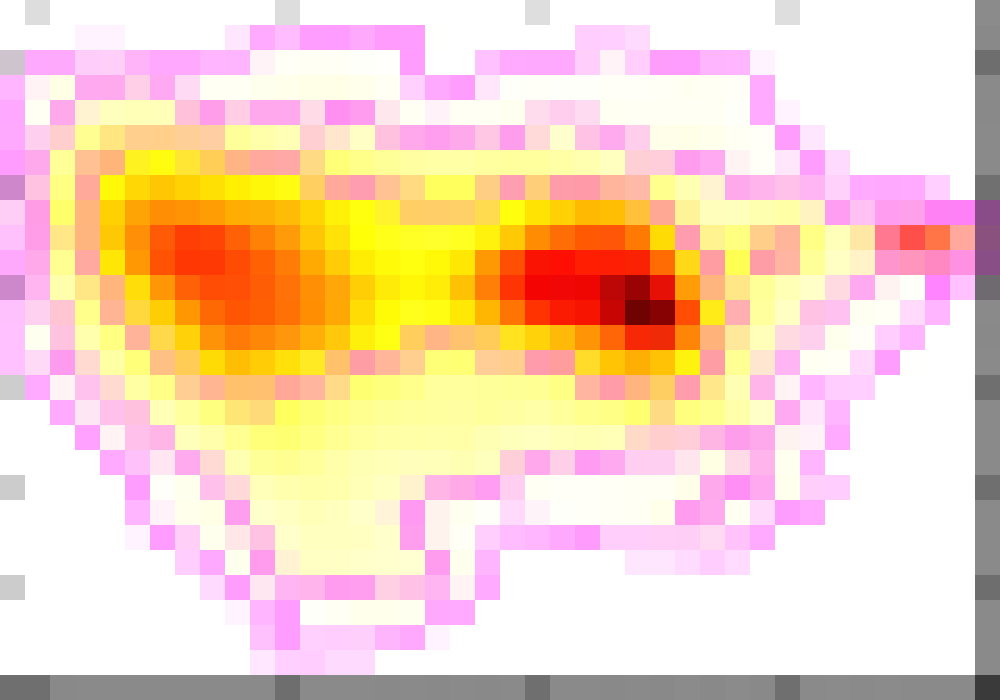}}&\raisebox{2.1pt}{\includegraphics[scale=0.1]{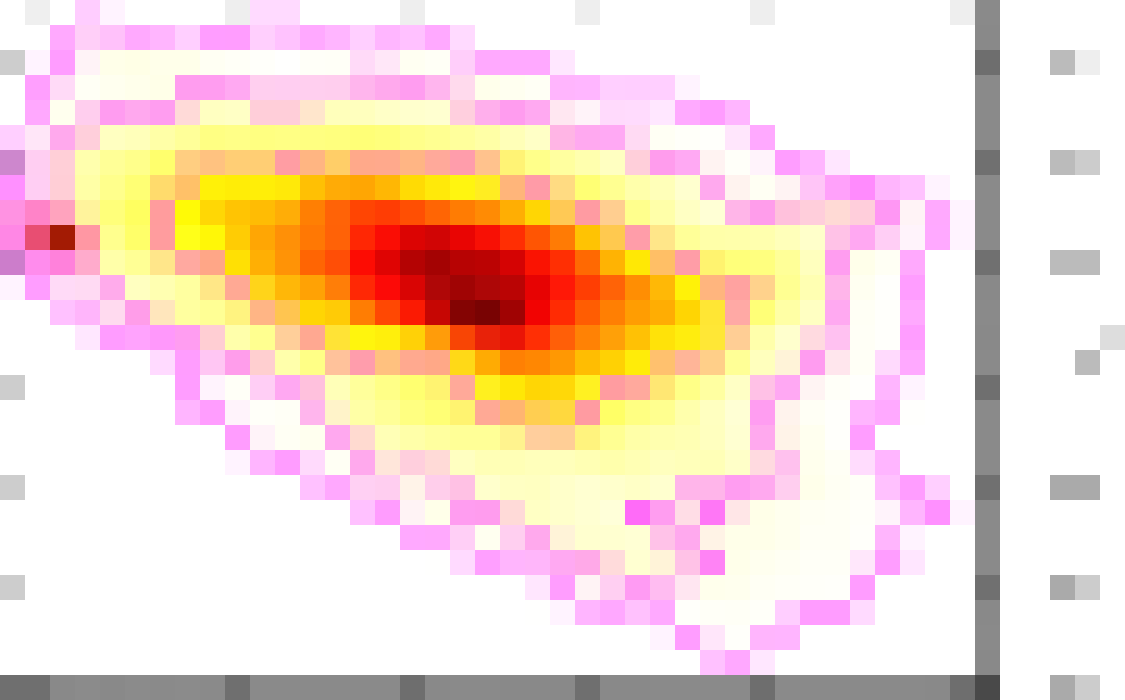}}\\
&&&&&&\hspace{-2.8pt}\includegraphics[scale=0.1]{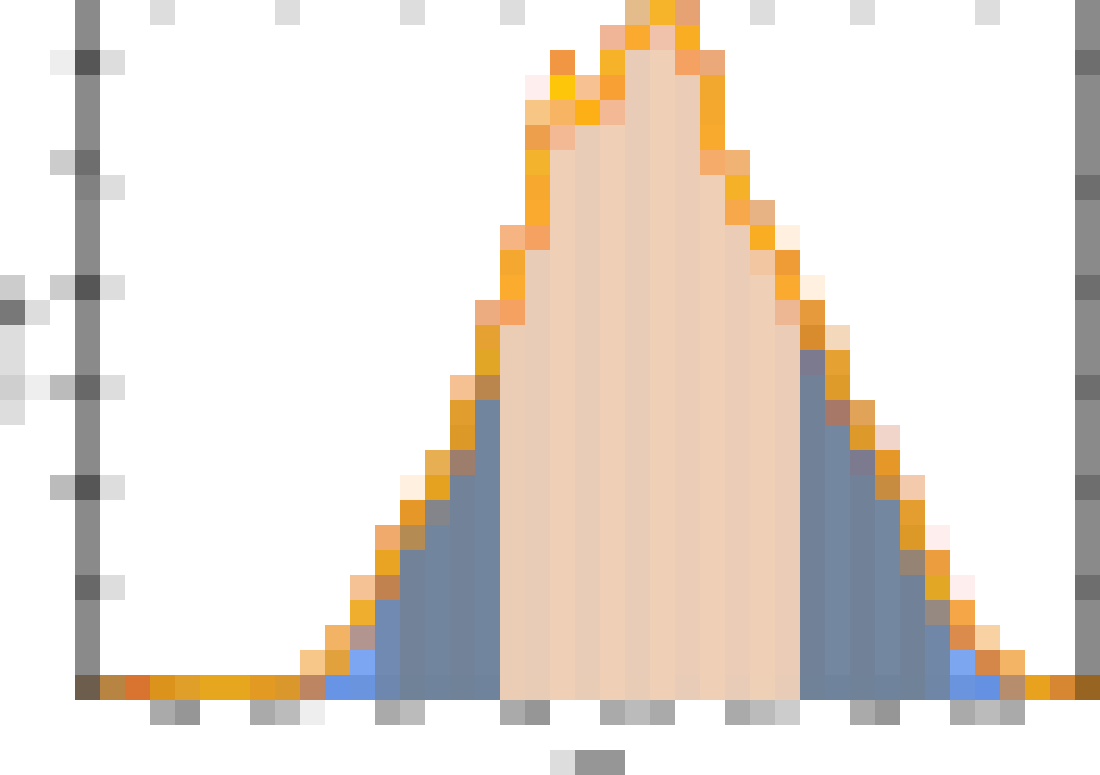}\vspace{-2.2pt}&\raisebox{2.2pt}{\includegraphics[scale=0.1]{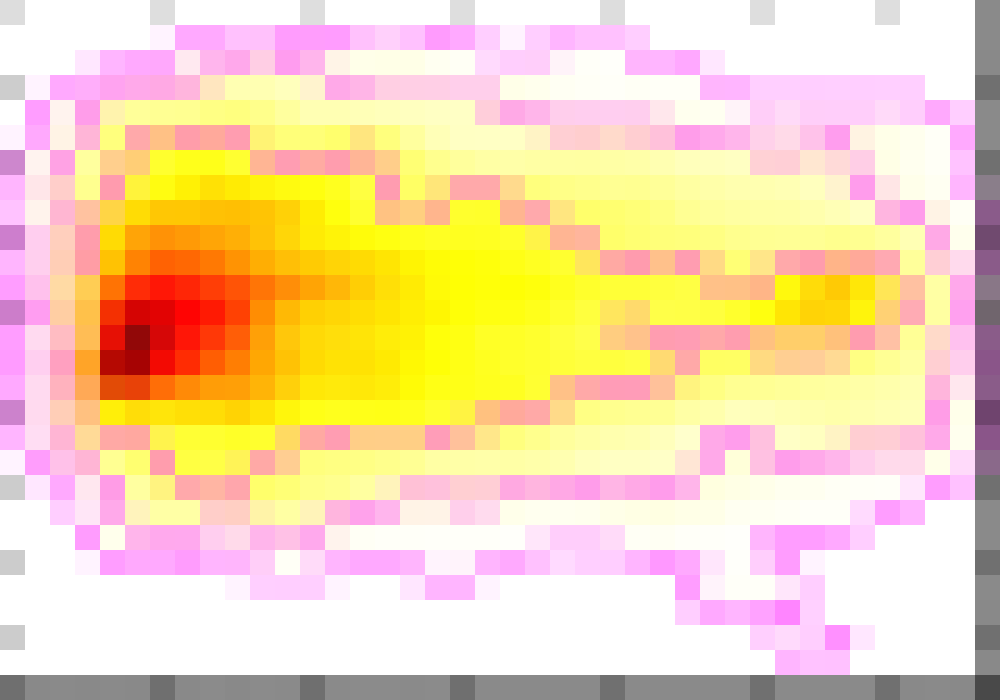}}&\raisebox{2.2pt}{\includegraphics[scale=0.1]{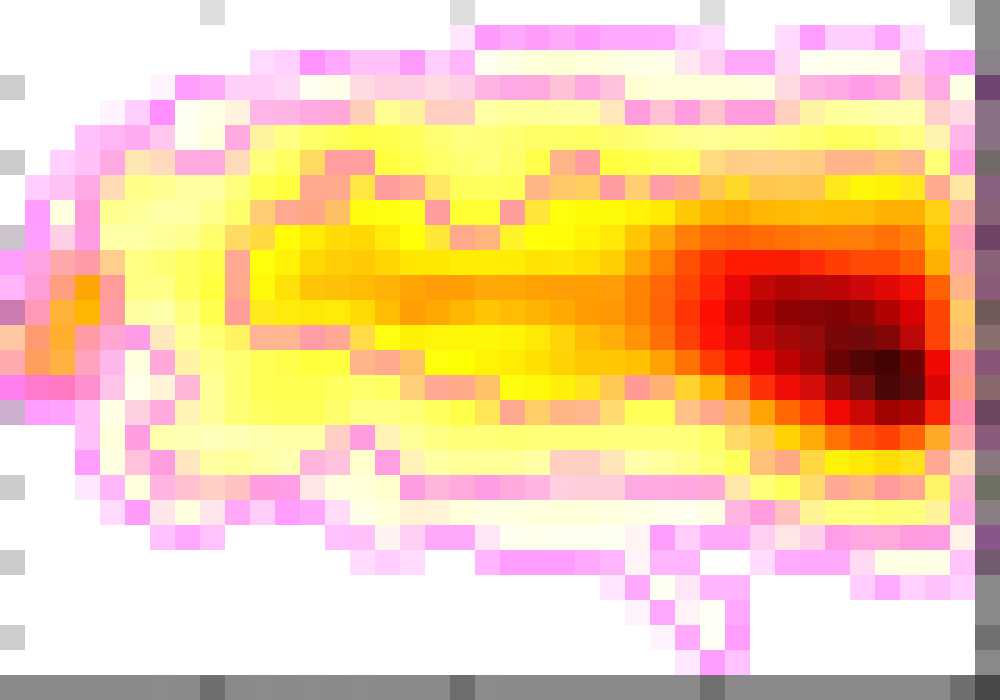}}&\raisebox{2.2pt}{\includegraphics[scale=0.1]{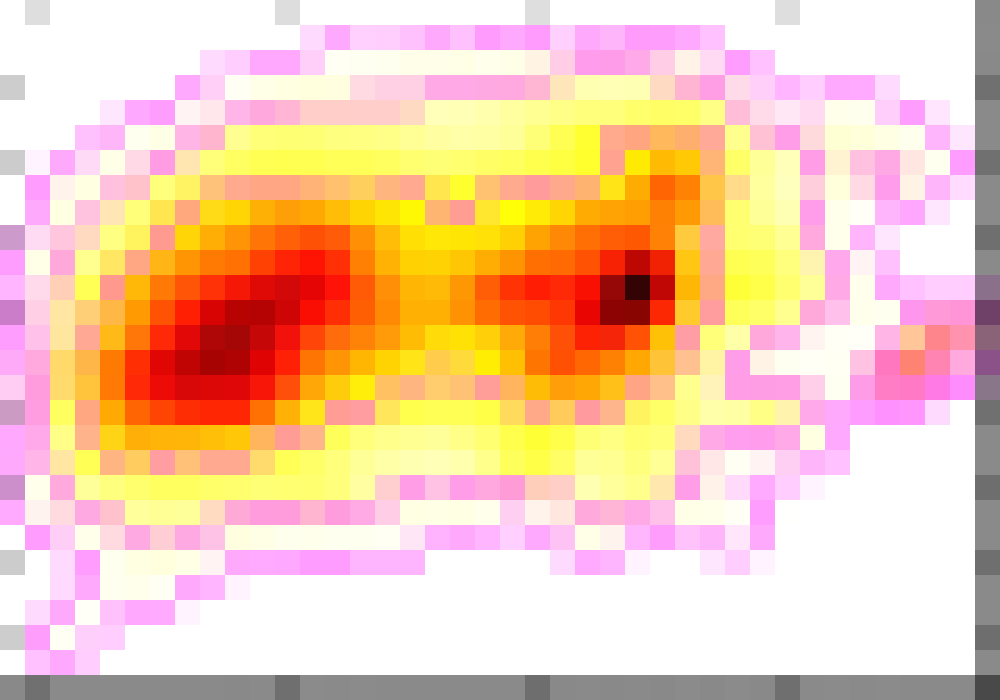}}&\raisebox{2.2pt}{\includegraphics[scale=0.1]{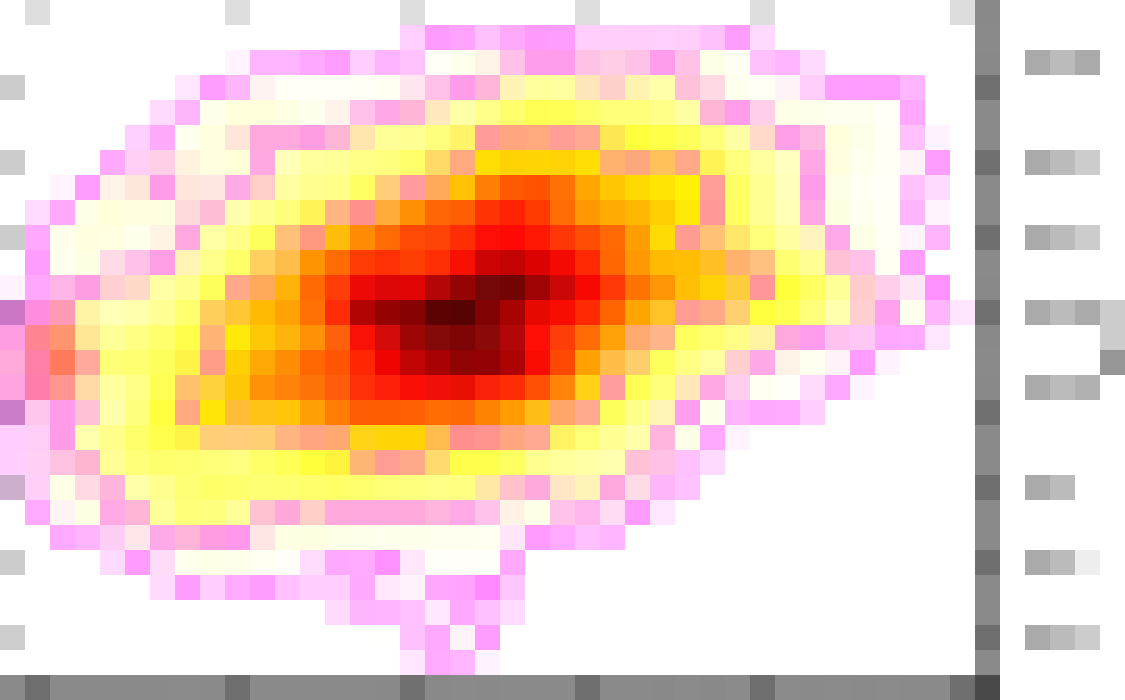}}\\
&&&&&&&\hspace{-4.1pt}\includegraphics[scale=0.1]{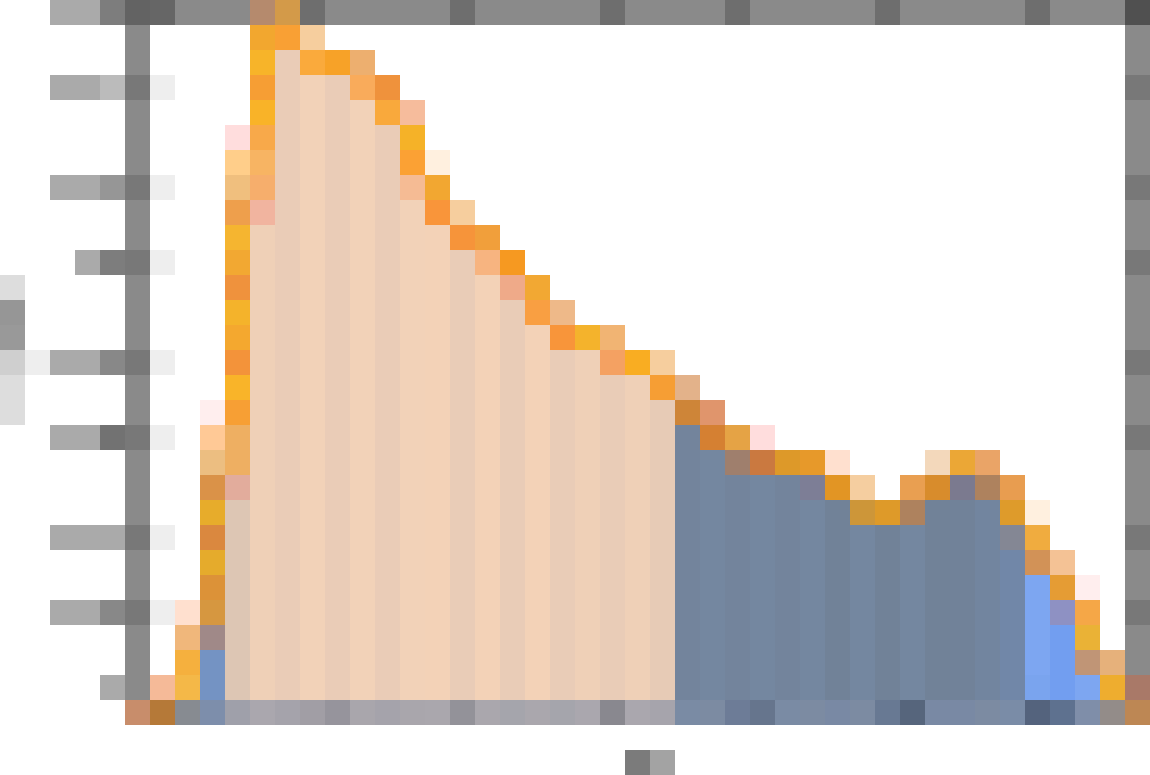}\vspace{-2.3pt}&\raisebox{2.3pt}{\includegraphics[scale=0.1]{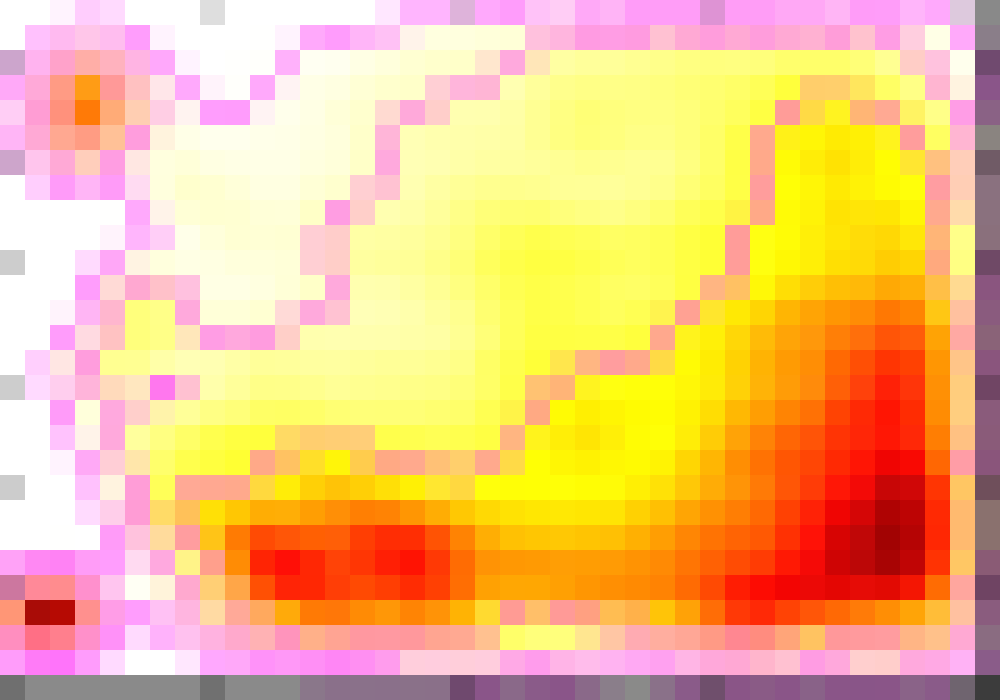}}&\raisebox{2.3pt}{\includegraphics[scale=0.1]{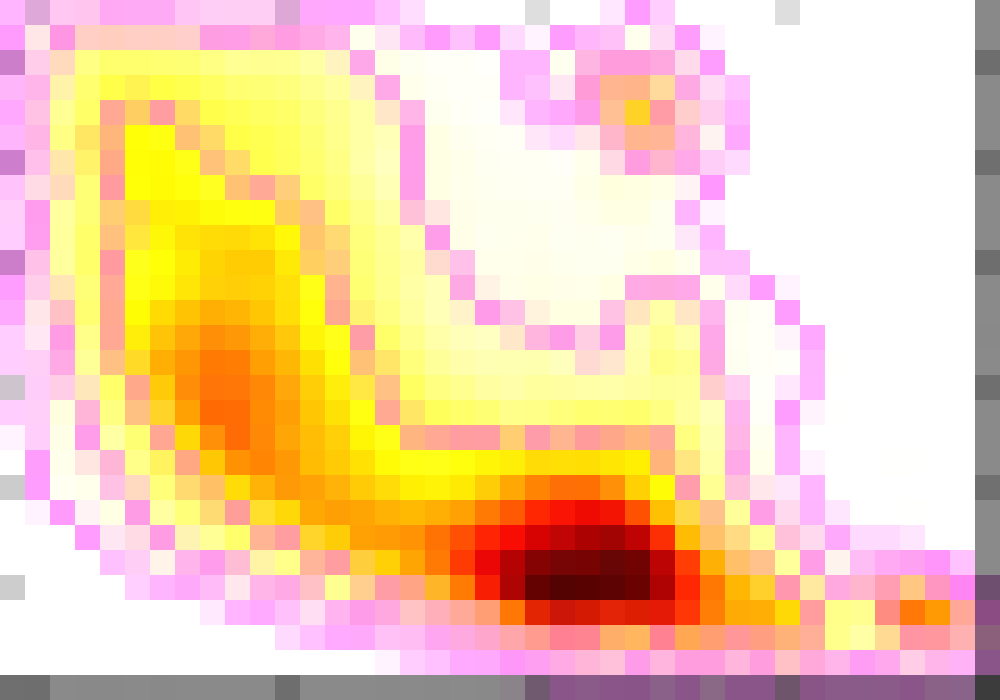}}&\raisebox{2.3pt}{\includegraphics[scale=0.1]{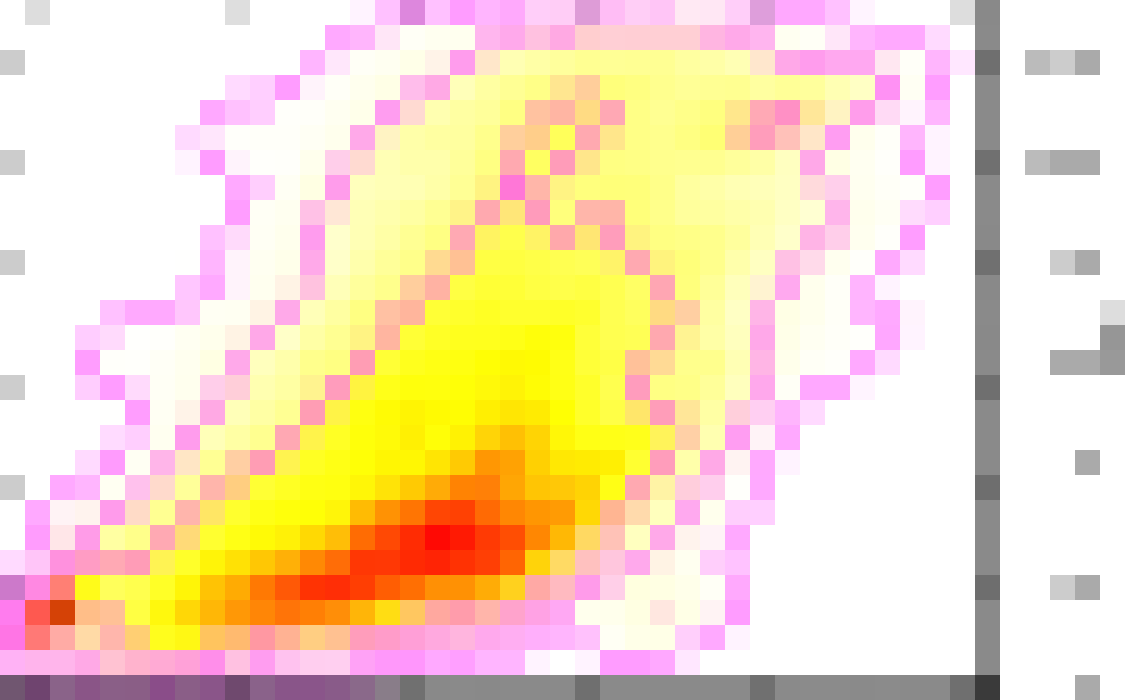}}\\
&&&&&&&&\hspace{-3.7pt}\includegraphics[scale=0.1]{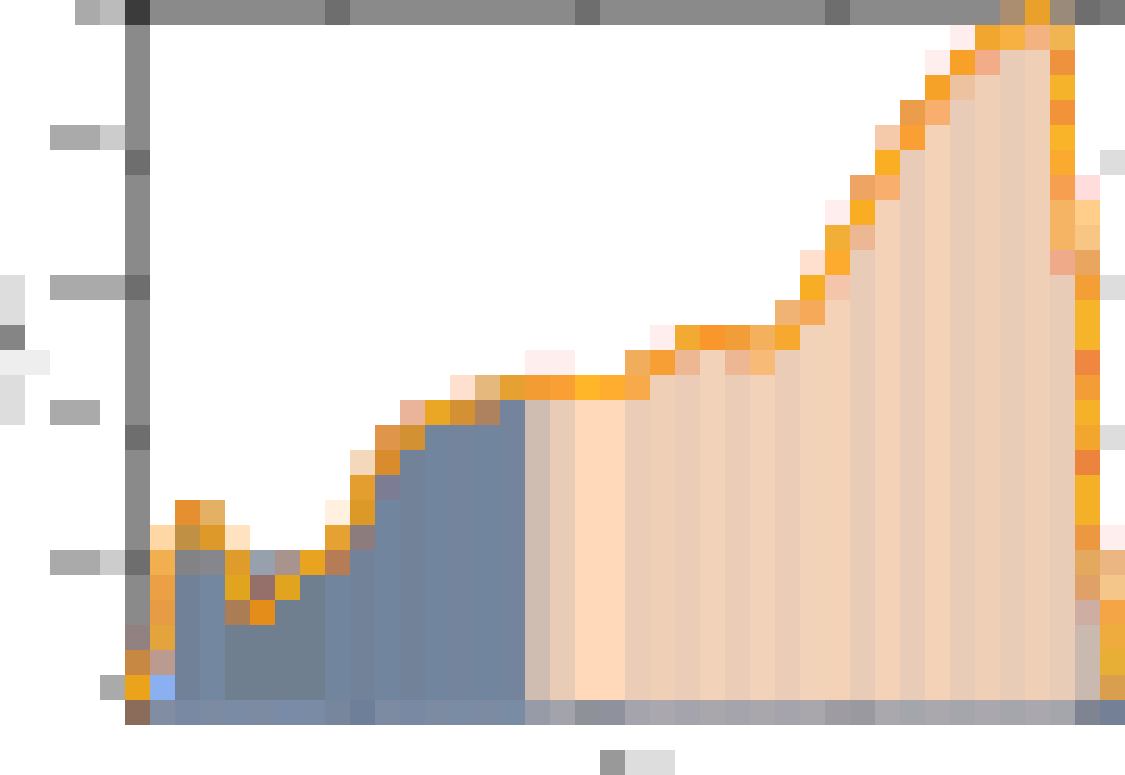}\vspace{-2.1pt}&\raisebox{2.1pt}{\includegraphics[scale=0.1]{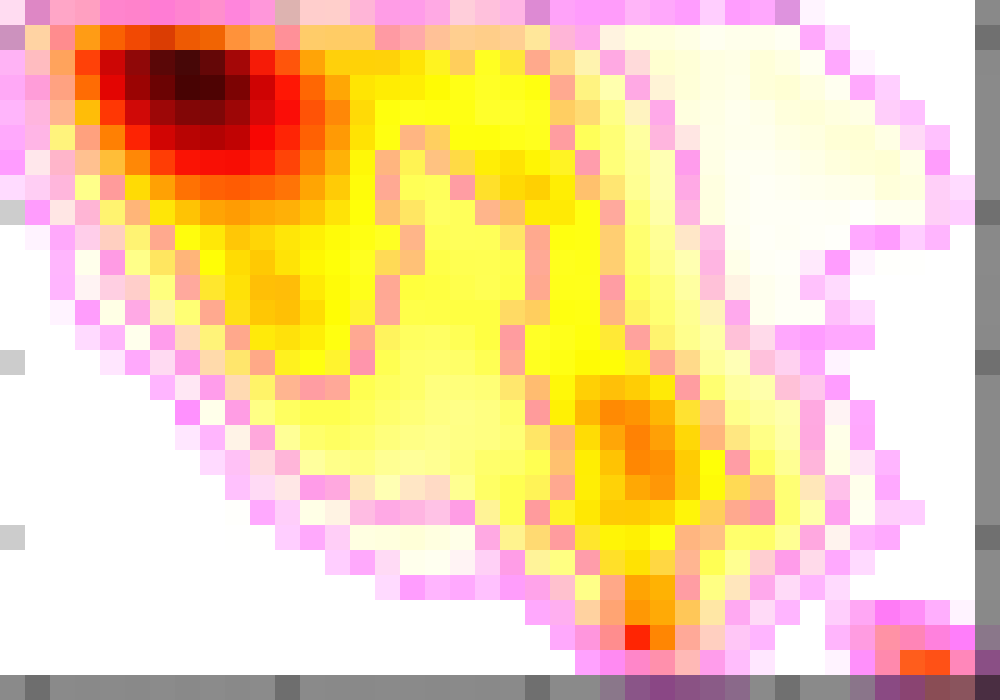}}&\raisebox{2.1pt}{\includegraphics[scale=0.1]{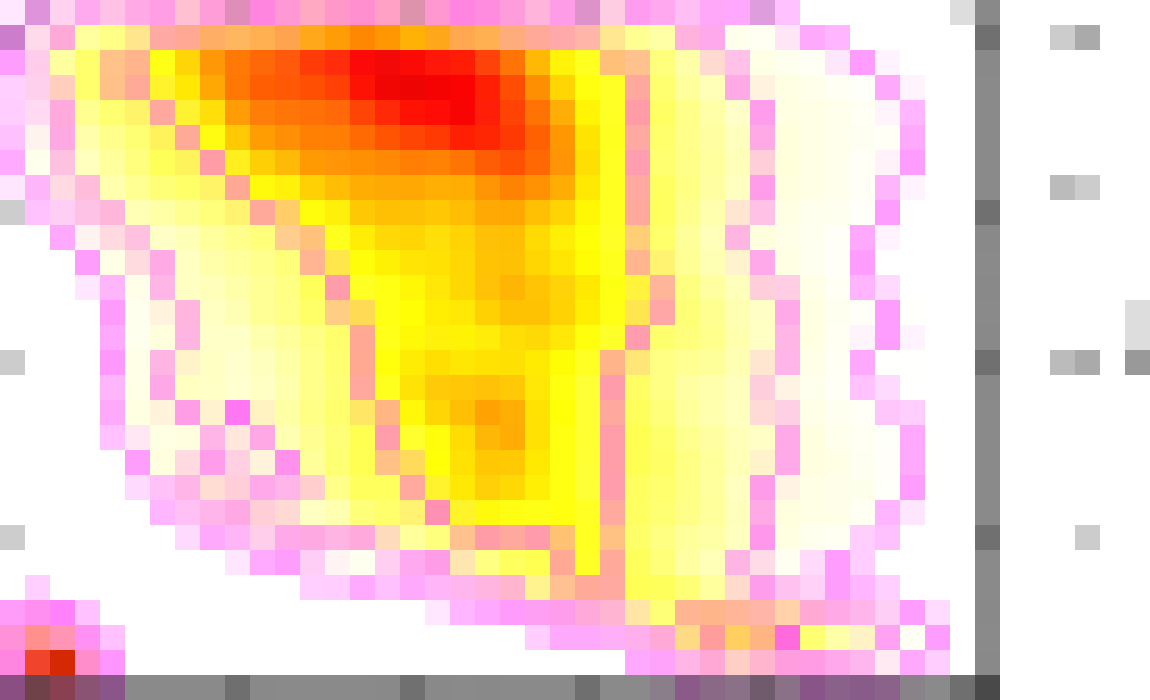}}\\
&&&&&&&&&\hspace{-3.3pt}\includegraphics[scale=0.1]{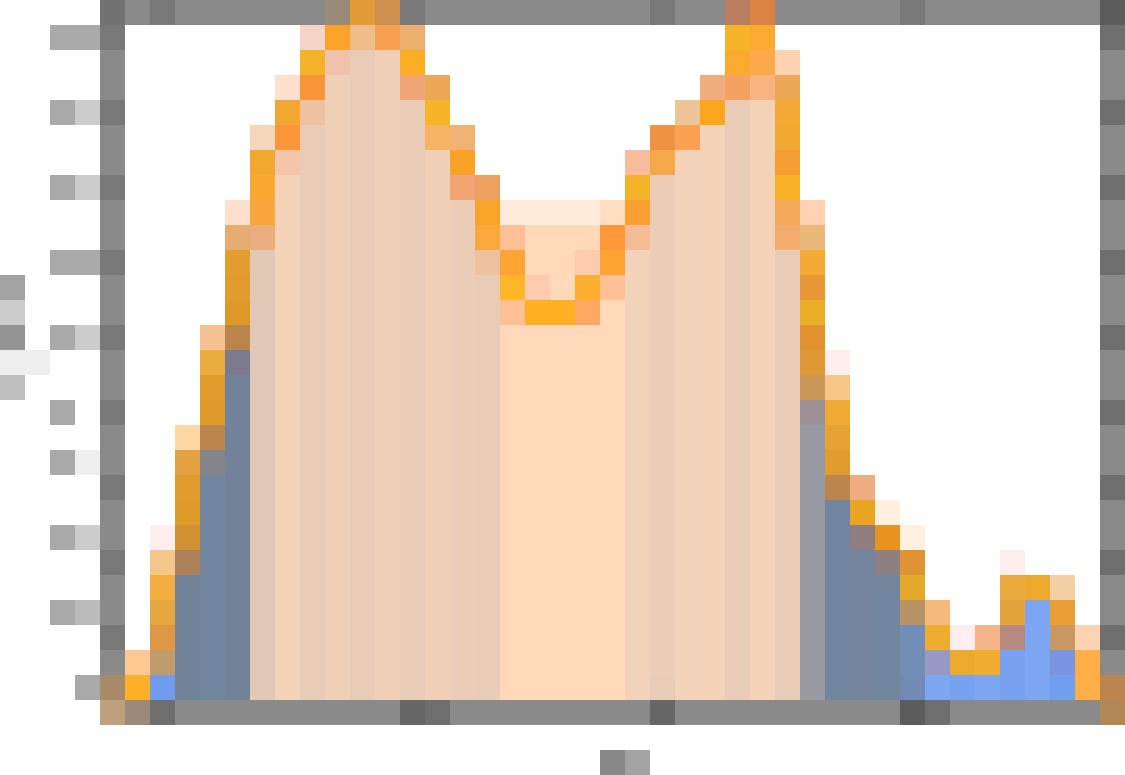}\vspace{-2.1pt}&\raisebox{2.1pt}{\includegraphics[scale=0.1]{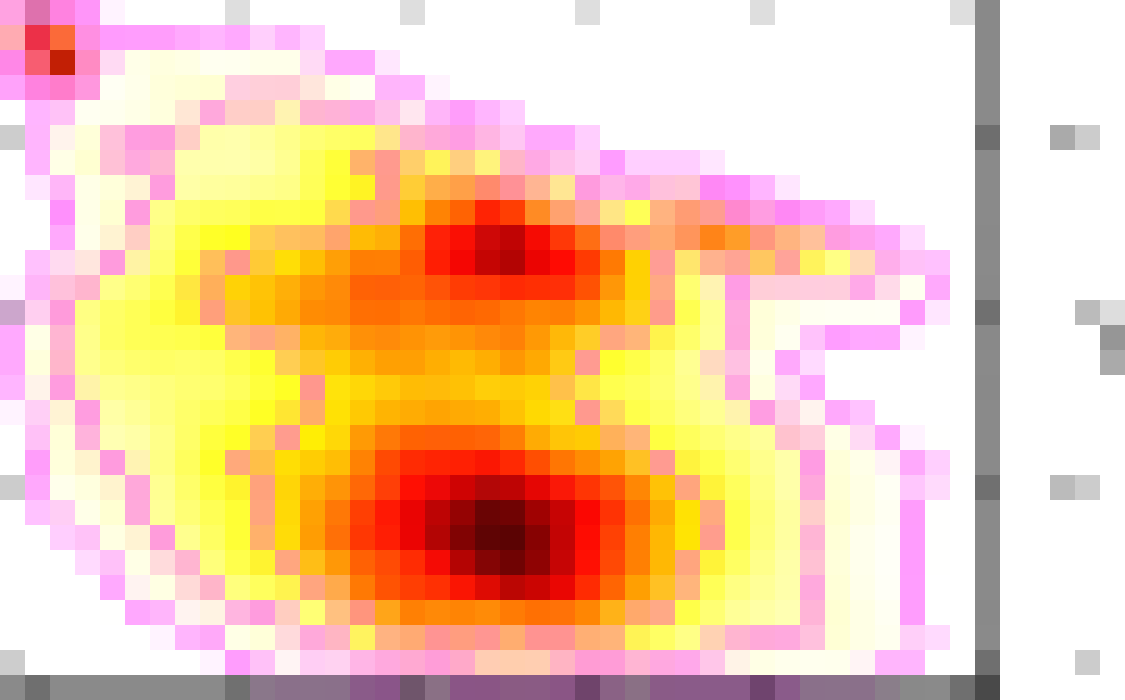}}\\
&&&&&&&&&&\hspace{-3.3pt}\includegraphics[scale=0.1]{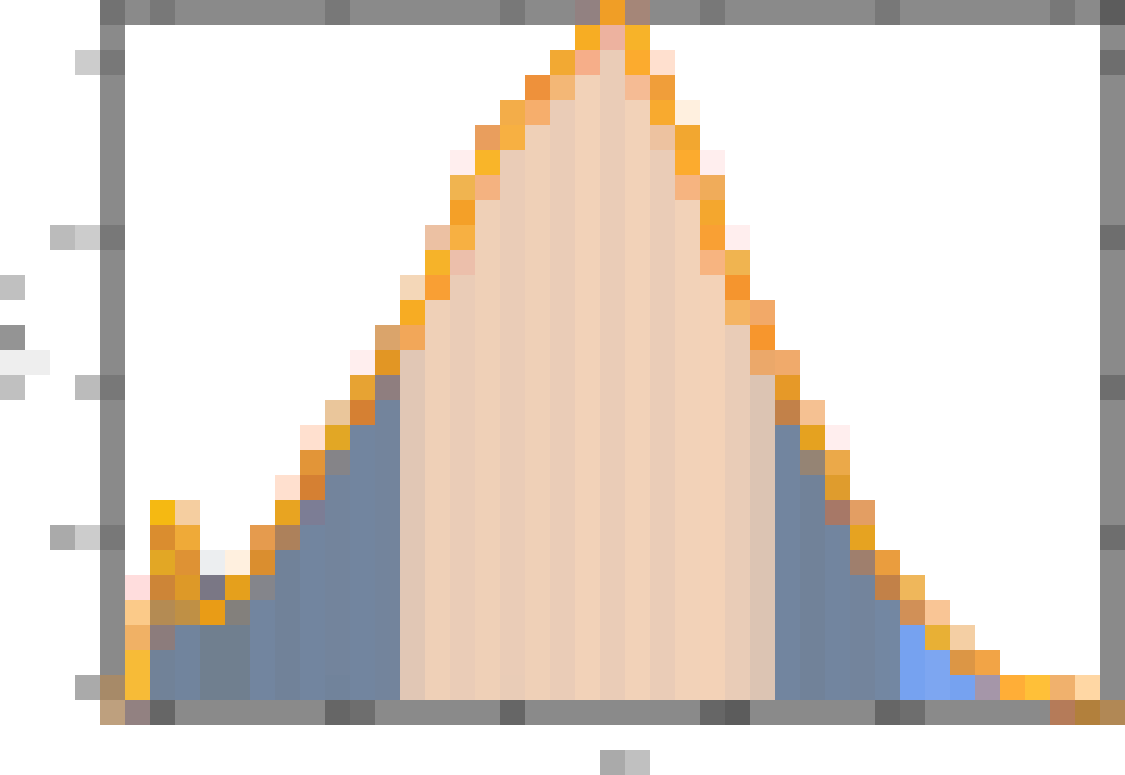}\vspace{-2.1pt}\\
}
\end{tabular}

 \caption{The \protect\PPD\ of parameters in our parametric \protect\HOD\ constrained to fit the observed stellar mass function of \protect\cite{li_distribution_2009}. The results shown are for the third cycle of our iterative procedure in which we constrain the \protect\HOD\ parameters using a covariance matrix derived from the previous set of \protect\HOD\ parameters. Panels on the diagonal show the marginalized \protect\PPD\ for each individual parameter, while off-diagonal panels show the \protect\PPD\ for pairs of parameters, marginalized over all other parameters.}
 \label{fig:ParameterConstraintsHOD}
\end{figure*}

We employ a Gelman-Rubin $\hat{R}$ statistic \citep{gelman_a._inference_1992} to determine when the chains have converged. When $\hat{R}\le 1.2$ we declare convergence and allow the chains to run for a further $10^6$ steps. Only these final $10^6$ steps are used to assess the \PPD\ of the \HOD\ parameters. The final \PPD\ is shown in Figure~\ref{fig:ParameterConstraintsHOD}.

\begin{table*}
\caption{Adopted priors for parameters of our \protect\HOD\ model. For $\alpha_{\rm sat}$, the slope of the satellite \protect\HOD\ at high masses, we adopt a prior consistent with the results of \protect\cite{kravtsov_dark_2004}. For all other parameters we adopted uniform priors spanning a wide range. Also shown are the widths, $\epsilon$, of the Cauchy distributions used to add a random component to each proposed step in our chains.}
 \label{tb:HODParameterPriors}
 \begin{center}
 \begin{tabular}{llll}
  \hline
  {\bf Parameter} & {\bf Units} & {\bf Prior} & \boldmath{$\epsilon$} \\
  \hline
  $\alpha_{\rm sat}$ & -- & $N(0.98,7.55\times 10^{-4},-\infty,+\infty)$ & $0.0002929$ \\
  $M_1$ & $M_\odot$ & $U_{\ln{}}(12.1,12.6)$ & $0.0005773$ \\
  $M_{\star,0}$ & $M_\odot$ & $U_{\ln{}}(10.5,10.8)$ & $0.0003486$ \\
  $\beta$ & -- & $U(0.35,0.5)$ & $0.0002719$ \\
  $\delta$ & -- & $U(0.4,0.65)$ & $0.0003668$ \\
  $\gamma$ & -- & $U(0.7,1.9)$ & $0.001382$ \\
  $\sigma_{\log M_\star}$ & -- &  $U(0.1,0.42)$ & $0.0001321$ \\
  $B_{\rm cut}$ & -- & $U(1.0,128.0)$ & $0.008308$ \\
  $B_{\rm sat}$ & -- & $U(1.0,20.0)$ & $0.01833$ \\
  $\beta_{\rm cut}$ & -- & $U(-2.0,0.0)$ & $0.003111$ \\
  $\beta_{\rm sat}$ & -- & $U(1.0,2.0)$ & $0.02055$ \\
  \hline
 \end{tabular}
 \end{center}
 \end{table*}

Finally, we take the maximum likelihood solution as our choice of parameters for the \HOD\ and use these in computing the covariance matrix of the \cite{li_distribution_2009} stellar mass function. This, of course, throws away all information about the \PPD\ of the \HOD\ parameters that the \BIE\ has determined. One could imagine instead sampling parameter sets from the \PPD\ and building a sample of covariance matrices representative of the uncertainties in the \HOD\ parameters. When constructing the likelihood function for our galaxy formation model one could then include this additional covariance in the data covariance matrix. Given the rather strong constraints on the \HOD\ parameters we do not feel that the substantial extra complexity and computation needed to achieve this is warranted, although we intend to revisit this point in future.

\section{Optimal Sampling of \boldmath{$z=0$} Halo Masses}\label{app:OptimalHaloMasses}

We wish to answer the following question: given some finite amount of computing time, what is the optimal distribution of halo masses to run when comparing to a given dataset? For example, is it better to run a volume limited sample (as one would get from an N-body simulation) or is it better to use equal numbers of halos per logarithmic interval of halo mass? The following section describes how to solve this optimization problem in the specific case of fitting to the stellar mass function.

We define:
\begin{description}
 \item [$n(M) \d \ln M$] is the dark matter halo mass function, i.e. the number of halos in the range $M$ to $M+M\d\ln  M$ per unit volume;
 \item [$\gamma(M) \d \ln M$] is the number of trees that we will simulate in the range $M$ to $M+M\d \ln M$;
 \item [$\alpha(M_\star)$] is the error on the observed stellar mass function at mass $M_\star$;
 \item [$P(N|M_\star,M;\delta \ln M_\star)$] is the conditional stellar mass distribution function of galaxies of stellar mass $M_\star$ in a bin of width $\delta \ln M_\star$ per halo of mass $M$;
 \item [$t(M)$] is the CPU time it takes to simulate a tree of mass $M$.
\end{description}
To clarify, $P(N|M_\star,M;\delta \ln M_\star)$ is the probability\footnote{To put it another way, $P(N|M_\star,M;\delta \ln M_\star)$ is closely related to the commonly used Halo Occupation Distribution.} to find $N$ galaxies of mass between $M_\star$ and $M_\star+M_\star\delta \ln M_\star$ in a halo of mass $M$. The usual conditional stellar mass function is simply the first moment of this distribution:
\begin{equation}
 \phi(M_\star;M) \delta \ln M_\star = \sum_{N=0}^\infty N P(N|M_\star,M;\delta \ln M_\star).
 \label{eq:cSMFdefinition}
\end{equation}
The model estimate of the stellar mass function $\Phi(M_\star)$ (defined per unit $\ln M_\star$) is
\begin{equation}
 \Phi(M_\star) = \int_0^\infty \phi(M_\star;M) {n(M) \over \gamma(M)} \gamma(M) \d \ln M,
\end{equation}
where the $n(M)/\gamma(M)$ term is the weight assigned to each tree realization---and therefore the weight assigned to each model galaxy when summing over a model realization to construct the stellar mass function. 

When computing a model likelihood, we must employ some statistic which defines how likely the model is given the data. Typically, for stellar mass functions we have an estimate of the variance in the data, $\alpha^2(M_\star)$, as a function of stellar mass (the full covariance matrix could easily be incorporated into this method if necessary). In that case, we can define a likelihood
\begin{equation}
 \ln \mathcal{L} = - {1 \over 2} \sum_i {[\phi_{{\rm obs},i} - \phi_i]^2 \over \alpha_i^2 + \sigma_i^2}
\end{equation}
where the sum is taken over all data points, $i$, and $\sigma_i^2$ is the variance in the model estimate and is given by
\begin{equation}
 \sigma^2(M_\star) = \langle [\phi(M_\star) - \bar{\phi}(M_\star)]^2 \rangle,
\end{equation}
where $\phi(M_\star)$ is the realization from a single model and $\bar{\phi}(M_\star)$ is the model expectation from an infinite number of merger tree realizations and the average is taken over all possible model realizations. Since the contributions from each merger tree are independent, 
\begin{equation}
 \sigma^2(M_\star) = \sum_i \zeta_i^2(M_\star;M)
\end{equation}
where $\zeta_i^2(M_\star;M)$ is the variance in the contribution to the stellar mass function from tree $i$. This in turn is given by
\begin{equation}
 \zeta^2(M_\star;M) = \psi^2(M_\star;M) \left[{n(M) \over \gamma(M)}\right]^2,
\end{equation}
where $\psi^2(M_\star;M)$ is the variance in the conditional stellar mass function. In the continuum limit this becomes
\begin{equation}
 \sigma^2(M_\star) = \int_0^\infty \psi^2(M_\star;M) \left[{n(M) \over \gamma(M)}\right]^2 \gamma(M) \d \ln M.
\end{equation}

Model variance artificially increases the likelihood of a given model. We would therefore like to minimize the increase in the likelihood due to the model variance:
\begin{equation}
\Delta 2 \ln \mathcal{L} = \sum_i {[\phi_{{\rm obs},i} - \phi_i]^2 \over \alpha_i^2} - {[\phi_{{\rm obs},i} - \phi_i]^2 \over \alpha_i^2 + \sigma_i^2} 
\end{equation}
Of course, we don't know the model prediction, $\phi_i$, in advance\footnote{Below, we will adopt a simple empirical model for $\phi(M_\star)$. However, it should not be used here since we will in actuality be computing the likelihood from the model itself.}. However, if we assume that a model exists which is a good fit to the data then we would expect that $[\phi_{{\rm obs},i} - \phi_i]^2 \approx \alpha_i^2$ on average. In that case, the increase in likelihood due to the model is minimized by minimizing the function\footnote{This can be seen intuitively: we are simply requiring that the variance in the model prediction is small compared the the variance in the data.}
\begin{equation}
 F[\gamma(M)] = \sum_i {\alpha_i^2 \over \alpha_i^2 + \sigma_i^2}.
 \label{eq:convergenceMeasure}
\end{equation}
If the bins all have the same $\delta \ln M_\star$ we can turn the sum into an integral
\begin{equation}
 F[\gamma(M)] = \int_0^\infty {\alpha(M_\star)^2 \over \alpha(M_\star)^2 + \sigma(M_\star)^2} \d \ln M_\star.
\end{equation}
Obviously, the answer is to make $\gamma(M)=\infty$, in which case $ F[\gamma(M)]=0$. However, we have finite computing resources. The total time to run our calculation is
\begin{equation}
 \tau = \int_0^\infty t(M) \gamma(M) \d \ln M.
\end{equation}
We therefore want to minimize $F[\gamma(M)]$ while keeping $\tau$ equal to some finite value. We can do this using a Lagrange multiplier and minimizing the function
\begin{eqnarray}
  F[\gamma(M)] &=& \int_0^\infty {\alpha(M_\star)^2 \over \alpha(M_\star)^2 + \sigma(M_\star)^2} \d \ln M_\star \nonumber \\ 
  & & + \int_0^\infty \lambda \gamma(M) t(M) \d \ln M.
\end{eqnarray}
Finding the functional derivative and setting it equal to zero gives:
\begin{equation}
 \gamma(M) = \sqrt{{\xi(M) \over \lambda t(M)}},
\end{equation}
in the limit where\footnote{This is the limit in which we would like our results to be.} $\sigma(M_\star) \ll \alpha(M_\star)$, and where
\begin{equation}
 \xi(M) = n^2(M) \int_{-\infty}^\infty {\psi^2(M_\star;M) \over \alpha^2(M_\star)} \d \ln M_\star.
\end{equation}
The values of $\lambda$ and $\delta \ln M_\star$, and the normalization of $t(M)$ are unimportant here since we merely want to find the optimal shape of the $\gamma(M)$ function---we can then scale it up or down to use the available time.

Figure~\ref{fig:optimalSamplingStellarMassFunction} shows the function $\gamma(M)$ obtained by adopting a model conditional stellar mass function which is a sum of central and satellite terms. Specifically, we use the model of \cite{leauthaud_new_2012} which is constrained to match observations from the COSMOS survey. In their model\footnote{This integral form of the conditional stellar mass function is convenient here since it allows for easy calculation of the number of galaxies expected in the finite-width bins of the observed stellar mass function.}:
\begin{eqnarray}
 \langle N_{\rm c}(M_\star|M)\rangle &\equiv& \int_{M_\star}^\infty \phi_{\rm c}(M_\star^\prime) \d \ln M_\star^\prime \nonumber \\
&=& {1 \over 2} \left[ 1 - \hbox{erf}\left( {\log_{10}M_\star - \log_{10} f_{\rm SHMR}(M) \over \sqrt{2}\sigma_{\log M_\star}} \right) \right]. \nonumber\\
\end{eqnarray}
Here, the function $f_{\rm SHMR}(M)$ is the solution of
\begin{equation}
 \log_{10}M = \log_{10}M_1 + \beta \log_{10}\left({M_\star \over M_{\star,0}}\right) + {(M_\star/M_{\star,0})^\delta \over 1 + (M_\star/M_{\star,0})^{-\gamma}} - {1/2}.
\end{equation}
For satellites,
\begin{eqnarray}
 \langle N_{\rm s}(M_\star|M)\rangle &\equiv& \int_{M_\star}^\infty \phi_{\rm s}(M_\star^\prime) \d \ln M_\star^\prime \nonumber \\
&=&  \langle N_{\rm c}(M_\star|M)\rangle \left({f^{-1}_{\rm SHMR}(M_\star) \over M_{\rm sat}}\right)^{\alpha_{\rm sat}} \nonumber \\
& & \times \exp\left(- {M_{\rm cut} \over f^{-1}_{\rm SHMR}(M_\star)} \right),
\end{eqnarray}
where
\begin{equation}
 {M_{\rm sat} \over 10^{12} M_\odot} = B_{\rm sat} \left({f^{-1}_{\rm SHMR}(M_\star) \over 10^{12} M_\odot}\right)^{\beta_{\rm sat}},
\end{equation}
and
\begin{equation}
 {M_{\rm cut} \over 10^{12} M_\odot} = B_{\rm cut} \left({f^{-1}_{\rm SHMR}(M_\star) \over 10^{12} M_\odot}\right)^{\beta_{\rm cut}}.
\end{equation}

We use the best fit parameters from the {\tt SIG\_MOD1} method of \cite{leauthaud_new_2012} for their $z_1$ sample, but apply a shift of $-0.2$ dex in masses to bring the fit into line with the $z=0.07$ mass function of \cite{li_distribution_2009}. The resulting parameter values are shown in Table~\ref{tb:z0SMFFitParameters}.

\begin{table}
\begin{center}
\caption{Parameters of the conditional stellar mass function fit.}
\label{tb:z0SMFFitParameters}
\begin{tabular}{lr@{.}lr@{.}l}
\hline
{\bf Parameter} & \multicolumn{2}{c}{\bf Value} \\
\hline
$\alpha_{\rm sat}$& 1&0 \\
$\log_{10} M_1$& 12&120 \\
$\log_{10} M_{\star,0}$& 10&516 \\
$\beta$& 0&430 \\
$\delta$& 0&5666 \\
$\gamma$& 1&53 \\
$\sigma_{\log M_\star}$& 0&206 \\
$B_{\rm cut}$& 0&744 \\
$B_{\rm sat}$& 8&00 \\
$\beta_{\rm cut}$& $-$0&13 \\
$\beta_{\rm sat}$& 0&859 \\
\hline
\end{tabular}
\end{center}
\end{table}

We assume that $P_{\rm s}(N|M_\star,M;\delta \ln M_\star)$ is a Poisson distribution while $P_{\rm c}(N|M_\star,M;\delta \ln M_\star)$ has a Bernoulli distribution, with each distribution's free parameter fixed by the constraint of eqn.~(\ref{eq:cSMFdefinition}), and the assumed forms for $\phi_{\rm c}$ and $\phi_{\rm s}$.

\begin{figure}
 \begin{center}
 \includegraphics[width=85mm,trim=0mm 0mm 0mm 3.5mm,clip]{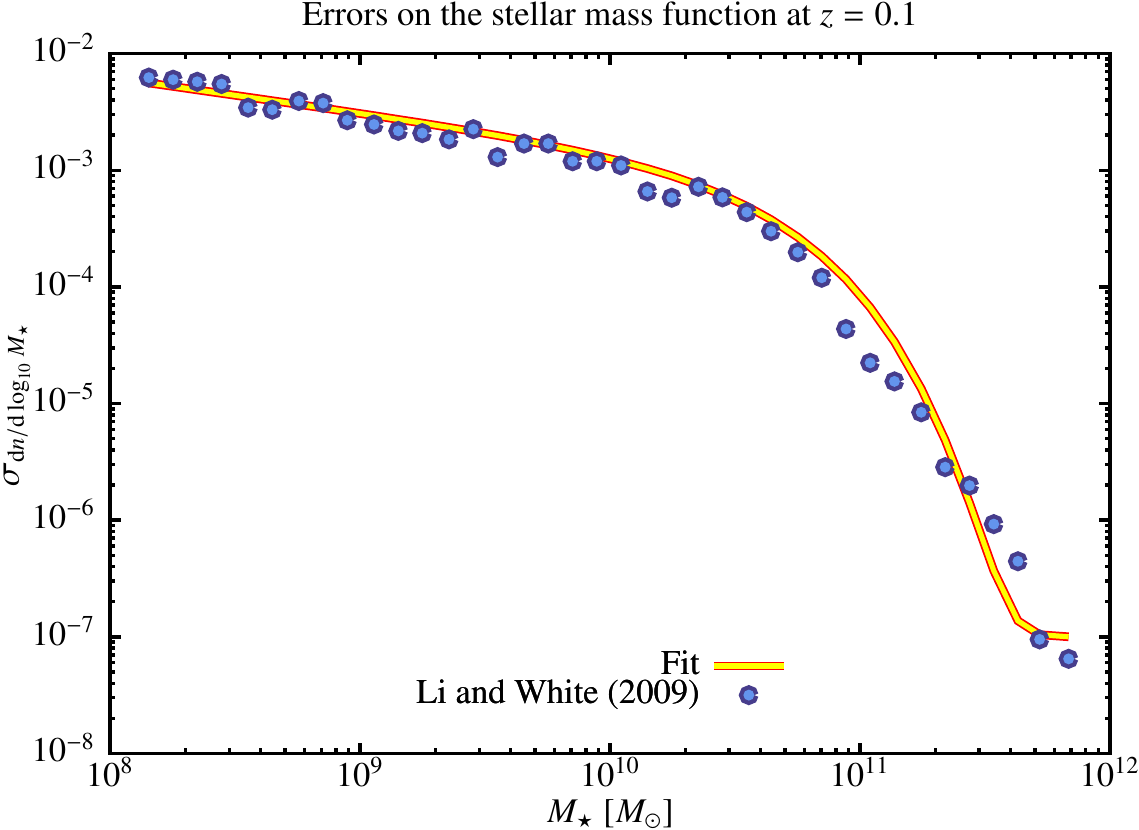}
 \end{center}
 \caption{Errors on the \protect\cite{li_distribution_2009} stellar mass function (points) and the fitting function (line) given by eqn.~(\protect\ref{eq:stellarMassFunctionErrorsFit}).}
 \label{fig:stellarMassFunctionErrors}
\end{figure}

\begin{figure*}
 \begin{tabular}{cc}
  \ifthenelse{\equal{\arabic{loRes}}{0}}
  {\includegraphics[width=85mm,trim=0mm 0mm 0mm 2.5mm,clip]{plots/accuracy/treeTiming.pdf}}{\includegraphics[width=85mm,trim=0mm 0mm 0mm 2.5mm,clip]{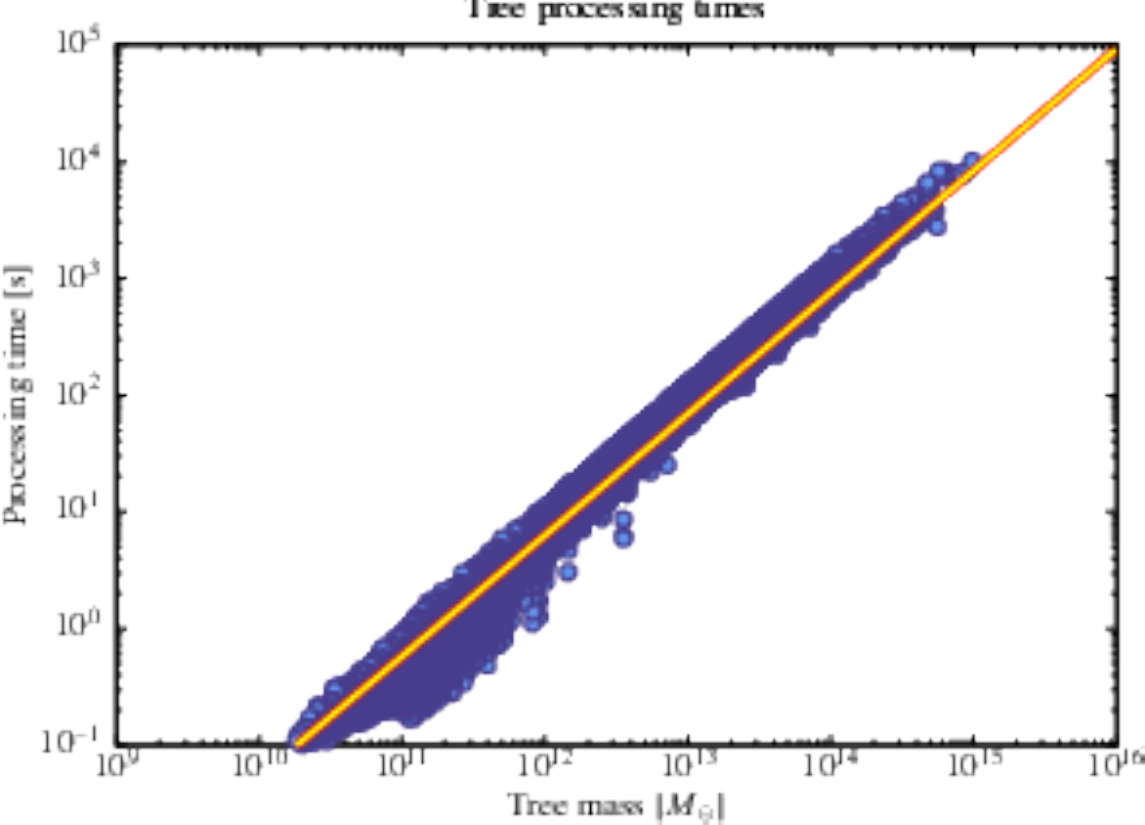}} &
  \includegraphics[width=85mm,trim=0mm 0mm 0mm 2.4mm,clip]{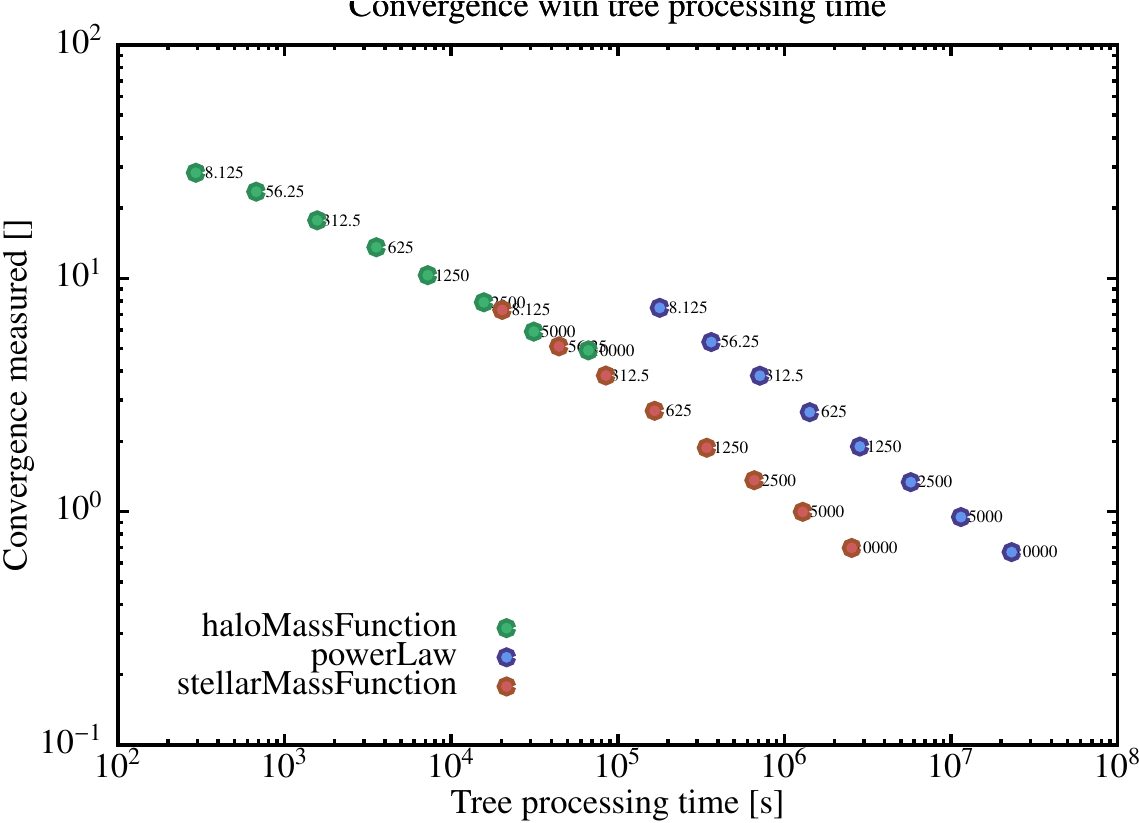}
 \end{tabular}
 \caption{\emph{Left panel:} The CPU time taken to a merger tree of given $z=0$ mass. Blue points show measured values for a large number of model merger trees. The yellow line indicates the best-fit $2^{\rm nd}$-order polynomial to this relation as described in eqn.~(\protect\ref{eq:TreeTiming}). \emph{Right panel:} The convergence measure, defined in eqn.~(\protect\ref{eq:convergenceMeasure}), as a function fo total time taken to process the set of merger trees. Results are shown for three different choices for sampling masses of present day dark matter halos as described in the text.}
 \label{fig:TreeTiming}
\end{figure*}

\begin{figure}
 \begin{center}
 \includegraphics[width=85mm,trim=0mm 0mm 0mm 3.5mm,clip]{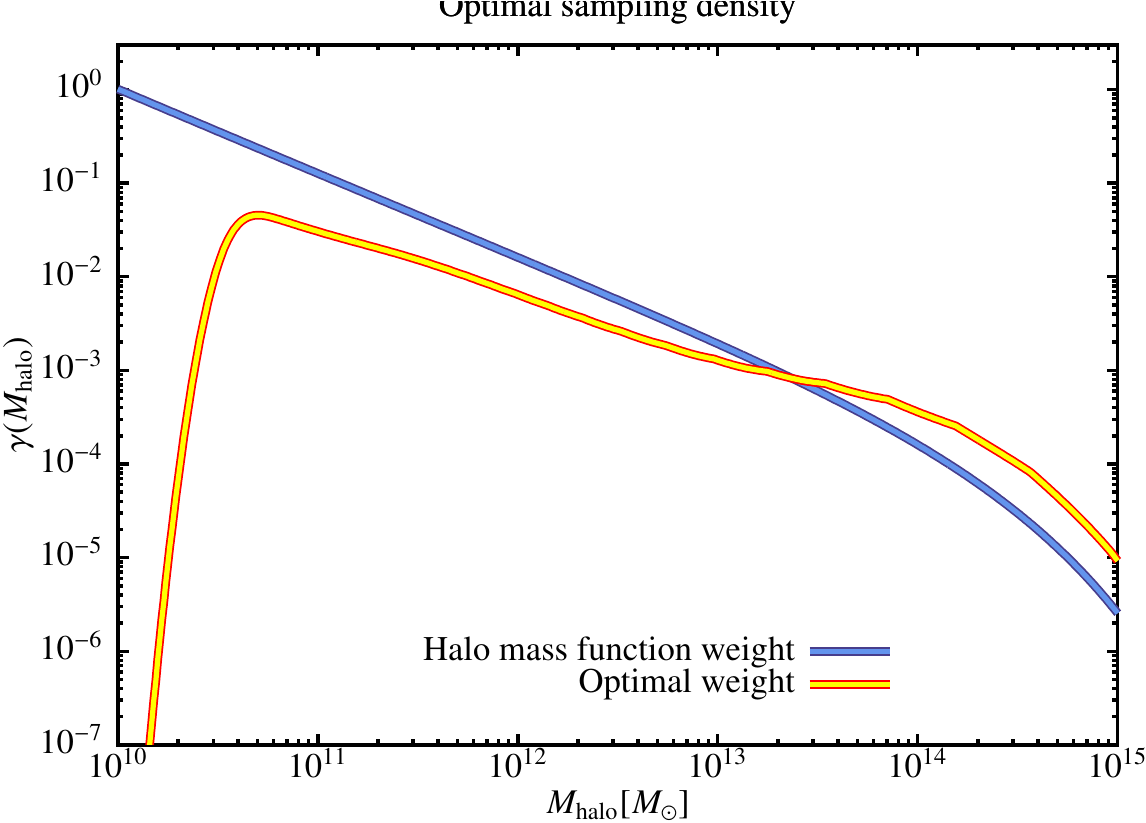}
 \end{center}
 \caption{Optimal weighting ($\gamma(M)$; yellow line) compared with weighting by the dark matter halo mass function (i.e. sampling halos at random from a representative volume; blue line). Sampling densities have been normalized to unit compute time.}
 \label{fig:optimalSamplingStellarMassFunction}
\end{figure}

The errors in the \cite{li_distribution_2009} observed stellar mass function are well fit by (see Fig.~\ref{fig:stellarMassFunctionErrors}):
\begin{eqnarray}
 \alpha(M_\star) &=& 10^{-3} \left({M_\star\over 4.5\times 10^{10}M_\odot}\right)^{-0.3} \exp\left(-{M_\star\over 4.5\times 10^{10}M_\odot}\right) \nonumber \\
 & & + 10^{-7},
 \label{eq:stellarMassFunctionErrorsFit}
\end{eqnarray}
and the tree processing time in our model can be described by:
\begin{equation}
 \log_{10} t(M) = \sum_{i=0}^2 C_i [ \log_{10} M ]^i
 \label{eq:TreeTiming}
\end{equation}
with $C_0=-16.25$, $C_1=1.767$ and $C_2=0.02883$ (see the left-hand panel of Fig.~\ref{fig:TreeTiming}).

The resulting optimal sampling density curve is shown in Fig.~\ref{fig:optimalSamplingStellarMassFunction} and is compared to weighting by the halo mass function (i.e. the result of sampling halos at random from a representative volume). Optimal sampling gives less weight to low mass halos (since a sufficient accuracy can be obtained without the need to run many tens of thousands of such halos) and to high mass halos which are computationally expensive. Nevertheless, the optimal sampling curve is quite close to the halo mass function, so we would expect that optimal sampling and sampling from the halo mass function should be similarly efficient. The right-hand panel of Fig.~\ref{fig:TreeTiming} shows the convergence measure $F[\gamma(M)]$ as a function of total tree processing time for three different methods of sampling the halo mass function. The optimal method achieves the smallest convergence measure in a given total processing time. It is clear that ``haloMassFunction'' (i.e. sampling halos in proportion to the halo mass function; green points) and ``stellarMassFunction'' (i.e. using the optimal weighting; red points) samplings give almost equally efficient results\footnote{The numbers next to each point indicate the number of mergers trees run on average per decade of halo mass. It can be seen that the ``stellarMassFunction'' sampling achieves the same convergence measure as ``haloMassFunction'' sampling using fewer merger trees. However, because ``stellarMassFunction'' sampling tends to sample more high mass trees, the total time required is about the same.}. Sampling using ``powerLaw'' ($\gamma(M) \propto [\log_{10} M]^{-1/2}$; blue points) is much less efficient. 

Clearly in this case our optimal sampling method did not improve over simply sampling from the halo mass function, although it does almost equally as well. This is likely because calculation of the optimal sampling requires a model for the conditional mass function of halos (eqn.~\ref{eq:cSMFdefinition}). If this model does not match the conditional mass function in our actual \SAM\ then the computed sampling will not be optimal. Once a viable \SAM\ is obtained it can be used to directly compute the conditional mass function, which should allow this method to more accurately compute the optimal sampling.

Given these results, for the present work, we choose to use sampling from the halo mass function in our \MCMC\ analysis\footnote{Note that the optimal sampling function here was computed assuming a fixed mass resolution in merger trees. In our \protect\MCMC\ analysis we instead use a variable mass resolution. We have computed an optimal sampling for that case also, and find that it performs equally well, but not better than sampling from the halo mass function.}. Additionally, we introduce upper and lower limits on the halo abundance, \emph{for the purposes of sampling halo mass only}, to ensure that a minimum number of high mass halos, and not too many low mass halos are simulated.

\section{Robustness and Convergence of Model Results}\label{app:Numerics}

For the purposes of these robustness and convergence tests (and for the computation of model discrepancy terms; see \S\ref{sec:ModelDiscrepancy}) we adopt an ``\emph{a priori}'' set of model parameters which were found to give a reasonable match (by eye, not in any strict statistical sense) to the observed stellar mass function. Obviously, prior to constraining our model to match the observational data we do not know the maximum likelihood set of parameters, and so must make a ``best guess'' for the purposes of evaluating these models---our ``\emph{a priori}'' model serves this purpose. Parameters of the ``\emph{a priori}'' model are given in Table~\ref{tb:APriorParameters}.

\begin{table}
 \begin{center}
 \begin{tabular}{lr@{.}l}
  \hline
  {\bf Parameter} & \multicolumn{2}{l}{\bf Value} \\
  \hline
  $\alpha_{\rm cool}$ & 0&35 \\
  $\beta_{\rm cool}$ & 0&50  \\
  $\Delta \log_{10}\mathcal{M}_{\rm cool}$ & 0&30  \\
  $\tau_{\rm cool}$ & 2&26 Gyr  \\
  $\mathcal{M}_{\rm cool}$ & 3&$0 \times 10^{11} M_\odot$ \\
  $\alpha_\star$ & 0&21 \\
  $\beta_\star$ & 2&40  \\
  $\tau_{\rm \star, min}$ & 3&50 Myr \\
  $\tau_\star$ & 0&68 Gyr  \\
  $\alpha_{\rm wind}$ & -6&68 \\
  $\beta_{\rm wind}$ & 4&78   \\
  $f_{\rm wind}$ & 0&0138 \\
  $\tau_{\rm wind, min}$ & 5&09 Myr \\
  $f_{\rm df}$ & 8&35  \\
  $V_{\rm reion}$ & 46&69 km/s \\
  $C_{1\ldots6}$ & 0&00   \\
  $\mu$ & -0&27    \\
  $\kappa$ & 0&0127   \\
  $H_1$ & -0&0198  \\
  $H_2$ & 0&00986  \\
  \hline
 \end{tabular}
 \end{center}
 \caption{Parameters of the ``\emph{a priori}'' model used for tests of model robustness and accuracy, and also for calculations of model discrepancy terms.}
 \label{tb:APriorParameters}
\end{table}

Before beginning our \MCMC\ analysis we test that our model is converged with respect to all numerical parameters. To test for convergence, we run an \emph{a priori} model multiple times, each time adjusting the value of a single numerical parameter, and compute the model stellar mass function (i.e. precisely the same statistic that we will use in our \MCMC\ study). Thereby, we generate a set of model stellar mass functions as a function of the values of model numerical parameters. For each sequence of models, in which the value of a single numerical parameter is varied, we define the \emph{most optimal} model as that with the lowest or highest value of the model parameter, depending on whether the most accurate results are expected for small or large values of the parameter.

We then define a convergence measure
\begin{equation}
 \chi^2 =  {\Delta \cdot [\mathcal{C}_{\rm (model)} + \mathcal{C}_{\rm (model,opt)}]^{-1} \cdot \Delta^{\rm T} \over N_{\rm eff}} 
\end{equation}
where $\Delta$ is the difference between a model evaluation of the stellar mass function and the that obtained using the same model but with the most optimal value of the parameter being tested, $\mathcal{C}_{\rm (model)}$ is the covariance matrix of the model,  $\mathcal{C}_{\rm (model,opt)}$ is the covariance matrix of the optimal model, and $N_{\rm eff}$ is an effective number of degrees of freedom. This measure should approach unity as convergence is achieved. The error on this convergence measure is
\begin{equation}
 {\sigma_{\chi^2}^2 \over \chi^2} = {2 \over N_{\rm eff}} + \left({\sigma_{N_{\rm eff}} \over N_{\rm eff}}\right)^2.
\end{equation}
The effective number of degrees of freedom is found by running many realizations of the ``\emph{a priori}'' model with different random seeds but otherwise unchanged parameters and taking the average of $\Delta \cdot [\mathcal{C}_{\rm (model)} + \mathcal{C}_{\rm (model,opt)}]^{-1} \cdot \Delta^{\rm T}$ over all realizations. This is necessary to account for the effects of significant multicolinearity in the model covariance matrix which result in this quantity being distributed as a $\chi^2$-distribution with fewer degrees of freedom than would be expected.

\begin{figure*}
 \begin{tabular}{cc}
  \includegraphics[width=85mm,trim=0mm 0mm 0mm 2.5mm,clip]{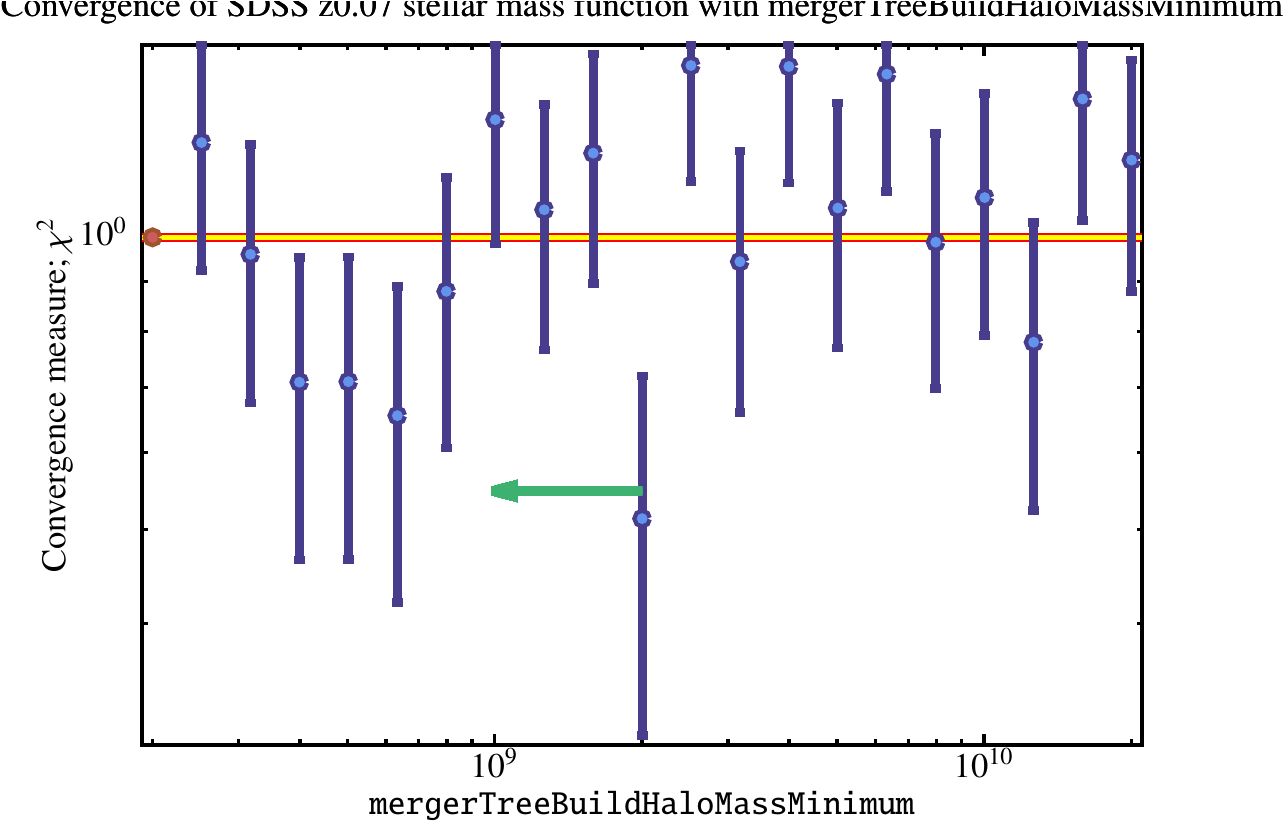} &
  \includegraphics[width=85mm,trim=0mm 0mm 0mm 2.5mm,clip]{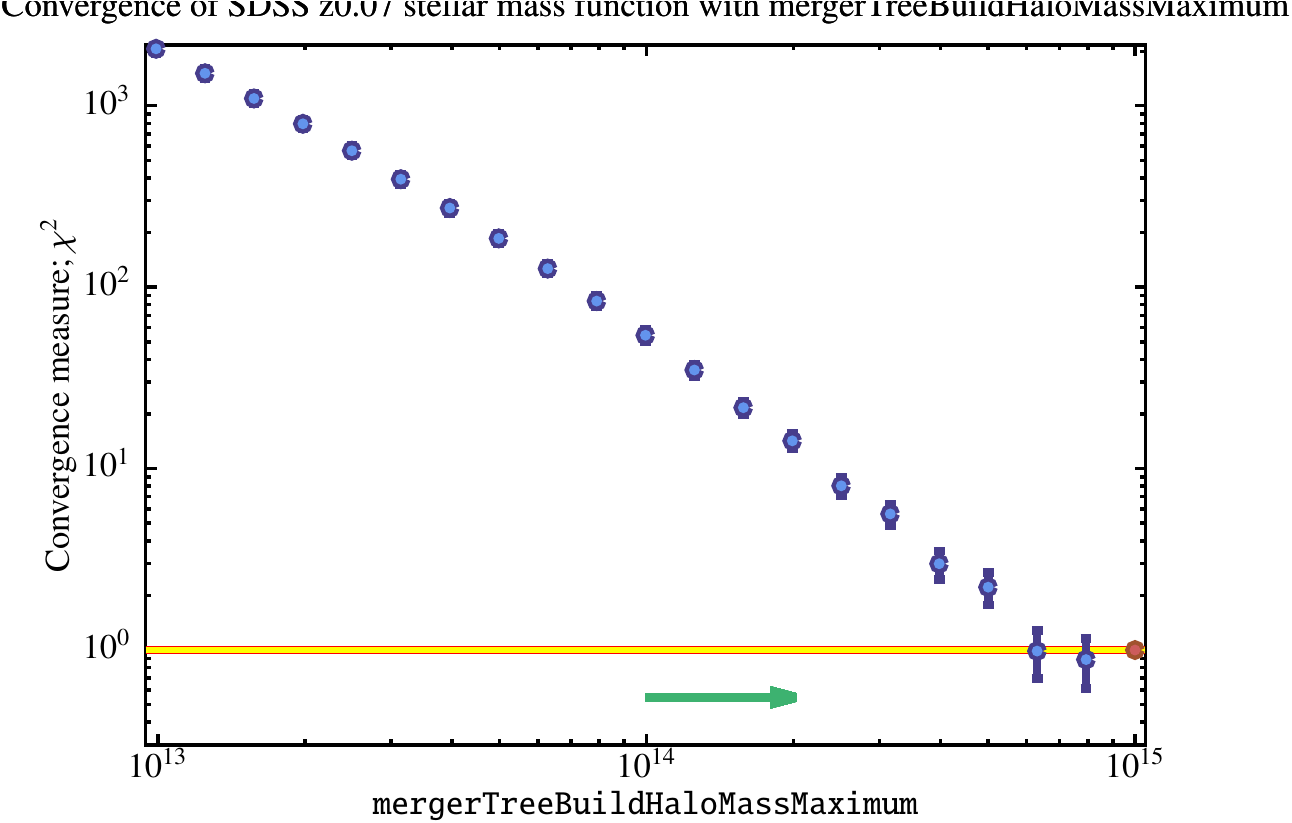} \\
  \includegraphics[width=85mm,trim=0mm 0mm 0mm 2.5mm,clip]{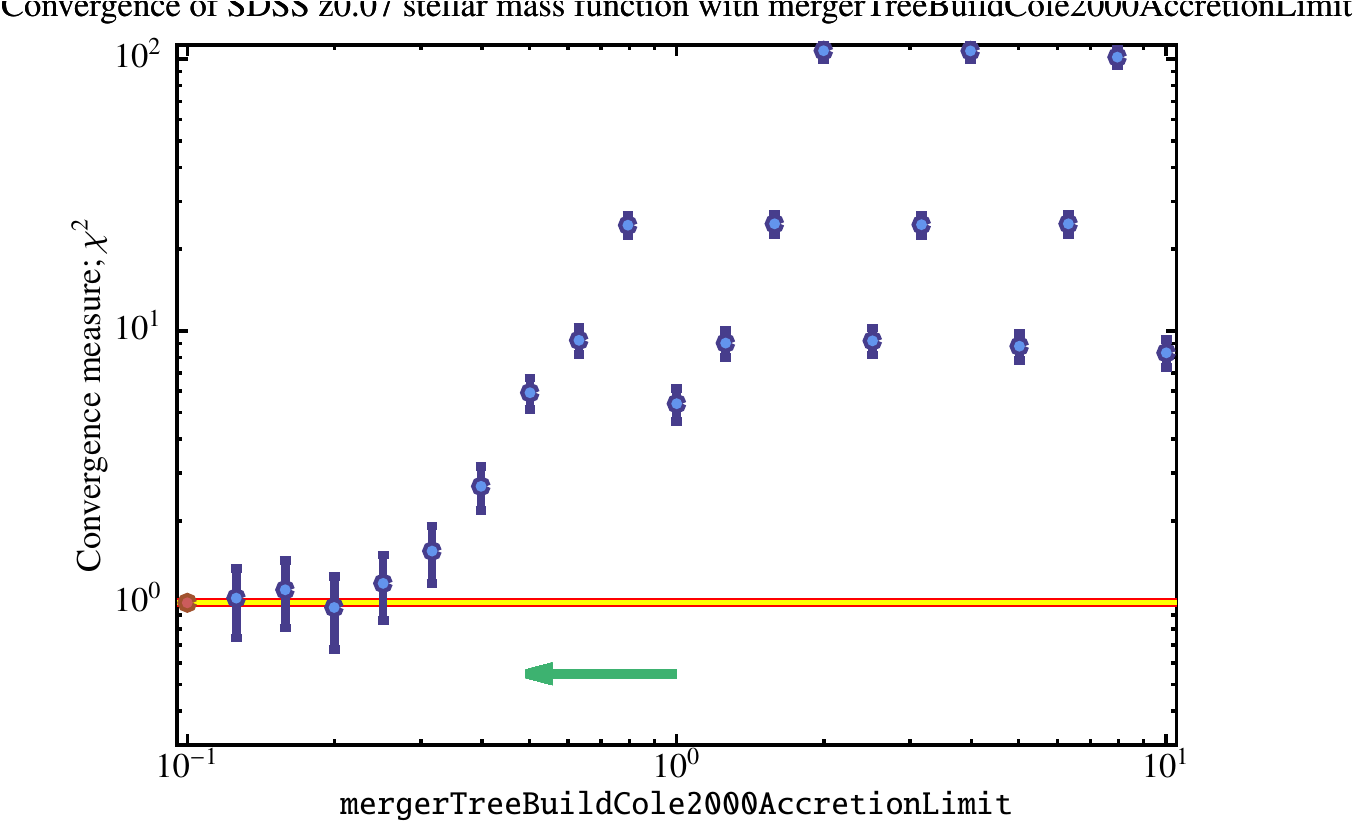} &
  \includegraphics[width=85mm,trim=0mm 0mm 0mm 2.5mm,clip]{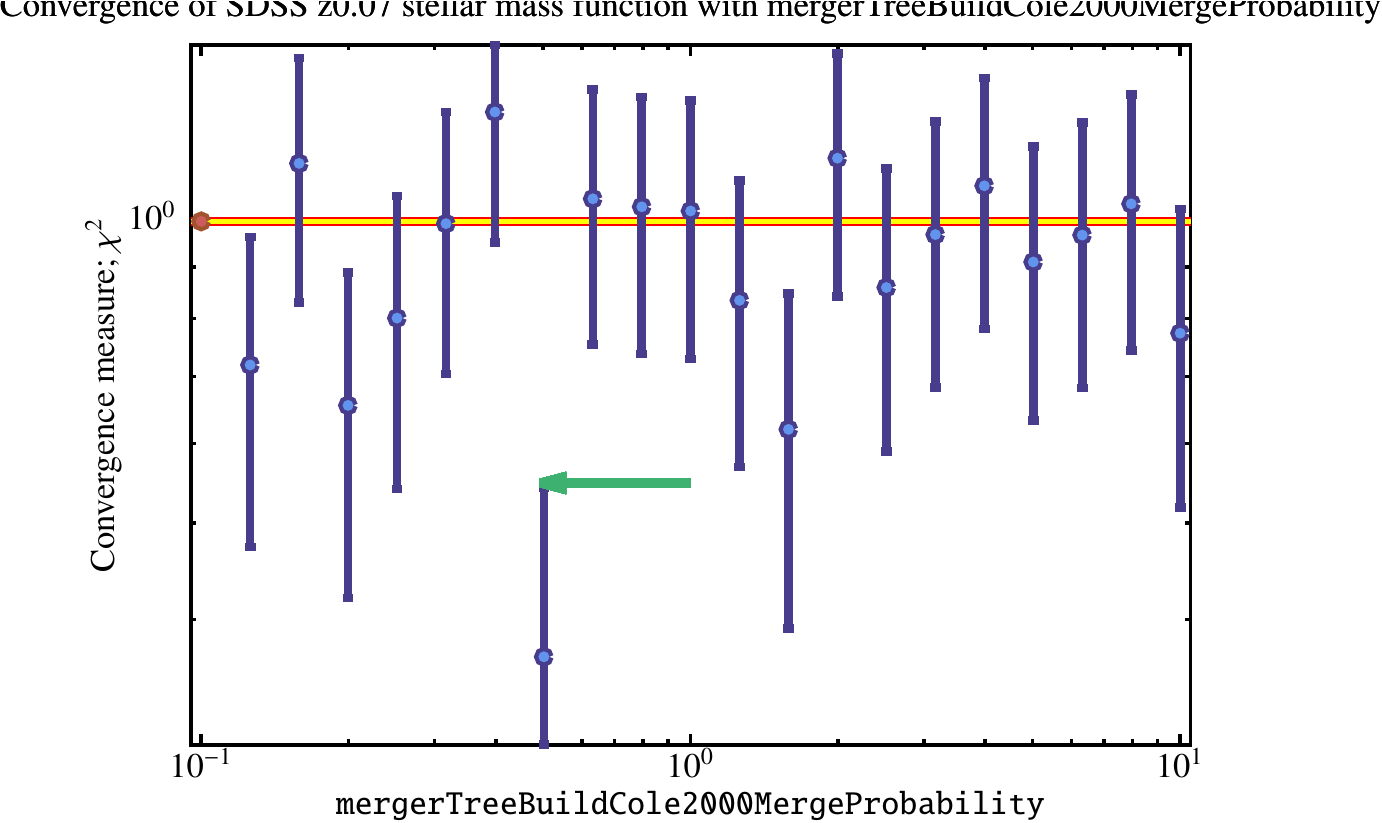} \\
  \includegraphics[width=85mm,trim=0mm 0mm 0mm 2.5mm,clip]{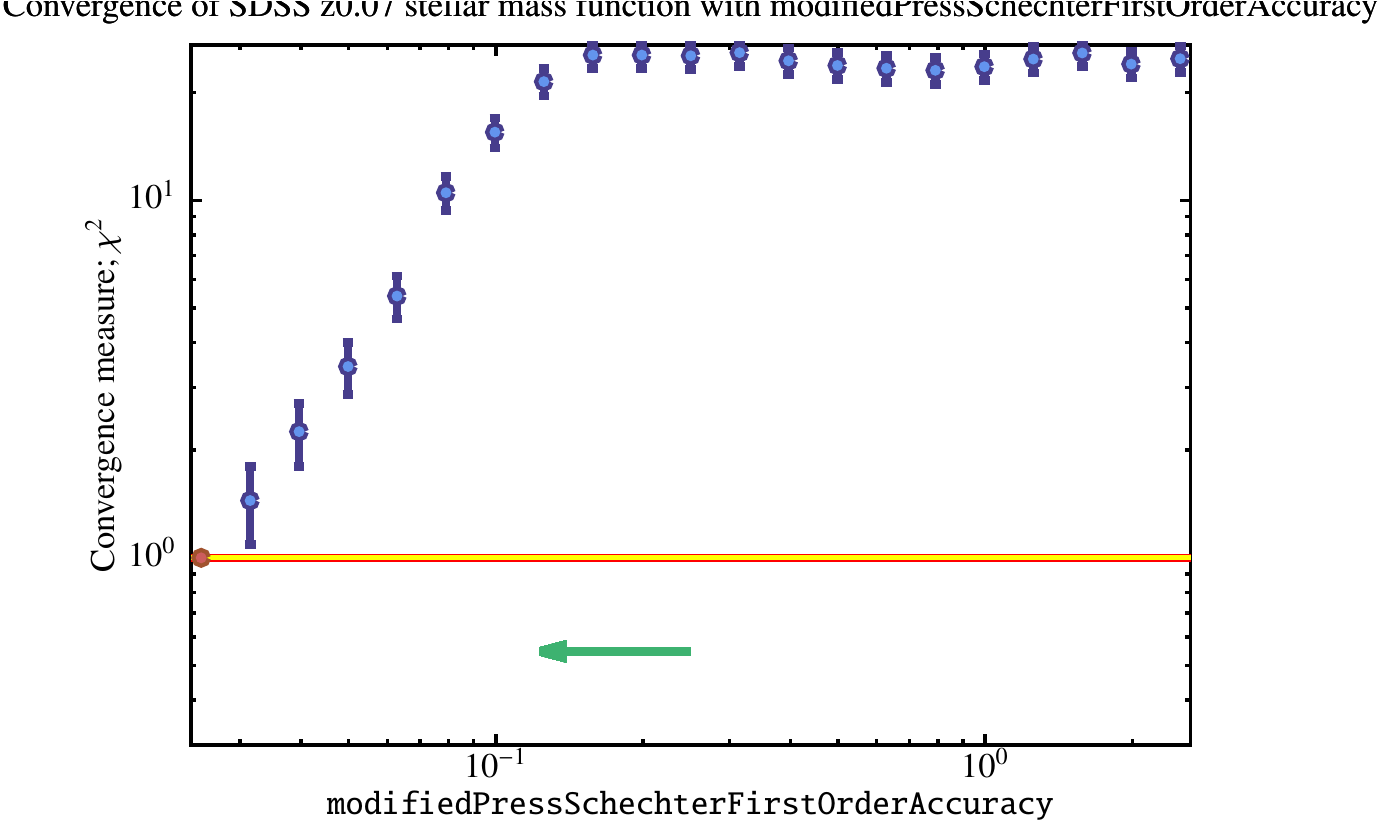} \\
  \includegraphics[width=85mm,trim=0mm 0mm 0mm 2.5mm,clip]{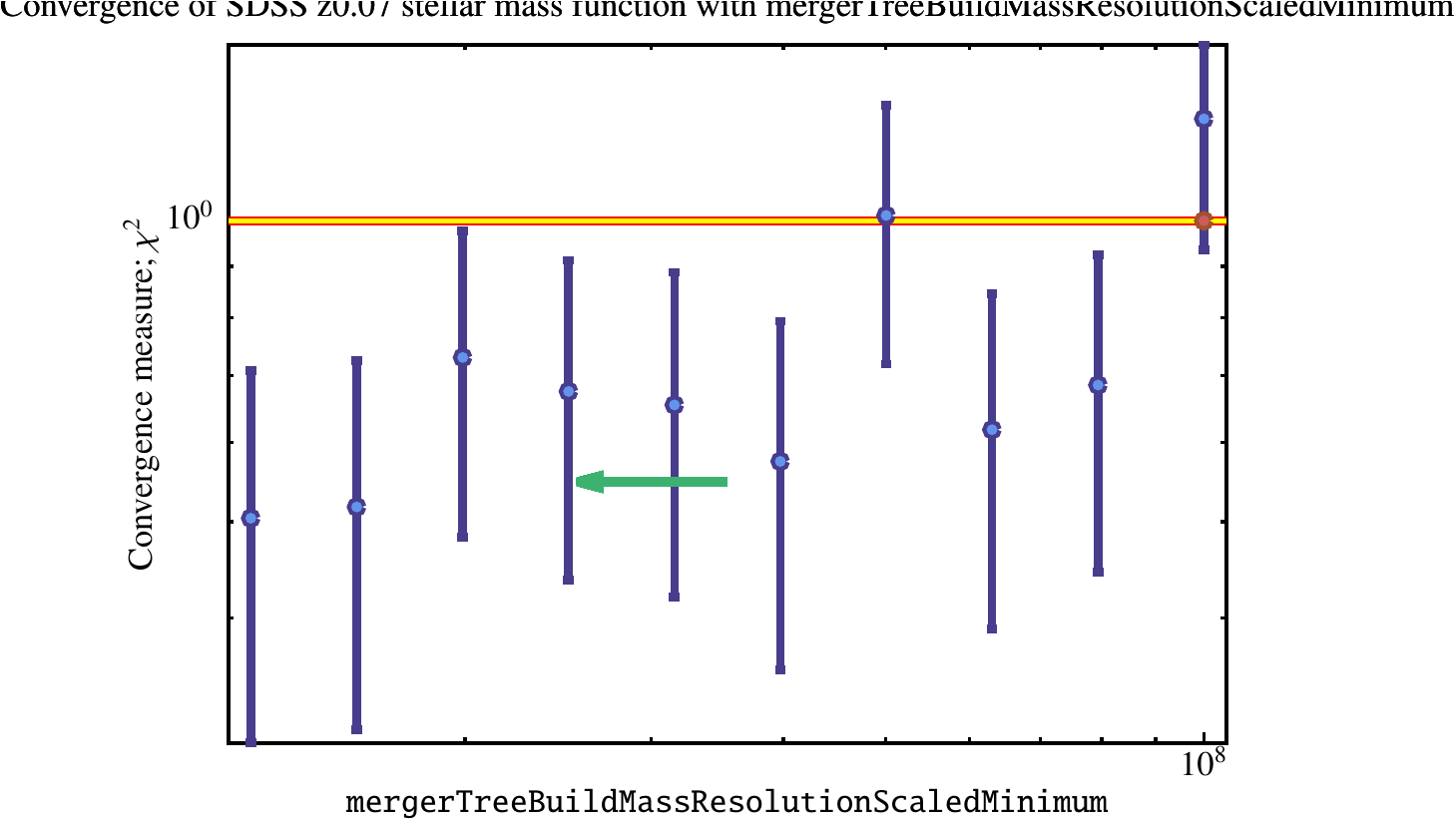} &
  \includegraphics[width=85mm,trim=0mm 0mm 0mm 2.5mm,clip]{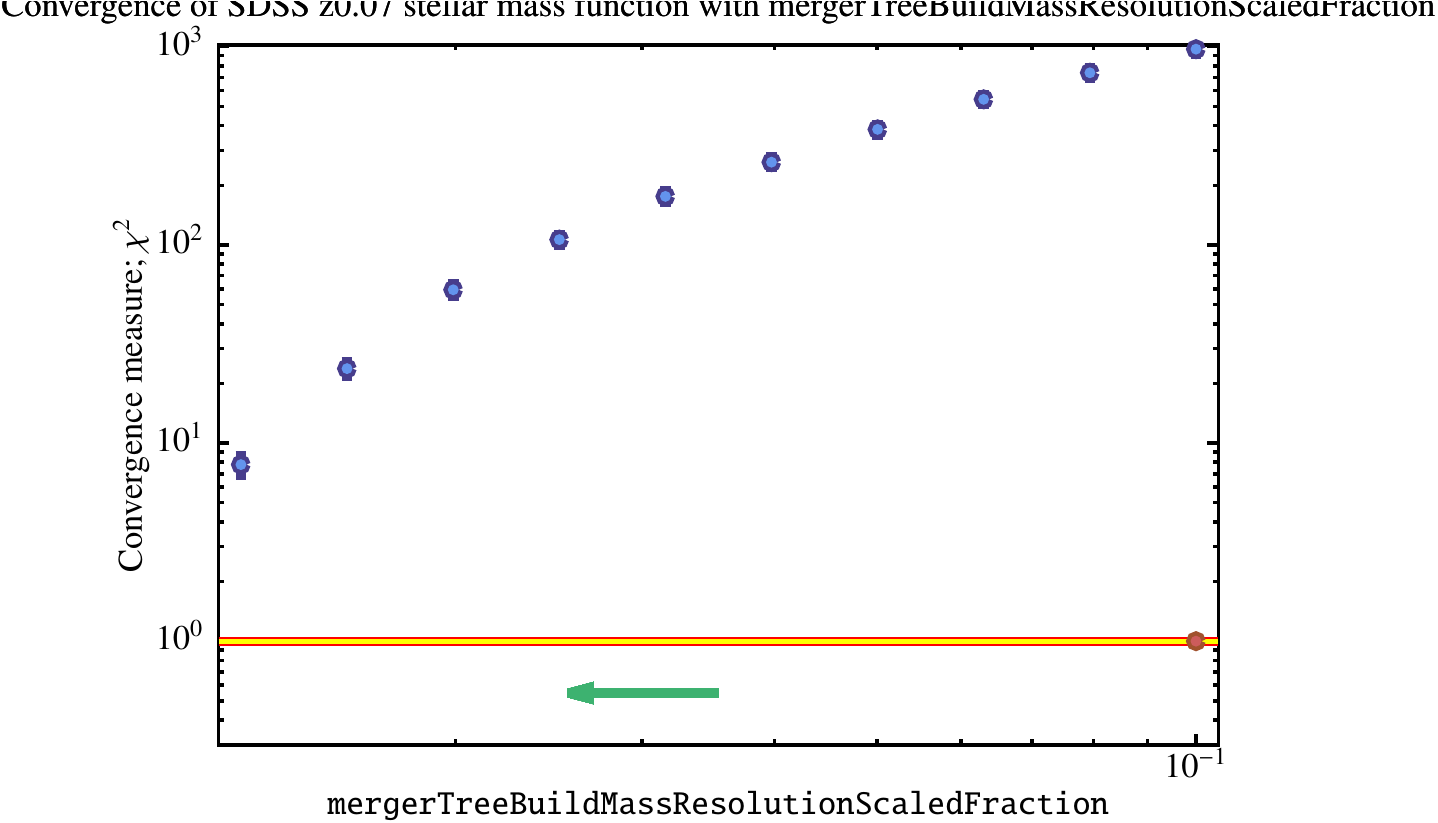} 
\end{tabular}
 \caption{Convergence diagrams for model parameters that control the construction of dark matter merger trees. The solid horizontal lines show $\chi^2=1$ (the expected value for converged models), the red point (shown at arbitrary $\chi^2$) indicates the value of the parameter used in our models for the \protect\MCMC\ analysis, the green arrow indicates the direction of increasing optimality of the parameter (i.e. moving in the direction of the arrow should lead to better converged models), and the blue points indicate the convergence measures obtained by running models while varying the parameter shown on the $x$-axis.}
 \label{fig:ConvergenceBuild}
\end{figure*}

\begin{figure*}
 \begin{tabular}{cc}
  \includegraphics[width=85mm,trim=0mm 0mm 0mm 2.5mm,clip]{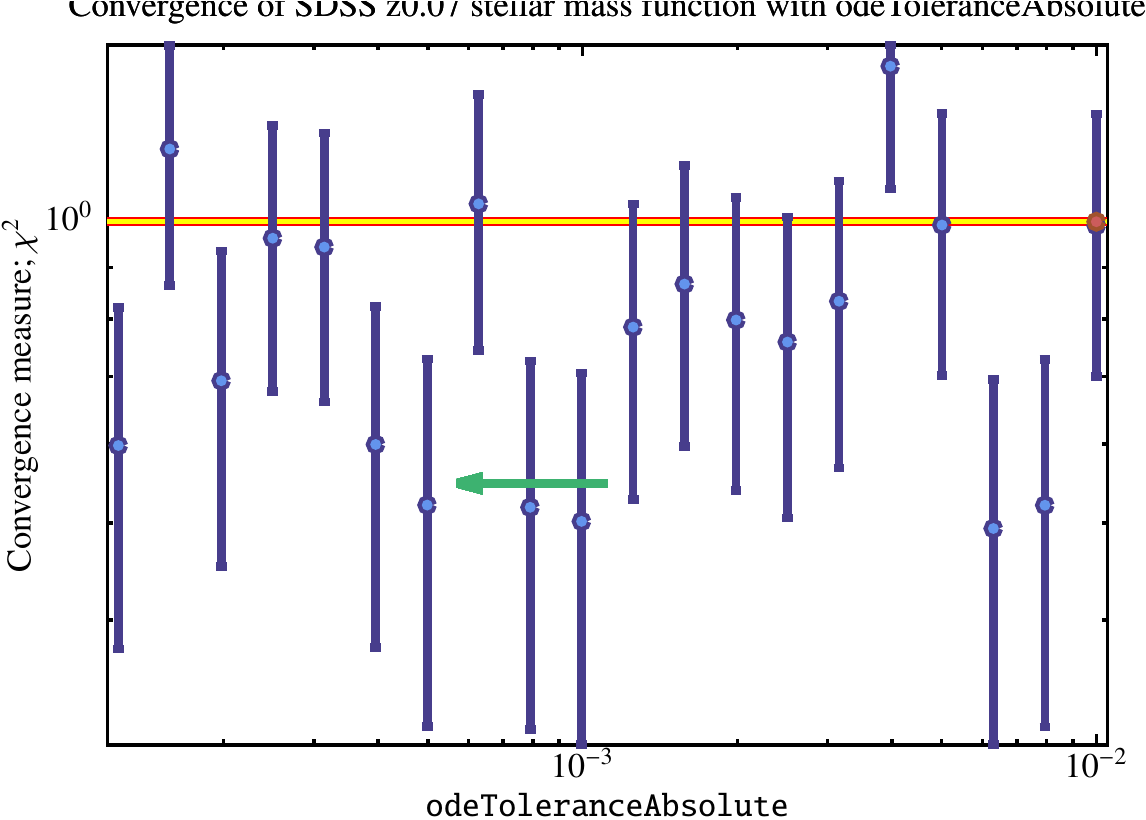} &
  \includegraphics[width=85mm,trim=0mm 0mm 0mm 2.5mm,clip]{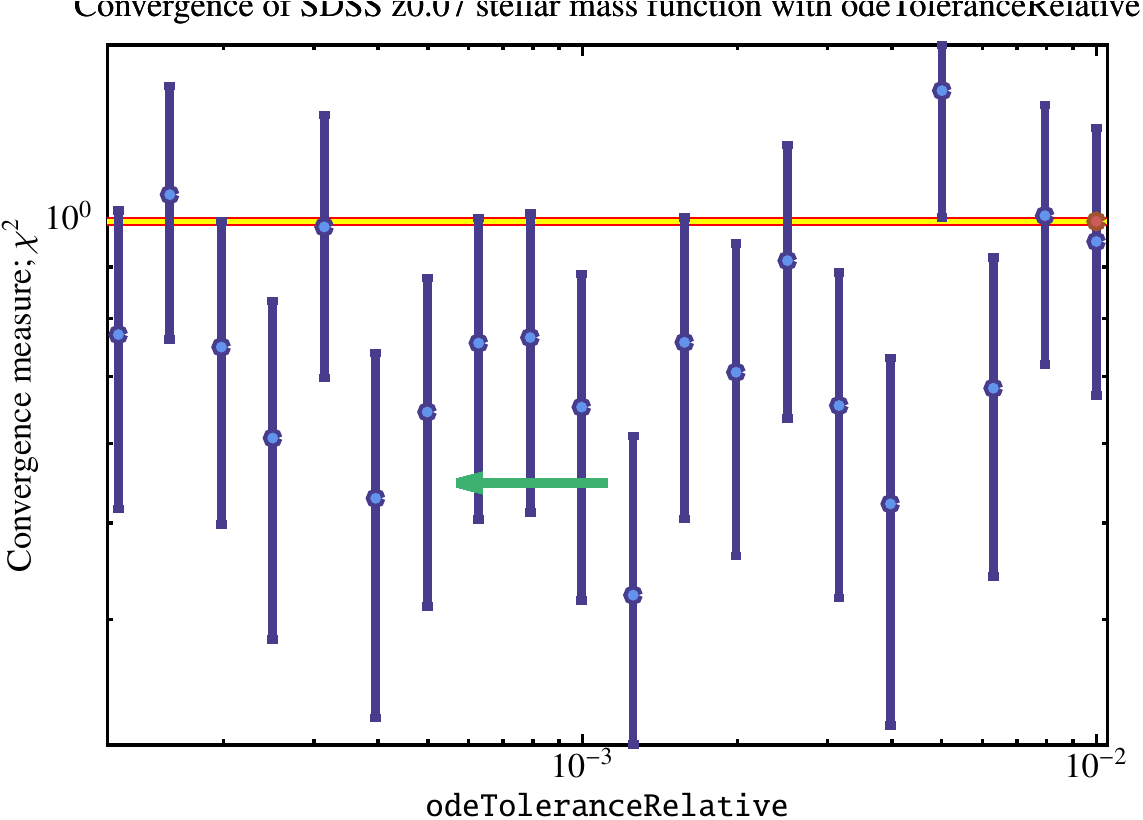} \\
  \includegraphics[width=85mm,trim=0mm 0mm 0mm 2.5mm,clip]{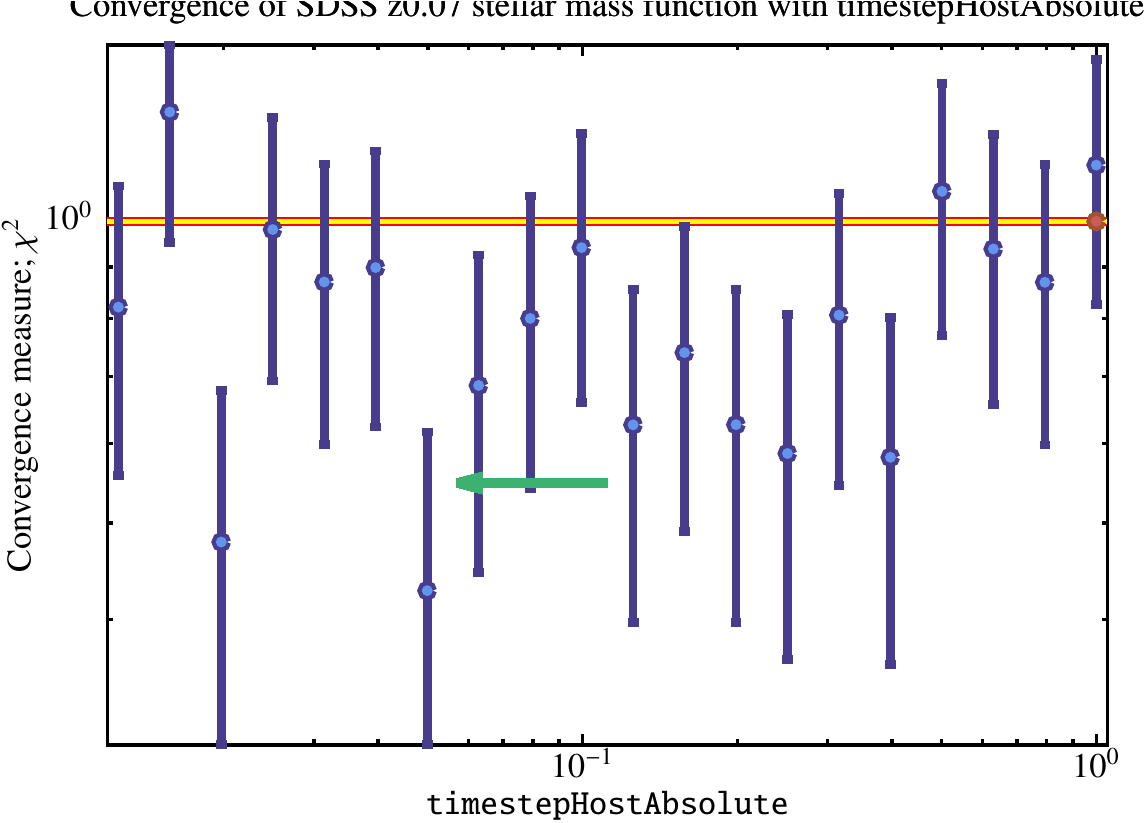} &
  \includegraphics[width=85mm,trim=0mm 0mm 0mm 2.5mm,clip]{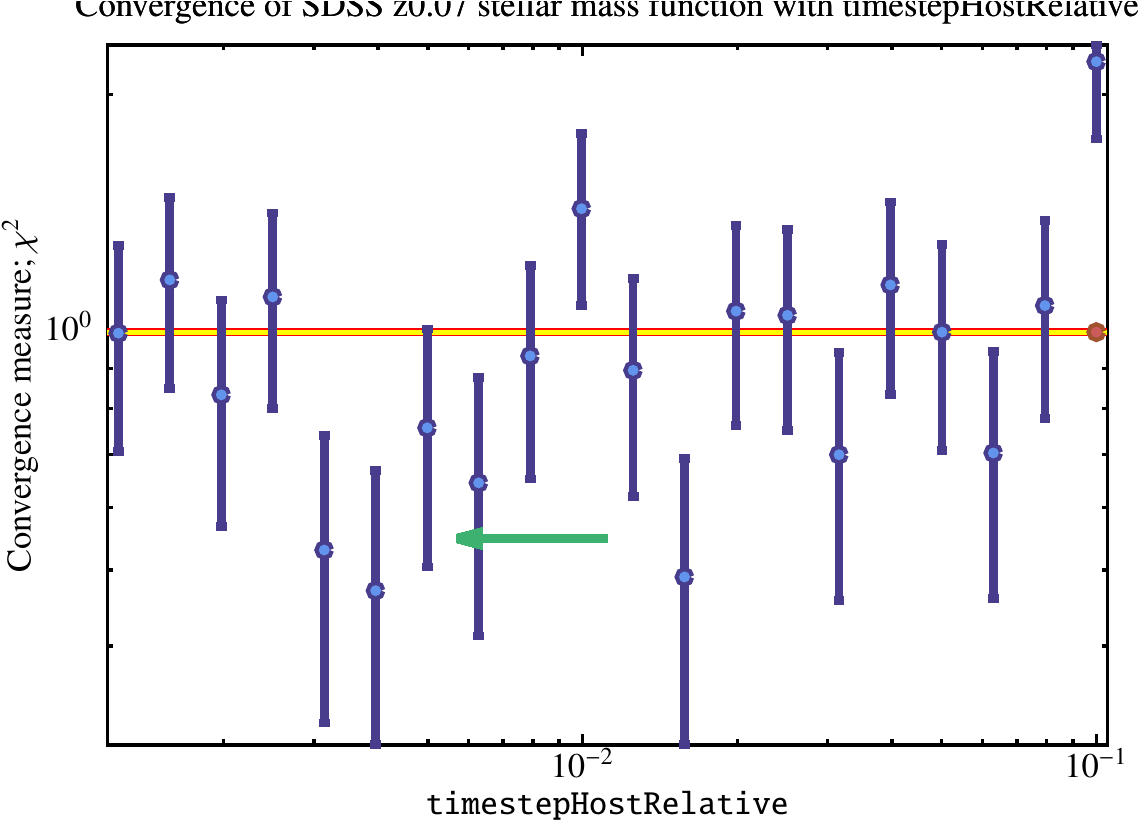} \\
  \includegraphics[width=85mm,trim=0mm 0mm 0mm 2.5mm,clip]{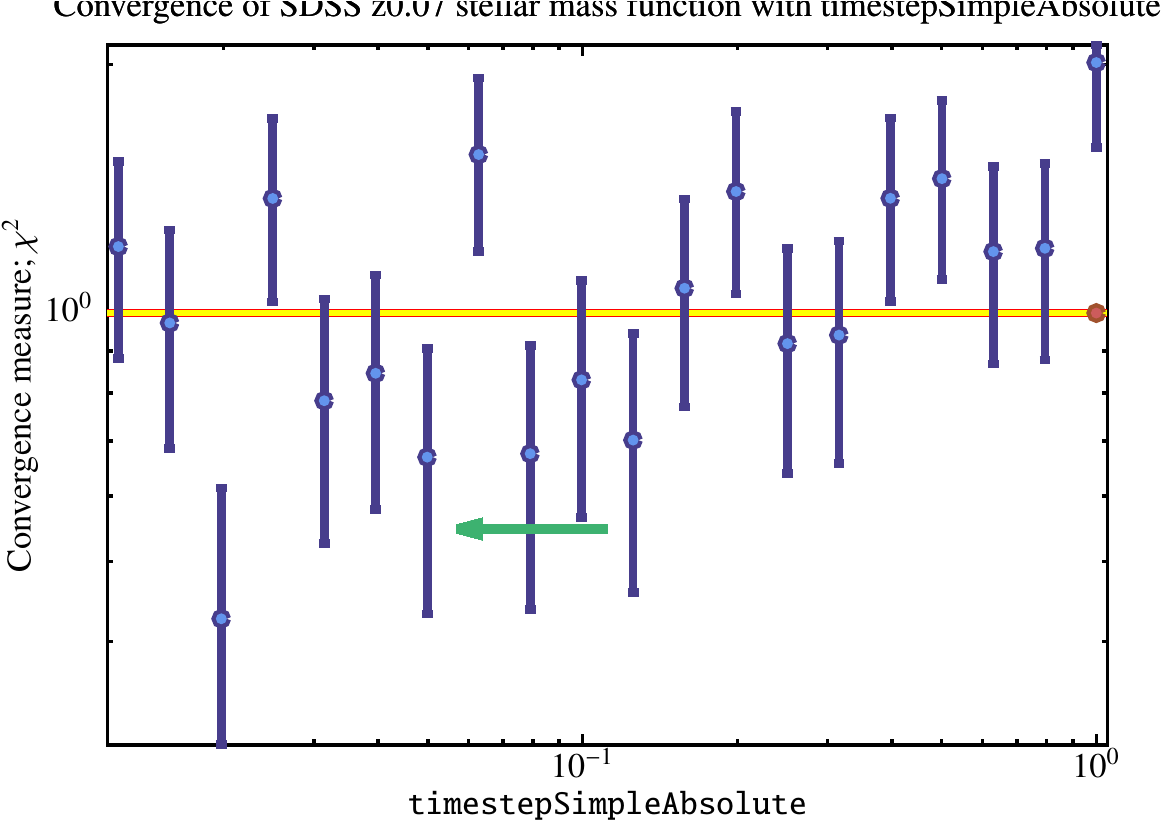} &
  \includegraphics[width=85mm,trim=0mm 0mm 0mm 2.5mm,clip]{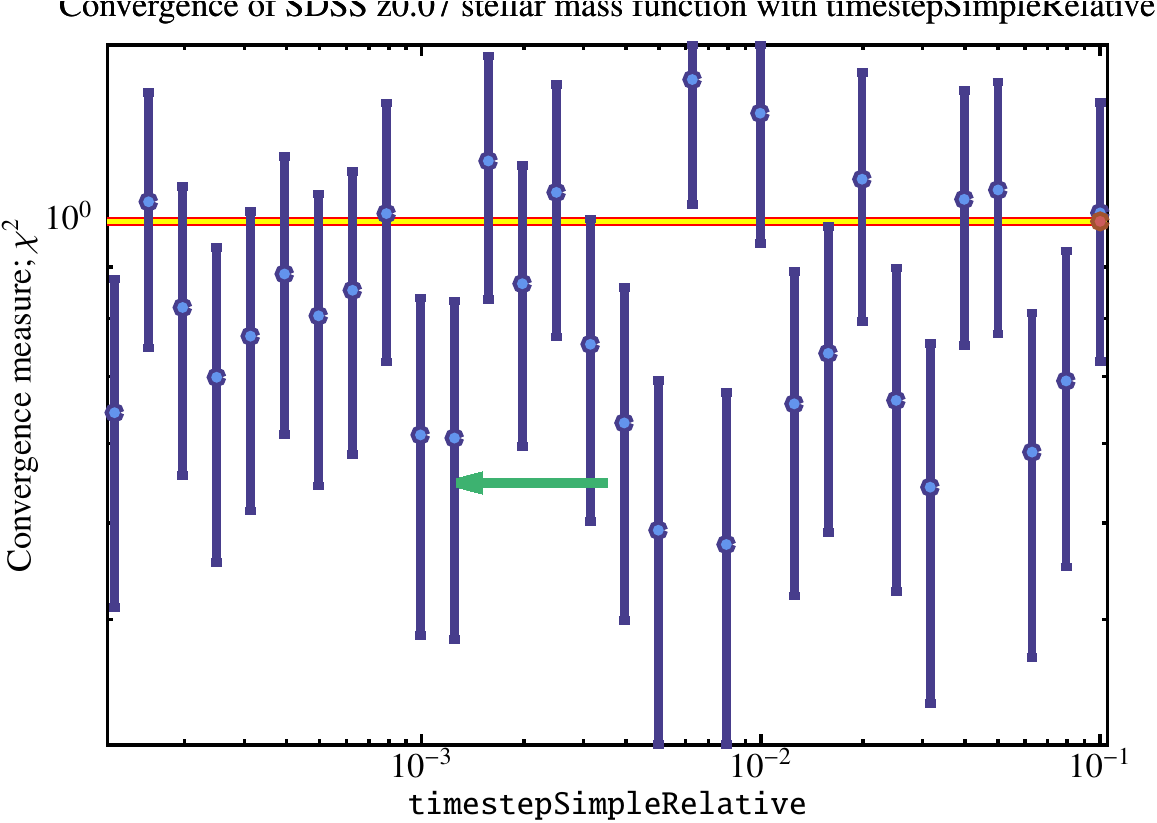}
 \end{tabular}
 \caption{Convergence diagrams for model parameters that control the time-stepping when evolving systems of \protect\ODEs\ through merger trees. The solid horizontal lines show $\chi^2=1$ (the expected value for converged models), the red point (shown at arbitrary $\chi^2$) indicates the value of the parameter used in our models for the \protect\MCMC\ analysis, the green arrow indicates the direction of increasing optimality of the parameter (i.e. moving in the direction of the arrow should lead to better converged models), and the blue points indicate the convergence measures obtained by running models while varying the parameter shown on the $x$-axis.}
 \label{fig:ConvergenceEvolve}
\end{figure*}

Figures~\ref{fig:ConvergenceBuild} and \ref{fig:ConvergenceEvolve} show convergence diagrams for all numerical parameters in our model. The solid horizontal line shows $\chi^2=1$ (the expected value for converged models), the red point (shown at arbitrary $\chi^2$) indicates the value of the parameter used in our models for the \MCMC\ analysis, the green arrow indicates the direction of increasing optimality of the parameter (i.e. moving in the direction of the arrow should lead to better converged models), and the blue points indicate the convergence measures obtained by running models while varying the parameter shown on the $x$-axis.

The top row of Fig.~\ref{fig:ConvergenceBuild} shows convergence in the parameters {\tt mergerTreeBuildHaloMassMinimum} and {\tt mergerTreeBuildHaloMassMaximum} which control the minimum and maximum masses halos at $z=0$ that are used as the roots of merger trees. The range spanned must be sufficient to capture all galaxies which contribute significantly to the stellar mass function. Clearly for both parameters the model is converged for our choice of values. In the case of {\tt mergerTreeBuildHaloMassMinimum} the model remains converged up to $2\times 10^{11}M_\odot$ (the highest value we consider). Nevertheless, we retain a value of $2\times10^8M_\odot$ for this parameter since, as the \MCMC\ algorithm explores parameter space it can sometimes produce models which form galaxies much more efficiently in lower mass halos, thereby requiring a lower value of {\tt mergerTreeBuildHaloMassMinimum} to achieve convergence. Given the available mass of baryons in $2\times 10^{8}M_\odot$ halos, this value of {\tt mergerTreeBuildHaloMassMinimum} is sufficient for convergence in any region of parameter space that the \MCMC\ algorithm might explore. For {\tt mergerTreeBuildHaloMassMaximum}, convergence is achieved above $6\times 10^{14}M_\odot$---higher mass trees are sufficiently rare that they do not contribute significantly to the stellar mass function.

The second and third rows of Fig.~\ref{fig:ConvergenceBuild} shows convergence in parameters which control the size of the timestep taken when constructing merger trees using the algorithm of \cite{parkinson_generating_2008}. The parameters {\tt mergerTreeBuildCole2000AccretionLimit} and {\tt mergerTreeBuildCole2000MergeProbability} correspond to the maximum allowed values of the quantities that \cite{parkinson_generating_2008} label $F$ and $P$ (their eqns.~4 and 5). Model results are clearly well-converged with respect to these parameters. The parameter {\tt modifiedPressSchechterFirstOrderAccuracy} corresponds to $\epsilon_1$ defined by \citeauthor{parkinson_generating_2008}~(\citeyear{parkinson_generating_2008}; their Appendix~A) and also limits timesteps in tree building\footnote{Specifically, it ensures that eqn.~(2) of \protect\cite{parkinson_generating_2008} is first-order accurate.}. Convergence in this parameter is just barely reached at our chosen value. Finally, the bottom row of Fig.~\ref{fig:ConvergenceBuild} shows parameters controlling the mass resolution of merger trees. Parameters {\tt mergerTreeBuildMassResolutionScaledMinimum} and {\tt mergerTreeBuildMassResolutionScaledFraction} correspond to $M_{\rm res,min}$ and $f_0$ of eqn.~(\ref{eq:TreeMassResolution}) respectively. {\tt mergerTreeBuildMassResolutionScaledMinimum} is well-converged at our chosen value, but {\tt mergerTreeBuildMassResolutionScaledFraction} is clearly far from being converged at our chosen value of $0.1$. However, we are forced to adopt this large value to permit models to be computed sufficiently rapidly. Knowing that this will cause of models to give discrepant answers, we include a model discrepancy term to account for this approximation.

Figure~\ref{fig:ConvergenceEvolve} shows convergence diagrams for six parameters which control time-stepping in our model. The top row shows {\tt odeToleranceAbsolute} and {\tt odeToleranceRelative}. In our model the \ODE\ solver uses adaptive timesteps to keep the error in variable $i$ below
\begin{equation}
 \delta_i = \epsilon_{\rm abs} s_i + \epsilon_{\rm rel} |y_i|,
\end{equation}
where $y_i$ is the value of the $i^{\rm th}$ variable, $s_i$ is a fixed value\footnote{More specifically, $s_i$ is fixed over a given timestep. It may change between timesteps due to possible dependencies on other variables.} for the $i^{\rm th}$ variable, and $\epsilon_{\rm abs}$ and $\epsilon_{\rm rel}$ correspond to {\tt odeToleranceAbsolute} and {\tt odeToleranceRelative} respectively. The model is clearly well-converged with respect to both of these parameters.

Nodes of merger trees in our model are evolved individually---that is, each node will evolve for some time before the code moves on to another node. This means that nodes are always slightly out-of-sync (they are, of course, brought in to sync at each requested output time). If galaxies in nodes are non-interacting this has no effect on the overall evolution of the tree and its constituent galaxies. However, galaxies do interact---specifically satellite galaxies interact with their host halo (and the central galaxy of that host) by by transferring their hot gas component to the host (``strangulation'') and by merging with the central galaxy. It is important therefore to prevent satellites and hosts from getting too far out of sync. The four parameters shown in the lower two rows of Fig.~\ref{fig:ConvergenceEvolve} limit how far satellites can get out of sync with their hosts before their evolution is halted until the host node ``catches up''. The reader is referred to \cite{benson_galacticus:_2012} for full details of these parameters. The convergence diagrams for these parameters controlling time-stepping of satellites and hosts all clearly show that the model is converged with respect to these parameters.

Convergence potentially depends on the parameters of the model being varied in the \MCMC\ analysis. Therefore, once the \MCMC\ analysis was completed we repeated this convergence study using the maximum likelihood set of parameters. The convergence characteristics of most parameters remain unchanged. There are two cases where the model is less well converged when using the maximum likelihood model compared to the \emph{a priori} model, both of which affect only the high-mass end of the galaxy stellar mass function. The first case is for the parameter {\tt modifiedPressSchechterFirstOrderAccuracy} which controls merger tree construction. With the \emph{a priori} model, this parameter was just barely converged at the value adopted in this work. With the maximum likelihood model this is no longer true, and {\tt modifiedPressSchechterFirstOrderAccuracy} would need to be decreased by a factor  of around 3 to restore convergence. The second case is the timestepping parameters {\tt timestepHostRelative} and {\tt timestepHostAbsolute}. These were set to $0.1$ and $1$~Gyr respectively in this work. For the maximum likelihood model to be converged would require either {\tt timestepHostRelative} to be reduced to $0.003$, or for {\tt timestepHostAbsolute} to be reduced to $0.03$~Gyr. This change in convergence characteristics reflects differences in the timescales for star formation and outflows in galaxies in the maximum likelihood and \emph{a priori} models.

We note that, while we can detect this lack of convergence in our model by running simulations using a very large number of merger trees (to reduce statistical noise), the resulting changes in the galaxy stellar mass function are small compared to the statistical noise in the models that we actually run in our \MCMC\ analysis (which utilize many fewer merger trees to keep calculation times sufficiently short).

Nevertheless, this suggests that, in future work, an iterative approach be adopted (see also \S\ref{sec:ModelDiscrepancyResults} where a similar approach is recommended for model discrepancy calculations), in which convergence is tested for the current maximum likelihood model, new constraints are introduced, the model parameter space is explored quickly using \MCMC, and convergence is re-tested using the new maximum likelihood model. This process can be iterated until convergence characteristics are no longer changing.

\section{Model Mass Function Covariance Matrix Model}\label{app:ModelCovariance}

Our \SAM\ uses a Monte Carlo approach to generate results, simulating a random sample of dark matter merger trees and then summing over their constituent galaxies. As such, the results from the model always have some random fluctuation around the ``true'' answer for a given set of input parameters. This additional covariance should be accounted for when evaluating model likelihoods. Ideally, the covariance would be made so small (by generating a sufficiently large number of Monte Carlo realizations) that it is entirely dominated by the covariance arising from the observational data, and so could be ignored. In practice, the \MCMC\ technique is so time consuming that we are always driven to generate ``just enough'' Monte Carlo merger tree realizations to keep the covariance non-dominant, but non-negligible.

It is therefore important to have a good model for the covariance of the model results. \glc\ works by sampling a set of tree ``root masses'' from the $z=0$ dark matter halo mass function. From each root, a tree is grown, within which the physics of galaxy formation is then solved. Root masses are sampled uniformly from the halo mass function. That is, the cumulative halo mass function, $N(M)$, is constructed between the maximum and minimum halo masses to be simulated. The number of root masses, $N_{\rm r}$, to be used in a model evaluation is then determined. Root masses are then chosen such that
\begin{equation}
 N(M_i) = N(M_{\rm min}) {i-1 \over N_{\rm r}-1}
\end{equation}
for $i=1\ldots N_{\rm r}$ (noting that $N(M_{\rm max})=0$ by construction). 

Consider first those galaxies which form in the main branch of each tree (i.e. those galaxies which are destined to become the central galaxy of the $z=0$ halo). Suppose that we simulate $N_k$ halos of root mass $M_k$ at $z=0$. In such halos the main branch galaxies will, at any time, have stellar masses drawn from some distribution $p_k(M_\star|t)$. The number of such galaxies contributing to bin $i$ of the mass function is therefore binomially distributed with success probability $p_{ik} = \int_{M_{i,\rm min}}^{M_{i,\rm max}} p_k(M_\star|t) \d M_\star$ and a sample size of $N_k$. The contribution to the covariance matrix from these main branch galaxies is therefore:
\begin{equation}
 \mathcal{C}_{ij} = \left\{ \begin{array}{ll} p_{ik}(1-p_{ik}) N_k w_k^2 & \hbox{ if } i = j, \\ -p_{ik} p_{jk} N_k w_k^2 & \hbox{ otherwise,} \end{array} \right.
\end{equation}
where $w_k$ is the weight to be assigned to each tree. To compute this covariance requires knowledge of the probabilities, $p_{ik}$. We estimate these directly from the model. To do this, we bin trees into narrow bins of root mass and assume that $p_{ik}$ does not vary significantly across the mass range of each bin. Using all realizations of trees that fall within a given bin, $k$, we can directly estimate $p_{ik}$.

In addition to the main branch galaxies, each tree will contain a number of other galaxies (these will be ``satellite'' galaxies at $z=0$, but at higher redshifts may still be central galaxies in their own halos). Previous studies \citep{kravtsov_dark_2004} have established that the number of satellites in halos is well described by a Poisson process.

To test these assumptions we run 1000 realizations of our \emph{a priori} model, each time using a different random seed such that each model is a statistically independent realization. For each model we construct the mass function. From this ensemble of model mass function estimates we directly measure the covariances of main branch galaxies, other galaxies, and the covariance between these two samples.

\begin{figure*}
 \begin{tabular}{cc}
 \multicolumn{2}{c}{Main branch galaxies} \\
 \includegraphics[width=85mm,trim=0mm 0mm 0mm 4mm,clip]{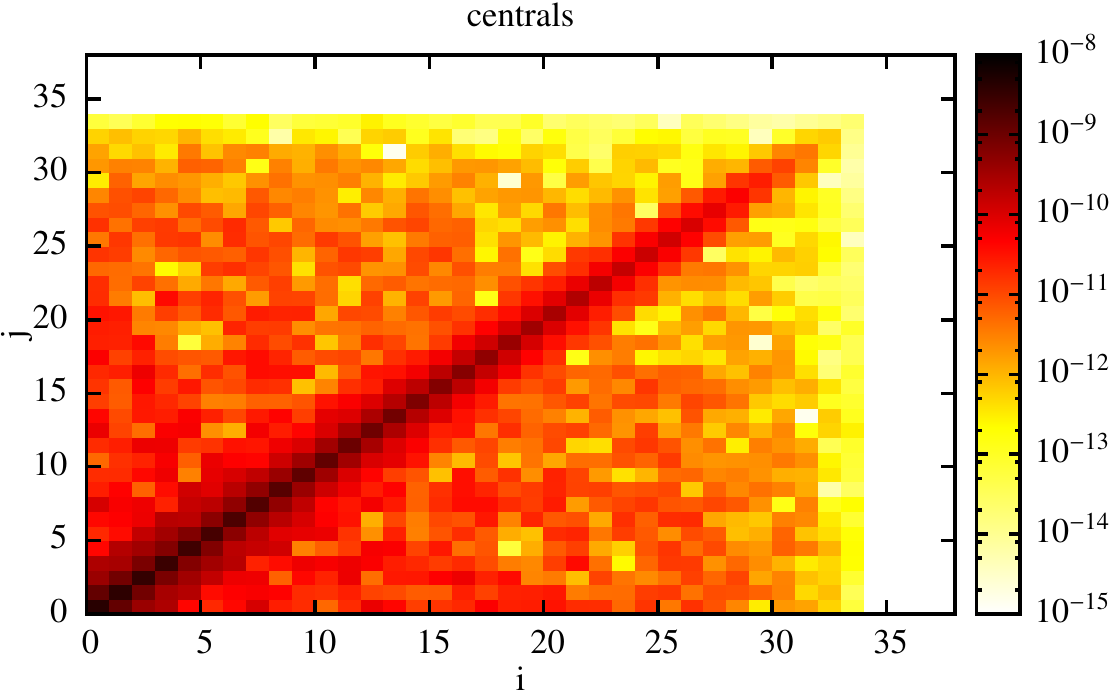} &
 \includegraphics[width=85mm,trim=0mm 0mm 0mm 4mm,clip]{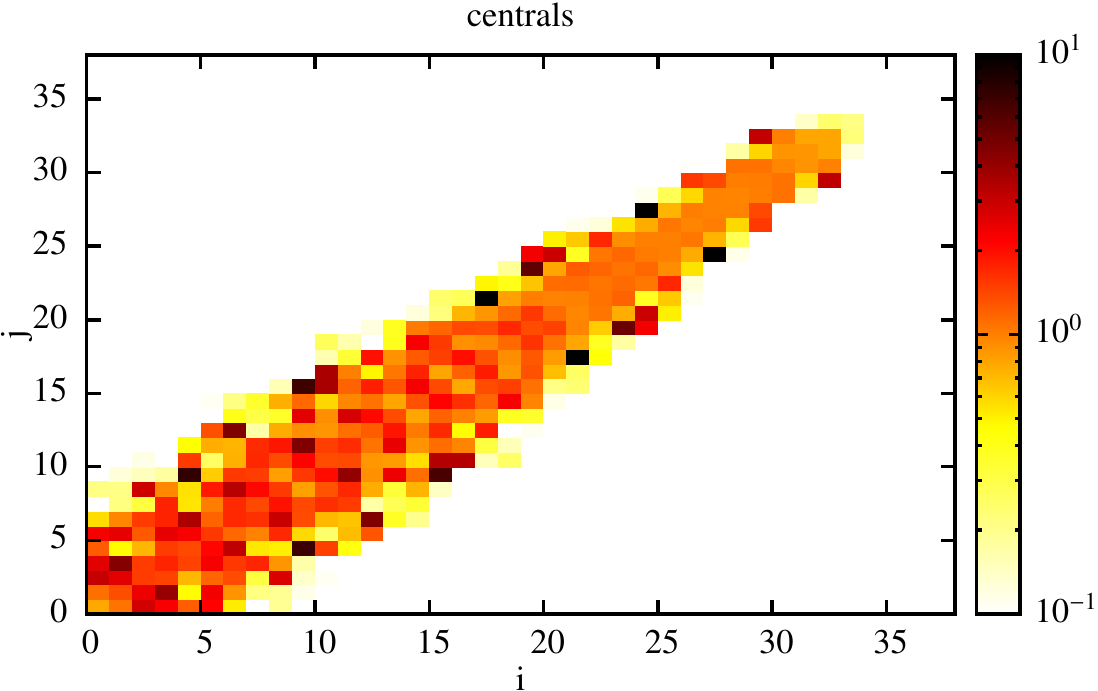} \\
 \multicolumn{2}{c}{Non-main branch galaxies} \\
 \includegraphics[width=85mm,trim=0mm 0mm 0mm 4mm,clip]{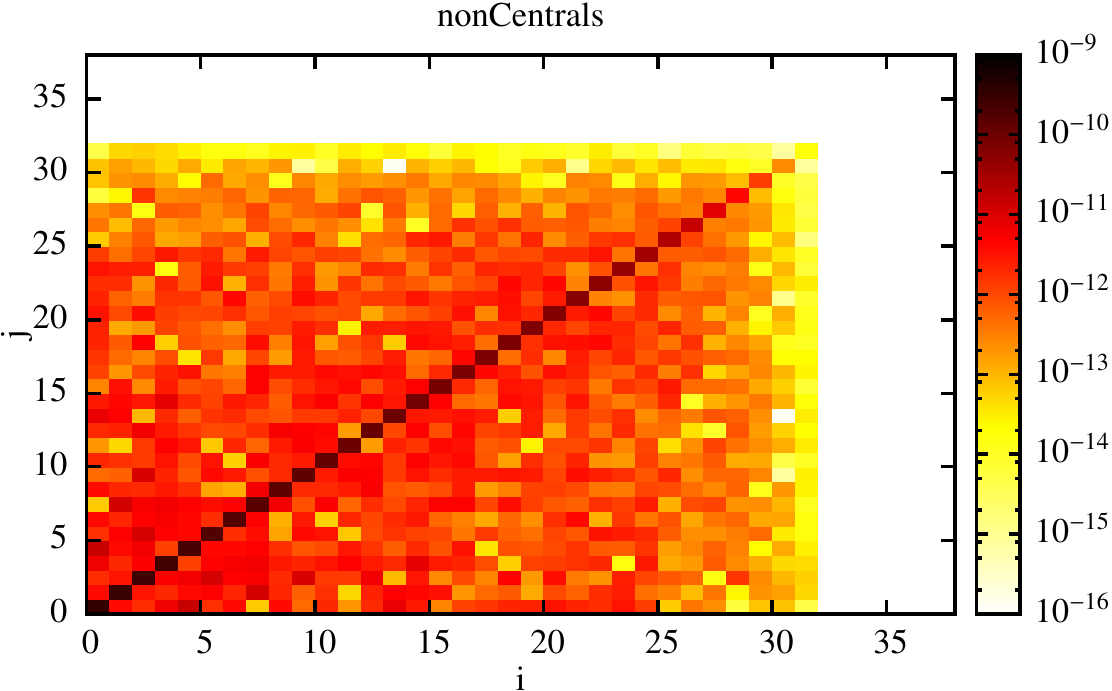} &
 \includegraphics[width=85mm,trim=0mm 0mm 0mm 4mm,clip]{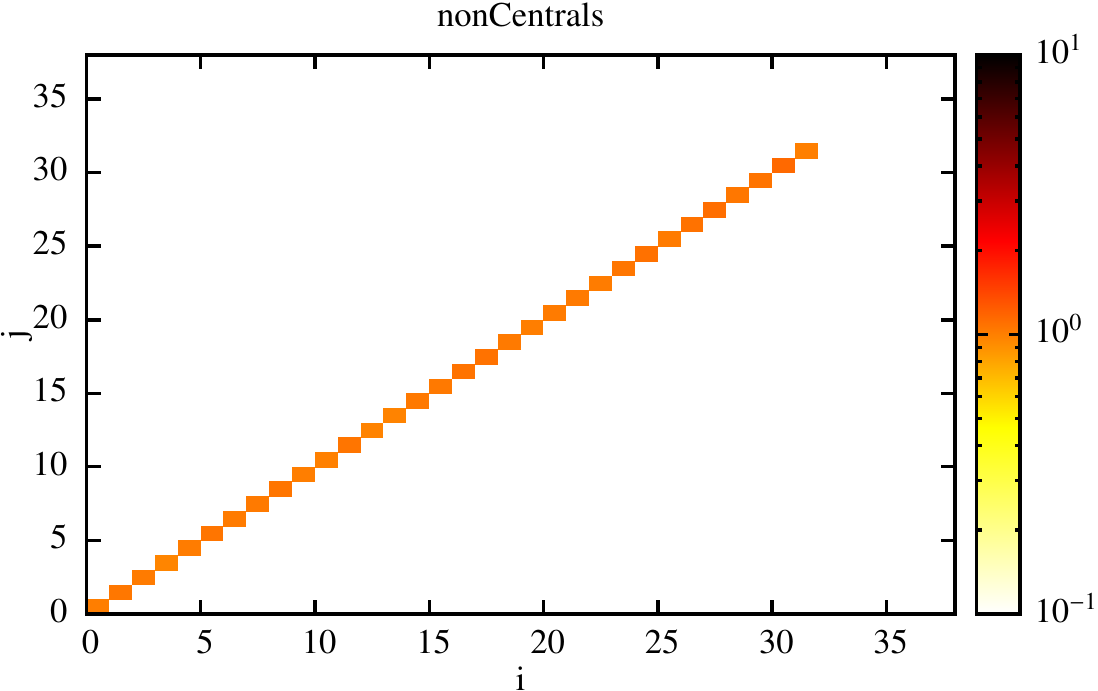} \\
 \multicolumn{2}{c}{Cross} \\
 \includegraphics[width=85mm,trim=0mm 0mm 0mm 4mm,clip]{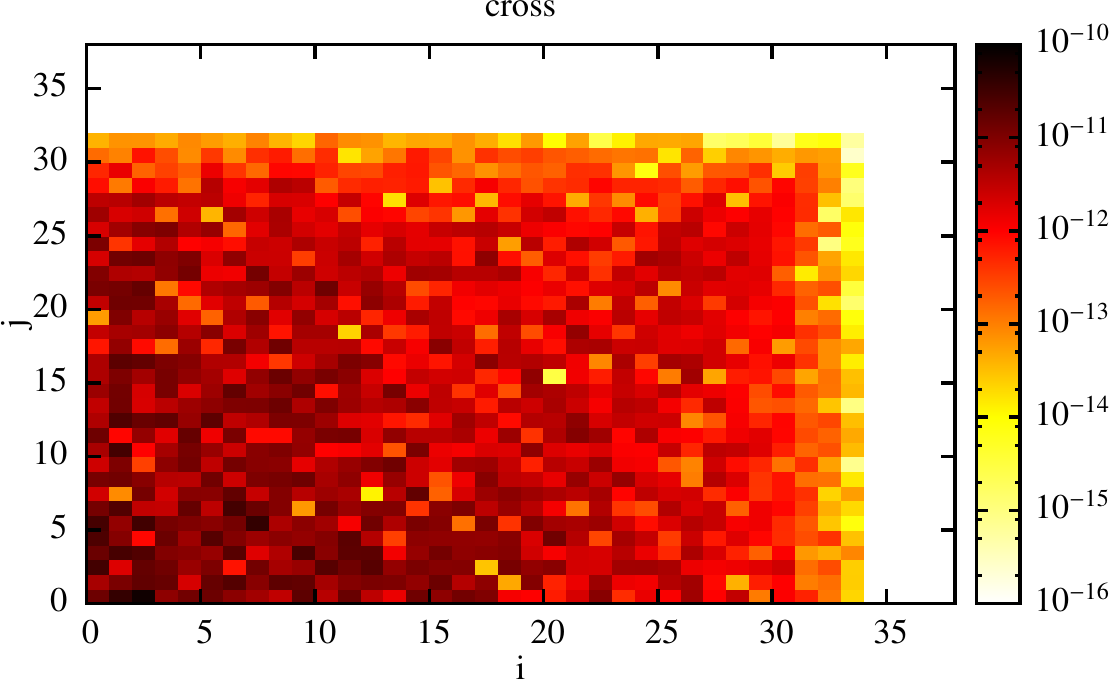} &
 \includegraphics[width=85mm,trim=0mm 0mm 0mm 4mm,clip]{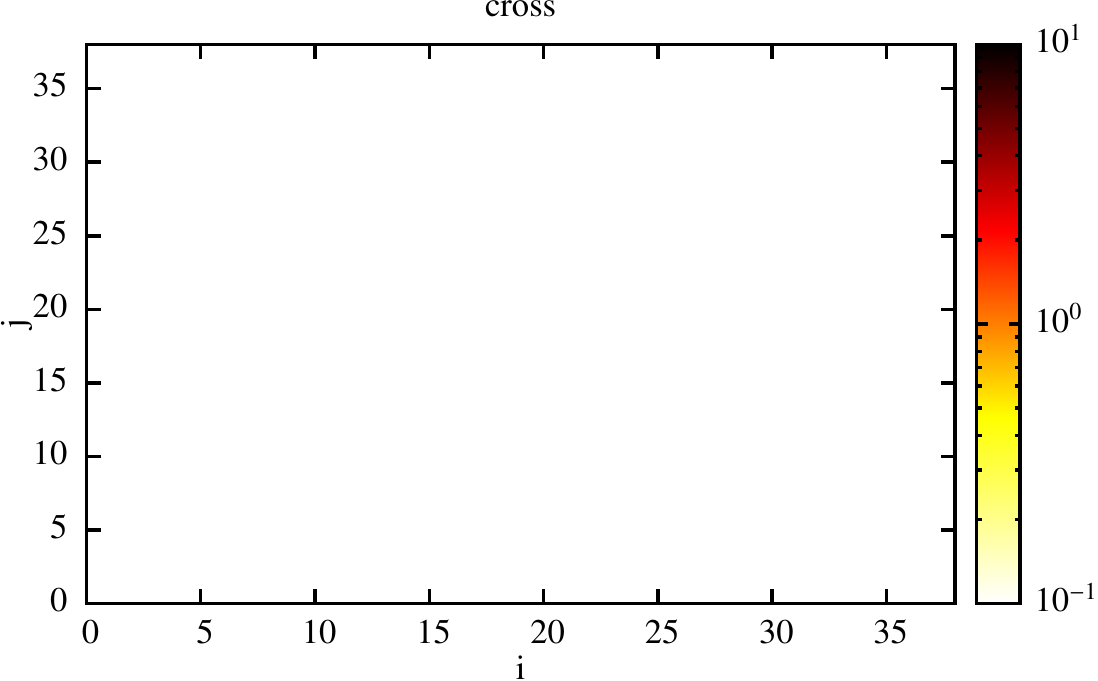}
 \end{tabular}
 \caption{\emph{Left column:} Covariance matrices measured from 1000 realizations of the model stellar mass function are shown for main branch galaxies, non-main branch galaxies, and the covariance between these two populations (i.e. considering the full covariance matrix as a $2 \times 2$ block matrix, these are the two diagonal blocks and the off-diagonal block respectively). \emph{Right column:} The ratio of model to measured covariances. White areas indicate zero model covariance.}
 \label{fig:modelCovariance}
\end{figure*}

\begin{figure}
 \includegraphics[width=85mm,trim=0mm 0mm 0mm 4mm,clip]{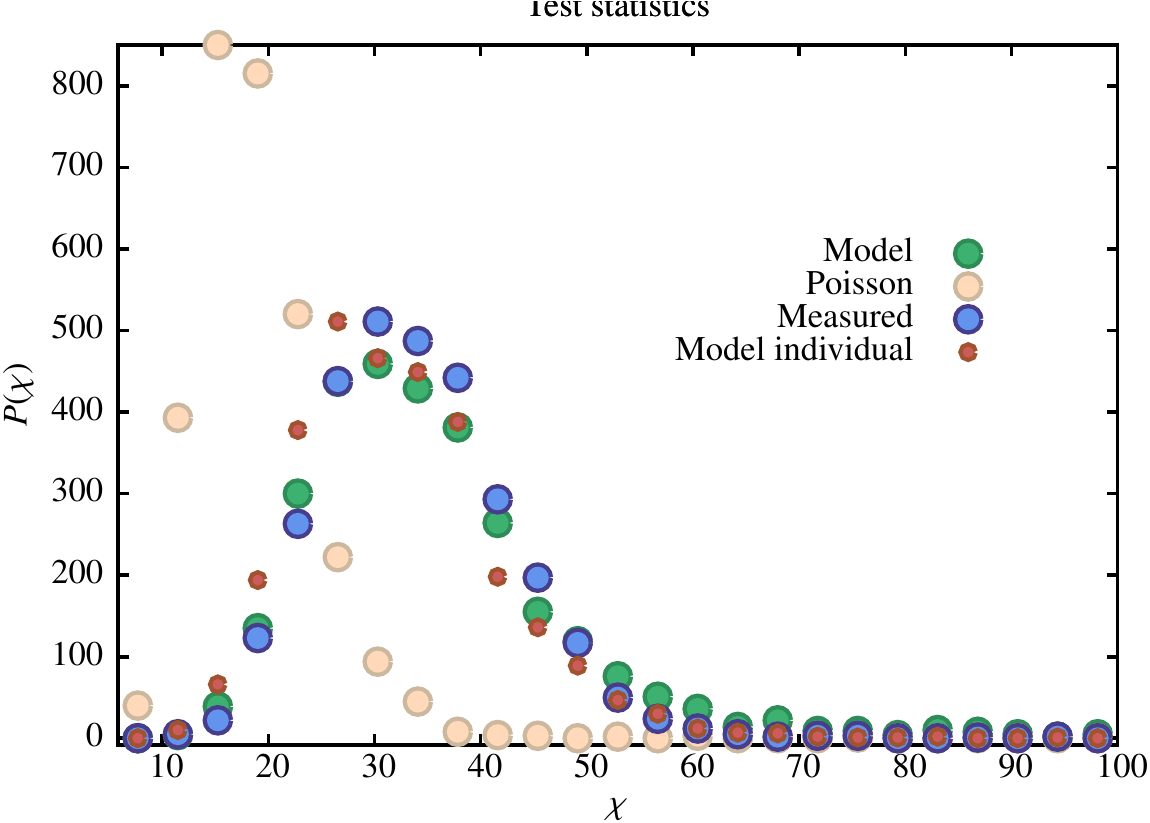}
 \caption{The test statistic, $\mathcal{T}$ (defined by eqn.~\protect\ref{eq:modelCovarianceTestStatistic}), for models of the model covariance matrix. Points show the distribution of test statistic obtained from 1000 model realizations. Large blue circles show $\mathcal{T}$ obtained using the covariance matrix measured directly from the 1000 model realizations (i.e. the true covariance matrix). Large green points show the test statistic obtained using the binomial model described in the text when the resulting covariance matrix is averaged over all 1000 realizations. Small red points are the same, but the covariance matrix used for each realization is estimated from that realization alone. Finally, the large pale pink points indicate $\mathcal{T}$ obtained when the covariance matrix is estimated assuming a Poisson process.}
 \label{fig:modelCovarianceTestStatistic}
\end{figure}

Figure~\ref{fig:modelCovariance} shows the measured covariance, and the ratio of model to measured covariance in each case. Our model (in which we assume no covariance between main branch and other galaxies) clearly performs quite well. To quantify this more precisely we construct a test statistic, $\mathcal{T}$, defined as:
\begin{equation}
 \mathcal{T} = \Delta \cdot \mathcal{C}^{-1} \cdot \Delta^{\rm T},
 \label{eq:modelCovarianceTestStatistic}
\end{equation}
where $\Delta$ is the difference between the mass function of a model realization and the mean mass function of all model realizations. If our model for the model covariance matrix is accurate, the distribution of $\mathcal{T}$ should be consistent with that obtained using the covariance matrix measured directly from the model realizations. Figure~\ref{fig:modelCovarianceTestStatistic} shows that this is indeed the case. For comparison, we show the results if we assume that main branch galaxies follow a Poisson process. Clearly this overestimates the covariance in the model and results in an incorrect $\mathcal{T}$ distribution.

\section{Merger Tree Construction}\label{app:MergerTreeConstruction}

Throughout this work we utilize merger trees built using the algorithm proposed by \cite{parkinson_generating_2008}, which those authors demonstrated provided a good match to several statistics of merger trees extracted from the Millennium Simulation. Furthermore, \cite{jiang_generating_2013} study several algorithms for constructing merger trees and compare their results with trees extracted from N-body simulations. They find that the \cite{parkinson_generating_2008} algorithm is the only algorithm which gives good agreement with the N-body merger trees for all statistics that they consider, and that (given the uncertainties in the construction of N-body merger trees) merger trees constructed using this algorithm are as accurate as those extracted from N-body simulations.

In this Appendix, we explore this algorithm further, comparing it to a set of merger trees extracted from the ``MillGas'' simulation (kindly provided for this study by the VIRGO Consortium). This simulation has Millennium-like resolution, cosmological parameters $(\Omega_{\rm M},\Omega_\Lambda,\Omega_{\rm b},H_0,\sigma_8,n_{\rm s})=(0.272,0.728,0.045,70.4\hbox{km/s/Mpc},0.81,1.0)$, but utilizes a smaller box of $125/h$~Mpc on a side. Despite the name, the version of MillGas used for this analysis is dark matter-only. Halo finding and merger tree construction is performed using {\sc subfind} and {\sc d-trees} respectively \citep{jiang_n-body_2014}.

The primary advantage of this simulation for purposes of our analysis is that a large number (976) of snapshots were stored from the simulation, allowing us to explore how the statistics of merger trees depend on the number of snapshots used in their construction. We find that the merger tree statistics are well-converged with 976 snapshots (as ecxpected; \citealt{benson_convergence_2012}).

Comparing the conditional mass functions (i.e. the distribution of halo masses, $M_{\rm a}$, at redshift $z_{\rm a}$, conditioned on those halos being progenitors of a halo of mass $M_{\rm b}$ at the later redshift $z_{\rm b}$), we find excellent agreement between N-body and Monte Carlo trees, as did \cite{parkinson_generating_2008}. In fact, we find that this agreement extends to the conditional mass functions of the 1$^{\rm st}$ through $4^{\rm th}$ most massive progenitors (\cite{parkinson_generating_2008} checked up to the 2$^{\rm nd}$ most massive progenitor), and for very small timesteps ($\Delta z=0.00122$ at $z=0$ for example).

One issue that we find does make a significant difference is whether we allow halos in the N-body merger trees to lose mass. As noted by \cite{helly_galaxy_2003}, halos in N-body merger trees can be less massive than their progenitors, while this is not possible by construction in the \cite{parkinson_generating_2008} algorithm. \cite{helly_galaxy_2003} were forced to confront this issue as their semi-analytic model required monotonically increasing halo masses. They therefore decided to force monotonicity by adding mass to halos which were less massive than the sum of their progenitor masses, making them equal to the sum of the progenitor masses. The N-body merger trees used by \cite{parkinson_generating_2008} were forced to have monotonically increasing halo masses.

While \glc\ is entirely capable of handling non-monotonic halo masses, there is evidence (John Helly, private communication) that some of this reduction in mass is a numerical artifact of the friends-of-friends algorithm which tends to artificially transfer mass from the second most massive progenitor just prior to a merger. Therefore, it is worth exploring the effects of enforcing monotonicity. 

\begin{figure*}
 \begin{tabular}{cc}
 \includegraphics[width=85mm,trim=0mm 0mm 0mm 4mm,clip]{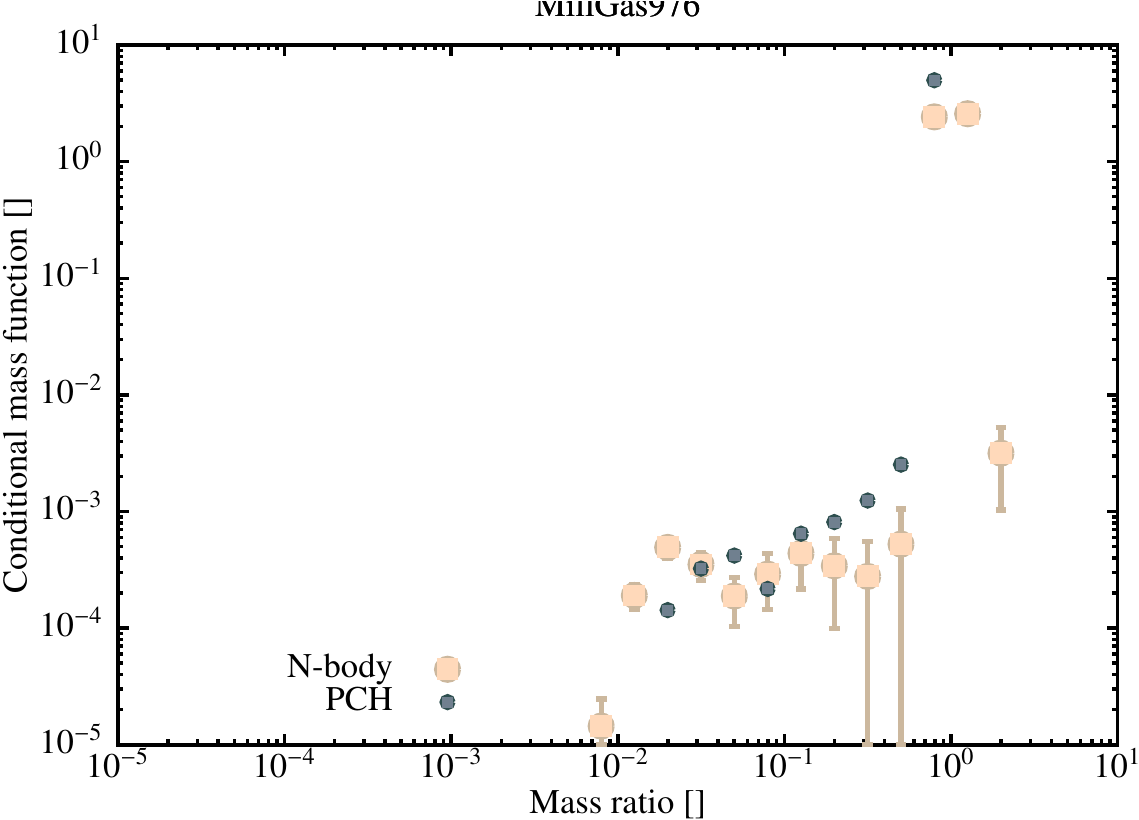} &
 \includegraphics[width=85mm,trim=0mm 0mm 0mm 4mm,clip]{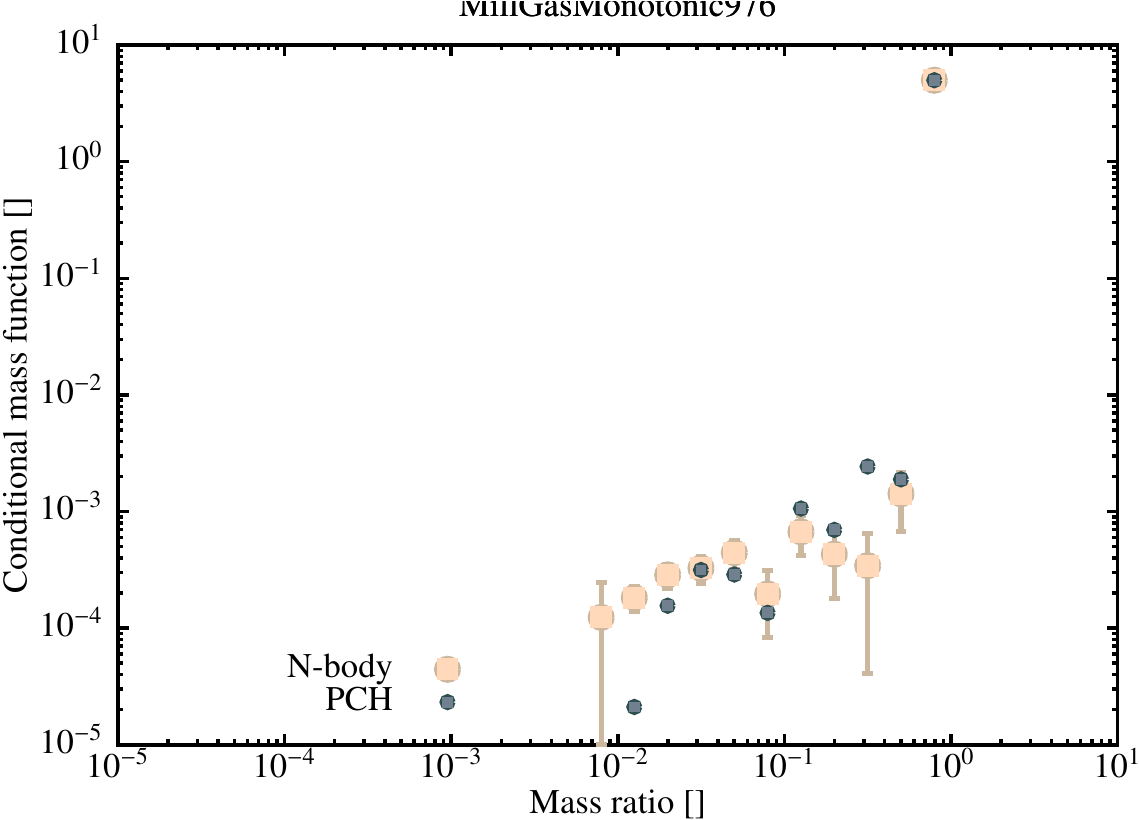} \\
 \includegraphics[width=85mm,trim=0mm 0mm 0mm 4mm,clip]{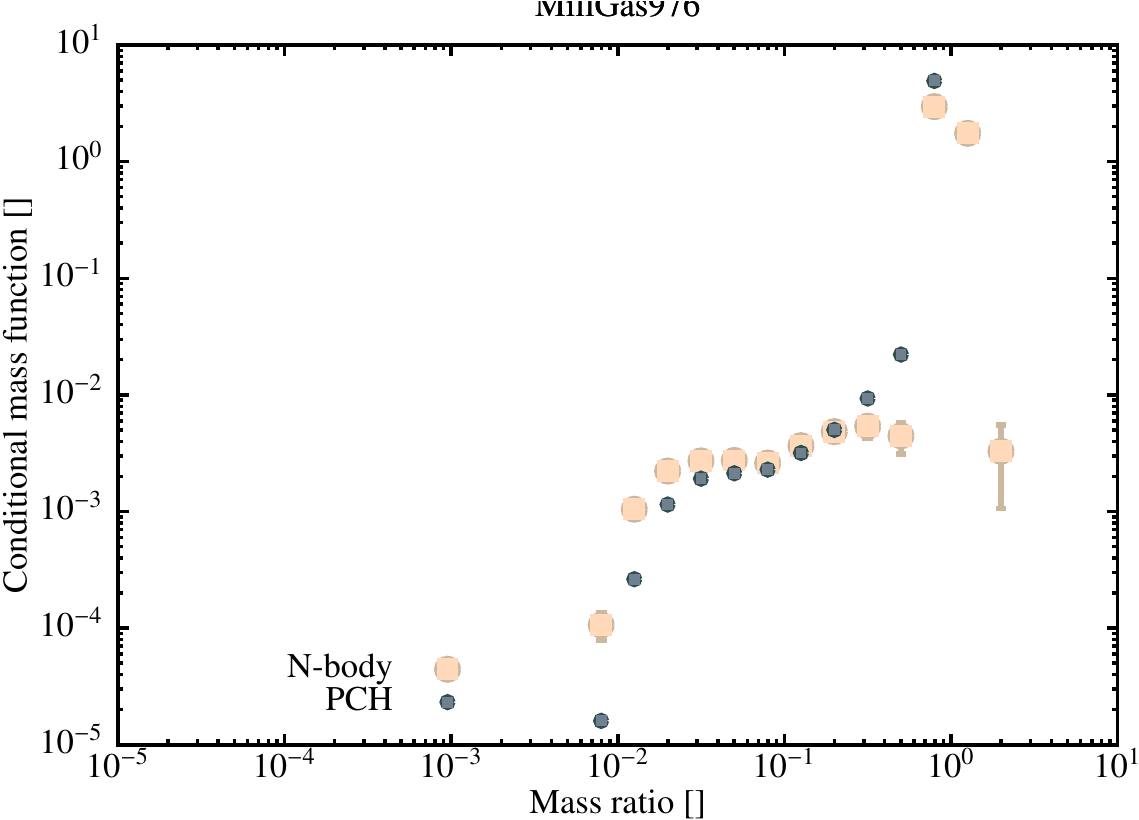} &
 \includegraphics[width=85mm,trim=0mm 0mm 0mm 4mm,clip]{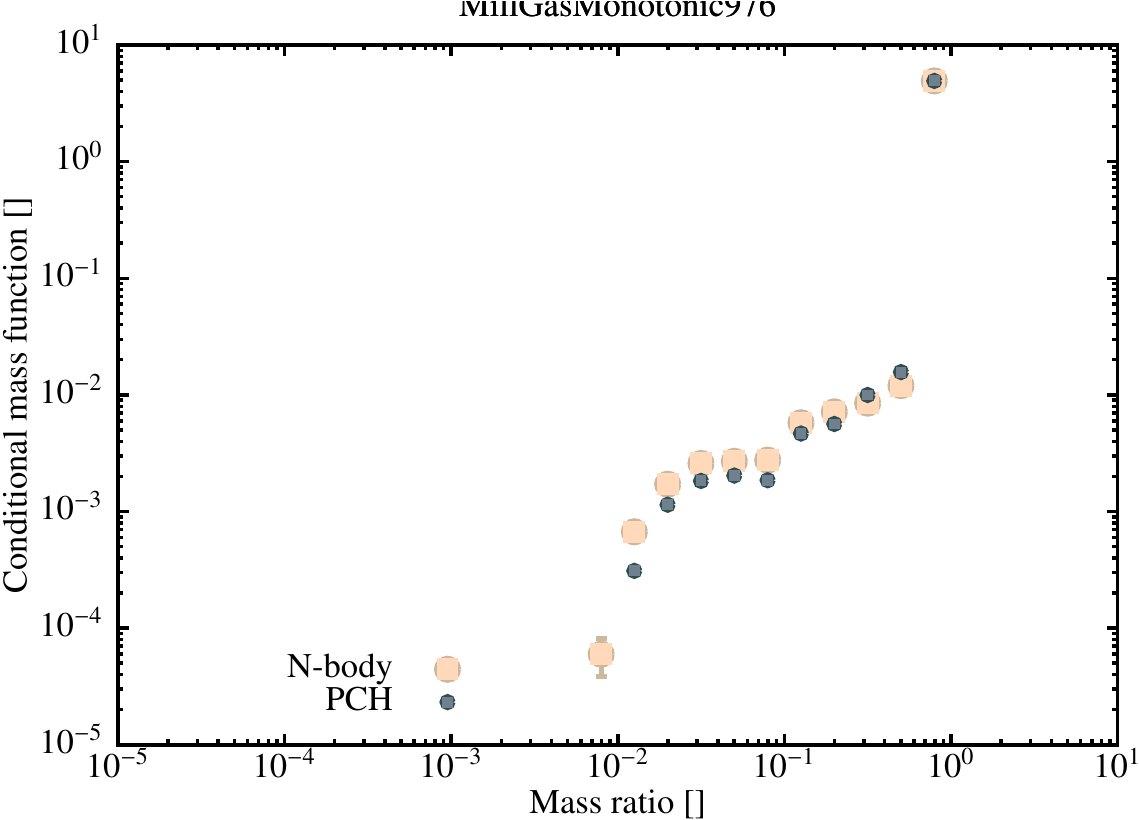} \\
 \includegraphics[width=85mm,trim=0mm 0mm 0mm 4mm,clip]{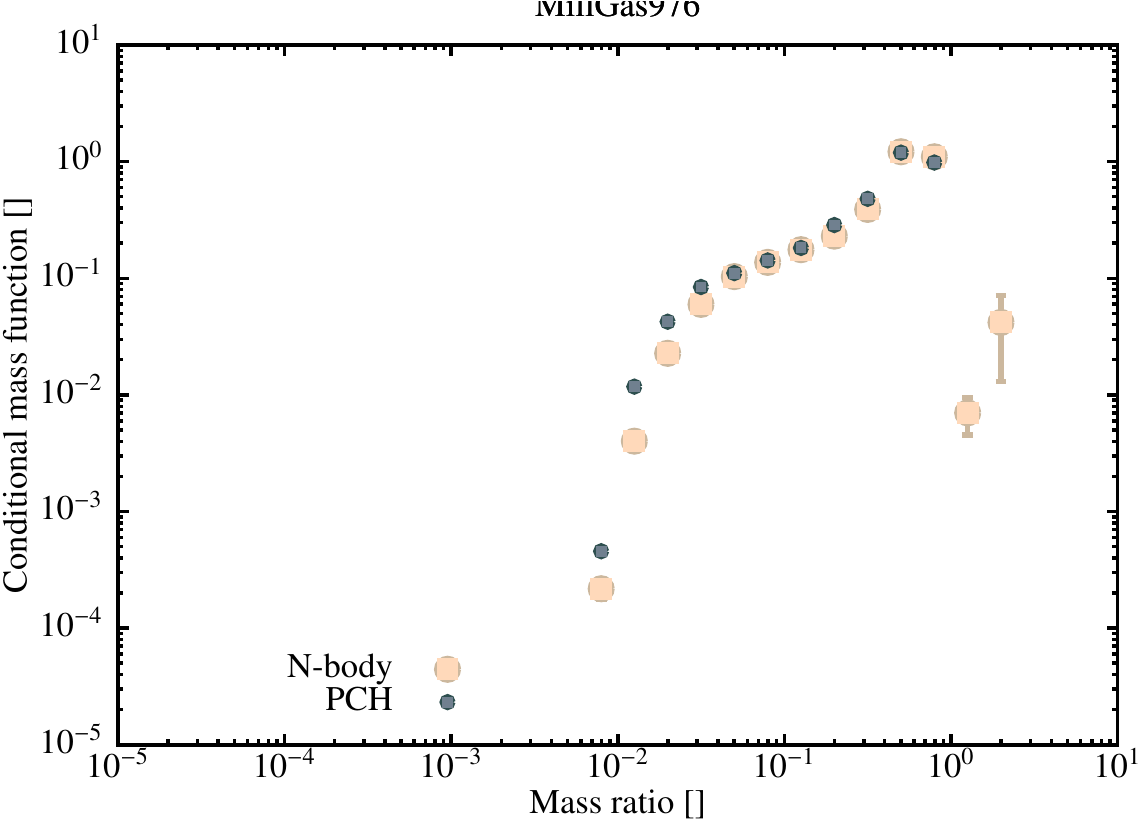} &
 \includegraphics[width=85mm,trim=0mm 0mm 0mm 4mm,clip]{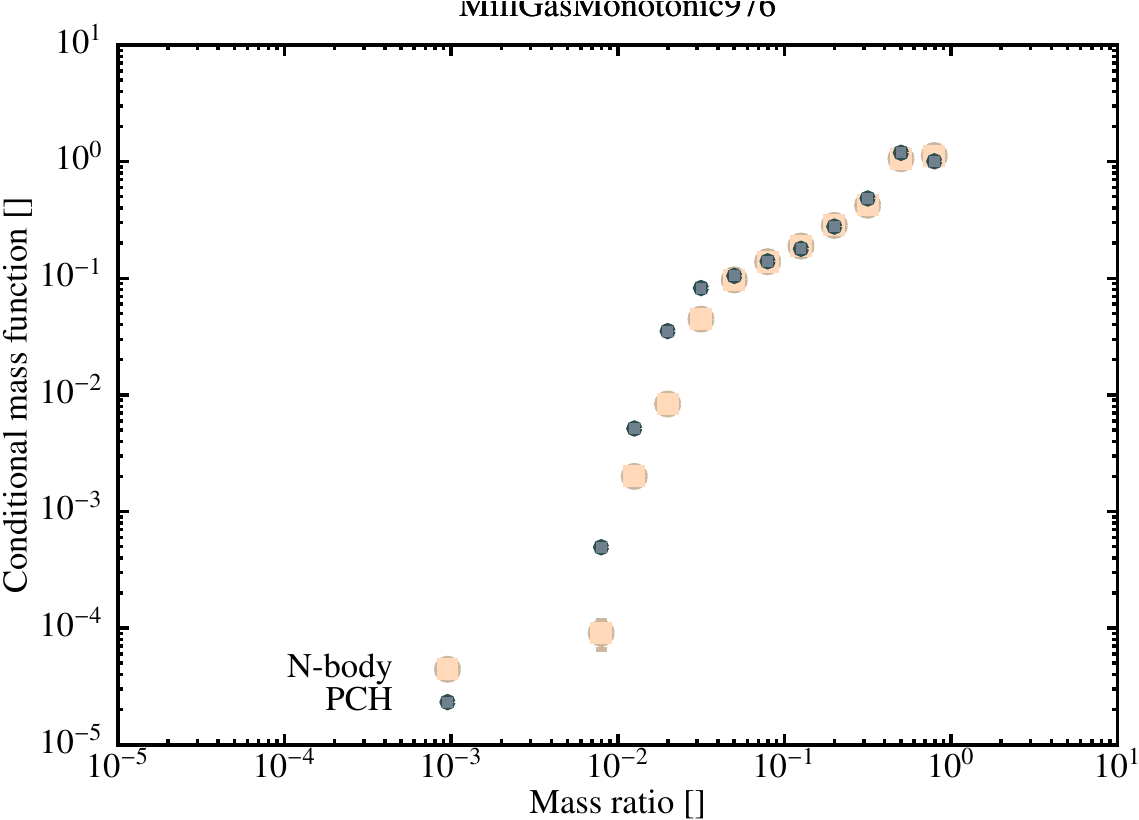}
 \end{tabular}
 \caption{Conditional mass functions from N-body and PCH merger trees for $M_{\rm b}=10^{12.22}M_\odot$, $z_{\rm b}=0$, $z_{\rm a}=0.00122$, $0.0199$, and $0.989$ (top, middle, and bottom rows respectively), as a function of mass ratio $M_{\rm a}/M_{\rm b}$. The left column shows the conditional mass function \emph{without} enforcing halo masses to grow monotonically, while the right column shows the conditional mass function \emph{with} monotonicity enforced.}
 \label{fig:ConditionalMassFunction}
\end{figure*}

Figure~\ref{fig:ConditionalMassFunction} shows the conditional mass function of N-body and PCH merger trees for $M_{\rm b}=10^{12.22}M_\odot$, $z_{\rm b}=0$, $z_{\rm a}=0.00122$, $0.0199$, and $0.989$ (top, middle, and bottom rows respectively), as a function of mass ratio $M_{\rm a}/M_{\rm b}$. Considering first the left column, in which the N-body trees have not been forced to be monotonically increasing in mass along each branch, the two show very good agreement, even for small time differences, with the following caveats:
\begin{itemize}
 \item For mass ratios above unity the PCH trees have no progenitors, while N-body trees can;
 \item There are small offsets in the mass ratio below which the conditional mass function is cut off.
\end{itemize}
This latter point is not surprising---the cut off arises from the finite mass resolution of the N-body simulations, which we attempt to match by imposing the same mass resolution limit on our Monte Carlo trees. However, there is no reason to expect that the details of the shape of the cut off will be matched---these will depend on precisely how resolution affects tree construction for Monte Carlo trees, and halo finding in the N-body simulations. It is also irrelevant for any careful calculation of galaxy properties which should, of course, check that changes in resolution make no difference to the results.

The right column of Figure~\ref{fig:ConditionalMassFunction} shows the same comparison but with monotonically increasing halo masses forced on the N-body merger trees. By construction, the N-body trees now have no progenitors with mass ratios above unity. The agreement with the Monte Carlo trees remains just as good, perhaps slightly better. For example, consider the point for $z_{\rm a}=0.0199$ (middle row) in the left column at mass ratio $0.5$. The N-body point lies significantly below the Monte Carlo results. However, once monotonically increasing halo masses are enforced these points agree extremely well.

\begin{figure*}
 \begin{tabular}{cc}
 \includegraphics[width=85mm,trim=0mm 0mm 0mm 4mm,clip]{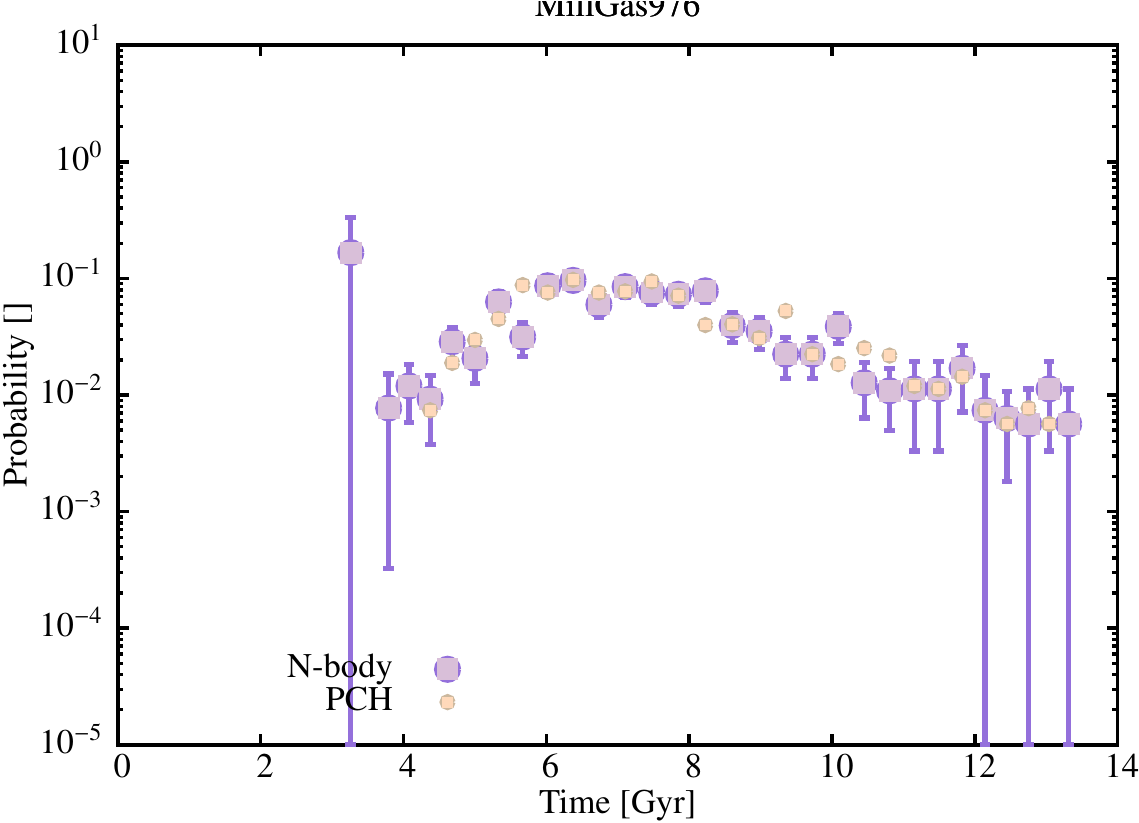} &
 \includegraphics[width=85mm,trim=0mm 0mm 0mm 4mm,clip]{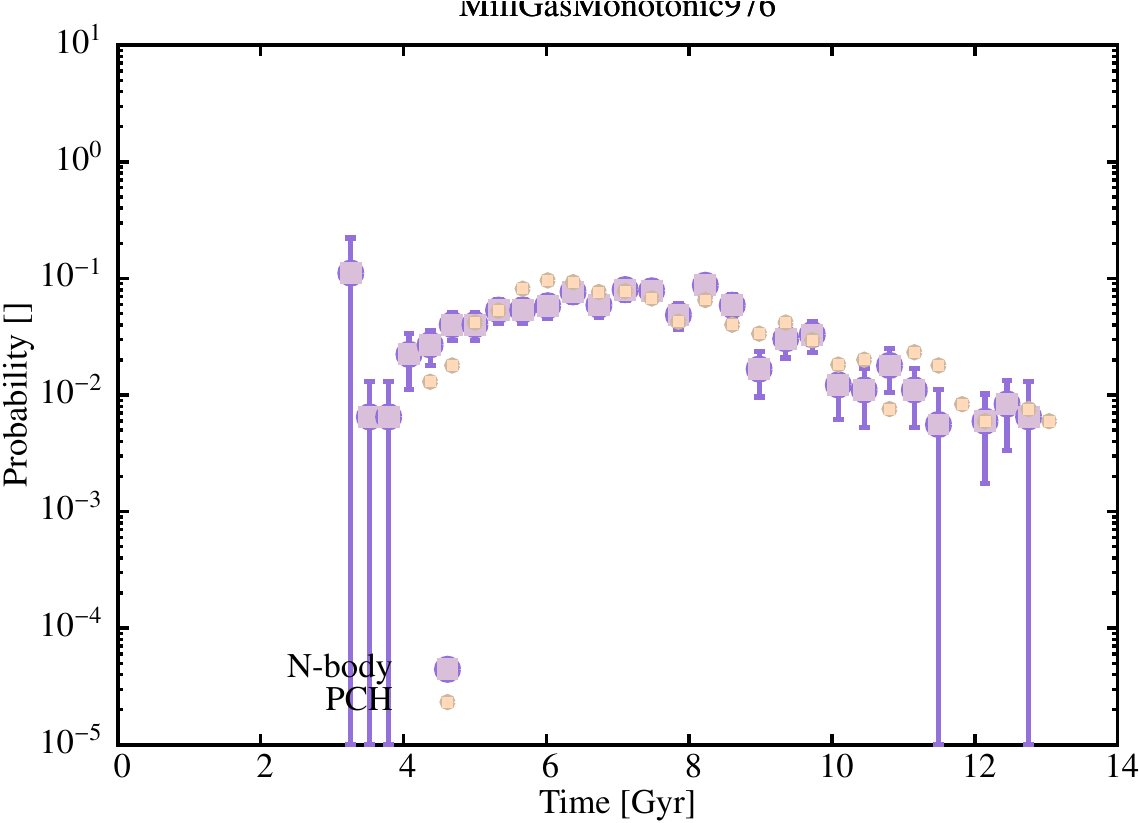} \\
 \includegraphics[width=85mm,trim=0mm 0mm 0mm 4mm,clip]{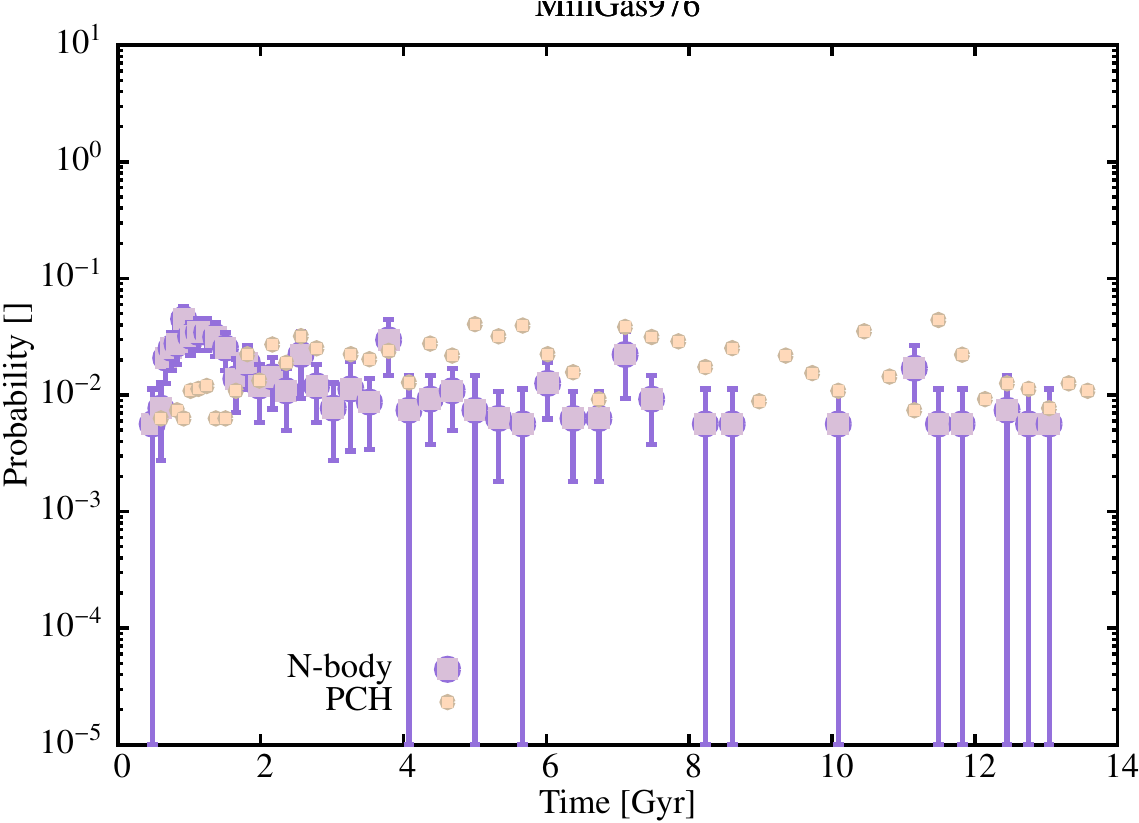} &
 \includegraphics[width=85mm,trim=0mm 0mm 0mm 4mm,clip]{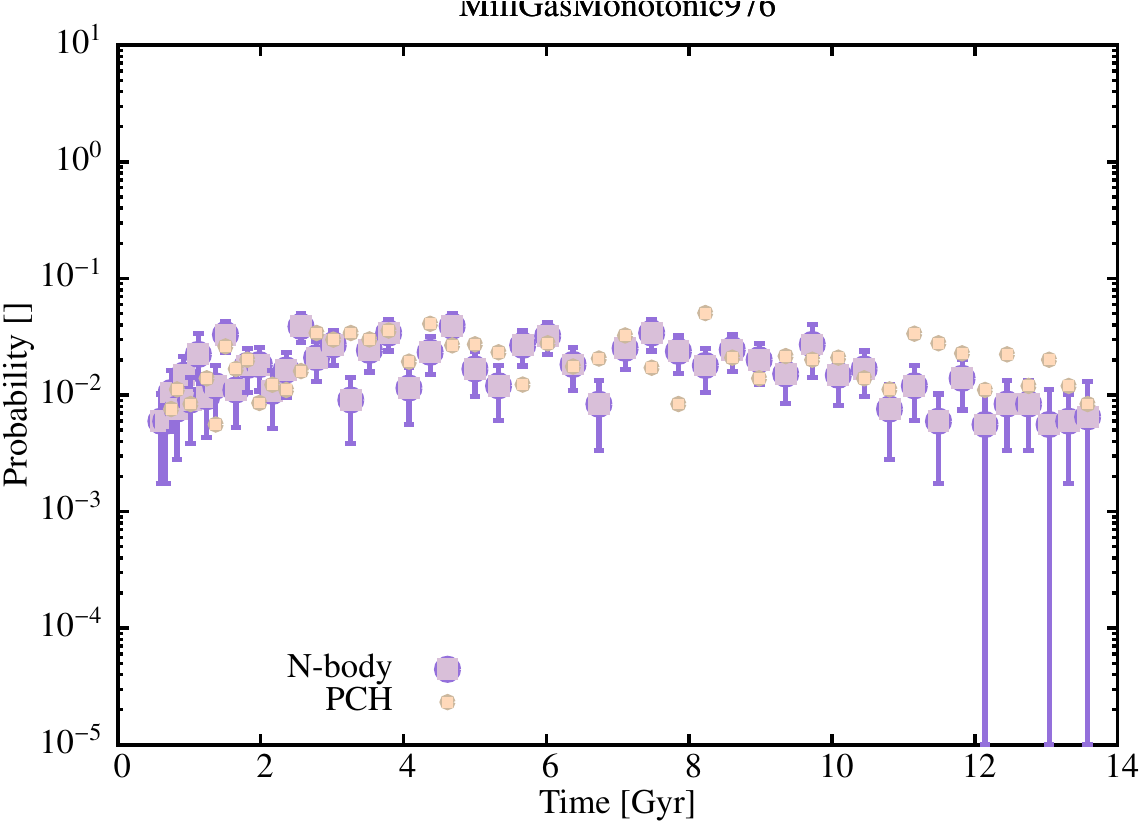} \\
 \includegraphics[width=85mm,trim=0mm 0mm 0mm 4mm,clip]{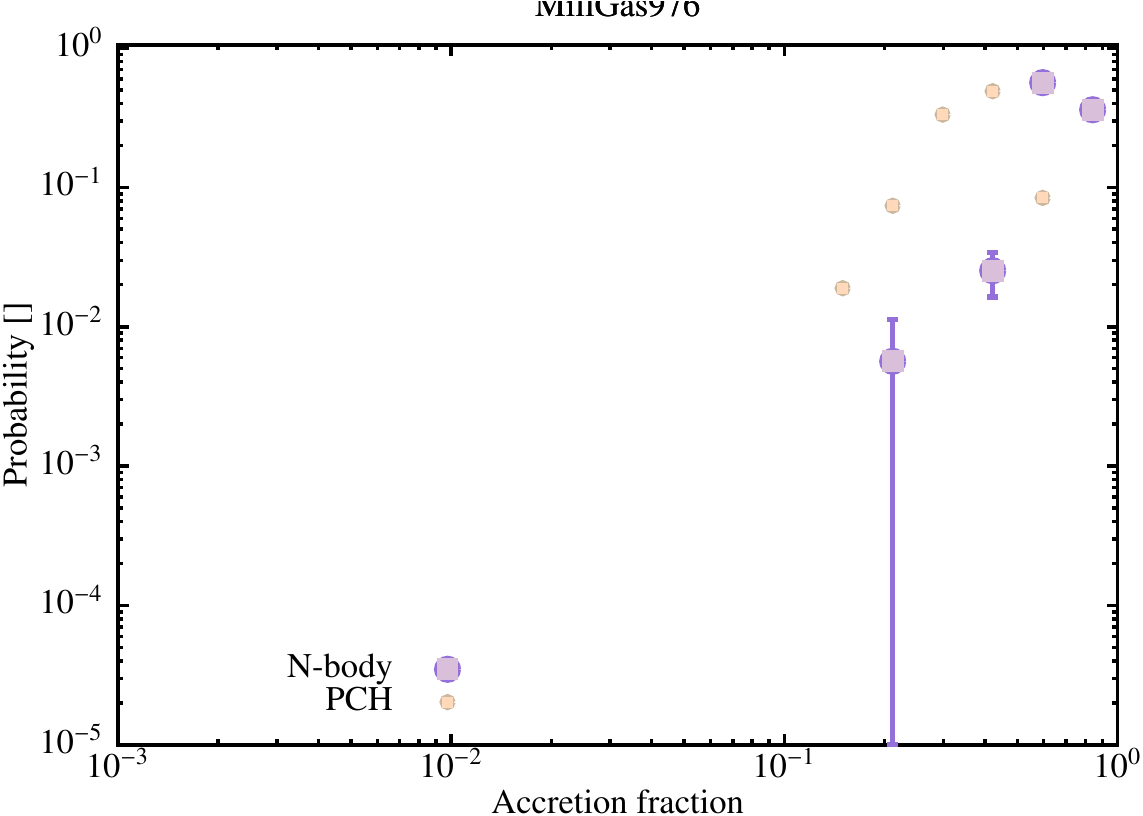} &
 \includegraphics[width=85mm,trim=0mm 0mm 0mm 4mm,clip]{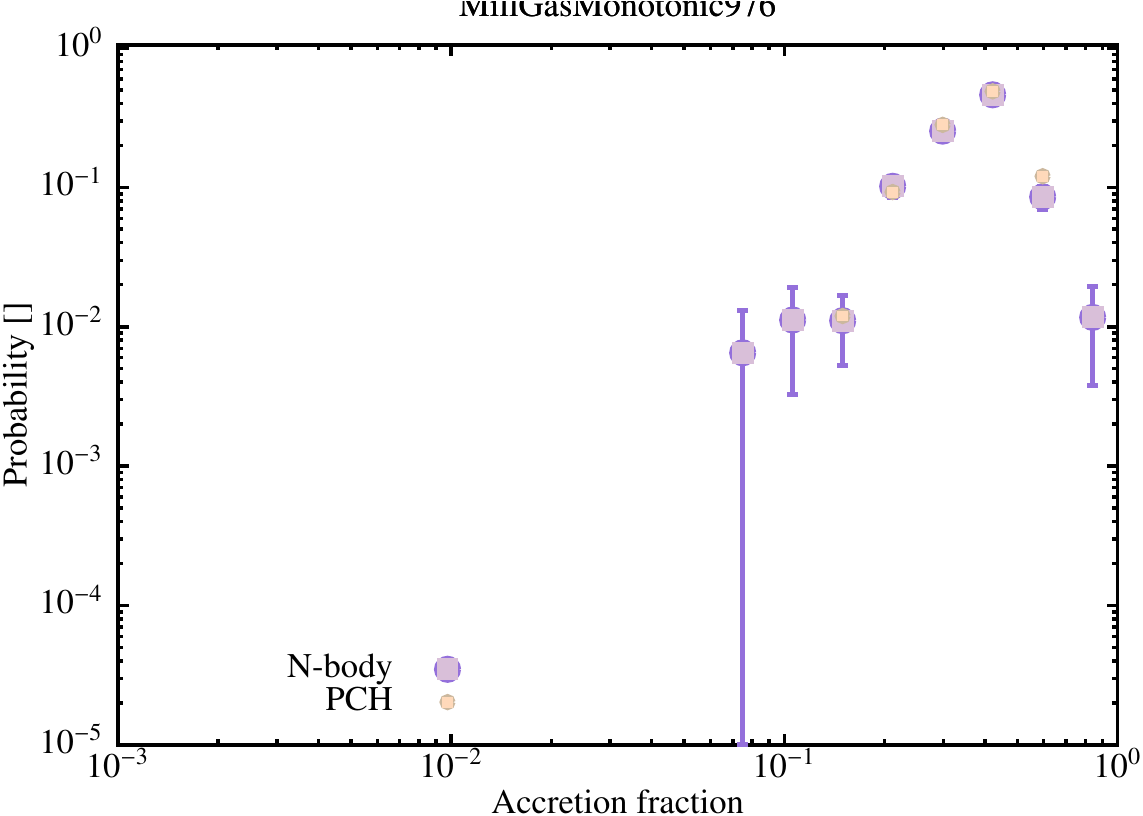}
 \end{tabular}
 \caption{Statistics of the merger trees of halos of mass $10^{13.5}M_\odot$ at $z=0$. The left column shows results for N-body merger trees \emph{without} enforcing monotonically increasing halo masses, while the right column shows results when this \emph{is} enforced. The top row shows the distribution of formation times (defined as the time at which 50\% of the final halo mass was assembled into a single halo), the middle row shows the distribution of last major merger times (defined as a merger with mass ratio of 4:1 or lower), and the bottom row shows the distribution of accretion fraction (i.e. the fraction of mass in the final halo which was gained via smooth accretion, and not via mergers with resolved halos).}
 \label{fig:NodeTiming}
\end{figure*}

To further explore the effects of enforcing montonically increasing halo masses we show in Figure~\ref{fig:NodeTiming} three other statistics for the merger trees of halos of mass $10^{13.5}M_\odot$ at $z=0$. In each case the left column shows the results for N-body trees \emph{without} enforcing monotonically growing halo masses, while the right column \emph{does} enforce this. The top row shows the distribution of formation times (defined as the time at which 50\% of the final halo mass was first assembled into a single halo). In this case, agreement between N-body and PCH merger trees is good irrespective of whether halo masses are forced to increase monotonically.

The middle row shows the distribution of last major merger times. This distribution is quite flat, but in the case where N-body trees are not forced to have monotonically increasing halo masses there is a clear difference with PCH trees, such that the N-body trees have last major merger times peaked toward early times. This is to be expected if the friends-of-friends algorithm artificially reduces the masses of halos just prior to merging. If we enforce halo masses to grow monotonically, the distribution for N-body trees now agrees very closely with that from PCH trees.

Finally, we examine the fraction of mass gained by the halo over its lifetime which can not be accounted for by the masses of merging halos (bottom row). In this case, N-body halos typically gain a large fraction of their mass from this ``smooth accretion'', much more so than the corresponding PCH halos, when halo masses are not required to increase monotonically with time. This is again what would be expected if the friends-of-friends algorithm artificially reduced the masses of halos just prior to merging. Enforcing monotonic growth results in distributions of accreted mass that are in excellent agreement between the two sets of merger trees. 

In conclusion, once monotonic growth of halo mass is enforced and the N-body trees have sufficiently high time resolution, there is no significant discrepancy between the N-body and PCH trees. This suggests the following:
\begin{enumerate}
 \item the PCH algorithm works extremely well in matching the N-body tree conditional mass functions, including distributions of first through fourth most massive progenitor, formation times, time of last major merger, and fraction of mass accreted sub-resolution;
 \item masses of the secondary progenitor are artificially reduced just prior to merging in N-body merger trees.
\end{enumerate}
This latter issue seems to be a result of the friends-of-friends algorithm (John Helly, private communication) and, as we have shown, can be mitigated by forcing the masses of halos in the N-body merger trees to always grow with time along each branch. For this reason, we choose to enforce monotonic growth on halos when computing the model discrepancy in \S\ref{sec:MonteCarloDiscrepancy}.

\end{document}